\documentclass[epj]{svjour}
\usepackage{graphics}
\usepackage{amsmath,latexsym,epsfig,epsf,rotate,amssymb}
\usepackage{graphicx}
\usepackage{url}
\usepackage{psfig}
\usepackage{eucal}
\usepackage{xspace}
\usepackage{color}
\usepackage{subfigure}

\usepackage{hyperref}
\hypersetup{colorlinks=true, linkcolor=blue, filecolor=blue,
pagecolor=blue, urlcolor=blue}

\begin{document}
\hugehead
%
\headnote{
\hfill 
\parbox{9cm}{\tt \normalsize BIHEP-TH-2009-005, BU-HEPP-09-08, CERN-PH-TH/2009-201,
DESY 09-092,\\ FNT/T 2009/03,
Freiburg-PHENO-09/07,\\ 
HEPTOOLS 09-018, IEKP-KA/2009-33,\\ LNF-09/14(P), LPSC 09/157,\\LPT-ORSAY-09-95, LTH 851, MZ-TH/09-38,\\PITHA-09/14, PSI-PR-09-14, \\SFB/CPP-09-53, WUB/09-07 \\}}

\title{Quest for precision in hadronic cross
sections at low energy:\\
  Monte Carlo tools vs. experimental data}

 \subtitle{Working Group on Radiative Corrections  and  Monte Carlo Generators for Low Energies}

\author{
S.~Actis \inst{38} \and 
A.~Arbuzov \inst{9,43} \and
G.~Balossini \inst{32,33} \and
P.~Beltrame \inst{13} \and 
C.~Bignamini \inst{32,33} \and
R.~Bonciani \inst{15} \and
C.~M.~Carloni Calame \inst{35} \and
V.~Cherepanov \inst{25,26} \and
M.~Czakon \inst{1} \and
H.~Czy\.z  \inst{19,44,47,48} \and 
A.~Denig \inst{22} \and 
S.~Eidelman \inst{25,26,45} \and 
G.~V.~Fedotovich \inst{25,26,43} \and 
A.~Ferroglia \inst{23} \and
J.~Gluza \inst{19} \and
A.~Grzeli{\'n}ska \inst{8}\and
M.~Gunia   \inst{19} \and
A.~Hafner  \inst{22} \and
F.~Ignatov \inst{25} \and
S.~Jadach  \inst{8} \and
F.~Jegerlehner \inst{3,19,41} \and
A.~Kalinowski \inst{29} \and
W.~Kluge \inst{17} \and 
A.~Korchin \inst{20} \and 
J.~H.~ K\"uhn \inst{18}\and
E.~A.~Kuraev \inst{9} \and
P.~Lukin \inst{25} \and
P.~Mastrolia \inst{14} \and
G.~Montagna \inst{32,33,42,48} \and
S.~E. M\"uller\inst{22,44} \and 
F.~Nguyen \inst{34,42} \and 
O.~Nicrosini \inst{33} \and
D.~Nomura \inst{36,46}  \and
G.~Pakhlova \inst{24} \and 
G.~Pancheri \inst{11} \and
M.~Passera \inst{28} \and
A.~Penin \inst{10} \and
F.~Piccinini \inst{33} \and
W.~P\l{a}czek \inst{7} \and
T.~Przedzinski \inst{6} \and 
E.~Remiddi \inst{4,5} \and
T.~Riemann \inst{41} \and
G.~Rodrigo \inst{37} \and
P.~Roig  \inst{27} \and
O.~Shekhovtsova \inst{11} \and 
C.~P. Shen \inst{16} \and
A.~L. Sibidanov \inst{25} \and
T.~Teubner \inst{21,46}  \and
L.~Trentadue \inst{30,31} \and
G.~Venanzoni \inst{11,47,48}  \and
J.~J.~~van der Bij \inst{12} \and
P.~Wang  \inst{2} \and
B.~F.~L.~Ward \inst{39} \and
Z.~Was \inst{8,45} \and
M.~Worek \inst{40,19} \and
C.~Z.~Yuan \inst{2}
}                     
%
%
\institute{
\small{Institut f\"ur Theoretische Physik E, RWTH Aachen Universit\"at,
D-52056 Aachen, Germany
\and
Institue of High Energy Physics, Chinese Academy of Sciences,
Beijing 100049, China 
\and
Institut f\"ur Physik 
Humboldt-Universit\"at zu Berlin,
D-12489 Berlin, Germany
\and
Dipartimento di Fisica dell'Universit\`a di Bologna, I-40126 Bologna, Italy
\and
INFN, Sezione di Bologna, 
I-40126 Bologna, Italy
\and
The Faculty of Physics, Astronomy and Applied Computer
Science,
Jagiellonian University, Reymonta 4, 30-059 Cracow, Poland
\and
Marian Smoluchowski Institute of Physics, Jagiellonian University, 
Reymonta 4, 30-059 Cracow, Poland
\and
Institute of Nuclear Physics 
Polish Academy of Sciences, PL-31342 Cracow, Poland 
\and
Bogoliubov Laboratory of Theoretical Physics, Joint Institute for Nuclear  
Research, 141980 Dubna, Russia
\and
Department of Physics, University of Alberta, 
 Edmonton, AB T6G 2J1, Canada
\and
Laboratori Nazionali di Frascati dell'INFN, I-00044 Frascati, Italy 
\and
Physikalisches Institut, Albert-Ludwigs-Universit\"at
Freiburg,  D-79104 Freiburg, Germany 
\and
CERN, Physics Department, CH-1211 Gen\`eve, Switzerland 
\and
CERN, Theory Department,
CH-1211 Gen\`eve, Switzerland 
\and
Laboratoire de Physique Subatomique et de Cosmologie,
Universit\'e Joseph Fourier/CNRS-IN2P3/INPG,\\
F-38026 Grenoble, France 
\and
University of Hawaii, Honolulu, Hawaii 96822, USA 
\and
Institut f\"ur Experimentelle Kernphysik, Universit\"at
Karlsruhe, D-76021 Karlsruhe, Germany 
\and
Institut f\"ur Theoretische Teilchenphysik,   
Universit\"at Karlsruhe, D-76128 Karlsruhe, Germany.
\and
Institute of Physics, University of Silesia, 
PL-40007 Katowice, Poland
\and
National Science Center ``Kharkov Institute of Physics and Technology", 
61108 Kharkov, Ukraine
\and
Department of Mathematical 
Sciences, University of Liverpool, Liverpool L69 3BX, U.K.
\and
Institut f\"ur Kernphysik, Johannes Gutenberg - 
Universit\"at Mainz, D-55128 Mainz, Germany  
\and
Institut f\"ur Physik (THEP), Johannes Gutenberg-Universit\"at,
D-55099 Mainz, Germany 
\and
Institute for Theoretical and Experimental Physics, 
Moscow, Russia
\and
Budker Institute of Nuclear Physics, 630090 Novosibirsk, Russia 
\and
Novosibirsk State University, 630090 Novosibirsk, Russia 
\and
Laboratoire de Physique Th\'eorique (UMR 8627),
Universit\'e de Paris-Sud XI, B\^{a}timent 210,
91405 Orsay Cedex, France
\and
INFN, Sezione di Padova, I-35131 Padova, Italy 
\and
LLR-Ecole Polytechnique, 91128 Palaiseau, France 
\and
Dipartimento di Fisica, Universit\`a di Parma, I-43100 Parma, Italy 
\and
INFN, Gruppo Collegato di Parma,  
I-43100 Parma, Italy
\and
Dipartimento di Fisica Nucleare e Teorica, Universit\`a di Pavia,
I-27100 Pavia, Italy  
\and 
INFN, Sezione di Pavia, I-27100 Pavia, Italy 
\and
Dipartimanto di Fisica dell'Universit\`a ``Roma Tre" and INFN 
Sezione di Roma Tre,  I-00146 Roma, Italy
\and 
School of Physics and Astronomy, University of Southampton, 
Southampton SO17 1BJ, U.K. 
\and 
Theory Center, KEK, Tsukuba, Ibaraki 
305-0801, Japan
\and
Instituto de Fisica Corpuscular (IFIC),
Centro mixto UVEG/CSIC,
Edificio Institutos de Investigacion, Apartado de Correos 22085,
E-46071 Valencia, Espanya 
\and
Paul Scherrer Institut, W\"urenlingen and Villigen, CH-5232 Villigen PSI,
Switzerland 
\and
Department of Physics, Baylor University, Waco, Texas 76798-7316, USA 
\and
Fachbereich C, Bergische Universit\"at Wuppertal,
D-42097 Wuppertal, Germany 
\and
Deutsches Elektronen-Synchrotron, DESY,
D-15738 Zeuthen, Germany 
\and
Section \ref{sec:1} conveners 
\and
Section \ref{sec:2} conveners 
\and
Section \ref{sec:3} conveners 
\and
Section \ref{sec:5} conveners 
\and
Section \ref{sec:4} conveners  
\and
Working group conveners 
\and
Corresponding authors: henryk.czyz@us.edu.pl, guido.montagna@pv.infn.it, graziano.venanzoni@lnf.infn.it 
} }
\date{Received: date / Revised version: date}
%

\abstract{
We present the achievements of the last years of the 
experimental and theoretical groups working on hadronic cross section
 measurements at the low energy $e^+ e^-$ colliders in Beijing, Frascati, Ithaca, Novosibirsk, Stanford and Tsukuba and on $\tau$ decays.
We sketch the prospects in these
  fields for the years to come. 
We emphasise the status and the precision 
  of the Monte Carlo generators used to analyse the hadronic cross section 
measurements obtained as well with energy scans as with radiative return, to
determine luminosities and $\tau$ decays. The radiative corrections 
fully or approximately
implemented in the various codes and the contribution of the vacuum
polarisation are discussed. 
 }
\maketitle




\tableofcontents
\noindent
\section{Introduction}
\label{intro}
 The systematic comparison of Standard Model ({\small SM})
predictions with precise experimental data served in the last 
decades as an invaluable tool to test the theory at the quantum
level. It has also provided stringent constraints on ``new
physics'' scenarios.  The (so far) remarkable agreement between the
measurements of the electroweak observables and their {\small
  SM} predictions is a striking experimental confirmation of the
theory, even if there are a few observables where the agreement is not
so satisfactory.  On the other hand, the Higgs boson has not yet been
observed, and there are clear phenomenological facts (dark matter, 
matter-antimatter asymmetry in the universe) as well as 
strong theoretical arguments hinting at the
presence of physics beyond the {\small SM}. New colliders, like the 
LHC or a future $e^+ e^-$ International Linear Collider
(ILC), will hopefully answer many questions, offering at
the same time great physics potential and a new challenge to provide
even more precise theoretical predictions.
 
Precision tests of the Standard Model require an appropriate inclusion
of higher order effects and the knowledge of very precise input parameters.
One of the basic input parameters is the fine-structure constant 
$\alpha$ , determined from the  anomalous 
 magnetic moment of the electron
 with an impressive accuracy of 0.37 parts per billion 
(ppb)~\cite{Hanneke:2008tm} 
relying on the validity of 
 perturbative QED~\cite{Gabrielse:2006gg}.
However, physics at nonzero squared momentum transfer $q^2$
is actually described by an effective electromagnetic
coupling $\alpha(q^2)$ rather than by the low-energy
constant $\alpha$ itself. The shift of the fine-structure
constant from the Thomson limit to high energy involves low
energy non-perturbative hadronic effects which spoil this
precision. In particular, the effective fine-structure
constant at the scale $M_{\scriptscriptstyle{Z}}$,
$\alpha(M_{\scriptscriptstyle{Z}}^2) = \alpha/[1-\Delta
\alpha(M_{\scriptscriptstyle{Z}}^2)]$, plays a crucial role
in basic {\small EW} radiative corrections of the {\small
  SM}. An important example is the {\small EW} mixing
parameter $\sin^2 \!\theta$, related to $\alpha$, the Fermi
coupling constant $G_F$ and $M_{\scriptscriptstyle{Z}}$ via
the Sirlin relation~\cite{Sirlin:1980nh,Sirlin:1989uf,Marciano:1980pb}
\begin{equation}
  \sin^2 \!\theta_{\scriptscriptstyle{S}} \cos^2 \!\theta_{\scriptscriptstyle{S}} = 
  \frac{\pi \alpha}
{\sqrt 2 G_F M_{\scriptscriptstyle{Z}}^2 (1-\Delta r_{\scriptscriptstyle{S}})},
\label{eq:sirlin}
\end{equation}
where the subscript $S$ identifies the renormalisation
scheme. $\Delta r_{\scriptscriptstyle{S}}$ incorporates the
universal correction $\Delta
\alpha(M_{\scriptscriptstyle{Z}}^2)$, large contributions
that depend quadratically on the top quark
mass $m_t$~\cite{Veltman:1977kh} (which led to its indirect determination
before this quark was discovered), plus all remaining
quantum effects.
In the {\small SM}, $\Delta r_{\scriptscriptstyle{S}}$
depends on various physical parameters, including
$M_{\scriptscriptstyle{H}}$, the mass of the Higgs boson. As
this is the only relevant unknown parameter in the {\small
  SM}, important indirect bounds on this missing ingredient
can be set by comparing the calculated quantity in
Eq.~(\ref{eq:sirlin}) with the experimental value of $\sin^2
\!\theta_{\scriptscriptstyle{S}}$ (e.g.\ the effective
{\small EW} mixing angle $\sin^2 \!\theta_{\rm eff}^{\rm
  lept}$ measured at LEP and SLC from the
on-resonance asymmetries) once
$\Delta\alpha(M_{\scriptscriptstyle{Z}}^2)$ and other
experimental inputs like $m_t$ are provided. It is important
to note that an error of $\delta
\Delta\alpha(M_{\scriptscriptstyle{Z}}^2) = 35 \times
10^{-5}$~\cite{Burkhardt:2005se} in the effective
electromagnetic coupling constant dominates the uncertainty
of the theoretical prediction of $\sin^2 \!\theta_{\rm
  eff}^{\rm lept}$, inducing an error $\delta(\sin^2
\!\theta_{\rm eff}^{\rm lept}) \sim 12 \times 10^{-5}$
(which is comparable with the experimental value
$\delta(\sin^2 \!\theta_{\rm eff}^{\rm
  lept})^{\scriptscriptstyle \rm EXP} = 16 \times 10^{-5}$
determined by LEP-I and SLD~\cite{leplsd:2005ema,LEPEWWG}) and
affecting the upper bound for $M_H$~\cite{leplsd:2005ema,LEPEWWG,Passera:2008jk}.  Moreover, as
measurements of the effective {\small EW} mixing angle at a
future linear collider may improve its precision by one
order of magnitude, a much smaller value of
$\delta\Delta\alpha(M_{\scriptscriptstyle{Z}}^2)$ will be
required (see below). It is therefore crucial to assess all
viable options to further reduce this uncertainty.

The shift $\Delta \alpha(M_{\scriptscriptstyle{Z}}^2)$ can
be split in two parts:
$\Delta\alpha(M_{\scriptscriptstyle{Z}}^2) =
\Delta\alpha_{\rm lep}(M_{\scriptscriptstyle{Z}}^2) + \Delta
\alpha_{\rm had}^{(5)}(M_{\scriptscriptstyle{Z}}^2)$. The
leptonic contribution is calculable in perturbation theory
and known up to three-loop accuracy: $\Delta\alpha_{\rm
  lep}(M_{\scriptscriptstyle{Z}}^2) = 3149.7686\times
10^{-5}$~\cite{Steinhauser:1998rq}. The hadronic contribution $\Delta
\alpha_{\rm had}^{(5)}(M_{\scriptscriptstyle{Z}}^2)$ of the
five light quarks ($u$, $d$, $s$, $c$, and $b$) can be
computed from hadronic $e^+ e^-$ annihilation data via the
dispersion relation~\cite{Cabibbo:1961sz}
\begin{equation} 
  \Delta \alpha_{\rm had}^{(5)}(M_{\scriptscriptstyle{Z}}^2) = 
-\left(\frac{\alpha M_{\scriptscriptstyle{Z}}^2}{3\pi}
  \right) \mbox{Re}\int_{m_\pi^2}^{\infty} {\rm d}s 
\frac{R(s)}{s(s- M_{\scriptscriptstyle{Z}}^2
  -i\epsilon)},
\label{eq:delta_alpha_had}
\end{equation}
where $R(s) = \sigma^{0}_{\rm had}(s)/(4\pi\alpha^2\!/3s)$ and
$\sigma^{0}_{\rm had}\!(s)$ is the total cross section for $e^+ e^-$
annihilation into any hadronic states, with 
vacuum polarisation and initial state {\small QED}
corrections subtracted off. The current
accuracy of this dispersion integral is of the order
of 1\%, dominated by the error of the hadronic cross section
measurements in the energy region below a few GeV~\cite{Eidelman:1995ny,Davier:1997kw,Burkhardt:2001xp,Burkhardt:2005se,Jegerlehner:2001wq,Jegerlehner:2003rx,Jegerlehner:2006ju,Jegerlehner:2008rs,Jegerlehner:2003qp,Jegerlehner:2003ip,Hagiwara:2003da,Hagiwara:2006jt}.
Table~\ref{tab:future} (from Ref.~\cite{Jegerlehner:2001wq}) shows that an
uncertainty $\delta \Delta\alpha_{\rm had}^{(5)} \sim 5 \times
10^{-5}$, needed for precision physics at a future linear collider,
requires the measurement of the hadronic cross section with a
precision of $O(1\%)$ from threshold up to the $\Upsilon$ peak.
\begin{table}[h]
\begin{center}
 \renewcommand{\arraystretch}{1.4}
 \setlength{\tabcolsep}{1.6mm}
{\footnotesize
\begin{tabular}{|c|c|c|c|}
\hline
$\delta \Delta\alpha_{\rm had}^{(5)} \!\times \! 10^{5} $ 
& $\delta(\sin^2 \!\theta_{\rm eff}^{\rm lept}) \! \times \! 10^{5}$  
& Request on $R$\\
\hline \hline
22   &  7.9 & Present \\
\hline
7   &   2.5 & $\!\delta R/R \sim  1\%$ up to ${J/\psi}$\\
\hline   
5   &   1.8 & $\delta R/R \sim  1\%$ up to $\Upsilon$\\
\hline   
\end{tabular}
}
\caption{\label{tab:future} 
Values of the uncertainties $\delta
\Delta\alpha_{\rm had}^{(5)}$ (first column) and the errors induced by these
uncertainties on the theoretical {\small SM} prediction for $\sin^2
\!\theta_{\rm eff}^{\rm lept}$ (second column). The third column indicates
the corresponding requirements for the $R$ measurement. From Ref.~\cite{Jegerlehner:2001wq}.}
\end{center}
\end{table}

Like the effective fine-structure constant at the scale
$M_{\scriptscriptstyle{Z}}$, the {\small SM} determination of the
anomalous magnetic moment of the muon $a_\mu$ is presently limited by the
evaluation of the hadronic vacuum polarisation effects, which cannot be 
computed perturbatively at low energies. 
However, using analyticity and unitarity, it was shown
long ago that this term can be computed from hadronic $e^+ e^-$
annihilation data via the dispersion integral~\cite{Gourdin:1969dm}:
\begin{eqnarray}
      a_{\mu}^{\mbox{$\scriptscriptstyle{\rm HLO}$}} &=& 
      \frac{1}{4\pi^3}
      \int^{\infty}_{m_{\pi}^2} {\rm d}s \, K(s) \sigma^{0}\!(s) \nonumber \\
                                                                  &=&
      \frac{\alpha^2}{3\pi^2}
      \int^{\infty}_{m_{\pi}^2} {\rm d}s \, K(s) R(s)/s \, .
\label{eq:amu_had}
\end{eqnarray}
The kernel function $K(s)$ decreases monotonically with increasing~$s$. This
integral is similar to the one entering the evaluation of the hadronic
contribution $\Delta \alpha_{\rm had}^{(5)}(M_{\scriptscriptstyle{Z}}^2)$ in
Eq.~(\ref{eq:delta_alpha_had}). Here, however, the weight function in the
integrand gives a stronger weight to low-energy data.
A recent compilation of $e^+e^-$ data gives~\cite{Davier:2009zi}:
\begin{equation}
a_\mu^{\scriptscriptstyle{\rm HLO}} = (695.5\pm4.1)\times 10^{-10} \, .
\end{equation}
Similar values are obtained by other
 groups~\cite{Hagiwara:2006jt,Jegerlehner:2008zz,Jegerlehner:2009ry,Eidelman:2009ft}.

By adding this contribution to the rest of the
{\small SM} contributions, a recent update of the
 SM prediction of $a_{\mu}$, which uses the hadronic light-by-light
result from~\cite{Prades:2009tw} gives \cite{Davier:2009zi,Prades:2009qp}:
\noindent $a_{\mu}^{\mbox{$\scriptscriptstyle{\rm SM}$}}= 116591834 (49)  
\times 10^{-11}$.
The difference between the experimental average~\cite{Bennett:2006fi},
$     a_{\mu}^{\mbox{$\scriptscriptstyle{\rm exp}$}}= 11659 2080 (63)  
\times 10^{-11}$
and the SM prediction is then
$\Delta a_{\mu} = a_{\mu}^{\mbox{$\scriptscriptstyle{\rm exp}$}}-
a_{\mu}^{\mbox{$\scriptscriptstyle{\rm SM}$}} = +246 (80) \times 10^{-11}$,
i.e. 3.1 standard deviations (adding all errors in quadrature).
Slightly higher discrepancies are obtained 
in 
Refs.~\cite{Hagiwara:2006jt,Jegerlehner:2009ry,Eidelman:2009ft}.
%
As in the case of $\alpha(M^2_Z)$, the uncertainty 
of the theoretical evaluation of
 $a_{\mu}^{\mbox{$\scriptscriptstyle{\rm SM}$}}$
 is still dominated by the  hadronic contribution at low energies, and
a reduction of the uncertainty is necessary in order to match 
the increased precision of  the proposed muon g-2 experiments 
at FNAL~\cite{Carey:2009zz} and J-PARC~\cite{Imazato:2004fy}.


The precise determination of the hadronic cross sections (accuracy $\lesssim 1\%$) 
requires an excellent control of higher order effects like Radiative
 Corrections (RC) and the 
non-perturbative hadronic contribution to the running of $\alpha$ 
(i.e. the vacuum polarisation, VP)
in Monte Carlo (MC) programs used for the analysis of the data.
Particularly in the last years, the increasing precision reached 
on the experimental side at the $e^+e^-$ colliders (VEPP-2M, DA$\mathrm{\Phi}$NE, BEPC, PEP-II and KEKB)
led to the development of dedicated high precision theoretical tools:
BabaYaga (and its successor BabaYaga@NLO) 
for the measurement of the luminosity, 
MCGPJ for the simulation of the exclusive 
QED channels, and 
PHOKHARA for the simulation of the process  with Initial State Radiation (ISR) $e^+e^-\to hadrons+\gamma$,
are examples of MC generators which include NLO corrections with per mill accuracy.
In parallel to these efforts, well-tested codes such as 
BHWIDE (developed for LEP/SLC colliders) were adopted.

Theoretical accuracies of these generators were estimated, whenever possible, 
 by evaluating  missing higher order contributions. From this point of view, the great pro\-gress in the calculation of two-loop corrections to the Bha\-bha scattering cross section was essential to 
establish the high
theoretical accuracy of the existing generators for the luminosity measurement.
However, usually only analytical or semi-analytical estimates
 of missing terms exist which don't take into account 
realistic experimental cuts.
In addition, MC event generators include different parametrisations for the VP
 which affect the prediction (and the precision) 
of the cross sections and also the RC are usually implemented differently. 

These arguments evidently imply the importance of 
comparisons of MC generators with a common set of input parameters 
and experimental cuts. 
Such  {\it tuned} comparisons, which started in the LEP era, are a key step for the validation of the generators, since they allow
 to check that the details entering the complex structure of the generators are under control and free of possible bugs.
This
was the main motivation for the 
{\it ``Working Group on Radiative Corrections  and  Monte Carlo Generators for Low Energies'' (Radio MontecarLow)},
which was formed a few years ago bringing together experts 
(theorists and experimentalists) working in the field of low energy $e^+e^-$ physics and partly also the $\tau$ community.

In addition to tuned comparisons, technical details of the MC
 generators, recent progress (like new calculations) and remaining
 open issues were also discussed in regular meetings.

This report is a summary of all these efforts:
it provides a self-contained and up-to-date description of the progress which 
occurred in the last years towards precision hadronic 
 phy\-sics at low energies, together with new results like comparisons and estimates of high order effects (e.g. of the 
pion pair correction to the Bhabha process) in the 
presence of realistic experimental cuts. 

The report is  divided into five sections: 
Sections~\ref{sec:1},~\ref{sec:2} and~\ref{sec:3} are devoted to the status of the MC tools for Luminosity,
the $R$-scan and Initial State Radiation (ISR).


Tau spectral functions of hadronic decays are also used to estimate  
$a_{\mu}^{\mbox{$\scriptscriptstyle{\rm HLO}$}}$, since they can be related to
 $e^+e^-$ annihilation cross section via isospin
symmetry~\cite{Alemany:1997tn,Davier:2002dy,Davier:2003pw,Davier:2009ag}.
The substantial difference
between the $e^+e^-$- and $\tau$-based determinations of
$a_{\mu}^{\mbox{$\scriptscriptstyle{\rm HLO}$}}$, even if isospin
violation corrections are taken into account, shows that further
common theoretical and experimental efforts are necessary
to understand this phenomenon. 
In
Section~\ref{sec:5}  the experimental status  and 
MC tools for tau decays are discussed. The recent improvements
of the generators  TAUOLA and PHOTOS are discussed and prospects
  for further developments are sketched.


Section~\ref{sec:4}  discusses vacuum polarisation at low energies, which is a
key ingredient for 
the high precision determination of the hadronic cross section, focusing on
the description and comparison of available parametrisations. 
Finally, Section~\ref{sec:6} contains a brief summary of the report.

\section{Luminosity}
\label{sec:1}

\newcommand{\dd}{\mbox{d}}
\newcommand{\eps}{\varepsilon}
\def\gsim{\mathrel{\raise.3ex\hbox{$>$\kern-.75em\lower1ex\hbox{$\sim$}}}}
\def\lsim{\mathrel{\raise.3ex\hbox{$<$\kern-.75em\lower1ex\hbox{$\sim$}}}}

\newcommand{\Obig}[1]{{\mathcal{O}\left (#1\right )}}
\newcommand{\ourcal}{\mathcal}
\newcommand{\oa}{${\ourcal O}(\alpha)$}
\newcommand{\myref}[1]{(\ref{#1})}
\newcommand{\bea}{\begin{eqnarray}}
\newcommand{\eea}{\end{eqnarray}}

\newcommand{\be}{\begin{equation}}
\newcommand{\ee}{\end{equation}}
\newcommand{\sss}{\@dottedtocline{2}{1.5em}{2.3em}}

The present Section addresses the most important experimental and theoretical issues
involved in the precision determination of the luminosity at meson factories. The luminosity
is the key ingredient underlying all the measurements and studies of the physics processes
discussed in the other Sections. Particular emphasis is put on the theoretical accuracy 
inherent to the event generators used in the experimental analyses, in comparison with the
most advanced perturbative calculations 
and experimental 
precision requirements. The effort done during the activity of the 
working group to perform tuned comparisons
between the predictions of the most accurate programs is described in detail. 
New calculations, leading to an update of the theoretical error associated with the 
prediction of the luminosity cross section, are also presented. 
The aim of the Section is to provide a self-contained and up-to-date description of the 
progress occurred during the last few years towards high-precision luminosity monitoring 
at flavour factories, as well as of the still open issues necessary for future advances. 

The structure of the Section is as follows. After an introduction on the motivation 
for precision luminosity measurements at meson factories 
(Section \ref{MOTIVATION}), the  leading-order (LO)
cross sections of the two QED processes of major interest, i.e. Bhabha scattering and 
photon pair production, are presented in Section 
\ref{LONLO}, together with the formulae for the
next-to-leading-order (NLO) 
photonic corrections to the above processes. The remarkable progress 
on the calculation of next-to-next-leading-order (NNLO) QED corrections to the Bhabha cross section, 
as occurred in the last few years, is reviewed in Section 
\ref{NNLO}. In particular, this Section 
presents new exact results on lepton and hadron pair corrections, 
taking into account realistic 
event selection criteria. Section \ref{MP} is devoted to the description 
of the theoretical methods used in the Monte Carlo (MC) generators for the simulation of 
multiple photon radiation. The matching of such contributions with NLO corrections is also described 
in Section \ref{MP}. The main features of the MC programs used by the experimental collaborations 
are summarised in Section 
\ref{MC}. Numerical results for 
the radiative corrections implemented into the MC generators 
are shown in Section \ref{NUMERICS} for both the Bhabha process and two-photon production. Tuned 
comparisons between the predictions of the most precise generators are presented and 
discussed in detail in 
Section \ref{TUNED}, considering the Bhabha process at different 
centre-of-mass (c.m.) energies and 
with realistic experimental cuts. The theoretical accuracy presently reached by the luminosity 
tools is addressed in Section \ref{TH}, where the most important sources of uncertainty are discussed
quantitatively. The estimate of the total error affecting the calculation
of the Bhabha cross section is given, as the main conclusion of the present work, 
in Section \ref{CONCLUSIONS}, updating and improving the robustness of
results available in the literature. 
Some remaining open issues are discussed in Section \ref{CONCLUSIONS} as well.

\begin{figure*}[hbtp]
\begin{center}
\mbox{
\resizebox{0.81\columnwidth}{0.86\columnwidth}{
\includegraphics{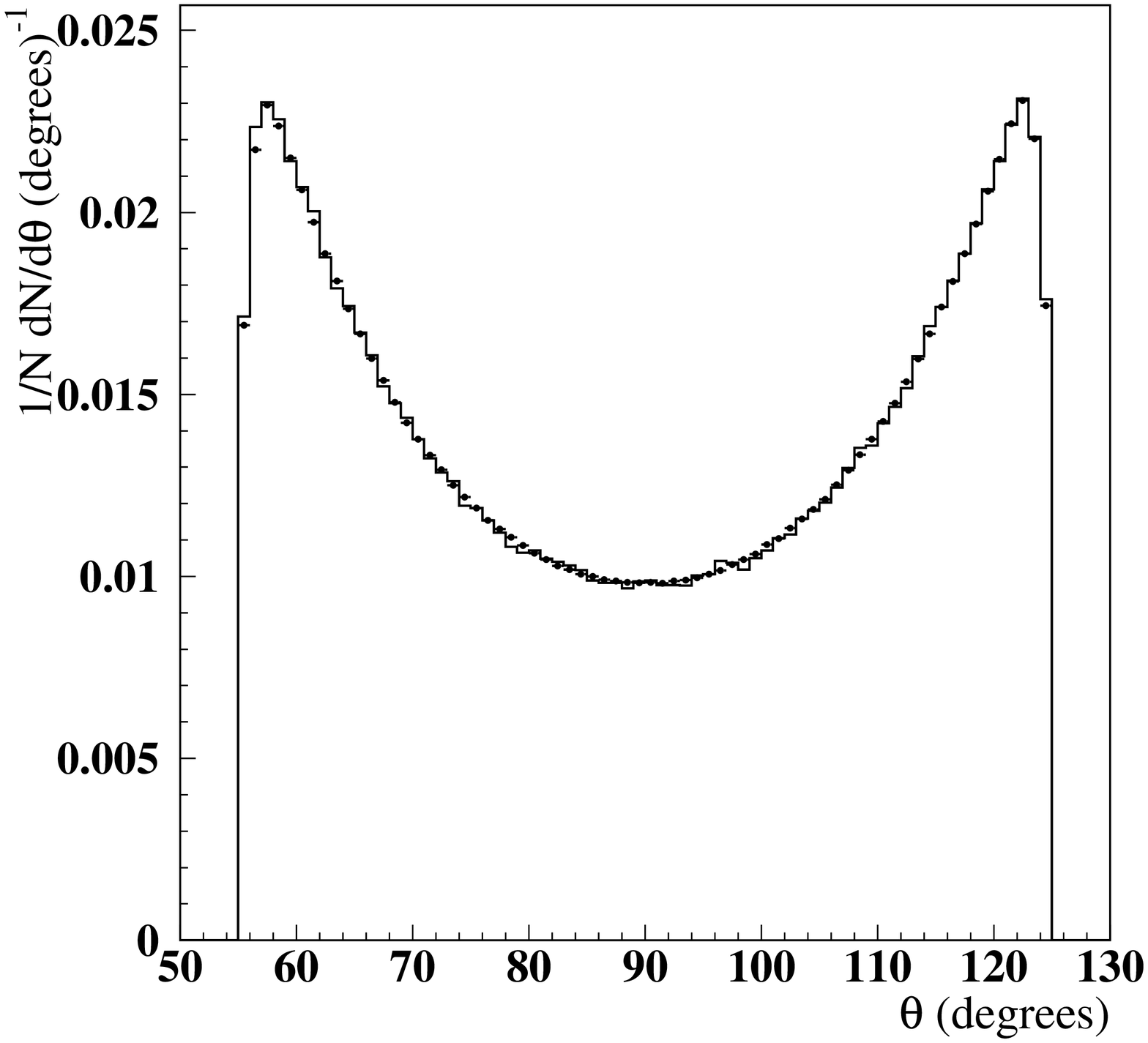}}
\hspace*{0.5cm}
\resizebox{0.85\columnwidth}{0.83\columnwidth}{
\includegraphics{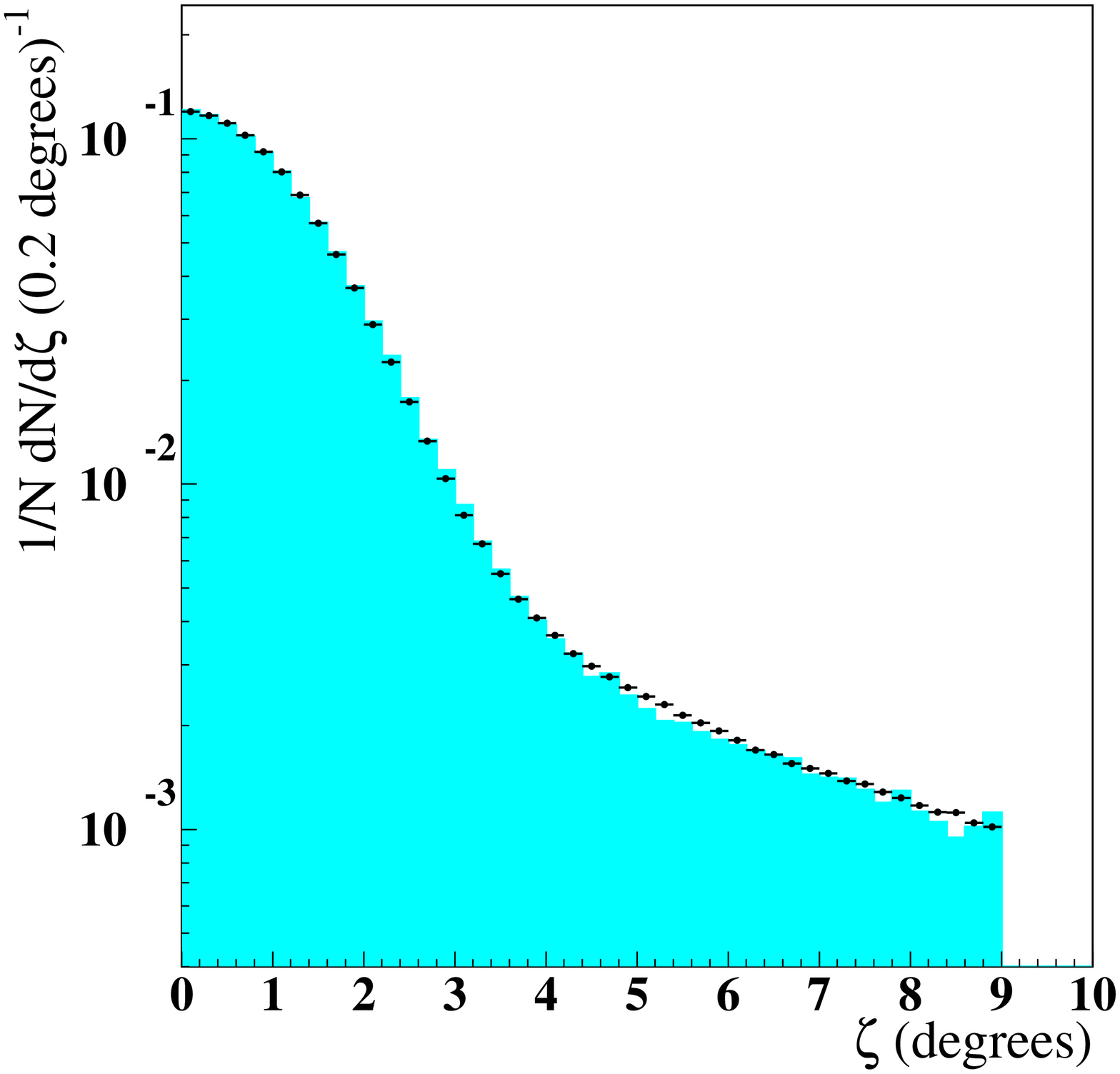}}
}
\caption{Comparison between large-angle Bhabha KLOE data
(points) and MC (histogram) distributions for the $e^\pm$ polar angle
$\theta$ (left) and for the acollinearity,
$\zeta=|\theta_{e^+}+\theta_{e^-}-180^\circ|$ (right), where
the flight direction of the $e^\pm$ is given by the 
position of clusters in the calorimeter. In each case, MC and data histograms are normalised
to unity. From~\cite{Ambrosino:2006te}.}
\label{figmotiv:1}
\end{center}
\end{figure*}

\subsection{Motivation \label{MOTIVATION}}

The luminosity of a collider is the normalisation constant between 
the event rate and the cross section of a given process. 
For an accurate measurement of the cross section
of an electron-positron ($e^+e^-$) annihilation 
process, the precise knowledge of the collider
luminosity is mandatory.

The luminosity depends on three factors:
beam-beam crossing
frequency, beam currents and the beam overlap area in the
crossing region. However, the last quantity is difficult to
determine accurately from the collider optics.
Thus, experiments prefer to determine the luminosity
by the counting rate of well
selected events whose cross section is known with good accuracy,
using the formula~\cite{Ambrosino:2006te}
\begin{equation}
\int\!\mathcal{L}\,{\rm d} t = \frac{N}{\epsilon \, \sigma} ,
\label{eq:1}
\end{equation}
where $N$ is the number of events of the chosen reference process, $\epsilon$ the 
experimental selection efficiency and $\sigma$ the theoretical cross section of the reference 
process. 
Therefore, the total luminosity
error will be given by the sum in quadrature of the fractional experimental and theoretical
uncertainties. 


Since the advent of low luminosity $e^+e^-$ colliders, a great
effort was devoted to obtain good precision in the cross section
of electromagnetic processes, extending the pioneering work of the
earlier days~\cite{Cabibbo:1961sz}. At the $e^+e^-$ colliders 
operating in the c.m. energy range $1\mbox{ GeV}<\sqrt{s}<3\mbox{ GeV}$,
such as ACO at Orsay, VEPP-II at Novosibirsk and Adone at Frascati, the
luminosity measurement was based on
Bhabha scattering~\cite{Bhabha:1936xxx,Barbiellini:1975}
with final-state electrons and positrons detected at small angles,
or single and double 
bremsstrahlung processes~\cite{Dehne:1974bc},
thanks to their high statistics. The electromagnetic cross
sections scale as $1/s$, while elastic
$e^+e^-$ scattering has a steep dependence on the polar angle,
$\sim 1/\theta^3$, thus providing a high rate for small values of
$\theta$.

Also at high-energy, accelerators running in the '90s around the $Z$ pole 
to perform precision tests of the Standard Model (SM), such 
as LEP at CERN and SLC at Stanford, the experiments used small-angle Bhabha
scattering events as a luminosity monitoring process. Indeed, for the very forward angular 
acceptances considered by the LEP/SLC collaborations, the Bhabha 
process is dominated by the electromagnetic interaction and, therefore, calculable, 
at least in principle, with very high accuracy.
At the end of the LEP and SLC operation,
a total (experimental plus theoretical) precision of 
one per mill (or better) was
achieved~\cite{Jadach:1996gu,Arbuzov:1995qd,Montagna:1996gw,Arbuzov:1996eq,Montagna:1998sp,Ward:1998ht,Jadach:2003zr}, 
thanks to the work of
different theoretical groups and 
the excellent performance of precision luminometers.

At current low- and intermediate-energy high-lu\-mi\-no\-si\-ty meson
factories, the small polar angle region is difficult
to access due to the presence of the low-beta insertions close to the
beam crossing region, while wide-angle Bhabha scattering produces a
large counting rate and can be exploited for a precise
measurement of the luminosity.

Therefore, also in this latter case of $e^\pm$ scattered at large angles,
e.g.  larger than 55$^\circ$ for the KLOE experiment~\cite{Ambrosino:2006te}
running at DA$\mathrm{\Phi}$NE in Frascati, and 
larger than 40$^\circ$ for the CLEO-c
experiment~\cite{cleo:2007zt} running
at CESR in Cornell, the main advantages of Bhabha scattering are preserved:
\begin{enumerate}
\item large statistics. For example at DA$\mathrm{\Phi}$NE, a statistical error
$\delta\mathcal{L}/\mathcal{L}\sim0.3\%$ is reached in
about two hours of data taking, even at the lowest luminosities;
\item high accuracy for the calculated cross section;
\item clean event topology of the signal and
small amount of background. 
\end{enumerate}

In Eq.~(\ref{eq:1}) the cross section is usually evaluated by inserting
event generators, which include radiative corrections at a high level of
precision, into
the MC code simulating the detector response. 
The code has to be developed to reproduce the detector performance
(geometrical acceptance, reconstruction efficiency and
resolution of the measured quantities) to a high level of confidence.

In most cases the major sources of the systematic errors of the
luminosity measurement are differences of efficiencies and resolutions
between data and MC.

In the case of KLOE, the largest experimental error of the
luminosity measurement is due to a different polar angle resolution
between data and MC which is observed at the edges of the accepted
interval for Bhabha scattering events. Fig.~1 shows a comparison
between large angle Bhabha KLOE data and MC, at left for the polar
angle and at right for the acollinearity $\zeta=|\theta_{e^+}+\theta_{e^-}-180^\circ|$. 
One observes a very good agreement between data and MC, but also 
differences (of about 0.3 \%) at the sharp interval edges. The analysis cut, 
$\zeta<9^\circ$, applied to the acollinearity distribution is very far
from the bulk of the distribution and does not introduce noteworthy
systematic errors.
\begin{figure}[h]
\begin{center}
\resizebox{0.4\textwidth}{!}{%
\includegraphics{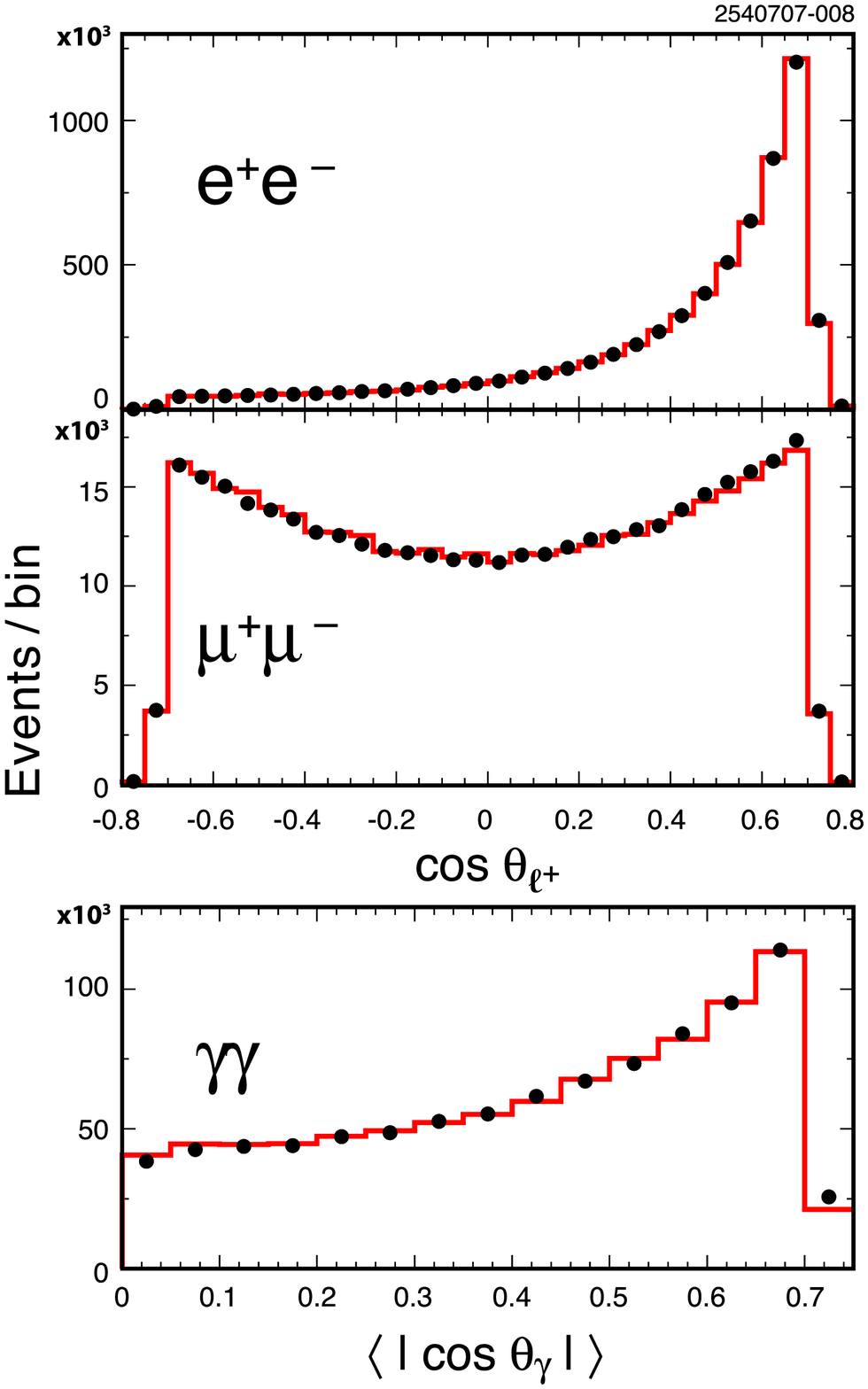}
}
\caption{Distributions of CLEO-c $\sqrt{s} = 3.774$ GeV
data (circles) and MC simulations (histograms) for the polar angle
of the positive lepton (upper two plots) in $e^+e^-$ and $\mu^+\mu^-$
events, and for the mean value of $|\cos\theta_\gamma|$ of the two photons
in $\gamma\gamma$ events (lower panel). MC histograms are normalised to the number of
data events. From~\cite{cleo:2007zt}.}
\label{figmotiv:2}
\end{center}
\end{figure}
Also in the CLEO-c luminosity measurement with Bhabha scattering
events, the detector modelling is the main
source of experimental error. In particular, uncertainties
include those due to finding and reconstruction of the electron
shower, in part due to the nature of the electron shower, as well as
the steep $e^\pm$ polar angle distribution.

The luminosity measured with Bhabha scattering events is often checked
by using other QED processes, such as $e^+e^-\to\mu^+\mu^-$
or $e^+e^-\to\gamma\gamma$. In KLOE, the luminosity measured
with $e^+e^-\to\gamma\gamma$ events differs by $0.3\%$
from the one determined from Bhabha events. In CLEO-c,
$e^+e^-\to\mu^+\mu^-$ events 
are also used, and the luminosity determined from $\gamma\gamma$ ($\mu^+\mu^-$)
is found to be $2.1\%$ ($0.6\%$) larger than that from Bhabha events.
Fig.~\ref{figmotiv:2} shows the CLEO-c data for the polar angle distributions 
of all three processes, compared with the corresponding MC predictions. 
The three QED processes are also used by the BaBar experiment at the 
PEP-II collider, Stanford, yielding a luminosity determination with an 
error of about 1\% \cite{andreas:2009xxxx}. 
Large-angle Bhabha scattering is the normalisation process adopted
by the CMD-2 and SND collaborations at VEPP-2M, Novosibirsk, while both
BES at BEPC in Beijing and Belle at KEKB in Tsukuba measure luminosity using 
the processes $e^+ e^- \to e^+ e^-$ and $e^+ e^- \to \gamma\gamma$ with 
the final-state particles detected at wide polar angles and an experimental 
accuracy of a few per cent. However, BES-III aims at reaching an error
of a few per mill in their luminosity measurement in the near future
\cite{Asner:2008nq}. 

The need of precision, namely better than $1\%$, 
and possibly redundant
measurements of the collider luminosity is of utmost importance to perform accurate measurements
of the $e^+ e^- \to {\rm hadrons}$ cross sections, which are the
key ingredient for evaluating the hadronic contribution
to the running of the electromagnetic coupling constant $\alpha$ and the muon anomaly $g-2$.

\subsection{LO cross sections and NLO corrections \label{LONLO}}

As remarked in Section \ref{MOTIVATION}, the processes of interest for
the luminosity 
measurement at meson factories are Bhabha scattering and electron-positron annihilation into two photons and muon pairs. Here we present the LO formulae for the 
cross section of the processes  $e^+ e^- \to e^+ e^-$ and $e^+ e^- \to \gamma\gamma$, as
well as the QED corrections to their cross sections in the NLO 
approximation of perturbation theory. The reaction $e^+ e^- \to \mu^+ \mu^-$ is 
discussed in Section \ref{sec:2}.

\subsubsection{LO cross sections \label{LO}}

\begin{figure}
\begin{center}
\resizebox{0.45\textwidth}{!}{%
\includegraphics{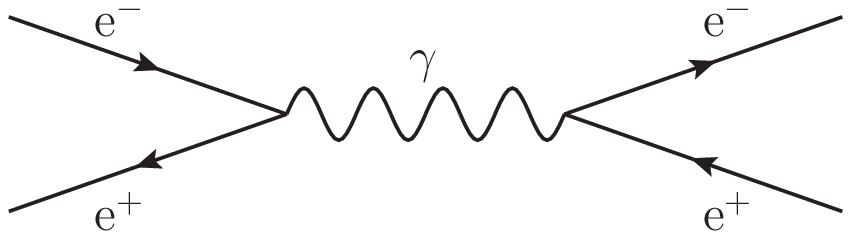}~\hspace{2.5cm}\includegraphics{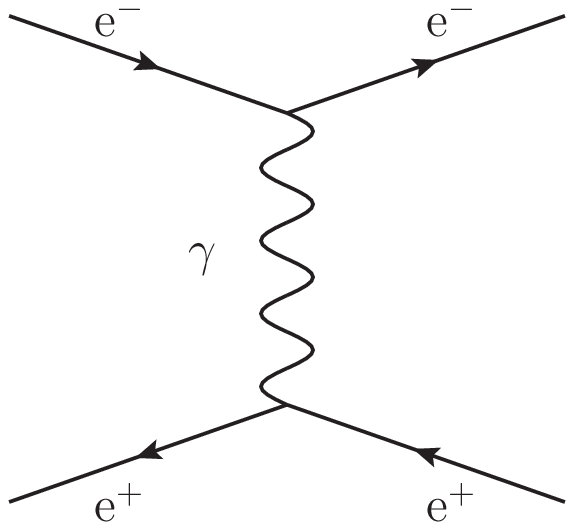}
}
\caption{LO Feynman diagrams for the Bhabha process in QED, corresponding 
to $s$-channel annihilation and $t$-channel scattering.}
\label{figlo:1}
\end{center}
\end{figure}

For the Bhabha scattering process
\begin{eqnarray}
e^-(p_-) + e^+(p_+)\to e^-(p_-') + e^+(p_+')
\end{eqnarray}
at Born level with 
simple one-photon exchange (see Fig.~\ref{figlo:1}) the differential cross section reads
\begin{eqnarray}
\frac{\dd{\sigma}^{\mathrm{Bhabha}}_0}{\dd\Omega_-} 
= \frac{\alpha^2}{4s}\;\left(\frac{3+c^2}{1-c}\right)^2 
+  O \left(\frac{m_e^2}{s}\right),
\end{eqnarray}
where 
\begin{eqnarray}
s &=& (p_-+p_+)^2, \qquad c = \cos\theta_-.
\end{eqnarray}
The angle $\theta_-$ is defined between the initial and final electron three-momenta,
$\dd\Omega_-=\dd\phi_-\dd\cos\theta_-$, and $\phi_-$ is the azimuthal angle of the
outgoing electron. The small mass correction terms suppressed by the ratio $m_e^2/s$ are
negligible for the energy range and the angular acceptances 
which are of interest here.

At meson factories the Bhabha scattering cross section is largely dominated by 
$t$-channel photon exchange, followed by $s$-$t$ interference and $s$-channel 
annihilation.
Furthermore, $Z$-boson exchange contributions and other electroweak
effects are suppressed at least by a factor $s/M_Z^2$. In particular,
for large-angle Bhabha scattering with a c.m. energy
$\sqrt{s} = 1$~GeV the $Z$ boson contribution amounts to about $-1\times 10^{-5}$.
For $ \sqrt{s} = 3$~GeV it amounts to $-1\times 10^{-4}$ 
and $-1 \times 10^{-3}$ for $\sqrt{s} = 10$~GeV. So only at $B$ factories the
electroweak effects should be taken into account at tree level, 
when aiming at a per mill precision level.

\begin{figure}
\begin{center}
\resizebox{0.35\textwidth}{!}{%
\includegraphics{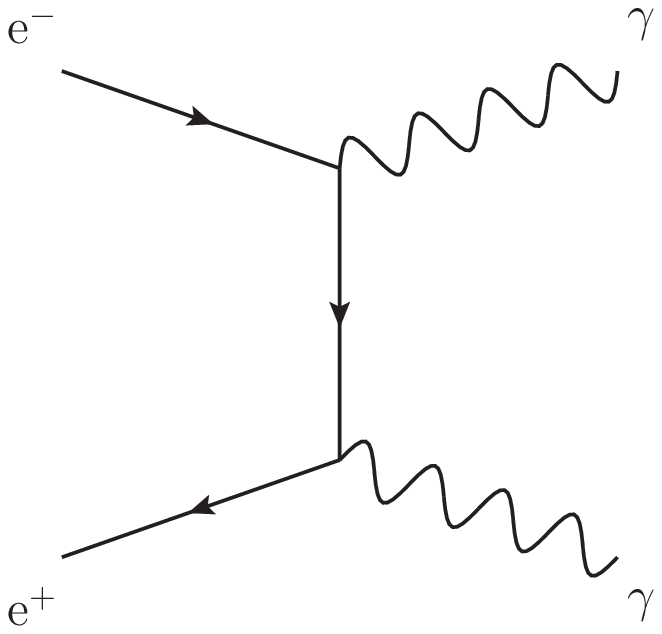}~\hspace{3cm}\includegraphics{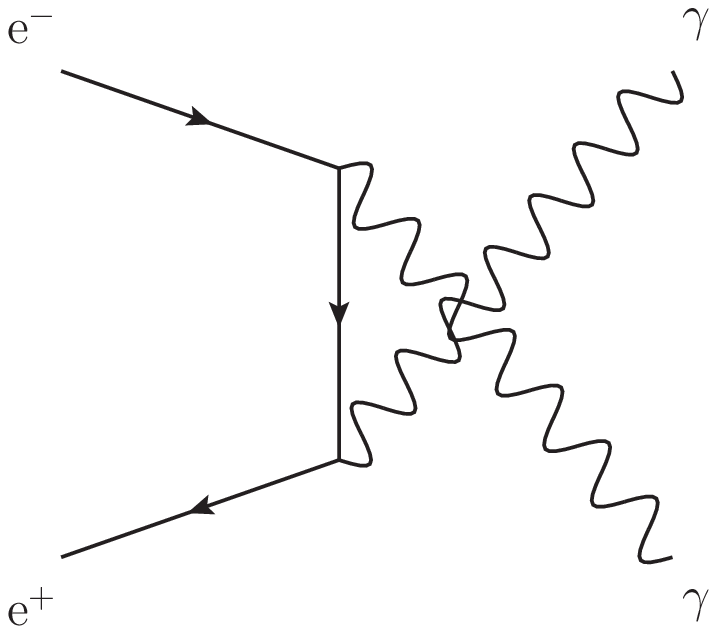}
}
\caption{LO Feynman diagrams for the process $e^+e^- \to \gamma\gamma$.}
\label{figlo:2}
\end{center}
\end{figure}

The LO differential cross section of 
the two-photon annihilation channel (see Fig.~\ref{figlo:2})
\begin{equation} \nonumber
e^+(p_+) + e^-(p_-) \to \gamma(q_1) + \gamma(q_2)
\end{equation}
can be obtained by a crossing relation from the Compton scattering
cross section computed by Brown and Feynman~\cite{Brown:1952eu}. It reads
\begin{eqnarray}
\frac{\dd\sigma^{\gamma\gamma}_0}{\dd\Omega_1}
= \frac{\alpha^2}{s}\left(\frac{1+c_1^2}{1-c_1^2}\right)
+ O \left(\frac{m_e^2}{s}\right),
\end{eqnarray}
where $\dd\Omega_1$ denotes the
differential solid angle of the first photon. It is assumed that both 
final photons are registered in a detector 
and that their polar angles with respect to the initial beam directions are
not small ($\theta_{1,2} \gg m_e/E$, where $E$ is the beam 
energy).

\subsubsection{NLO corrections \label{NLO}}

The complete set of NLO radiative 
corrections, emerging at  $O (\alpha)$ of perturbation theory, 
to Bhabha scattering and two-photon annihilation can be split into gauge-invariant subsets: 
QED corrections, due to emission of real photons off the charged leptons and 
exchange of virtual photons between them, and purely weak contributions arising from 
the electroweak sector of the SM. 

The complete $O (\alpha)$ QED corrections to Bhabha
scattering are known since a long time~\cite{Redhead:1953xxx,Polovin:1956xxx}. 
The first complete NLO prediction in the electroweak SM was performed in \cite{Consoli:1979xw},
followed by  \cite{Bohm:1984yt} and several others. At NNLO, 
the leading virtual weak corrections from the top quark were derived first in \cite{Bardin:1990xe} and are available in the 
fitting programs ZFITTER \cite{Bardin:1999yd,Arbuzov:2005ma} and 
TOPAZ0 \cite{Montagna:1993py,Montagna:1993ai,Montagna:1998kp}, extensively used by the 
experimentalists for the 
extraction of the electroweak parameters at LEP/SLC. 
The weak NNLO corrections in the SM are also known for the $\rho$-parameter
\cite{%
Djouadi:1987gn,Djouadi:1988di,Kniehl:1988ie,vanderBij:1986hy,Barbieri:1993dq,Fleischer:1993ub,%
Fleischer:1994cb,Boughezal:2006xk,Chetyrkin:2006bj,Schroder:2005db,Chetyrkin:1995ix,%
Chetyrkin:1995js,Barbieri:1992nz,vanderBij:2000cg,Faisst:2003px,Boughezal:2004ef,%
Boughezal:2005eb}
and the weak mixing angle
 \cite{Awramik:2003rn,Awramik:2004ge,Hollik:2005va,Hollik:2006ma,Awramik:2006uz,Awramik:2006ar}, as well as corrections  from Sudakov logarithms
\cite{Kuhn:1999nn,Kuhn:2000hx,Kuhn:2001hz,Feucht:2004rp,Jantzen:2005xi,Jantzen:2005az,%
Denner:2006jr,Denner:2008hw}. 
Both NLO and NNLO weak effects are 
negligible at low energies and are not 
 implemented yet in numerical packages for Bhabha scattering at meson factories.
In pure QED, the situation is considerably different due to the remarkable progress 
made on NNLO 
corrections in recent years, as emphasised and discussed in detail in 
Section \ref{NNLO}.

\begin{figure}
\begin{center}
\resizebox{0.5\textwidth}{!}{
\includegraphics{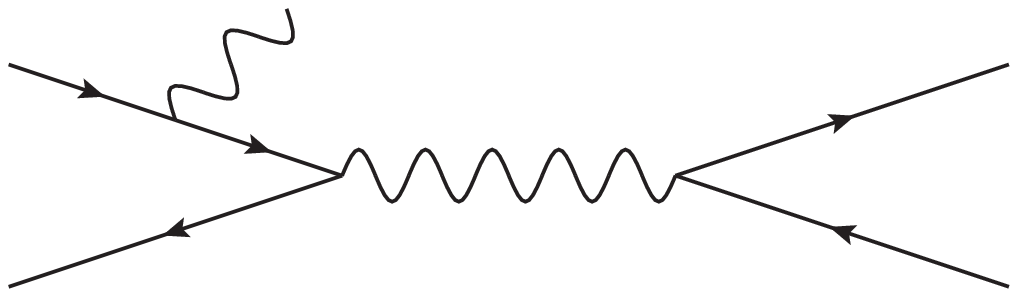}~\hspace{0.75cm}\includegraphics{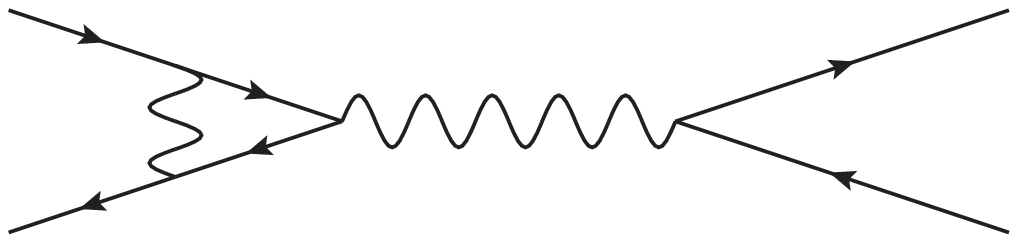}
}
\caption{Examples of Feynman diagrams for real and virtual NLO QED 
initial-state corrections to the $s$-channel contribution of the Bhabha process.}
\label{fignlo:1}
\end{center}
\end{figure}

As usual, the photonic corrections can be split into two parts according to
their kinematics. The first part preserves the Born-like kinematics and contains the effects
due to one-loop amplitudes (virtual corrections) and single soft-photon emission. 
Examples of Feynman diagrams giving rise to such corrections are represented in 
Fig.~\ref{fignlo:1}. The energy of a soft photon is assumed not to exceed an energy 
$\Delta E$, where
$E$ is the beam energy and the auxiliary parameter $\Delta \ll 1$ should be chosen
in such a way that the validity of the soft-photon approximation is guaranteed. 
The second contribution is due to hard photon emission, 
i.e. to single bremsstrahlung with photon energy above $\Delta E$ and 
corresponds to the radiative process $e^+ e^- \to e^+ e^- \gamma$.

\begin{figure}
\begin{center}
\resizebox{0.5\textwidth}{!}{%
\includegraphics{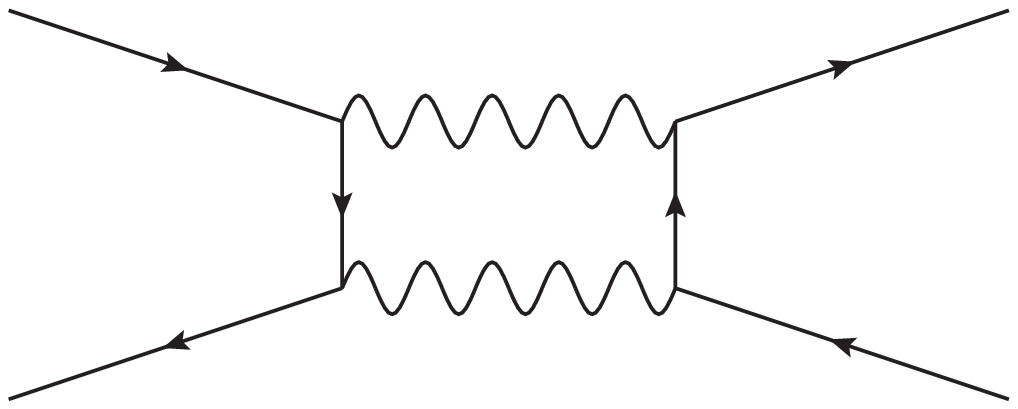}~\hspace{0.75cm}\includegraphics{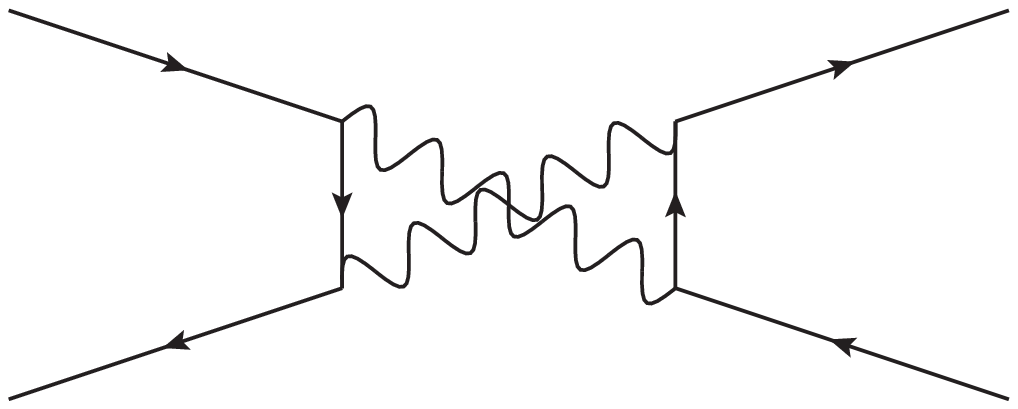}
}
\caption{Feynman diagrams for the NLO QED box corrections to the $s$-channel contribution of the Bhabha process.}
\label{fignlo:2}
\end{center}
\end{figure}

Following~\cite{Berends:1973jb,Beenakker:1990mb},
the soft plus virtual (SV) correction can be cast into the form 
\begin{eqnarray}
\frac{\dd\sigma^{\mathrm{Bhabha}}_{B+S+V}}{\dd\Omega_-}
&=& \frac{\dd{\sigma}^{\mathrm{Bhabha}}_{0}}{\dd\Omega_-}
\biggl\{1+ \frac{2\alpha}{\pi}(L-1)\left[2\ln\Delta
+ \frac{3}{2}\right] \nonumber \\ \label{bsv}
&-& \frac{8\alpha}{\pi}\ln(\mbox{ctg}\frac{\theta}{2})
\ln\Delta + \frac{\alpha}{\pi}K^{\mathrm{Bhabha}}_{SV} \biggr\},
\end{eqnarray}
where the factor $K^{\mathrm{Bhabha}}_{SV}$ is given by
\begin{eqnarray}
&& K^{\mathrm{Bhabha}}_{SV} = -1-2\mathrm{Li}_2(\sin^2\frac{\theta}{2}) 
+ 2\mathrm{Li}_2(\cos^2\frac{\theta}{2})
\nonumber \\ 
&& \quad + \frac{1}{(3+c^2)^2}\biggl[\frac{\pi^2}{3}(2c^4 - 3c^3 - 15c)
+ 2(2c^4 - 3c^3 + 9c^2 
\nonumber \\ 
&& \quad + 3c + 21)\ln^2(\sin\frac{\theta}{2})
- 4(c^4+c^2-2c)\ln^2(\cos\frac{\theta}{2}) 
\nonumber \\ \nonumber
&& \quad - 4(c^3+4c^2+5c+6)\ln^2(\mbox{tg}\frac{\theta}{2})
+ 2(c^3-3c^2+7c
\nonumber \\
&& \quad -5)\ln(\cos\frac{\theta}{2}) 
+2(3c^3+9c^2+5c+31)\ln(\sin\frac{\theta}{2}) \biggr],
\label{eq:svbh}
\end{eqnarray}
and depends on the scattering angle, due to the contribution from 
initial-final-state interference and box diagrams (see Fig.~\ref{fignlo:2}). 
It is worth noticing that the SV correction contains a leading logarithmic 
(LL) part enhanced by
the collinear logarithm $L = \ln(s/m_e^2)$. Among the virtual corrections 
there is also a numerically important effect due to
vacuum polarisation in the photon propagator. Its contribution 
is omitted in Eq.~(\ref{eq:svbh}) but can be taken into account
in the standard way by insertion of the resummed vacuum polarisation
operators in the photon propagators of the Born-level Bhabha amplitudes.

The differential cross section of the single hard 
brems\-strah\-lung process
\begin{eqnarray*}
e^+(p_+) + e^-(p_-) \to e^+(p_+') + e^-(p_-') + \gamma(k)
\end{eqnarray*}
for scattering angles up to corrections of order $m_e/E$ reads
\begin{eqnarray} \label{bere}
\dd\sigma^{\mathrm{Bhabha}}_{\mathrm{hard}} &=& \frac{\alpha^3}{2\pi^2s}\;
R_{e\bar{e}\gamma}\dd \Gamma_{e\bar{e}\gamma} , 
\\ \nonumber 
\dd\Gamma_{e\bar{e}\gamma} &=& \frac{\dd^3p_+'\dd^3p_-'\dd^3k}{\eps_+'\eps_-'k^0}
\delta^{(4)}(p_++p_--p_+'-p_-'-k), 
\\ \nonumber
R_{e\bar{e}\gamma} &=& \frac{WT}{4}
- \frac{m_e^2}{(\chi_+')^2}\left(\frac{s}{t}+\frac{t}{s}+1\right)^2
\\ \nonumber 
&-& \frac{m_e^2}{(\chi_-')^2}\left(\frac{s}{t_1}+\frac{t_1}{s}+1\right)^2
- \frac{m_e^2}{\chi_+^2}\left(\frac{s_1}{t}+\frac{t}{s_1}+1\right)^2
\\ \nonumber
&-& \frac{m_e^2}{\chi_-^2}\left(\frac{s_1}{t_1}+\frac{t_1}{s_1}+1\right)^2,
\end{eqnarray}
where
\begin{eqnarray}
W &=& \frac{s}{\chi_+\chi_-} + \frac{s_1}{\chi_+'\chi_-'}
- \frac{t_1}{\chi_+'\chi_+} \nonumber
- \frac{t}{\chi_-'\chi_-}
+ \frac{u}{\chi_+'\chi_-} + \frac{u_1}{\chi_-'\chi_+}\, , \\ \nonumber
T &=& \frac{ss_1(s^2+s_1^2) + tt_1(t^2+t_1^2)+uu_1(u^2+u_1^2)}{ss_1tt_1}\, ,
\end{eqnarray}
and the invariants are defined as
\begin{eqnarray}
\nonumber
&& s_1 = 2p_-'p_+',\qquad t=-2p_-p_-',\qquad
t_1=-2p_+p_+', \\ \nonumber
&& u = -2p_-p_+',\quad u_1=-2p_+p_-',\quad \chi_{\pm}=kp_{\pm},\quad
\chi_{\pm}'=kp_{\pm}'.
\end{eqnarray}

NLO QED radiative corrections to the two-photon annihilation channel were 
obtained in~\cite{terentyev:1969xxx,Berends:1973tm,Eidelman:1978rw,Berends:1981uq}, 
while weak corrections were computed in~\cite{bohm:1986bs}. 

In the one-loop approximation the part of the differential cross section with
the Born-like kinematics reads
\begin{eqnarray}
\label{bsvgg}
&& \frac{\dd\sigma^{\gamma\gamma}_{B+S+V}}{\dd\Omega_1} 
= \frac{\dd{\sigma}^{\gamma\gamma}_0}{\dd\Omega_1}
\Biggl\{1 + \frac{\alpha}{\pi}
\biggl[(L-1)\biggl(2\ln\Delta+\frac{3}{2}\biggr)
\nonumber \\ 
&&\qquad\qquad\quad + \, K^{\gamma\gamma}_{SV}\biggr]\Biggr\}, 
\nonumber \\ 
&& K^{\gamma\gamma}_{SV} = \frac{\pi^2}{3}+\frac{1-c_1^2}{2(1+c_1^2)}\biggl[\biggl(1
+ \frac{3}{2}\,\frac{1+c_1}{1-c_1}\biggr)\ln\frac{1-c_1}{2} 
\nonumber\\
&& \quad + \biggl(1+\frac{1-c_1}{1+c_1}+\frac{1}{2}\frac{1+c_1}{1-c_1}\biggr)
\ln^2\frac{1-c_1}{2}
+ (c_1\to -c_1)\biggr], \nonumber \\
&& c_1 = \cos\theta_1,\qquad
\theta_1=\widehat{\vec{q}_1\vec{p}}_-\, .
\end{eqnarray}

In addition, the three-photon production process
\begin{eqnarray*}
e^+(p_+)\ +\ e^-(p_-)\ \to\ \gamma(q_1)\ +\ \gamma(q_2)\ +\ \gamma(q_3)
\end{eqnarray*}
must be included. Its cross section is given by
\begin{eqnarray} \label{3gamma}
&& \dd\sigma^{e^+e^-\to 3\gamma} = \frac{\alpha^3}{8\pi^2 s}
R_{3\gamma}\,\dd\Gamma_{3\gamma}\, , \\ \nonumber
&& R_{3\gamma} = s\;\frac{\chi_3^2+(\chi_3')^2}{\chi_1\chi_2\chi_1'\chi_2'}
- 2m_e^2\biggl[\frac{\chi_1^2+\chi_2^2}{\chi_1\chi_2(\chi_3')^2}
+ \frac{(\chi_1')^2+(\chi_2')^2}{\chi_1'\chi_2'\chi_3^2}\biggr] 
\\ \nonumber 
&&\quad \quad + \, \mbox{(cyclic permutations)}, 
\\ \nonumber
&& \dd \Gamma_{3\gamma} = \frac{\dd^3q_1\dd^3q_2\dd^3q_3}{q_1^0q_2^0q_3^0}
\delta^{(4)}(p_++p_--q_1-q_2-q_3),
\end{eqnarray}
where
\begin{eqnarray*}
\chi_i=q_ip_-,\qquad \chi_i'=q_ip_+,\qquad i=1,2,3\, .
\end{eqnarray*}
The process has to be treated as a radiative correction to
the two-photon production. The energy of the third photon should 
exceed the soft-photon energy threshold $\Delta E$. In practice, the tree photon
contribution, as well as the radiative Bhabha process 
$e^+ e^- \to e^+ e^- \gamma$, should be simulated with the help of a MC event generator
 in order to take into account the proper experimental criteria of a given event selection.

In addition to the corrections discussed above, also the effect of vacuum polarisation, due
to the insertion of fermion loops inside the photon propagators, must be included in the precise 
calculation of the Bhabha scattering cross section. Its
theoretical treatment, which faces the non-trivial 
problem of the non-perturbative contribution due to hadrons, 
is addressed in detail in Section \ref{sec:4}. However, 
numerical results for such a correction are presented in Section \ref{NUMERICS} 
and Section \ref{TH}.

\begin{figure}[h]
\begin{center}
\resizebox{0.475\textwidth}{!}{%
\includegraphics{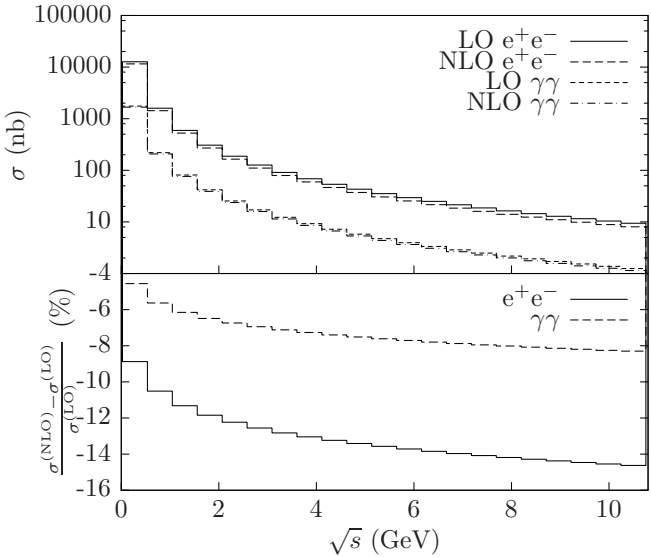}
}
\caption{Cross sections of the processes $e^+e^- \to e^+e^-$ and $e^+e^- \to \gamma\gamma$ 
in LO and NLO approximation as a function of the c.m. energy at meson factories (upper panel).
In the lower panel, the relative contribution due to the NLO QED corrections 
(in per cent) to the two processes is shown.}
\label{fignlo:3}
\end{center}
\end{figure}

In Fig.~\ref{fignlo:3} the cross sections of the Bhabha and two-photon production processes in LO and
NLO approximation are shown as a function of the c.m. energy between $\sqrt{s} \simeq 2\, m_{\pi}$ and
$\sqrt{s} \simeq 10$~GeV (upper panel). The results were obtained imposing the following cuts 
for the Bhabha process:
\begin{eqnarray}
&& \theta_{\pm}^{\textrm{min}}  = 45^{\circ}\,, 
    \qquad \quad \, \, \, \theta_{\pm}^{\textrm{max}} = 135^{\circ}\,,  \nonumber \\
&& E_{\pm}^{\textrm{min}} = 0.3\sqrt{s}\,,  \qquad \xi_{\textrm{max}} =10^{\circ}\,,
\label{eq:cutsmf}
\end{eqnarray}
where $\theta_{\pm}^{\textrm{min,max}}$ are the angular acceptance cuts, 
$E_{\pm}^{\textrm{min}}$ are the minimum energy thresholds for the detection of the 
final-state electron/positron and $\xi_{\textrm{max}}$ is the maximum $e^+ e^-$ acollinearity.
For the photon pair production processes we used correspondingly:
\begin{eqnarray}
&& \theta_{\gamma}^{\textrm{min}}  = 45^{\circ}\,, 
\qquad \quad \, \, \, \theta_{\gamma}^{\textrm{max}} = 135^{\circ}\,,  \nonumber \\
&& E_{\gamma}^{\textrm{min}} = 0.3\sqrt{s}\,,  \qquad \xi_{\textrm{max}} =10^{\circ}\,,
\label{eq:cutsmfg}
\end{eqnarray}
where, as in Eq.~(\ref{eq:cutsmf}), $\theta_{\gamma}^{\textrm{min,max}}$ are the angular acceptance cuts, $E_{\gamma}^{\textrm{min}}$ is the minimum energy threshold for the detection of at 
least two photons and $\xi_{\textrm{max}}$ is the maximum acollinearity between the most energetic and next-to-most energetic photon.

The cross sections display the typical $1/s$ QED behaviour. The relative 
effect of NLO corrections is shown in the lower panel. It can be seen that the NLO corrections are 
 largely negative and increase with increasing
c.m. energy, because of the growing importance of the 
collinear logarithm $L= \ln(s/m_e^2)$. The corrections to $e^+e^- \to \gamma\gamma$ are 
about one half of those to Bhabha scattering, because of the absence of final-state radiation 
effects in photon pair production.

\subsection{NNLO corrections to the Bhabha scattering cross section \label{NNLO}}

Beyond the NLO corrections discussed in the previous Section, in recent years a
significant effort was devoted to the calculation of the perturbative corrections  to
the Bhabha process at NNLO in QED.

The calculation of the full NNLO corrections to the Bhabha scattering cross section 
requires three types of ingredients: {\it i)} the two-loop matrix elements for the 
$e^{+}e^{-} \to e^{+}e^{-}$ process; {\it ii)} the one-loop matrix elements for the 
$e^{+}e^{-} \to e^{+}e^{-} \gamma$ process, both in the case in which the additional 
photon is soft or hard; 
 {\it iii)} the tree-level matrix elements for $e^{+}e^{-} \to
e^{+}e^{-}\gamma\gamma$, with two soft or two hard photons, or one soft and one
hard photon. Also the process $e^{+}e^{-} \to e^{+}e^{-} e^{+}e^{-}$, with 
one of the two $e^{+}e^{-}$ pairs remaining 
undetected, contributes to the Bhabha signature at NNLO. 
%
Depending on the kinematics, other final states like, e.g., $e^{+}e^{-} \mu^{+}\mu^{-}$
or those with hadrons are also possible.


The advent of new calculational techniques and a deeper understanding of the IR
structure of unbroken gauge theories, such as QED or QCD, made the calculation of
the complete set of two-loop QED corrections possible. The history of this
calculation will be presented in Section~\ref{virtual}.
%

Some remarks on the one-loop matrix elements with
three particles in the final state are in order now. 
The diagrams involving the emission of a soft photon are known and they were included
in the calculations of the two-loop matrix elements, in order to remove the IR soft
divergences. However, although the contributions due to a hard collinear  photon are
taken  into account in logarithmic accuracy by the MC generators, a full  calculation
of the diagrams involving a hard photon in a general phase-space configuration is still
missing. In Section~\ref{Penta}, we shall comment on the possible strategies which can
be  adopted in order to calculate these corrections.\footnote{As emphasised in Section \ref{TH} and 
Section \ref{CONCLUSIONS}, the complete calculation of this class of corrections became available 
\cite{Actis:2009uq} during the completion of the present work.}

As a general comment, it must be noticed that the fixed-order corrections 
calculated up to NNLO are taken into account at the LL, and, 
partially, next-to-leading-log (NLL) level in the most precise MC generators, which include, as 
will be discussed in Section 
\ref{MP} and Section \ref{MC}, the logarithmically enhanced contributions of soft and
collinear photons at all orders in perturbation theory.


Concerning the tree level graphs with four particles in the final state, the production of a 
soft $e^{+}e^{-}$ pair was considered in the literature by the authors of \cite{Arbuzov:1995vj} 
by following the evaluation of pair production \cite{Arbuzov:1995cn,Arbuzov:1995vi} within the calculation of the 
$O (\alpha^2 L)$ single-logarithmic accurate small-angle Bhabha cross section 
\cite{Arbuzov:1995qd}, 
and it is included in the two-loop calculation (see
Section~\ref{ElLoop}). New results on lepton and hadron pair corrections, which are at present 
approximately included in the available Bhabha codes, are presented in 
Section~\ref{4pfinal}.




\subsubsection{Virtual corrections for the $e^{+}e^{-} \to e^{+}e^{-}$ process \label{virtual}}

The calculation of the virtual two-loop QED corrections to the Bhabha scattering
differential cross section was carried out in the last 10 years.  This calculation was
made possible by an improvement of the techniques employed in  the evaluation of
multi-loop Feynman diagrams. An essential tool used to manage the calculation is the 
Laporta algorithm \cite{Laporta:1996mq,Laporta:2001dd,Tkachov:1981wb,Chetyrkin:1981qh}, which enables one to reduce a generic combination
of dimensionally-regularised scalar integrals to a combination of a small set of
independent integrals called the ``Master Integrals'' (MIs) of the problem under
consideration. The calculation of the MIs is then pursued by means of a variety of 
methods. Particularly important are the differential equations method \cite{Kotikov:1991kg,Kotikov:1991hm,Kotikov:1991pm,Remiddi:1997ny,Caffo:1998yd,Caffo:1998du,Argeri:2007up} and
the Mellin-Barnes techniques 
\cite{Smirnov:2004,Friot:2005cu,Usyukina:1975yg,Smirnov:1999gc,Tausk:1999vh,%
Smirnov:1999wz,Smirnov:2001cm,Heinrich:2004iq,Czakon:2005rk,Gluza:2007rt}.
Both methods proved to be very useful in the evaluation of virtual corrections to Bhabha 
scattering because they are especially effective in problems with a small number of different 
kinematic parameters. They both allow one to obtain an analytic expression for the 
integrals, which must be written in terms of a  suitable functional basis. A basis
which was extensively employed in the calculation of multi-loop Feynman diagrams of the 
type discussed here is represented by the Harmonic Polylogarithms \cite{Goncharov:1998,Broadhurst:1998rz,Remiddi:1999ew,Gehrmann:2001pz,Gehrmann:2001jv,Maitre:2005uu,Maitre:2007kp,Vollinga:2004sn,Weinzierl:2007cx} and their 
generalisations.
Another fundamental achievement which enabled one to complete the calculation of  the QED
two-loop corrections was an improved  understanding of the IR structure of QED. In
particular, the relation between the collinear logarithms in which the electron mass
$m_e$ plays  the role of a natural cut-off and the corresponding  poles  in the
dimensionally regularised  massless theory was extensively investigated in
\cite{Penin:2005kf,Penin:2005eh,Mitov:2006xs,Becher:2007cu}.

\begin{figure}
\begin{center}
\resizebox{0.45\textwidth}{!}{%
\includegraphics{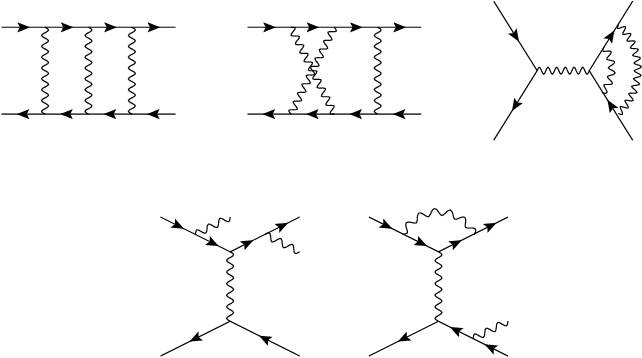}
}
\caption{\small{ Some of the diagrams belonging to the class of the ``photonic'' NNLO
corrections to the Bhabha scattering differential cross section. The additional
	photons in the final state are soft. \label{fig0}}}
\end{center}	
\end{figure}

\begin{figure}
\begin{center}
\resizebox{0.45\textwidth}{!}{%
\includegraphics{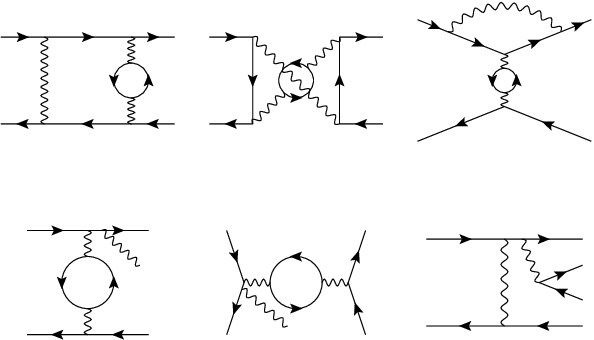}
}
\caption{\small{ Some of the diagrams belonging to the class of the ``electron loop'' 
NNLO corrections. The additional photons or electron-positron pair in the final state 
are soft. \label{fig1}}}
\end{center}
\end{figure}

\begin{figure}
\begin{center}
\resizebox{0.45\textwidth}{!}{%
\includegraphics{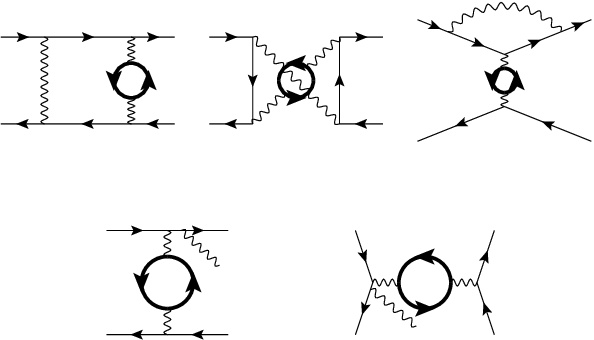}
}
\vspace*{4mm}
\caption{\small{ Some of the diagrams belonging to the class of the ``heavy fermion'' 
NNLO corrections. The additional photons in the final state are soft.
\label{fig2}}}
\end{center}
\end{figure}

\begin{figure}
\begin{center}
\resizebox{0.45\textwidth}{!}{%
\includegraphics{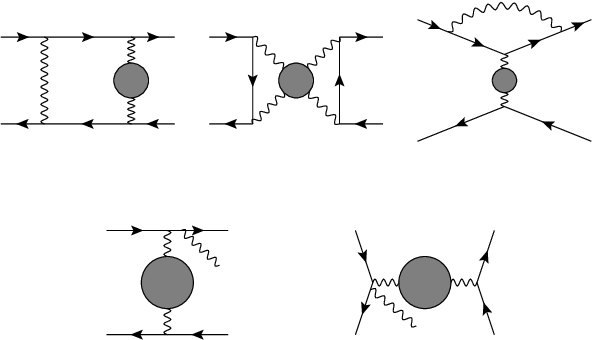}
}
\caption{\small{ Some of the diagrams belonging to the class of the ``hadronic'' 
corrections. The additional photons in the final state are soft.
\label{fig3}}}
\end{center}
\end{figure}

The first complete diagrammatic calculation of the two-loop QED virtual corrections  to
Bhabha scattering can be found in \cite{Bern:2000ie}. However, this result was obtained  in
the fully massless approximation ($m_e = 0$) by employing dimensional regularisation
(DR) to regulate  both soft and collinear divergences.
Today, the complete set of two-loop corrections to Bhabha scattering in pure QED have
been evaluated using $m_e$ as a collinear regulator, as required in order to include
these fixed-order calculations in available Monte Carlo event generators.
The Feynman diagrams involved in the calculation can be divided in three gauge-independent 
sets: {\em i)} diagrams without fermion loops (``photonic'' diagrams), {\em
ii)} diagrams involving a closed electron loop, and {\em iii)} diagrams involving a
closed loop of hadrons or a fermion heavier than the electron. Some of the diagrams
belonging to the aforementioned sets are shown in Figs.~\ref{fig0}--\ref{fig3}. 
These three sets are discussed in more detail below. \\

\noindent {\em{Photonic corrections \label{Phot}}} \\
\vspace*{-2mm}

\noindent A large part of the NNLO photonic corrections can be evaluated in a closed
analytic form, retaining  the full dependence on $m_e$ \cite{Bonciani:2005im}, by using the Laporta
algorithm for the reduction of the Feynman diagrams to a combination of MIs, and then
the differential equations method for their analytic evaluation.
With this technique it is possible to calculate, for instance, the NNLO corrections to
the form factors \cite{Bonciani:2003te,Bonciani:2003hc,Bonciani:2003ai,Czakon:2004wm}.
However, a calculation of the two-loop photonic boxes retaining the full dependence on
$m_e$ seems to be beyond the reach of this method. This is due to the fact that the number of
MIs belonging to the same topology is, in some cases, large.
Therefore, one must solve analytically large systems of first-order
ordinary linear differential equations; this is not possible in general.
Alternatively, in order to calculate the different MIs involved, one could use the
Mellin-Barnes techniques, as shown in \cite{Smirnov:2001cm,Heinrich:2004iq,Czakon:2004wm,Czakon:2006pa,Broadhurst:1993mw,Davydychev:2003mv}, or a combination 
of both methods. The calculation is very complicated and a full result is not 
available yet.\footnote{For the planar double box diagrams, all the MIs are known 
\cite{Czakon:2006pa} for small $m_e$, while the MIs for the non-planar double box diagrams 
are not completed.} However, the full dependence on $m_e$ is not phenomenologically relevant.
In fact, the physical problem exhibits a well defined mass hierarchy. The mass of the
electron is always very small compared to the other kinematic invariants and can be
safely neglected everywhere, with the exception of the terms in which it acts as a
collinear regulator. The ratio of the photonic NNLO corrections to the Born cross
section is given by 
%
\be
\frac{{\rm d} \sigma^{(2,\mbox{\tiny{PH}})}}{{\rm d} \sigma^{(\mbox{\tiny{Born}})}}  =
\left( \frac{\alpha}{\pi} \right)^2 \sum_{i=0}^{2} \delta^{(\mbox{\tiny{PH}},i)} 
\left(L_e \right)^i +
O \left(\frac{m_e^2}{s}, \frac{m_e^2}{t}\right) \, ,
\label{exp-photonic}
\ee
where $L_e=\ln{(s/m_e^2)}$ and the coefficients $\delta^{(\mbox{\tiny{PH}},i)}$
contain infrared logarithms and are functions of the scattering angle $\theta$. The approximation given by
Eq.~(\ref{exp-photonic}) is sufficient for a phenomenological description of the
process.\footnote{It can be shown that the terms suppressed by a positive power of
$m_e^2/s$ do not play any phenomenological role already at very low c.m. energies,
$\sqrt{s} \sim 10$ MeV. Moreover, the terms $m_e^2/t$ (or $m_e^2/u$) become important
in the extremely forward (backward) region, unreachable for the experimental setup.}
%
The coefficients of the double and single collinear logarithm in
Eq.~(\ref{exp-photonic}), $\delta^{(\mbox{\tiny{PH}},2)}$ and
$\delta^{(\mbox{\tiny{PH}},1)}$,  were obtained in \cite{Arbuzov:1998du,Glover:2001ev}. However, the
precision required for luminosity  measurements at $e^+ e^-$ colliders demands the
calculation of the  non-logarithmic coefficient, $\delta^{(\mbox{\tiny{PH}},0)}$.  The
latter  was  obtained in \cite{Penin:2005kf,Penin:2005eh} by reconstructing the  differential cross section
in the $s \gg m_e^2 \neq 0$ limit from the dimensionally regularised massless 
approximation \cite{Bern:2000ie}. The main idea of the method developed in \cite{Penin:2005kf,Penin:2005eh} is
outlined below: As far as the leading term in the small electron mass expansion is
considered, the difference between the massive and the dimensionally regularised
massless Bhabha scattering can be viewed as a difference between two regularisation
schemes for the infrared divergences. With the known massless two-loop result at hand,
the calculation of the massive one is reduced to constructing the {\it infrared
matching} term which relates the two above  mentioned regularisation schemes. To
perform the matching an auxiliary amplitude is constructed, which has the same
structure of the infrared singularities but is sufficiently simple to be evaluated at
least at the leading order in the small mass expansion. The particular form of the auxiliary
amplitude is dictated by the general theory of infrared singularities in QED and
involves the exponent of the one-loop correction as well as the two-loop corrections to
the logarithm of the electron form factor. The difference between the full and
the auxiliary amplitudes is infrared finite.  It can be evaluated by using dimensional
regularisation for each amplitude and then taking the limit of four space-time
dimensions. The infrared divergences, which induce the asymptotic dependence of the
virtual corrections on the electron and photon masses, are absorbed into the auxiliary
amplitude while the technically most nontrivial calculation of the full amplitude  is
performed in the massless approximation. The matching of the massive and massless
results is then necessary only for the auxiliary amplitude and is straightforward.
Thus the two-loop massless result for the scattering amplitude along  with the two-loop
massive electron form factor \cite{Burgers:1985qg} are sufficient to obtain the two-loop photonic
correction to the differential cross section in the small electron mass limit.

A method based on a similar principle was subsequently developed in
\cite{Mitov:2006xs,Becher:2007cu}; the authors of \cite{Becher:2007cu} confirmed  the result of 
\cite{Penin:2005kf,Penin:2005eh} for the NNLO photonic corrections to the Bhabha  scattering differential cross section.  \\

%
%
%

\noindent {\em{Electron loop corrections \label{ElLoop}}} \\
\vspace*{-2mm}

\noindent The NNLO electron loop corrections arise from the interference of two-loop 
Feynman diagrams with the tree-level amplitude as well as from the interference of
one-loop diagrams, as long as one of the diagrams contributing to each term involves a
closed electron loop. This set of corrections presents a single two-loop box
topology and is therefore technically less challenging to evaluate  with respect to
the photonic correction set.  The calculation of the electron loop corrections was
completed a few years ago \cite{Bonciani:2003cj0,Bonciani:2004gi,Bonciani:2004qt,Actis:2007gi}; the final result retains the full
dependence of the differential cross section on the electron mass $m_e$.  The MIs
involved in the calculation were identified by means of the Laporta algorithm and
evaluated with the differential equation method. As expected, after UV renormalisation
the differential cross section contained only  residual IR  poles which were removed by
adding the contribution of the  soft photon emission diagrams.  The resulting NNLO
differential cross section could be conveniently written in terms of  1- and
2-dimensional Harmonic Polylogarithms (HPLs) of maximum weight three. Expanding the
cross section in the limit $s, |t| \gg m_e^2$, the ratio of the NNLO corrections to the
Born cross section can be  written as in Eq.~(\ref{exp-photonic}):
\be
\frac{{\rm d} \sigma^{(2,\mbox{\tiny{EL}})}}{{\rm d} \sigma^{(\mbox{\tiny{Born}})}}  =
\left( \frac{\alpha}{\pi} \right)^2 \sum_{i=0}^{3} \delta^{(\mbox{\tiny{EL},i})} 
\left(L_e\right)^i +
O \left(\frac{m_e^2}{s}, \frac{m_e^2}{t}\right) \, .
\label{exp-electron}
\ee
Note that the series now contains a cubic collinear logarithm. This logarithm
appears, with an opposite sign, in the corrections due to the production of an
electron-positron pair (the soft-pair production was considered in
\cite{Arbuzov:1995vj}). When the two contributions are considered together in the full
NNLO, the cubic collinear logarithms cancel. Therefore, the physical cross section includes 
at most a double logarithm, as in Eq.~(\ref{exp-photonic}).

The explicit expression of all the coefficients $\delta^{(\mbox{\tiny{EL},i})}$,
obtained by expanding the results of \cite{Bonciani:2003cj0,Bonciani:2004gi,Bonciani:2004qt}, was confirmed by two different
groups \cite{Becher:2007cu,Actis:2007gi}. In \cite{Becher:2007cu} the small electron mass expansion was
performed within the soft-collinear effective theory (SCET)  framework, while  the
analysis in \cite{Actis:2007gi} employed  the asymptotic expansion of the  MIs.  \\

\noindent {\em{Heavy-flavor and hadronic corrections \label{HFLoop}}} \\
\vspace*{-2mm}

\noindent Finally, we consider the corrections originating from two-loop Feynman
diagrams involving a heavy flavour fermion loop.\footnote{Here by ``heavy flavour'' we
mean a muon or a $\tau$-lepton, as well as a heavy quark, like the top, the $b$- or the
$c$-quark, depending on the c.m. energy range that we are considering.} Since this set
of corrections  involves one more mass scale with respect to the corrections analysed
in the previous sections, a direct diagrammatic calculation is in principle a more
challenging task.
Recently, in \cite{Becher:2007cu} the authors applied their technique based on SCET to Bhabha
scattering and obtained the heavy flavour NNLO corrections in the limit in which $s,
|t|, |u| \gg m_f^2 \gg m_e^2$, where $m_f^2$ is the mass of the heavy fermion running
in the loop. Their result was very soon confirmed in \cite{Actis:2007gi} by means of a method
based on the asymptotic expansion of Mellin-Barnes representations of the MIs involved in the calculation.
However, the results obtained in the approximation $s, |t|, |u| \gg m_f^2 \gg m_e^2$
cannot be applied to the case in which $\sqrt{s} < m_f$ (as in the case of a tau
loop at $\sqrt{s} \sim 1$ GeV), and they apply only to a relatively narrow angular
region perpendicular to the beam direction when $\sqrt{s}$ is not very much larger than
$m_f$ (as in the case of top-quark loops at the ILC). It was therefore necessary to
calculate the heavy flavour corrections to Bhabha scattering assuming only that the
electron mass is much smaller than the other scales in the process, but retaining the
full dependence on  the heavy mass, $s, |t|, |u|, m_f^2 \gg m_e^2$.

The calculation was carried out in two different ways: in \cite{Bonciani:2007eh,Bonciani:2008ep} it was
done analytically, while in \cite{Actis:2007,Actis:2008br} it was done numerically with dispersion relations.


The technical problem of the diagrammatic calculation of Feynman integrals with four
scales can be simplified by considering carefully, once more, the structure of the
collinear singularities of the  heavy-flavour  corrections. The ratio of the NNLO heavy
flavour corrections to the Born cross section is given by
\be \label{exp-heavyflavor}
\frac{{\rm d} \sigma^{(2,\mbox{\tiny{HF}})}}{{\rm d} \sigma^{(\mbox{\tiny{Born}})}}  =
\left( \frac{\alpha}{\pi} \right)^2 \sum_{i=0}^{1} \delta^{(\mbox{\tiny{HF},i})} 
\left(L_e\right)^i +
O \left(\frac{m_e^2}{s}, \frac{m_e^2}{t}\right) \, ,
\ee
where now the coefficients $\delta^{(i)}$ are functions of the scattering angle
$\theta$ and, in general, of the mass of the heavy fermions involved in the virtual
corrections.
It is possible to prove that, in a physical gauge, all the collinear singularities
factorise and can be absorbed in the external field renormalisation \cite{Frenkel:1976bj}. This
observation has two consequences in the case at hand. The first one is that box
diagrams are free of collinear divergences in a physical gauge; since the sum of all
boxes forms a gauge independent block, it can be concluded that the sum of all box
diagrams is free of collinear divergences in any gauge. The second consequence is that
the single collinear logarithm in Eq.~(\ref{exp-heavyflavor}) arises from vertex
corrections only. Moreover, if one chooses on-shell UV renormalisation conditions, the
irreducible two-loop vertex graphs are free of collinear singularities.
%
Therefore, among all the two-loop diagrams contributing to the NNLO heavy flavour
corrections to Bhabha scattering, only the reducible vertex corrections are
logarithmically divergent in the $m_e \to 0$ limit.\footnote{Additional collinear
logarithms arise also from the interference of one-loop diagrams in which at least one
vertex is present.} The latter are easily evaluated even if they depend on two
different masses.
By exploiting these two facts, one can obtain the NNLO heavy-flavour corrections to the
Bhabha scattering differential cross section assuming only that $s, |t|, |u|, m_f^2 \gg
m_e^2$. In particular, one can set $m_e =0$ from the 
beginning in all the two-loop diagrams
with the exception of the reducible ones. This procedure allows one  to effectively
eliminate one mass scale from the two-loop boxes, so that these graphs can be  evaluated
with the techniques already employed in the diagrammatic calculation of the electron
loop corrections.\footnote{{The necessary MIs can be found in \cite{Bonciani:2008ep,Fleischer:1998nb,Aglietti:2003yc,Aglietti:2004tq}.}}
In the case in which the heavy flavour fermion is a quark, it is straightforward to
modify the calculation of the two-loop self-energy diagrams to obtain the mixed QED-QCD
corrections to Bhabha scattering \cite{Bonciani:2008ep}.


An alternative approach to the calculation of the heavy flavour corrections to
Bhabha scattering is based on dispersion relations. 
This method also applies to hadronic corrections.
The  hadronic and heavy fermion corrections to the Bhabha-scattering cross section
can be obtained by appropriately 
inserting the renormalised irreducible photon
vacuum-polarisation function  $\Pi$  in the photon propagator:
\begin{equation}
\label{1stReplace}
\frac{g_{\mu\nu}}{q^2+i\, \delta} \, \to \,
\frac{g_{\mu\alpha}}{q^2+i\, \delta} \,
\left( q^2\, g^{\alpha\beta} - q^\alpha\, q^\beta \right) \,\Pi(q^2)\,
\frac{g_{\beta\nu}}{q^2+i\, \delta}.
\end{equation}
The vacuum polarisation $\Pi$ 
can be represented by a once-subtracted dispersion integral \cite{Cabibbo:1961sz},
\begin{equation}
\label{DispInt}
\Pi(q^2) =
- \frac{q^2}{\pi} \, 
  \int_{4 M^2}^{\infty} \, {\rm d} z \, 
  \frac{\textrm{Im} \, \Pi(z)}{z} \, 
  \frac{1}{q^2-z+i\, \delta} .
\end{equation}
The contributions to $\Pi$ may then be determined from a (properly normalised) production cross 
section by the optical theorem~\cite{Cutkosky:1960sp},
\begin{eqnarray}
\label{Rhad0}
\textrm{Im} \, \Pi_{\rm had}(z)&=& 
- \frac{\alpha}{3} \, R(z).
\end{eqnarray}
In this way, the hadronic vacuum polarisation may be obtained 
from the experimental data for  $R$: 
\begin{eqnarray}
\label{Rhad}
R(z) &=& 
\frac{\sigma^0_{\rm had}(z)}
     {(4 \pi \alpha^2)\slash (3z)} \, ,
\end{eqnarray}
where $\sigma^0_{\rm had}(z)\equiv \sigma(\{e^+e^-\to\gamma^\star\to \textrm{hadrons}\};z)$.
In the low-energy region the inclusive experimental data may be used \cite{Davier:2002dy,rintpl:2008AA}.
Around a narrow hadronic resonance with mass $M_{\rm res}$ and width $\Gamma^{e^+ e^-}_{\rm res}$ one may use the relation
\begin{equation}
R_{\rm res}(z)= \frac{9 \pi}{\alpha^2} M_{\rm res} \Gamma^{e^+ e^-}_{\rm res}
\delta(z-M^2_{\rm res}),
\end{equation}
and in the remaining regions the perturbative QCD prediction \cite{Harlander:2002ur}.
Contributions to $\Pi$ arising from leptons and heavy  quarks with mass $m_f$, charge $Q_f$ and colour $C_f$ can be 
computed directly in perturbation theory. In the lowest order it reads
\begin{eqnarray}
R_f(z;m_f)&=&
 Q_f^2\, C_f \,\left(1+2\, \frac{m_f^2}{z}\right)\,
        \sqrt{1-4\,\frac{m_f^2}{z}}.
\end{eqnarray}
As a result of the above formulas, the massless photon propagator gets replaced by a massive 
propagator, whose effective mass $z$ is subsequently integrated over:
\begin{equation}
\label{PropReplace}
\frac{g_{\mu\nu}}{q^2+i \delta}
\to
\frac{\alpha}{3 \pi}
\int_{4 M^2}^{\infty} 
\frac{{\rm d}z~R_{\rm tot}(z)}{z(q^2-z+i \delta)}
\left(
g_{\mu\nu} - \frac{q_\mu q_\nu}{q^2+i \delta}
\right)\, ,
\end{equation}
where $R_{\rm tot}(z)$ contains hadronic and leptonic contributions.

For self-energy corrections to Bhabha scattering at one-loop order,  the dispersion relation approach  was first employed 
in \cite{Berends:1976zn}.
Two-loop applications of this technique, prior to Bhabha scattering, are the evaluation of the had\-ronic 
vertex correction \cite{Kniehl:1988id} and of two-loop had\-ronic corrections to the lifetime of the muon \cite{vanRitbergen:1998hn}. The approach was also applied to the evaluation of the two-loop form factors in QED in \cite{Barbieri:1972as,Barbieri:1972hn,Mastrolia:2003yz}.

The fermionic and hadronic corrections to Bhabha scattering at one-loop accuracy come only from  the self-energy diagram; see for details Section~\ref{sec:4}.
At two-loop level  there are  
reducible and  irreducible  self-energy contributions, vertices and boxes.
The reducible corrections are easily treat\-ed. 
For the evaluation of the irreducible two-loop diagrams, it is advantageous that they are one-loop diagrams with self-energy insertions because the application of the dispersion technique as described here is possible.

The kernel function for the irreducible two-loop vertex was derived in \cite{Kniehl:1988id} and verified e.g. in \cite{Actis:2008br}. The three kernel functions for the two-loop box functions were first obtained in \cite{Actis:2007pn,Actis:2007,Actis:2008br} and verified in \cite{Kuhn:2008zs}. 
A complete collection of all the relevant formulae may be found  in \cite{Actis:2008br}, and the corresponding Fortran code bhbhnnlohf is publicly available 
at the web page \cite{ACGR:bhbhnnlohf} \\ 
{\tt www-zeuthen.desy.de/theory/research/bhabha/} .

In \cite{Actis:2008br}, the dependence of the various heavy fermion NNLO corrections on $\ln(s/m_f^2)$ for $s,|t|,|u| \gg m_f^2$ was studied. The irreducible vertex behaves (before a combination with real pair emission terms) like $\ln^3(s/m_f^2)$ \cite{Kniehl:1988id}, while the sum of the various infrared divergent diagrams as a whole behaves like $\ln(s/m_f^2)\ln(s/m_e^2)$.
This is in accordance with 
Eq.~(\ref{exp-heavyflavor}), but the limit plays no effective role at the energies studied here.

\begin{figure}[t]
\begin{center}
\vspace*{7mm}
\begin{center}
\scalebox{0.33}{\includegraphics{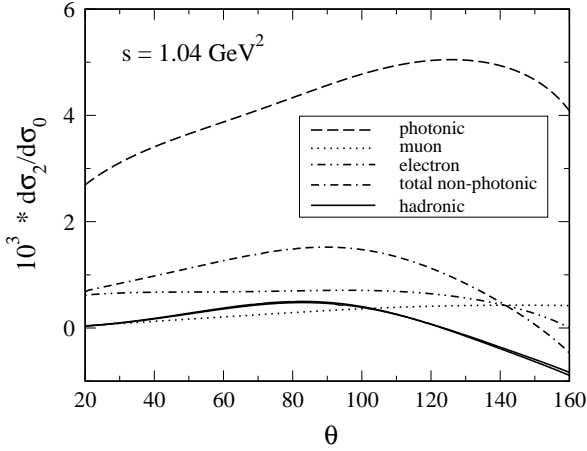}}
\end{center}
\caption[]
{Two-loop photonic and non-photonic corrections to Bhabha scattering at $\sqrt{s}=1.02$ GeV, normalised
to the QED tree-level cross section, as a function of the electron polar angle; no cuts; the parameterisations of $R_{\rm had}$ from \cite{Burkhardt:1981jk} and \cite{Davier:2002dy,rintpl:2008AA,Harlander:2002ur} are very close to each other. }
\label{1gevnoSE}
\end{center}
\end{figure}

\begin{figure}[t]
\begin{center}
\vspace*{7mm}
\begin{center}
\scalebox{0.33}{\includegraphics{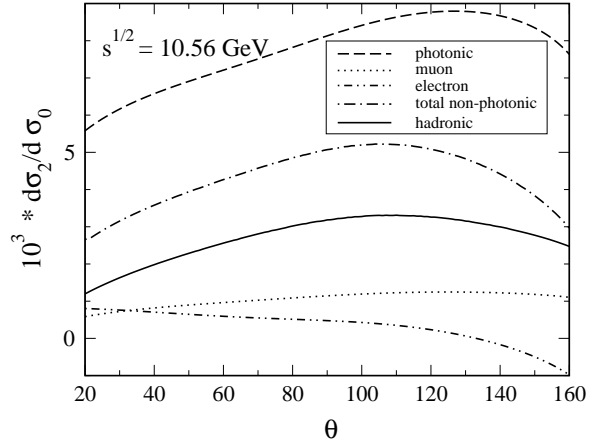}}
\end{center}
\caption[]
{Two-loop photonic and non-photonic corrections to Bhabha scattering 
at $\sqrt{s}=10.56$ GeV, normalised
to the QED tree-level cross section, 
as a function of the electron polar angle; no cuts; the parameterisations of $R_{\rm had}$ is from \cite{Burkhardt:1981jk}.}
\label{10gevnoSE}
\end{center}
\end{figure}

As a result of the efforts of recent years we now have at least two completely independent calculations for all the non-photonic virtual two-loop contributions. 
The net result, as a ratio of the NNLO corrections to the Born cross section in per mill, is shown in 
Fig.~\ref{1gevnoSE} for KLOE and in Fig.~\ref{10gevnoSE} 
for BaBar/Belle.\footnote{The pure self-energy corrections deserve a special 
discussion and are thus omitted in the plots.}
While the non-photonic corrections stay at one per mill or less for  KLOE, they reach 
a few per mill at the BaBar/Belle energy range. The NNLO 
photonic corrections are the dominant contributions and amount 
to some per mill, both at $\mathrm{\phi}$ and $B$ factories. However, as already emphasised, 
the bulk of both photonic and non-photonic 
corrections is incorporated into the generators used by the experimental collaborations. Hence, 
the consistent comparison between the results of NNLO calculations and the MC predictions 
at the same perturbative level enables one to assess the theoretical accuracy of the luminosity tools, 
as will be discussed quantitatively in Section \ref{TH}.

\subsubsection{Fixed-order calculation of the hard photon emission at one loop
\label{Penta}}

The one-loop matrix element for the process $e^{+}e^{-} \to e^{+}e^{-} \gamma$ is one
of the contributions to the complete set of NNLO corrections to Bhabha scattering. Its
evaluation requires the nontrivial computation of one-loop tensor integrals associated
with pentagon diagrams.

According to the standard Passarino-Veltman (PV) approach \cite{Passarino:1978jh},
one-loop tensor integrals can be expressed in terms of 
MIs with trivial numerators that are independent of the loop variable,
each multiplied by a Lorentz structure depending only on combinations of the external
momenta and the metric tensor.
The achievement of the complete PV-reduction amounts to solving a nontrivial system of
equations. 
Due to its size, it is reasonable to replace the analytic techniques by
numerical tools.
It is difficult to implement the PV-reduction numerically, since it gives rise to Gram
determinants. The latter naturally arise in the procedure of inverting a system and they 
can vanish at special phase space points. This fact requires a proper modification of the 
reduction algorithm \cite{Bern:1996je,Dixon:1996wi,Binoth:1999sp,Denner:2002ii,Binoth:2005ff,Denner:2005nn,Denner:2006fy}. A viable solution for the complete algebraic reduction of 
tensor-pentagon (and tensor-hexagon) integrals was formulated in \cite{Davydychev:1991va,Tarasov:1996br,Fleischer:1999hq}, by exploiting the algebra of signed minors \cite{Melrose:1965kb}.
In this approach the cancellation of powers of inverse Gram determinants was performed recently
in \cite{Diakonidis:2008dt,Diakonidis:2008ij}.


Alternatively, the computation of the one-loop five-point amplitude $e^{+}e^{-} \to e^{+}e^{-} \gamma$
can be performed by using generalised-unitarity cutting rules (see
\cite{Bern:2007dw} for a detailed compilation of references). In the following we
propose two ways to achieve the result, via an analytical and via a 
semi-numerical method. The application of generalised cutting rules as an on-shell
method of calculation is based on two fundamental properties of scattering amplitudes:
{\it i)} analyticity, according to which any amplitude is determined by its
singularity structure \cite{Landau:1959fi,Mandelstam:1958xc,Mandelstam:1959bc,Cutkosky:1960sp,Eden:1966}; and {\it ii)} unitarity, according to which
the residues at the singularities are determined by products of simpler amplitudes. 
Turning these properties into a tool for computing scattering amplitudes is possible
because of the underlying representation of the amplitude in terms of Feynman integrals
and their PV-reduction, which grants the existence of a representation of any one-loop
amplitudes as linear combination of MIs, each multiplied by a rational coefficient. In
the case of $e^{+}e^{-} \to e^{+}e^{-} \gamma$, pentagon-integrals 
may be
expressed, through PV-reduction, by a linear combination of 17 MIs (including 3 boxes,
8 triangles, 5 bubbles and 1 tadpole).
 Since the required MIs are analytically known
\cite{'tHooft:1978xw,Bern:1992em,Bern:1993kr,Tarasov:1996br,Binoth:1999sp,Duplancic:2003tv,Ellis:2007qk},
the determination of their coefficients is needed for reconstructing the amplitude
as a whole.
Matching the generalised cuts of the amplitude with the cuts
of the MIs provides an efficient way to extract their (rational) coefficients from the amplitude itself. 
In general the fulfilment of multiple-cut conditions requires loop momenta with
complex components. The effect of the cut conditions is to freeze some or all of its
components, depending on the number of the cuts. With the
quadruple-cut \cite{Britto:2004nc} the loop momentum is completely frozen, yielding
the algebraic determination of the coefficients of $n$-point functions with $n\ge4$. In
cases where fewer than four denominators are cut, like triple-cut
\cite{MastroliaTriple,FordeTriBub,BjerrumBohr:2007vu}, double-cut
\cite{Britto:2005ha,Britto:2006sj,Anastasiou:2006jv,Anastasiou:2006gt,Britto:2006fc,FordeTriBub} and single-cut
\cite{NigelGlover:2008ur}, the loop momentum is not frozen: the free components are
left over as phase-space integration variables. 

For each multiple-cut, the
evaluation of the phase-space integral would generate, in general, logarithms and a
non-logarithmic term. The coefficient of a given $n$-point MI finally appears in the
non-logarithmic term of the corresponding $n$-particle cut, where all the internal lines
are on-shell (while the logarithms correspond to the cuts of higher-point MIs which
share that same cut). Therefore all the coefficients of MIs can be determined in a
{\it top-down} algorithm, starting from the quadruple-cuts for the extraction of the
four-point coefficients, and following with the triple-, double- and single-cuts for the
coefficients of three-, two- and one-point, respectively. 
The coefficient of an $n$-point MI ($n \ge 2$) can also be obtained by specialising the generating formulas given in \cite{Britto:2008vq} for general one-loop amplitudes to the case at hands.


Instead of the analytic evaluation of the multiple-cut phase-space integrals, it is
worth considering the feasibility of computing the process $e^{+}e^{-} \to e^{+}e^{-}
\gamma$ with a semi-numerical technique by now known as OPP-reduction
\cite{Ossola:2006us,Ossola:2007bb}, based on the decomposition of the numerator of any
one-loop integrand in terms of its denominators \cite{delAguila:2004nf,Pittau:2004bc,Pittau:1996ez,Pittau:1997mv}. Within this approach
the coefficients of the MIs can be found simply by solving a system of numerical
equations, avoiding any explicit integration. The OPP-reduction algorithm
exploits the polynomial structures of the integrand when evaluated at values of the
loop-mo\-men\-tum fulfilling multiple cut-conditions: {\it i)} for each $n$-point MI one
considers the $n$-particle cut obtained by setting all the propagating lines on-shell;
{\it ii)} such a cut is associated with a polynomial in terms of the free components of
the loop-momentum, which corresponds to the numerator of the integrand evaluated at the
solution of the on-shell conditions; {\it iii)} the constant-term of that polynomial
{\it is} the coefficient of the MI. \\ Hence the difficult task of evaluating one-loop
Feynman integrals is reduced to the much simpler problem of polynomial fitting,
recently optimised by using a projection technique based on the Discrete Fourier
Transform \cite{Mastrolia:2008jb}.

In general the result of a dimensional-regulated amplitude in the 4-dimensional limit,
with $D$ ($=4-2\epsilon$) the regulating parameter, is expected to contain
(poly)logarithms, often referred to as the cut-constructible term, and a pure 
rational term. In a later paper \cite{Ossola:2008xq}, which completed the OPP-method, the rising of the rational term was attributed to two potential
sources (of UV-divergent integrals): one, defined as $R_1$, due to the $D$-dimensional
completion of the 4-dimensional contribution of the numerator; a second one, called
$R_2$, due to the $(-2\epsilon)$-dimensional algebra of Dirac-matrices. Therefore
in the OPP-approach the calculation of the one-loop amplitude $e^{+}e^{-} \to
e^{+}e^{-} \gamma$ can proceed through two computational stages:
\begin{enumerate}
\item
the coefficients of the MIs that are responsible both for the cut-constructible and
for the $R_1$-rational terms can be determined by applying the OPP-reduction
discussed above \cite{Ossola:2006us,Ossola:2007bb,Mastrolia:2008jb};
\item 
the $R_2$-rational term can be computed by using additional tree-level-like
diagrammatic rules, very much resembling the computation of the counter terms needed for
the renormalisation of UV-divergences \cite{Ossola:2008xq}.
\end{enumerate}

The numerical influence of the radiative loop diagrams, including the pentagon diagrams, is expected 
not to be  
particularly large. However, the calculation of such corrections would greatly help to 
assess the physical precision of existing luminosity programs.\footnote{As already remarked, the exact calculation of one-loop
corrections to hard photon emission in Bhabha scattering became available~\cite{Actis:2009uq}
 during the completion of the report, exactly according to the
methods described in the present Section.}


\subsubsection{Pair corrections \label{4pfinal}}
As was mentioned  in the paragraph on virtual heavy fla\-vour and hadronic corrections of Section~\ref{virtual}, these virtual corrections have to be combined with real corrections in order to get physically sensible results.
The virtual NNLO electron, muon, tau and pion corrections have to be combined with the emission of real electron, muon, tau and pion pairs, respectively.
The real pair production cross sections are finite, but cut dependent. 
We consider here the pion pair production as it is the dominant part of the hadronic corrections and 
can serve as an estimate of the role of the whole set of hadronic corrections. The description of all relevant hadronic contributions is a much more involved task and will not be covered in this review. 
As was first explicitly shown for Bhabha scattering in \cite{Arbuzov:1995vj} for electron pairs, and also discussed in \cite{Actis:2008br}, there appear exact cancellations of terms of the order $\ln^3(s/m_e^2)$ or $\ln^3(s/m_f^2)$, so that the leading terms are at most of order $\ln^2(s/m_e^2), \ln^2(s/m_f^2)$.

\begin{table}[t]
\caption{\label{tab-pairs}%
The NNLO lepton and pion pair corrections to the Bhabha scattering Born cross section $\sigma_B$:
 virtual corrections $\sigma_v$ , soft and hard real photon emissions $\sigma_s, \sigma_h$, and pair emission contributions $\sigma_{pairs}$. The total pair correction cross sections are obtained from the sum $\sigma_{s+v+h} + 
\sigma_{pairs}$. All cross sections, 
according to the cuts given in the text, are given in nanobarns.}
\begin{center}
\begin{tabular}{cllllll}
\hline 
\multicolumn{6}{c}{\bf{Electron pair corrections}}
\\
       &$\sigma_{B}$ & $\sigma_{h}$ & $\sigma_{v+s}$ & $\sigma_{v+s+h}$ & $\sigma_{pairs}$ 
\\
\hline
  KLOE & 529.469 & 9.502  &  -11.567 & -2.065 & 0.271 
\\
BaBar & 6.744 & 0.246 & -0.271 & -0.025 & 0.017 
\\ \hline
\multicolumn{6}{c}{
\bf{Muon pair corrections}} 
\\
& $\sigma_{B}$ & $\sigma_{h}$ & $\sigma_{v+s}$ & $\sigma_{v+s+h}$ & $\sigma_{pairs}$ 
\\
\hline
KLOE & 529.469 & 1.494  & -1.736  & -0.241 &  --
\\
BaBar & 6.744  & 0.091 & -0.095 & -0.004 & 0.0005 \\
\hline
\multicolumn{6}{c}{
\bf{Tau pair corrections}} 
\\
& $\sigma_{B}$ & $\sigma_{h}$ & $\sigma_{v+s}$ & $\sigma_{v+s+h}$ & $\sigma_{pairs}$ 
\\
\hline
KLOE&  529.469  & 0.020  & -0.023  & -0.003 & --
\\
BaBar &  6.744 & 0.016 & -0.017 & -0.0007 & $< 10^{-7}$ \\
\hline
\multicolumn{6}{c}{
\bf{Pion pair corrections}} 
\\
& $\sigma_{B}$ & $\sigma_{h}$ & $\sigma_{v+s}$ & $\sigma_{v+s+h}$ & $\sigma_{pairs}$ 
\\
\hline
KLOE & 529.469  & 1.174 &  -1.360 & -0.186 & --
\\
BaBar & 6.744  & 0.062  & -0.065  & -0.003 & 0.00003
\\
\hline
\end{tabular}
\end{center}
\end{table}

In Table~\ref{tab-pairs} we show NNLO 
lepton and pion pair contributions with typical kinematical cuts for the KLOE and BaBar experiments.
Besides contributions from unresolved pair emissions $\sigma_{pairs}$,  we also add unresolved real hard photon emission contributions $\sigma_{h}$.
The corrections  $\sigma_{pairs}$ from fermions have been calculated with the Fortran 
package HELAC-PHEGAS
 \cite{Kanaki:2000ey,Papadopoulos:2000tt,Papadopoulos:2005ky,Cafarella:2007pc}, 
the real pion corrections with EKHARA 
\cite{Czyz:2005ab,Czyz:2006dm},  
the NNLO hard photonic corrections  $\sigma_{h}$ with a program \cite{chunp} based 
on the  generator  BHAGEN-1PH \cite{Caffo:1996mi}.
The latter depend, technically, on the soft photon cut-off $E_{\gamma}^{\min} = \omega$.
After adding up with  $\sigma_{v+s}$, the sum of the two $\sigma_{v+s+h}$ is independent of that; in fact here we use $\omega/E_{\rm beam} = 10^{-4}$.
In order to cover also pion pair corrections $\sigma_{v+s}$ is determined with an updated version of the Fortran package bhbhnnlohf \cite{Actis:2008br,ACGR:bhbhnnlohf}. 
The cuts 
applied in Table~\ref{tab-pairs} for the  KLOE experiment are
\begin{itemize}
\item $\sqrt{s} = 1.02 ~\textnormal{GeV}$\,,
\item  $E_{\rm min} = 0.4 ~\textnormal{GeV}$\,, 
\item $55^{\circ} < \theta_{\pm}<125^{\circ}$\,,
\item  $\xi_{\rm max} = 9^{\circ}$\,,
\end{itemize}
and for the BaBar experiment
\begin{itemize}
\item $\sqrt{s} = 10.56 ~\textnormal{GeV}$\,,
\item  $|\cos(\theta_{\pm})|<0.7 
~~\textnormal{and} 
\\
|\cos(\theta_{+})|<0.65 
~~\textnormal{or} ~~|\cos(\theta_{-})|<0.65$\,,  
\item $|\vec{p}_{+}|/E_{\rm beam} > 0.75 ~~\textnormal{and}~~ 
|\vec{p}_{-}|/E_{\rm beam} > 0.5 ~~\textnormal{or}$ \\
$|\vec{p}_{-}|/E_{\rm beam} > 0.75 ~~\textnormal{and}~~ 
|\vec{p}_{+}|/E_{\rm beam} > 0.5$\,, 
\item $\xi_{\rm max}^{3d} =30^{\circ}$\,.
\end{itemize}
Here $E_{\rm min}$  is the energy threshold for the final-state electron/positron, 
$\theta_{\pm}$ are the electron/positron polar angles and $\xi_{max}$ is
the maximum allowed polar angle acollinearity:
\begin{eqnarray}
\xi = \left|\theta_{+}+\theta_{-} -180^{\circ}\right| ,
\end{eqnarray}
and $\xi_{\rm max}^{3d}$ is the 
maximum allowed three dimensional a\-col\-linea\-ri\-ty: 
\begin{eqnarray}
\xi^{3d} = \left|\arccos\left(\frac{p_{+}\cdot p_{-}}{(|\vec{p}_{-}|
|\vec{p}_{+}|}\right) \times \frac{180^{\circ}}{\pi}  -180^{\circ}\right|.
\end{eqnarray}
For $e^{+}e^{-}\rightarrow e^{+}e^{-}\mu^{+}\mu^{-}$,  
cuts are applied only to the $e^{+}e^{-}$ pair.
In the case of  $e^{+}e^{-}\rightarrow 
e^{+}e^{-}e^{+}e^{-}$, all possible $e^{\pm}e^{\mp}$ combinations are 
checked and if at least one pair fulfils the cuts the 
event is accepted. 

At KLOE the electron pair corrections contribute about $3\times 10^{-3}$ and at BaBar
about $1\times 10^{-3}$, while all the other contributions of pair production are even smaller. Like in 
small-angle Bhabha scattering at LEP/SLC the pair corrections \cite{Montagna:1998vb} 
are largely dominated by the electron pair contribution.

\subsection{Multiple photon effects and matching with NLO corrections 
\label{MP}}

\subsubsection{Universal methods for leading logarithmic corrections 
\label{HO}}

From inspection of Eqs.~(\ref{bsv}) and (\ref{bsvgg}) for the SV NLO QED 
corrections to the cross section of the Bhabha scattering and $e^+ e^- \to \gamma\gamma$
process, it can be seen that large logarithms $L=\ln(s/m_e^2)$, due to
collinear photon emission, are present. Similar large logarithmic terms arise after 
integration of the hard photon contributions from the kinematical domains of photon 
emission at small angles with respect to charged particles. For the  energy range of 
meson factories
 the logarithm is large numerically, i.e. $L \sim 15$ at the $\mathrm{\phi}$ factories and 
 $L \sim 20$ at the $B$ factories, and the corresponding terms give 
the bulk of the total radiative correction. These contributions represent also the dominant
part of the NNLO effects discussed in Section \ref{NNLO}. Therefore, to achieve the required theoretical accuracy, the logarithmically enhanced contributions due to emission of soft and collinear photons must be taken into account at all orders in perturbation theory. 
The methods for the calculation of higher-order (HO) QED corrections on the 
basis of the generators employed nowadays at flavour factories were already 
widely and successfully used in the 90s at LEP/SLC for electroweak tests of the SM.  
They were adopted for the calculation of both the small-angle
Bhabha scattering cross section (necessary for the high-precision luminosity
measurement) and $Z$-boson observables. Hence, the theory accounting for 
the control of HO QED corrections at meson factories can be considered 
particularly robust, having passed the very stringent tests of the LEP/SLC era. 
 
The most popular and standard methods to keep multiple photon effects under control are the QED Structure Function (SF)
approach \cite{Kuraev:1985hb,Altarelli:1986kq,Nicrosini:1986sm,Nicrosini:1987sw}
and Yennie-Frautschi-Suura (YFS) exponentiation \cite{Yennie:1961ad}. 
The former is used in all the versions of the generator BabaYaga 
\cite{CarloniCalame:2000pz,CarloniCalame:2003yt,Balossini:2006wc} and 
MCGPJ \cite{Arbuzov:2005pt}
(albeit according to different realisations), while the latter is the theoretical recipe adopted 
in BHWIDE \cite{Jadach:1995nk}. Actually, analytical QED SFs $D(x,Q^2)$, valid in the strictly collinear approximation, 
are implemented in MCGPJ, whereas BabaYaga is based on
a MC Parton Shower (PS) algorithm to reconstruct $D(x,Q^2)$ numerically.  \\

\noindent {\em{The Structure Function approach \label{SF}}} \\
\vspace*{-2mm}
 
Let us consider the annihilation process $e^- e^+ \to  X$, where $X$ is some given final
state and $\sigma_0 (s)$ its LO cross section. 
Initial-state (IS) QED radiative corrections can be described according to the following 
picture.  Before arriving at the annihilation point, the incoming electron (positron) of
four-momentum $p_{-(+)}$ radiates real and virtual photons. These 
photons, due to the dynamical features of QED, are mainly  radiated along the direction of
motion of the radiating particles, and their effect is mainly to reduce 
the  original four-momentum of the incoming electron (positron) to $ x_{1(2)} p_{-(+)}$. 
After this pre-emission,  the hard scattering process $e^- (x_1 p_-) e^+ (x_2 p_+) \to  X$
takes place, at a reduced  squared c.m. energy ${\hat s}  = x_1 x_2 s$. The 
resulting cross section, corrected for IS QED radiation, can be represented in the form~\cite{Kuraev:1985hb,Altarelli:1986kq,Nicrosini:1986sm}
\begin{equation}
\sigma (s) = \int_0^1 {\rm d} x_1 {\rm d} x_2  D(x_1, s) D(x_2, s) \sigma_0 (x_1 x_2 s) 
\Theta({\rm cuts}), 
\label{eq:masterd}
\end{equation}
where $D(x,s)$ is the electron SF, representing the probability that an
incoming electron (positron) radiates a collinear photon, retaining a fraction $x$ of its
original momentum at the energy scale $Q^2 = s$, and $\Theta({\rm cuts})$ 
stands for a rejection algorithm taking care of experimental cuts. When considering  photonic radiation
only the non-singlet part of the SF is of interest. If the running of the QED
coupling constant  is neglected, the non-singlet part of the SF is the solution 
of the following Renormalisation Group (RG) equation, analogous to the 
Dokshitzer-Gribov-Li\-pa\-tov-Altarelli-Parisi (DGLAP) equation of QCD 
\cite{Gribov:1972rt,Altarelli:1977zs,Dokshitzer:1977sg}:
\begin{equation}
s  {{\partial} \over {\partial s}} D(x,s) = {{\alpha} \over {2 \pi}} \int_x^1 {{{\rm d}z} \over {z}}
P_+(z) D \left( {{x} \over {z}} , s \right) ,  
\label{eq:apeqdiff}
\end{equation}
where $P_+(z) $ is the regularised Altarelli-Parisi (AP) splitting function for the process ${\rm electron} \to {\rm electron} + {\rm photon}$, given by 
\begin{eqnarray}
&& P_+(z) = P(z) - \delta (1-z) \int_0^1 {\rm d}x P(x) , \nonumber \\
&& P(z) = {{1 + z^2} \over {1-z}} . 
\label{eq:apvert}
\end{eqnarray}
Equation (\ref{eq:apeqdiff}) can be also transformed into an integral equation, subject to the
boundary  condition $D(x, m_e^2) = \delta(1-x)$: 
\begin{equation}
D(x,s)  = \delta (1-x) + {{\alpha} \over {2 \pi}} \int_{m_e^2}^{s} {{{\rm d} Q^2} \over {Q^2}}
\int_x^1  {{{\rm d}z} \over {z}} P_+(z) D \left( {{x} \over {z}} , Q^2  \right).  
\label{eq:apeq}
\end{equation}

Equation~(\ref{eq:apeq}) can be solved exactly by means of numerical methods, such as the inverse
Mellin transform method. However, this derivation of $D(x,s)$ turns out be problematic 
in view of phenomenological applications. Therefore, approximate (but very accurate)
analytical representations of the solution of the evolution equation are of major interest
for practical purposes. This type of solution was the one typically adopted in the context
of LEP/SLC phenomenology.  A first analytical solution  can be obtained in the 
soft photon approximation, i.e. in the limit $x \simeq 1$. This solution, also known as 
Gribov-Lipatov (GL) approximation, exponentiates the
large logarithmic contributions of infrared and collinear origin at all perturbative orders, but 
it does not take into account hard-photon (collinear) effects. This drawback can be 
overcome by solving the evolution equation iteratively. At the $n$-th step of the iteration, 
one obtains the $O (\alpha^n)$  contribution to the SF for any value
of $x$. By combining the GL  solution with the iterative one, in which 
the soft-photon part has been eliminated in order to avoid 
double counting, one can build a hybrid solution of the evolution equation. It exploits 
all the positive features of the two kinds of solutions and is not affected by the limitations 
intrinsic to each of them. Two classes of hybrid solutions, namely
the additive and factorised ones, are known in the literature, and both were adopted 
for applications to LEP/SLC precision physics. A typical additive solution, where
the GL approximation $D_{GL} (x,s)$ is supplemented by finite-order terms present in the
iterative solution, is given by \cite{Cacciari:1992pz}
\begin{eqnarray}
&& D_{A} (x,s)  = \sum_{i=0}^3 d_A^{(i)} (x,s) , \nonumber \\
&& d_A^{(0)} (x,s) = {{\exp \left[ {{{1} \over {2}} \beta \left( {{3} \over {4}} - \gamma_E \right)} \right] } 
\over { \Gamma \left( 1 + {{1} \over {2}} \beta \right) }} {{1} \over {2}} \beta (1 - x)^{{{1} \over {2}} \beta - 1}, \nonumber \\
&& d_A^{(1)} (x,s) = - {1 \over 4 }  \beta (1+x) , \nonumber \\
&& d_A^{(2)} (x,s) = {1 \over 32 } \beta^2 \left[  (1+x) \left( -4 \ln (1-x) + 3 \ln x 
\right)  \right. \nonumber \\
&& \qquad \qquad \left. - 4 {{\ln x } \over  {1-x}} - \, 5 - x  \right] ,  \nonumber \\ 
&&  d_A^{(3)} (x,s) = {1 \over 384 } \beta^3 \left\{ (1+x) \left[ 18 \zeta (2) - 6 \hbox{\rm Li}_2 (x) 
\right. \right.  \nonumber \\
&& \qquad \qquad \left. -12  \ln^2 (1-x) \right]  + {1 \over {1-x}} \left[ - {3 \over 2} (1 + 8 x + 3 x^2) \ln x 
\right. \nonumber \\
&& \qquad \qquad + {1 \over 2} (1 + 7 x^2 ) \ln^2 x -12 (1+ x^2) \ln x \ln (1-x) \nonumber \\
&& \qquad \qquad \left. \left. 
 - 6 (x+5) (1-x) \ln (1-x)  \right. \right.  \nonumber \\ 
&&  \qquad \qquad \left. \left. - {1 \over 4} (39 - 24 x - 15 x^2 ) \right] \right\} ,
\label{eq:strucfuna}
\end{eqnarray}
where $\Gamma$ is the Euler gamma-function, $\gamma_E 
\approx 0.5772$ the Euler-Mascheroni constant, 
$\zeta $ the Riemann $\zeta$-function and $\beta$ is the large collinear factor
\begin{eqnarray}
\beta = {{2 \alpha} \over {\pi}} \left[ \ln \left( {{s} \over {m_e^2}} \right) 
- 1 \right]  . 
\end{eqnarray}

Explicit examples of factorised solutions, which are obtained by multiplying the GL solution 
by finite-order terms in such a way that, order by order, the iterative contributions are exactly recovered,
can be found in \cite{Skrzypek:1992vk}. For the calculation of HO corrections with a per mill accuracy
analytical SFs in additive and factorised form containing up to $O (\alpha^3)$ 
finite-order terms are sufficient and in excellent agreement. They also agree 
with an accuracy much better than 0.1 with the exact numerical solution of the QED evolution equation. 
Explicit solutions 
up to the fifth order in $\alpha$ were calculated in~\cite{Przybycien:1992qe,Arbuzov:1999cq}. 


The RG method described above was applied in~\cite{Arbuzov:1997pj}
for the treatment of LL QED radiative corrections to various processes of 
interest for physics at meson factories. Such a formulation 
was later implemented in the generator MCGPJ. 
For example, according to~\cite{Arbuzov:1997pj}, the Bhabha scattering cross section, 
accounting for LL terms in all orders, $O (\alpha^nL^n),\ n=1,2,\ldots$,
of perturbation theory, is given by
\begin{eqnarray} \label{master}
&& \dd \sigma^{\mathrm{Bhabha}}_{\mathrm{LLA}} = \!\!\!\!
\sum\limits_{a,b,c,d=e^\pm,\gamma}^{} \int^{1}_{\bar{z}_1}\dd z_1 
\int^{1}_{\bar{z}_2} \dd z_2 
D^{\mathrm{str}}_{ae^-} (z_1) D^{\mathrm{str}}_{be^+} (z_2)
\nonumber \\ 
&& \quad \times \dd \sigma_{0}^{ab\to cd}(z_1,z_2)  
\int^{1}_{\bar{y}_1}\! \frac{\dd y_1}{Y_1}D^{\mathrm{frg}}_{e^-c} (\frac{y_1}{Y_1}) 
\int^{1}_{\bar{y}_2}\!\! \frac{\dd y_2}{Y_2}D^{\mathrm{frg}}_{e^+d} (\frac{y_2}{Y_2})
\nonumber \\ 
&& \quad + \,  O \left(\alpha^2L,\alpha\frac{m_e^2}{s}\right)\,.
\label{eq:sfmcgpj}
\end{eqnarray}
Here $\dd\sigma_{0}^{ab\to cd}(z_1,z_2)$ is the differential LO cross 
section of the process $ab\to cd$, with energy fractions of the incoming particles
being scaled by factors $z_1$ and $z_2$ with respect to the initial
electron and positron, respectively. In the notation of~\cite{Arbuzov:1997pj},
the electron SF $D^{\mathrm{str}}_{ab}(z)$ is distinguished from the 
electron fragmentation function $D^{\mathrm{frg}}_{ab}(z)$ to point out the 
role played by IS radiation (described by $D^{\mathrm{str}}_{ab}(z)$) 
with respect to the one due to final-state radiation (described by
$D^{\mathrm{frg}}_{ab}(z)$). However, because of their probabilistic 
meaning, the electron structure and fragmentation functions coincide. 
In Eq.~(\ref{eq:sfmcgpj}) the quantities $Y_{1,2}$ are the energy fractions of particles
$c$ and $d$ with respect to the beam energy. Explicit expressions for 
$Y_{1,2}=Y_{1,2}(z_1,z_2,\cos\theta)$ and other details on the kinematics can
be found in~\cite{Arbuzov:1997pj}.
The lower limits of the integrals, $\bar{z}_{1,2}$ and 
$\bar{y}_{1,2}$, should be defined according to the experimental conditions of
particle detection and kinematical constraints.  For the case of 
the $e^+e^-\to \gamma\gamma$ process 
one has to change the master formula~(\ref{master}) by picking up the two-photon final state.
Formally this can be done by just choosing the proper fragmentation functions,
$D^{\mathrm{frg}}_{\gamma c}$ and $D^{\mathrm{frg}}_{\gamma d}$.

The photonic part of the non-singlet electron structure (fragmentation) 
function in $O (\alpha^nL^n)$ considered in~\cite{Arbuzov:1997pj} reads
\begin{eqnarray} \nonumber 
&& D_{ee}^{NS,\gamma}(z)=\delta(1-z)+\sum_{i=1}^{n}\left(\frac{\alpha}{2\pi}(L-1)\right)^i 
\frac{1}{i!}\left[P^{(0)}_{ee}(z)\right]^{\otimes i} \!\!\!\! ,
\\ \nonumber
&& D_{\gamma e}(z)= \frac{\alpha}{2\pi}(L-1)P_{\gamma e}(z) + O (\alpha^2L^2),
\\ \nonumber
&& D_{e \gamma}(z)= \frac{\alpha}{2\pi}L P_{e\gamma}(z) + O (\alpha^2L^2),
\\ \nonumber
&& P^{(0)}_{ee}(z)=\left[\frac{1+z^2}{1-z}\right]_{+}
\\ \nonumber
&&\quad = \lim\limits_{\Delta \to 0}\left\{\delta(1-z)(2\ln\Delta
+ \frac{3}{2}) + \Theta(1-z-\Delta)\frac{1+z^2}{1-z}\right\}, 
\\ 
&& \left[P^{(0)}_{ee}(z)\right]^{\otimes i} = \int\limits_{z}^{1}\frac{\dd t}{t}
P^{(i-1)}_{ee}(t)P^{(0)}_{ee}
\left(\frac{z}{t}\right),
\\ \nonumber
&& P_{\gamma e}(z) = z^2 + (1-z)^2, \quad
P_{e\gamma}(z) = \frac{1+(1-z)^2}{z}\, . 
\end{eqnarray}
Starting from the second order in $\alpha$ there appear also non-singlet  and singlet
$e^+e^-$ pair contributions to the structure function:
\begin{eqnarray} 
&& D_{ee}^{NS,e^+e^-}(z) = \frac{1}{3}\left(\frac{\alpha}{2\pi}L\right)^2 P^{(1)}_{ee}(z)
+  O (\alpha^3L^3),
\nonumber \\ \nonumber 
&& D_{ee}^{S,e^+e^-}(z) = \frac{1}{2!}\left(\frac{\alpha}{2\pi}L\right)^2R(z)
+  O (\alpha^3L^3),
\\  \nonumber 
&& R(z) = P_{e \gamma}\otimes P_{\gamma e}(z)
= \frac{1-z}{3z}(4+7z+4z^2) \nonumber \\
&& \qquad \qquad \qquad \qquad \quad + 2(1+z)\ln z.
\end{eqnarray}
Note that radiation of a real pair, {\it i.e.} appearance of additional 
electrons and positrons in the final state, require the application of nontrivial 
conditions of experimental particle registration. Unambiguously, that can be done
only within a MC event generator based on 
four-particle matrix elements, 
as already discussed in Section \ref{NNLO}.

In the same way as in QCD, the LL cross sections depend on the choice of
the factorisation scale $Q^2$ in the argument of the large logarithm $L=\ln(Q^2/m_e^2)$, 
which is not fixed a priori by the theory. However, the scale should be taken of the order 
of the characteristic  energy transfer in the 
process under consideration. Typical choices are $Q^2=s$, $Q^2=-t$ and $Q^2=s t/u$.
The first one is good for annihilation channels like $e^+e^-\to \mu^+\mu^-$, the
second one is optimal for small-angle Bhabha scattering where the $t$-channel exchange
dominates, see~\cite{Arbuzov:2006mu}. 
The last choice allows to exponentiate the leading contribution due to initial-final state 
interference \cite{Greco:1990on} and is particularly suited for large-angle Bhabha scattering 
in QED. The option $Q^2=s t/u$ is adopted in all the versions of the generator 
BabaYaga. Reduction of the scale dependence can be achieved by taking into account
next-to-leading corrections in $O (\alpha^nL^{n-1})$, next-to-next-to-leading
ones in  $O (\alpha^nL^{n-2})$ {\it etc.} \\
 
 \noindent {\em{The Parton Shower algorithm \label{PS}}} \\
\vspace*{-2mm}

The PS  algorithm  is a method for providing a MC iterative
solution of the evolution equation and, at the same time, for generating the four-momenta of the
electron and photon at a given step of the iteration. It  was developed   
within the context of QCD 
and later applied in QED too.

In order to implement the algorithm, it
is first necessary to assume the existence of an upper limit for the energy fraction 
$x$ in such a way that the AP splitting function is regularised by writing
\begin{equation}
P_+(z) = \theta(x_+ - z) P(z) - \delta (1-z) \int_0^{x_+} {\rm d}x P(x) . 
\label{eq:apvertmod}
\end{equation}
Of course, in the limit $x_+ \to 1$, Eq.~(\ref{eq:apvertmod}) recovers the usual definition of
the AP splitting function given in Eq.~(\ref{eq:apvert}). By inserting the modified AP vertex into
Eq.~(\ref{eq:apeqdiff}), one obtains
\begin{eqnarray}
s {{\partial} \over {\partial s}}  D(x,s) = && {{\alpha} \over {2 \pi}} \int_x^{x_+} 
{{{\rm d}z} \over {z}} P(z) D \left( {{x} \over {z}} , s \right) \nonumber\\
&& - {{\alpha} \over {2 \pi}} D(x,s) \int_x^{x_+} {\rm d}z P(z) . 
\label{eq:apeqmod}
\end{eqnarray}
Separating the variables and introducing the Sudakov form factor
\begin{equation}
\Pi (s_1, s_2) = \exp \left[ 
- {{\alpha} \over {2 \pi}} \int_{s_2}^{s_1} {{{\rm d} s'} \over {s'}} \int_0^{x_+} {\rm d}z P(z) 
\right] , 
\label{eq:sudakov}
\end{equation}
which is the probability that the electron evolves from virtuality $-s_2$ to $-s_1$ without
emitting photons of energy fraction larger than $1-x_+ \equiv \epsilon$ ($\epsilon \ll 1$), 
Eq.~(\ref{eq:apeqmod}) can be recast into the integral form
\begin{eqnarray}
D(x,s) && = \Pi (s,m_e^2) D(x, m_e^2) \nonumber\\
&& \quad \! \! \! \! \! \! + {{\alpha} \over {2 \pi}} \int_{m_e^2}^{s} 
{{{\rm d}s'} \over {s'}} \Pi (s,s') \int_{x}^{x_+} {{{\rm d}z} \over {z}} P(z) 
D \left( {{x} \over {z}} , s' \right) . \nonumber\\\
\label{eq:apeqmodi}
\end{eqnarray}
The formal iterative solution of Eq.~(\ref{eq:apeqmodi}) can be represented by the infinite series
\begin{eqnarray}
&& D(x,s) = \sum_{n=0}^{\infty} \prod_{i=1}^{n} \left\{ 
\int_{m_e^2}^{s_{i-1}} {{{\rm d} s_i} \over {s_i}} \Pi (s_{i-1}, s_i)  \right. \nonumber \\
&& \! \! \! \left. \times {{\alpha} \over {2 \pi}}
\int_{x/(z_1 \cdots z_{i-1})}^{x_+} {{{\rm d} z_i} \over {z_i}} P(z_i) \right\}
\Pi (s_n, m_e^2) D \left( {{x} \over {z_1 \cdots z_n}} , m_e^2 \right) . \nonumber \\
\label{eq:dmc}
\end{eqnarray}
The particular form of Eq.~(\ref{eq:dmc}) allows to exploit a MC method for building the
solution iteratively. The steps of the algorithm are as follows: 
\begin{itemize}
\item[1 --]{set $Q^2 = m_e^2$, and fix $x=1$ according to the boundary condition $D(x,m_e^2) =
\delta(1-x)$; }
\item[2 --]{generate a random number $\xi$ in the interval $[0,1]$; }
\item[3 --]{if $\xi < \Pi (s,Q^2)$} stop the evolution; otherwise
\item[4 --]{compute $Q'^2$ as solution of the equation $\xi = \Pi (Q'^2, Q^2)$}; 
\item[5 --]{generate  a random number $z$ according to the probability density $P(z)$  in the
interval $[0,x_+]$; }
\item[6 --]{substitute $x \to xz$ and $Q^2 \to Q'^2$}; go to 2. 
\end{itemize}

\begin{figure}
\begin{center}
\resizebox{0.45\textwidth}{!}{%
\includegraphics{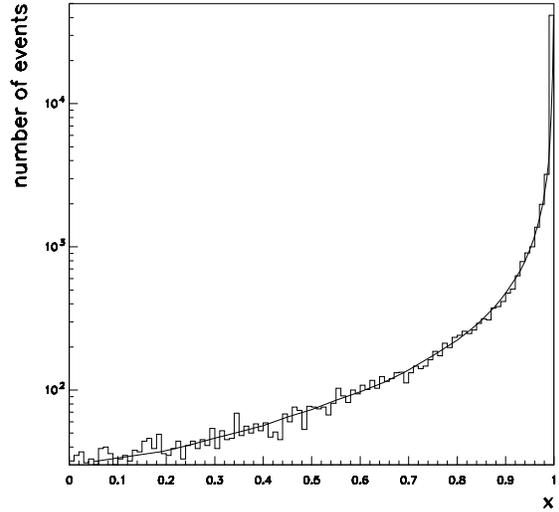}
}
\caption{Comparison for the $x$ distribution of the electron SF 
as obtained by means of a numerical solution of the QED evolution equation 
(solid line) and the PS algorithm (histogram). From \cite{CarloniCalame:2000pz}.}
\label{figmp:1}
\end{center}
\end{figure}

\begin{figure}
\begin{center}
\resizebox{0.5\textwidth}{!}{%
\includegraphics{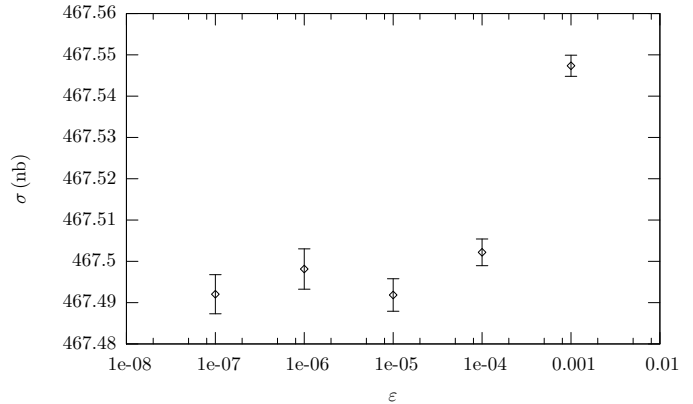}
}
\caption{QED corrected Bhabha cross section at DA$\mathrm{\Phi}$NE as a function of the infrared regulator 
 $\varepsilon$ of the PS approach, according to 
 the setup of Eq.~(\ref{eq:cutsmf}). The error bars correspond to $1\sigma$ MC errors. From 
 \cite{Balossini:2006wc}.}
\label{figmp:2}
\end{center}
\end{figure}

The $x$ distribution of the electron SF as obtained by means of the PS algorithm and 
a numerical solution (based on the inverse Mellin transform method) of the QED evolution
equation is shown in Fig.~\ref{figmp:1}. Perfect agreement is seen. Once $D(x,s)$ has been reconstructed 
by the algorithm, the master formula of 
Eq.~(\ref{eq:masterd}) can be used for the calculation of LL corrections 
to the cross section of interest. This cross section must be independent of the soft-hard 
photon separator $\epsilon$ in the limit of small values for $\epsilon$. This can be clearly seen 
in Fig.~\ref{figmp:2}, where the QED corrected Bhabha cross section as a function of the fictitious 
parameter $\varepsilon$ is shown for DA$\mathrm{\Phi}$NE energies  with the
cuts of Eq.~(\ref{eq:cutsmf}), but for an angular acceptance 
$\theta_{\pm}$ of $55^\circ \div 125^\circ$. The cross section reaches a plateau for $\epsilon$ smaller 
than $10^{-4}$. 

The main advantage of the PS algorithm with respect to the 
analytical solutions 
of the electron evolution is the possibility of going beyond the
strictly collinear approximation and generating transverse momentum $p_\perp$
of electrons and photons at each branching. In fact, the kinematics 
of the branching process $e(p) \to e'(p') + \gamma(q)$ can be written as
\begin{eqnarray}  
&& p=(E, \vec{0}, p_z)\,, \nonumber\\
&& p'=(zE,  \vec{p}_\perp,  p'_z)\,, \nonumber\\
&& q=((1-z)E,  - \vec{p}_\perp,  q_z)\,.
\label{eq:kinealt} 
\end{eqnarray}
Once the variables $p^2$,  ${p'}^2$ and $z$ are generated by the PS algorithm, 
the on-shell condition $q^2=0$, together with the longitudinal momentum 
con\-ser\-va\-tion, al\-lows to obtain an expression for the $p_{\perp}$
variable:
\begin{equation}
p^{2}_{\perp}=(1-z)(zp^2-p^{\prime 2}), \ \ 
\label{eq:kinep}
\end{equation}
valid at first order in $p^2 / E^2 \ll 1$, $p^2_\perp / E^2 \ll 1$. 

However, due to the approximations inherent to Eq.~(\ref{eq:kinep}), this PS approach can lead to an incorrect behaviour of the reconstruction of the exclusive photon kinematics.
First of all, since
within the PS algorithm the generation of $p^{\prime 2}$ and $z$ are
independent, it can happen that in some branchings the $p^2_{\perp}$ as given
by Eq.~(\ref{eq:kinep}) is negative. In order to avoid this problem, the
introduction of any kinematical cut on the $p^2$ or $z$ generation (or the
regeneration of the whole event) would prevent the correct reconstruction of
the SF $x$ distribution, which is important for a precise cross section
calculation. Furthermore, in the PS scheme, each fermion produces its photon
cascade independently of the other ones, missing the effects due to the
interference of radiation coming from different charged particles. As far as
inclusive cross sections
%
(i.e. cross sections with no cuts imposed on the generated photons)
%
are concerned, these effects are largely integrated out. 
However, as shown in \cite{CarloniCalame:2001ny}, they become important when more exclusive variables distributions are considered. 
 
The first problem can be overcome by 
%
choosing the generated $p_\perp$ of the photons different from Eq.~(\ref{eq:kinep}).
%
For example, one can choose to extract the photon 
$\cos\vartheta_\gamma$ according to the universal leading poles $1/p\cdot k$
present in the matrix element for photon emission. %
%
Namely, one can generate $\cos\vartheta_\gamma$ as
\begin{equation}
\cos\vartheta_\gamma \propto \frac{1}{1-\beta\cos\vartheta_\gamma} \ \ ,
\label{cthg_LL}
\end{equation} 
where $\beta$ is the speed of the emitting particle. In this way, photon
energy and angle are generated independently, different from Eq.~(\ref{eq:kinep}). 
The nice feature of this prescription is that
$p^2_{\perp}=E^2_\gamma\sin^2\vartheta_\gamma$ is always well defined, and the 
$x$ distribution reproduces exactly the SF, because  
no further kinematical cuts have to be imposed to avoid unphysical events. At this stage, the
PS is used only to generate the
energies and multiplicity of the photons.
The problem of including the radiation interference is still unsolved, because
the variables of photons emitted by a fermion are still uncorrelated with those of the
other charged particles. The issue of including photon interference can be successfully worked out 
looking at the YFS formula \cite{Yennie:1961ad}:
\begin{equation}
{\rm d}\sigma_n\approx {\rm d}\sigma_0\frac{e^{2n}}{n!}\prod_{l=1}^n\frac{{\rm d}^{\it 3}
\mathbf{k}_l}{(2\pi)^32k^0_l}\sum_{i,j=1}^N\eta_i\eta_j\frac{-p_i\cdot p_j}
{(p_i\cdot k_l)(p_j\cdot k_l)} \ \ .
\label{yfse}
\end{equation}  
It 
gives the differential cross section
${\rm d}\sigma_n$ for the emission of $n$ photons, whose momenta are 
$k_1,\cdots,k_n$, from a
kernel process described by ${\rm d}\sigma_0$ and involving $N$ fermions, whose 
momenta are $p_1,\cdots,p_N$. In Eq.~(\ref{yfse}) $\eta_i$ is a charge
factor, which is $+1$ for incoming $e^-$ or outgoing $e^+$ and $-1$ for
incoming $e^+$ or outgoing $e^-$. Note that Eq.~(\ref{yfse}) is valid in the
soft limit ($k_i\to 0$). The important point is that it also accounts for
coherence effects. From the YFS formula it is straightforward to read out the
angular spectrum of the $l^{th}$ photon:
\begin{equation}
\cos\vartheta_l\propto-\sum_{i,j=1}^N\eta_i\eta_j
\frac{1-\beta_i\beta_j\cos\vartheta_{ij}}{(1-\beta_i\cos\vartheta_{il})
(1-\beta_j\cos\vartheta_{jl})} \ \ .
\label{cthYFS}
\end{equation}

It is worth noticing that in the LL prescription the same quantity can be written as
\begin{equation}
\cos\vartheta_l\propto\sum_{i=1}^N\frac{1}{1-\beta_i\cos\vartheta_{il}},
\label{cthLL}
\end{equation}
whose terms are of course contained in Eq.~(\ref{cthYFS}).

In order to consider also coherence effects in the angular distribution of the
photons, one can generate $\cos\vartheta_\gamma$ according to Eq.~(\ref{cthYFS}), rather 
than to Eq.~(\ref{cthLL}). This recipe \cite{CarloniCalame:2001ny}
is adopted in BabaYaga v3.5 and BabaYaga@NLO. \\

\noindent {\em{Yennie-Frautschi-Suura exponentiation \label{YFS}}} \\
\vspace*{-2mm}


The YFS  exponentiation procedure, 
implemented in the code BHWIDE, is 
a technique for summing up all the infrared (IR) singularities present in any 
process 
accompanied by photonic radiation~\cite{Yennie:1961ad}. It is inherently exclusive, i.e. all the 
summations of the IR singular
contributions are done before any phase-space integration over the virtual or real photon
four-momenta are performed. 
The method was mainly developed by S. Jadach, B.F.L. Ward and collaborators to 
realise precision MC tools.
In the following, the general ideas underlying the procedure are summarised.  

Let us consider the scattering process $e^+ (p_1) e^- (p_2) \to f_1 (q_1) \cdots f_n (q_n)$, 
where $f_1 (q_1) \cdots f_n (q_n)$ represents a given arbitrary final state, and let  
${\cal M}_0$ be its tree-level matrix element. 
By using standard Feynman-diagram techniques, it is possible
to show that the  same process, when accompanied by $l$ 
additional real photons radiated by the
IS particles, and under the assumption that the $l$ additional photons are soft,
i.e. their energy is much smaller that any energy scale involved in the process, can be
described by the factorised matrix element built up by the LO one,  ${\cal M}_0$,
times the product of $l$ eikonal currents, namely 
\begin{equation} 
{\cal M} \simeq {\cal M}_0 \prod_{i=1}^{l} \left[ e \left( 
{{\varepsilon_i (k_i) \cdot p_2} \over {k_i \cdot p_2}} - 
{{\varepsilon_i (k_i) \cdot p_1} \over {k_i \cdot p_1}}
\right) \right] , 
\label{eq:lrphot}
\end{equation}
where $e$ is the electron charge, $k_i$ are the momenta of the photons and 
$\varepsilon_i (k_i)$ their polarisation vectors. 
Taking the square of the matrix element 
in Eq.~(\ref{eq:lrphot}) and multiplying by the proper flux factor and the Lorentz-invariant 
phase space volume, 
the cross section for the process  
$e^+ (p_1) e^- (p_2) \to f_1 (q_1) \cdots f_n (q_n) + \, l \, {\rm real} \, {\rm photons}$ can be written as
\begin{eqnarray}
{\rm d} \sigma^{(l)}_{r} = && {\rm d} \sigma_0 {{1} \over {l!}} \prod_{i=1}^{l} \left[
k_i {\rm d} k_i {\rm d} \cos \vartheta_i {\rm d} \varphi_i {{1} \over {2 (2 \pi)^3}} \right. \nonumber \\
&& \left. \times\,\sum_{\varepsilon_i} e^2 \left( 
{{\varepsilon_i (k_i) \cdot p_2} \over {k_i \cdot p_2}} - 
{{\varepsilon_i (k_i) \cdot p_1} \over {k_i \cdot p_1}}
\right)^2 \right] . 
\label{eq:lreal}
\end{eqnarray}
By summing over the number of final-state photons, one obtains the cross section for the
original process accompanied by an arbitrary number of real photons, namely
\begin{eqnarray}
{\rm d} \sigma^{(\infty)}_{r} =&& \sum_{l=0}^{\infty} {\rm d} \sigma^{(l)}_{r} \nonumber \\
=&& {\rm d} \sigma_0  \exp \left[
k {\rm d} k {\rm d} \cos \vartheta {\rm d} \varphi {{1} \over {2 (2 \pi)^3}} \right. \nonumber\\
&& \left. \times\,\sum_{\varepsilon} e^2 \left( 
{{\varepsilon (k) \cdot p_2} \over {k \cdot p_2}} - 
{{\varepsilon (k) \cdot p_1} \over {k \cdot p_1}}
\right)^2 \right] . 
\label{eq:expr}
\end{eqnarray}
Equation~(\ref{eq:expr}), being limited to real radiation only, is IR divergent once the phase
space integrations are performed down to zero photon energy. This problem, as is well known,
finds its solution in the matching between real and virtual photonic radiation. 
Equation~(\ref{eq:expr}) already shows the key feature of exclusive exponentiation, i.e. summing
up all the perturbative contributions before performing any phase space integration.

In order to get meaningful radiative corrections it is necessary to consider, besides IS real photon corrections, also IS virtual photon 
corrections, i.e. the corrections due to additional  internal photon lines connecting the
IS electron and positron. 
For a vertex-type 
amplitude, the result can be written as
\begin{eqnarray}
{\cal M}_{V_1} = - i {{e^2} \over {(2  \pi)^4}} \int {\rm d}^4 k && \!\! \!\! \!\! {{1} \over {k^2  + i \varepsilon}}
{\bar v} (p_1) \gamma^\mu 
{{- ( {\rlap / {p_1}} + {\rlap / {k}} ) + m } \over {2 p_1 \cdot k + k^2 + i 
\varepsilon}}  \nonumber \\
&\times &  \!\! \Gamma 
{{ ( {\rlap / {p_2}} + {\rlap / {k}} ) + m } \over {2 p_2 \cdot k + k^2 + i \varepsilon}}
\gamma_\mu u(p_2) , 
\label{eq:virt}
\end{eqnarray}
where $\Gamma$ stands for the Dirac structure of the LO process, in such a
way that ${\cal M}_0 = {\bar v} (p_1) \Gamma u(p_2)$. The soft-photon part  of the amplitude
can be extracted by taking $k^\mu \simeq 0 $ in all the numerators. In this approximation, the
amplitude of Eq.~(\ref{eq:virt}) becomes
\begin{eqnarray} 
&& {\cal M}_{V_1} = {\cal M}_0 \times V , \nonumber \\
&& V = {{2 i \alpha} \over {(2 \pi)^3}} \int {\rm d}^4 k                            
{{4 p_1 \cdot p_2} \over {(2 p_1 \cdot k + k^2 + i \varepsilon) 
(2 p_2 \cdot k + k^2 + i \varepsilon)}}   \nonumber\\
&& \qquad \qquad \qquad \quad \times {{1} \over {k^2 + i \varepsilon}} .
\label{eq:onevirt}
\end{eqnarray}
It can be seen that, as in the real case, the IR virtual correction
factorises off the LO matrix element so that it is universal, i.e. independent
of the details of the process under consideration, and divergent in the IR portion of the
phase space.  

The correction given by $n$  soft virtual photons can be seen to factorise with an additional factor $1/n!$, namely 
\begin{equation}
{\cal M}_{V_n}  = {\cal M}_0  \times {{1} \over {n!}} V^n , 
\label{eq:nvirt}
\end{equation}
so that by summing over all  the additional soft virtual photons one obtains
\begin{equation}
{\cal  M}_V = {\cal M}_0 \times \exp [V] . 
\label{eq:virtexp} 
\end{equation}

As  already noticed both the real  and virtual factors are IR divergent. In order to obtain
meaningful expressions one has to adopt  some regularisation procedure. One 
possibility is to give the photon a (small) mass $\lambda$ and to modify
Eqs.~(\ref{eq:lreal}) and (\ref{eq:onevirt}) accordingly. Once all the expressions are
properly regularised, one can write down a YFS master formula that takes  into account
real and virtual photonic corrections to the LO process. In virtue of the
factorisation properties discussed above, the master formula can be obtained from
Eq.~(\ref{eq:expr}) with the substitution ${\rm d} \sigma_0 \to {\rm d} \sigma_0 \vert \exp (V) \vert^2$, i.e.
\begin{eqnarray}
{\rm d} \sigma &&  = {\rm d} \sigma_0  \vert \exp (V) \vert^2 \exp \left[
k {\rm d} k {\rm d} \cos \vartheta {\rm d} \varphi {{1} \over {2 (2 \pi)^3}} \right. \nonumber\\
&& \left. \times\,\sum_{\varepsilon} e^2 \left( 
{{\varepsilon (k) \cdot p_2} \over {k \cdot p_2}} - 
{{\varepsilon (k) \cdot p_1} \over {k \cdot p_1}}
\right)^2 \right] . 
\label{eq:yfsmaster}
\end{eqnarray} 
As a last step it is possible to analytically perform the IR cancellation between virtual 
and very soft  real photons. Actually, since very soft real photons do not affect the
kinematics of the process, the real photon exponent can be split into a contribution coming
from photons with energy less than a cutoff $k_{min}$ plus a contribution from photons
with energy above it. The first contribution can be integrated over all its phase
space and can then be combined with the virtual exponent. After this step it is possible to remove the regularising photon mass by taking the limit 
$\lambda \to 0$, so that Eq.~(\ref{eq:yfsmaster}) becomes
\begin{eqnarray}
{\rm d} \sigma  && = {\rm d} \sigma_0   \exp (Y)  \exp \left[
k {\rm d} k {\rm d} \Theta (k - k_{min})\cos \vartheta {\rm d} \varphi {{1} \over {2 (2 \pi)^3}} \right. \nonumber\\
&& \left. \times\,\sum_{\varepsilon} e^2 \left( 
{{\varepsilon (k) \cdot p_2} \over {k \cdot p_2}} - 
{{\varepsilon (k) \cdot p_1} \over {k \cdot p_1}}
\right)^2 \right] , 
\label{eq:yfsfinal}
\end{eqnarray} 
where  $Y$ is given  by 
\begin{eqnarray}
Y && = 2 V + \int  k {\rm d} k {\rm d} \Theta ( k_{min} - k )
\cos \vartheta {\rm d} \varphi {{1} \over {2 (2 \pi)^3}} \nonumber\\
&& \times\,\sum_{\varepsilon} e^2 \left( 
{{\varepsilon (k) \cdot p_2} \over {k \cdot p_2}} - 
{{\varepsilon (k) \cdot p_1} \over {k \cdot p_1}}
\right)^2  . 
\end{eqnarray}
The explicit form of $Y$  can be derived by performing all the details of the calculation, and
reads
\begin{eqnarray}
Y &=& \beta \ln {{k_{min}} \over {E}} + \delta_{YFS} , \nonumber \\
\delta_{YFS} &=& {{1} \over {4}} \beta + {{\alpha} \over {\pi}} \left( 
{{\pi^2}  \over {3}} - {{1} \over {2}} \right) . 
\end{eqnarray}

\subsubsection{Matching NLO and higher-order corrections 
\label{NLO-HO}}

As will be shown numerically in Section \ref{NUMERICS}, NLO corrections must be combined with multiple
photon emission effects to achieve a theoretical accuracy at the per mill level. 
This combination, technically known as {\em matching}, is a fundamental ingredient of 
the most precise generators used for luminosity monitoring, i.e. BabaYaga@NLO,
BHWIDE  and MCGPJ. Although the matching is implemented according to
different theoretical details, some general aspects are common to all the recipes
and must be emphasised:
\begin{enumerate}

\item It is possible to match NLO and HO corrections consistently, avoiding double 
counting of LL contributions at order $\alpha$ and preserving the advantages of
resummation of soft and collinear effects beyond $O (\alpha)$.

\item The convolution of NLO corrections with HO terms allows 
to include the dominant part of NNLO corrections, given by 
infrared-enhanced $\alpha^2 L$ sub-leading contributions. This was argued 
and demonstrated analytically and numerically in \cite{Montagna:1996gw} through comparison with the available 
$O (\alpha^2)$ corrections to $s$-channel processes and $t$-channel
Bhabha scattering. Such an aspect of the matching procedure is crucial to
settle the theoretical accuracy of the generators by means of explicit comparisons
with the exact NNLO perturbative corrections discussed in Section \ref{NNLO}, and 
will be addressed in Section \ref{TH}.

\item BabaYaga@NLO and BHWIDE implement a fully factorised 
matching recipe, while MCGPJ includes some terms in additive form, 
as will be visible in
the formulae reported below.
\end{enumerate}

\vspace{-0.2cm}
In the following we summarise the basic features of the matching
procedure as implemented in the codes MCGPJ, 
BabaYaga@NLO and BHWIDE.

The matching approach realised in the MC event generator MCGPJ was 
developed in~\cite{Arbuzov:2005pt}. In particular, Bha\-bha scattering with complete 
$O (\alpha)$ and HO LL photonic corrections can written as
\begin{eqnarray}
&& \frac{\dd\sigma^{e^+e^-\to e^+e^-(\gamma)}}{\dd\Omega_-}
= \int\limits_{\bar{z}_1}^{1}\dd z_1\;\int\limits_{\bar{z}_2}^{1}\dd z_2\;
D_{ee}^{NS,\gamma}(z_1) D_{ee}^{NS,\gamma}(z_2)
\nonumber \\ && \quad \times 
\frac{\dd\hat{\sigma}_{0}^{\mathrm{Bhabha}}(z_1,z_2)}{\dd\Omega_-}
\left(1+\frac{\alpha}{\pi}K_{SV}\right)\Theta({\mathrm{cuts}}) 
\nonumber \\ \nonumber && \quad
\times \int\limits^{Y_1}_{y_{\mathrm{th}}}\frac{\dd y_1}{Y_1}\;
\int\limits^{Y_2}_{y_{\mathrm{th}}}\frac{\dd y_2}{Y_2}\;
D_{ee}^{NS,\gamma}(\frac{y_1}{Y_1})D_{ee}^{NS,\gamma}(\frac{y_2}{Y_2})
\\ \nonumber && \quad 
+ \frac{\alpha}{\pi}\int\limits_{\Delta}^{1}\frac{\dd x}{x}
\Biggl\{\biggl[\left(1-x+\frac{x^2}{2}\right)\ln\frac{\theta_0^2(1-x)^2}{4}
+ \frac{x^2}{2}\biggr] 
\\ \nonumber && \quad \times
2 \frac{\dd\sigma_0^{\mathrm{Bhabha}}}{\dd \Omega_-}
+ \biggl[\left(1-x+\frac{x^2}{2}\right)\ln\frac{\theta_0^2}{4}
+ \frac{x^2}{2}\biggr]
\\ \nonumber && \quad \times
\Biggl[ \frac{\dd\hat{\sigma}_0^{\mathrm{Bhabha}}(1-x,1)}{\dd \Omega_-}
+ \frac{\dd\hat{\sigma}_0^{\mathrm{Bhabha}}(1,1-x)}{\dd \Omega_-}
\Biggr]\Biggr\}\Theta({\mathrm{cuts}})
\\ && \quad 
- \frac{\alpha^2}{4s}
\left(\frac{3+c^2}{1-c}\right)^2
\frac{8\alpha}{\pi}\ln(\mbox{ctg}\frac{\theta}{2})
\ln\frac{\Delta\eps}{\eps}
\nonumber \\ && \quad
 + \frac{\alpha^3}{2\pi^2s}\!\!\!\!\!
\int\limits_{\stackrel{k^0>\Delta\eps}{\theta_i >\theta_0}}
\!\! \frac{WT}{4}\Theta({\mathrm{cuts}}) \frac{\dd \Gamma_{e\bar{e}\gamma}}{\dd \Omega_-}\, .
 \label{bhabha_nlo}
\end{eqnarray}
Here the step functions $\Theta({\mathrm{cuts}})$ stand for the particular 
cuts applied.
The auxiliary parameter $\theta_0$ defines cones around the directions of 
the motion of the charged particles 
in which the emission of hard photons is approximated by the factorised form
by convolution of collinear radiation factors~\cite{Arbuzov:2007wu} 
with the Born cross section.
The dependence on the parameters $\Delta$ and $\theta_0$ cancels out in the sum with the last term 
of Eq.~(\ref{bhabha_nlo}),  where the photon energy and emission angles with respect to all charged
particles are limited from below $(k^0 > \Delta\eps, \theta_i > \theta_0)$. 
Taking into account vacuum polarisation, the Born level Bhabha cross section with reduced 
energies of the incoming electron and positron can be cast in the form
\begin{eqnarray}
&& \frac{\dd\hat{\sigma}_{0}^{\mathrm{Bhabha}}(z_1,z_2)}{\dd\Omega_-} = 
\frac{4\alpha^2}{sa^2}\biggl\{\frac{1}{|1-\Pi(\hat{t})|^2}\;
\frac{a^2+z_2^2(1+c)^2}{2z_1^2(1-c)^2}
\nonumber \\ && \quad
+ \frac{1}{|1-\Pi(\hat{s})|^2}\;
\frac{z_1^2(1-c)^2+z_2^2(1+c)^2}{2a^2} 
\nonumber \\ && \quad
- \mbox{Re}\;\frac{1}{(1-\Pi(\hat{t}))(1-\Pi(\hat{s}))^{*}}\;
\frac{z_2^2(1+c)^2}{az_1(1-c)} \biggr\}\dd\Omega_-\, ,
\nonumber \\ &&
\hat{s} = z_1z_2s, \quad \hat{t} = -\frac{sz_1^2z_2(1-c)}{z_1+z_2-(z_1-z_2)c}\,,
\end{eqnarray}
where $\Pi(Q^2)$ is the photon self-energy correction. Note that in the  cross section 
above the cosine of the scattering angle, $c$, is
given for the original c.m. reference frame of the colliding beams.

For the two-photon production channel, a similar representation is used in MCGPJ:
\begin{eqnarray}
&& \dd\sigma^{e^+e^-\to\gamma\gamma(\gamma)}
= \int\limits_{\bar{z}_1}^{1}\dd z_1 D_{ee}^{NS,\gamma}(z_1)
\int\limits_{\bar{z}_2}^{1}\dd z_2 D_{ee}^{NS,\gamma}(z_2)
\nonumber \\ && \quad \times 
\dd\hat{\sigma}_0^{\gamma\gamma}(z_1,z_2)
\left(1+\frac{\alpha}{\pi}K_{SV}^{\gamma\gamma}\right) 
+\ \frac{\alpha}{\pi}\int\limits_{\Delta}^{1}\frac{\dd x}{x}
\nonumber \\  && \quad \times 
\left[\left(1-x+\frac{x^2}{2}\right)\ln\frac{\theta^2_0}{4}
+\frac{x^2}{2}\right]
\biggl[\dd\hat{\sigma}_0(1-x,1) 
\nonumber \\  && \quad 
+ \dd\hat{\sigma}_0(1,1-x)\biggr] 
+\ \frac{1}{3}\,\frac{4\alpha^3}{\pi^2s^2}  \!\!\!\!\int\limits_{\stackrel{z_i\geq\Delta}
{\pi-\theta_0\geq\theta_i\geq\theta_0}}\!\!
\dd \Gamma_{3\gamma}
\nonumber \\  && \quad \times 
\biggl[\frac{z_3^2(1+c_3^2)}{z_1^2z_2^2(1-c_1^2)(1-c_2^2)}
+ \mbox{two cyclic permutations}\biggr], 
\nonumber \\ 
&& z_i=\frac{q_i^0}{\eps},\quad c_i=\cos\theta_i,\quad
\theta_i=\widehat{\vec{p}_-\vec{q}}_i \, ,
\label{eq:agg}
\end{eqnarray}
where 
the cross section with reduced energies has the form
\begin{eqnarray} \nonumber
\frac{\dd\hat{\sigma}_0^{\gamma\gamma}(z_1,z_2)}{\dd\Omega_1}
= \frac{2\alpha^2}{s}\frac{z_1^2(1-c_1)^2+z_2^2(1+c_1)^2}
{(1-c_1^2)(z_1+z_2+(z_2-z_1)c_1)^2} \, ,
\end{eqnarray}
and the factor $1/3$ in the last term of Eq.~(\ref{eq:agg}) takes into
account the identity of the final-state photons.
The sum of the last two terms does not
depend on $\Delta$ and $\theta_0$. 

Concerning BabaYaga@NLO, the matching starts from the observation that 
Eq.~(\ref{eq:masterd}) for the QED corrected all-order cross section 
can be rewritten in terms of the PS ingredients as
\begin{equation}
{\rm d}\sigma^{\infty}_{LL}=
{\Pi}(Q^2,\varepsilon)~
\sum_{n=0}^\infty \frac{1}{n!}~|{\ourcal M}_{n,LL}|^2~{\rm d}\Phi_n\,.
\label{generalLL}
\end{equation}
By construction, the expansion of Eq.~(\ref{generalLL}) at \oa\ does not coincide with the exact \oa\ result. In fact
\bea
{\rm d}\sigma^{\alpha}_{LL}&=&
\left[
1-\frac{\alpha}{2\pi}~I_+~\ln\frac{Q^2}{m^2}
\right] |{\ourcal M}_0|^2 {\rm d}\Phi_0+
|{\ourcal M}_{1,LL} |^2 {\rm d}\Phi_1 \nonumber\\
&\equiv&
\left[
1+C_{\alpha,LL}
\right] |{\ourcal M}_0|^2 {\rm d}\Phi_0
+
|{\ourcal M}_{1,LL} |^2 {\rm d}\Phi_1\,,
\label{LL1}
\eea
where $I_+  \equiv \int_0^{1-\epsilon} P(z) {\rm d}z$, whereas the exact NLO cross section can always be cast in the form
\be
{\rm d}\sigma^{\alpha}
=
\left[
1+C_{\alpha}
\right] |{\ourcal M}_0|^2 {\rm d}\Phi_0
+
|{\ourcal M}_{1} |^2 {\rm d}\Phi_1\,.
\label{exact1}
\ee
The coefficients $C_{\alpha}$ contain the complete \oa\  virtual 
and soft-bremsstrahlung corrections in units of
the squared Born amplitude,
and $|{\ourcal M}_1|^2$ is the exact squared matrix element with the
emission of one hard photon.
We remark that $C_{\alpha,LL}$ has the same logarithmic structure as
$C_\alpha$ and that $|{\ourcal M}_{1,LL}|^2$ has the same singular
behaviour as $|{\ourcal M}_1|^2$.

In order to match the LL and NLO calculations,
the following correction factors, which are by construction
infrared safe and free of collinear logarithms, are introduced: 
\be
F_{SV}~=~
1+\left(C_\alpha-C_{\alpha,LL}\right),~~~
F_H~=~
1+\frac{|{\ourcal M}_1|^2-|{\ourcal M}_{1,LL}|^2}{|{\ourcal M}_{1,LL}|^2}\,.
\label{FSVH}
\ee
With them the exact \oa\ cross section can be expressed, up
to terms of ${\ourcal O}(\alpha^2)$,
in terms of its LL approximation as
\be
{\rm d}\sigma^\alpha~=~
F_{SV} (1+C_{\alpha,LL} ) |{\ourcal M}_0|^2 {\rm d}\Phi_0
~+~
F_H |{\ourcal M}_{1,LL}|^2 {\rm d}\Phi_1 .
\label{matchedalpha}
\ee
Driven by Eq.~\myref{matchedalpha}, Eq.~\myref{generalLL} can be improved
by writing the resummed matched cross section as
\begin{eqnarray}
{\rm d}\sigma^{\infty}_{\rm matched} = &&
F_{SV}~\Pi(Q^2,\varepsilon) \nonumber\\
&& \times\,\sum_{n=0}^\infty \frac{1}{n!}~
\left( \prod_{i=0}^n F_{H,i}\right)~
|{\ourcal M}_{n,LL}|^2~
{\rm d}\Phi_n .
\label{matchedinfty}
\end{eqnarray}
The correction factors $F_{H,i}$ follow from the definition \myref{FSVH}
for each photon emission.
The \oa\  expansion of Eq.~\myref{matchedinfty} now coincides with
the exact NLO cross section of Eq.~\myref{exact1}, and
all HO LL contributions
are the same as in Eq.~\myref{generalLL}. This formulation is implemented in BabaYaga@NLO for both Bhabha scattering and
photon pair production, using, of course, the appropriate SV and hard 
bremsstrah\-lung formulae. This matching formulation has also been 
applied to the study of Drell-Yan-like processes, 
by combining the complete $O (\alpha)$ electroweak corrections 
with QED shower evolution in the 
generator HORACE 
\cite{CarloniCalame:2003ux,CarloniCalame:2005vc,CarloniCalame:2006zq,CarloniCalame:2007cd}. 

As far as BHWIDE is concerned, this MC event generator realises
the process 
\begin{equation}
e^+(p_1) + e^-(q_1)\; \longrightarrow\; e^+(p_2) + e^-(q_2) \;
+ \gamma_1(k_1) + \ldots + \gamma_n(k_n)
\label{process} 
\end{equation}
via the YFS exponentiated cross section formula
\begin{eqnarray}
{\rm d}\sigma= && e^{2\alpha\mathrm{Re}B+2\alpha
\tilde B}\,\sum_{n=0}^\infty{1\over n!}\int\prod_{j=1}^n{{\rm d}^3k_j\over k_j^0
}\int{{\rm d}^4y\over(2\pi)^4} \nonumber\\
&& \times\,e^{iy(p_1+q_1-p_2-q_2-\sum_jk_j)+D}
\bar\beta_n(k_1,\dots,k_n)\,{{\rm d}^3p_2{\rm d}^3q_2\over p_2^0q_2^0}, \nonumber\\
\label{eqone}
\end{eqnarray}
where the
real infrared function $\tilde B$ and the virtual infrared function $B$ are
given in~\cite{Jadach:1995nk}. 
Here we note the usual connections
\begin{eqnarray}
2\alpha\tilde B && = \int^{k\le K_{max}}{{\rm d}^3k\over k_0}\, \tilde S(k)\,,
\nonumber\\
D && =\int {\rm d}^3k\,{\tilde S(k)\over k^0}
     \left(e^{-iy\cdot k}-\theta(K_{\rm max}-k)\right)
\label{eqtwo}
\end{eqnarray} 
for the standard YFS infrared real emission factor
\begin{equation}
\tilde S(k)= {\alpha\over4\pi^2}\left[Q_fQ_{f'}
\left({p_1\over p_1\cdot k}-{q_1
\over q_1\cdot k}\right)^2+\ldots\right]\,,
\label{eqthree}
\end{equation} 
and where $Q_f$ is the electric charge of $f$ in units of the positron charge. 
In Eq.~(\ref{eqthree}) the ``$\ldots$'' 
represent the remaining terms in $\tilde S(k)$, obtained from the given one
by respective of $Q_f$, $p_1$, 
$Q_{f'}$, $q_1$ with corresponding values for the
other pairs of the external charged legs according to the YFS
prescription of Ref.~\cite{Yennie:1961ad,Mahanthappa:1962ex} (wherein due attention is taken to obtain
the correct relative sign of each of the terms in $\tilde S(k)$ according to
this latter prescription). The explicit representation is given by 
\begin{eqnarray}
&& 2\alpha\mathrm{Re} B(p_1,q_1,p_2,q_2) + 2\alpha\tilde{B}(p_1,q_1,p_2,q_2;k_m)
 =  \nonumber\\
 && R_1(p_1,q_1;k_m) + R_1(p_2,q_2;k_m)
+ R_2(p_1,p_2;k_m) + \nonumber\\
&& R_2(q_1,q_2;k_m) - R_2(p_1,q_2;k_m) - R_2(q_1,p_2;k_m)\,,
\label{YFSff}
\end{eqnarray}
with
\begin{equation}
R_1(p,q;k_m)  = R_2(p,q;k_m) + \left(\frac{\alpha}{\pi}\right) 
                                \frac{\pi^2}{2}
\label{R1}
\end{equation}
and
\begin{eqnarray}
R_2(p,q;k_m)  && = \frac{\alpha}{\pi} \left\{ \bigg(\ln\frac{2pq}{m_e^2} -1 \bigg)
                 \ln\frac{k_m^2}{p^0q^0} + \frac{1}{2}\ln\frac{2pq}{m_e^2} \right. \nonumber \\
                && \left. -\frac{1}{2}\ln^2\frac{p^0}{q^0} 
               -\frac{1}{4}\ln^2\frac{(\Delta+\delta)^2}{4p^0q^0} 
                -\frac{1}{4}\ln^2\frac{(\Delta-\delta)^2}{4p^0q^0} \right. \nonumber \\
                && \left. -\mathrm{Re}\, {\rm Li}_2\left(\frac{\Delta+\omega}{\Delta+\delta}\right) 
                -\mathrm{Re}\, {\rm Li}_2\left(\frac{\Delta+\omega}{\Delta-\delta}\right) \right. \nonumber \\     
                && \left.  -\mathrm{Re}\, {\rm Li}_2\left(\frac{\Delta-\omega}{\Delta+\delta}\right) 
                -\mathrm{Re}\, {\rm Li}_2\left(\frac{\Delta-\omega}{\Delta-\delta}\right) \right. \nonumber\\
                && \left. +\frac{\pi^2}{3} -1 \right\},
 \label{R2}
\end{eqnarray}   
where $\Delta=\sqrt{2pq+(p^0-q^0)^2}$, $\omega=p^0+q^0$, $\delta=p^0-q^0$,
and $k_m$ is a soft photon cut-off in the c.m. system 
($E_{\gamma}^{\rm soft}<k_m\ll E_{\rm beam}$).

The YFS hard photon residuals $\bar\beta_i$ in Eq.~(\ref{eqone}), $i=0,1$,
to $O (\alpha)$ are given exactly 
in Ref.~\cite{Jadach:1995nk} for BHWIDE. Therefore this 
event generator calculates the YFS exponentiated exact
$O (\alpha)$ cross section for $e^+e^-\rightarrow e^+ e^- + n(\gamma)$
with multiple initial, initial-final and final state radiation, 
using a corresponding MC realisation
of Eq.~(\ref{eqone}) in the wide angle regime. The library for $O (\alpha)$ electroweak corrections, relevant
for higher energies, is taken from \cite{Beenakker:1990mb,Bohm:1986fg}.\par
The result (\ref{eqone}) is an exact rearrangement of the loop expansion for the
respective cross section and is independent of the dummy parameter $K_{\rm max}$. To derive this, one may proceed as follows.
Let the amplitude for the emission of $n$ real photons in the
Bhabha process be 
\begin{equation}
{\cal M}^{(n)} 
 = \sum_{\ell}M^{(n)}_{\ell},
\label{subp1}
\end{equation}
where $M^{(n)}_{\ell}$ is the contribution to 
${\cal M}^{(n)}$ from Feynman diagrams with ${\ell}$ virtual loops.
The key result in the YFS theory of Ref.~\cite{Yennie:1961ad,Mahanthappa:1962ex} on virtual corrections
is that we may rewrite Eq.~(\ref{subp1}) as the exact representation
\begin{equation}
{\cal M}^{(n)} = e^{\alpha B}\sum_{j=0}^\infty{\sf m}^{(n)}_j,
\label{yfsrepv}
\end{equation}
where we have defined
\begin{equation}
\alpha B = \int {d^4k\over (k^2-\lambda^2+i\epsilon)}S(k) ,
\label{vbfn}
\end{equation}
with the virtual infrared emission factor given by
\begin{eqnarray}
S(k) && = \frac{-i\alpha}{8\pi^2}\sum_{i'<j}Z_{i'}\theta_{i'}Z_j\theta_j\left(\frac{(2\bar{p}_{i'}\theta_{i'}-k)_\mu}{k^2-2k\bar{p}_{i'}\theta_{i'}+i\epsilon} \right. \nonumber\\
&& \left. +\frac{(2\bar{p}_{j}\theta_{j}+k)_\mu}{k^2+2k\bar{p}_{j}\theta_{j}+i\epsilon}\right)^2.
\label{vbfn1}
\end{eqnarray}
Here, $\lambda$ is an infrared regulator mass, and following Refs.~\cite{Yennie:1961ad,Mahanthappa:1962ex} we identify the sign of the $j$-th external line charge here as $Z_j=Q_j$ and $\theta_j=+(-)$ for outgoing (incoming) 4-momentum $\bar{p}_j$, so that here $\bar{p}_1=p_1,\;\bar{p}_2=q_1,\;\bar{p}_3=p_2,\;\bar{p}_4=q_2$,$\;Z_1=+1,\; \theta_1=-,\; Z_2=-1,\; \theta_2=-,\; Z_3=+1,\;\theta_3=+,\; Z_4=-1,\;\theta_4=+$. The amplitudes $\{{\sf m}^{(n)}_j\}$ are free of all virtual infrared divergences. 

Using the result (\ref{yfsrepv}) for ${\cal M}^{(n)}$, we get the attendant differential cross section by the standard methods as  
\begin{eqnarray}
  {\rm d}\hat\sigma^n && = {e^{2\alpha \mathrm{Re} B}\over {n !}}\int\prod_{l=1}^n
{{\rm d}^3k_l\over (k_l^2+\lambda^2)^{1/2}} \nonumber\\
&& \times \bar\rho^{(n)}(p_1,q_1,p_2,q_2,k_1,\cdots,k_n)\,
{{\rm d}^3p_2{\rm d}^3q_2\over p^0_2 q^0_2} \nonumber\\
&& \times \delta^{(4)}\left(p_1+q_1-p_2-q_2-\sum_{i=1}^nk_i\right) ,
\label{diff1}
\end{eqnarray}
where we have defined
\begin{equation}
\bar\rho^{(n)}(p_1,q_1,p_2,q_2,k_1,\cdots,k_n)=
\sum_{\rm spin} \left|\sum_{j=0}^\infty{\sf m}^{(n)}_j\right|^2 ,
\label{diff2}
\end{equation}
in the incoming $e^+e^-$ c.m. system.
Here we have absorbed the remaining kinematical factors for
the initial state flux and spin averaging into the
normalisation of the amplitudes ${\cal M}^{(n)}$ for pedagogical reasons,
so that the $\bar\rho^{(n)}$ are averaged over initial spins
and summed over final spins. We then use the key result of 
Ref.~\cite{Yennie:1961ad,Mahanthappa:1962ex} on real corrections to write the exact result 
\begin{eqnarray}
&&\bar\rho^{(n)}(p_1,q_1,p_2,q_2,k_1,\cdots,k_n)=\prod_{i=1}^n\tilde{S}(k_i)\bar\beta_0+\cdots  + \nonumber\\
&& \sum_{i=1}^n\tilde{S}(k_i)\bar\beta_{n-1}(k_1,\ldots,k_{i-1},k_{i+1},\ldots,k_n) \nonumber\\
&& + \bar\beta_n(k_1,\ldots,k_n), 
\label{realem}
\end{eqnarray}
where the hard photon 
residuals $\bar\beta_j$ are determined recursively~\cite{Yennie:1961ad,Mahanthappa:1962ex}
and are free of all virtual and all real infrared singularities to all orders in $\alpha$. Introducing the result (\ref{realem}) into Eq.~(\ref{diff1}) and summing
over the number of real photons $n$ leads directly to master formula (\ref{eqone}). We see that it allows for exact exclusive treatment of hard photonic effects on
an event-by-event basis.

\subsection{Monte Carlo generators \label{MC}}

To measure the luminosity, event generators, rather than analytical calculations, are 
mandatory to provide theoretical results of real experimental interest.
The software tools used in early measurements of 
the luminosity at flavour factories (and sometimes 
still used in recent experimental publications) include generators such as BHAGENF
 \cite{Drago:1997px}, 
BabaYaga v3.5 \cite{CarloniCalame:2003yt} and BKQED \cite{Berends:1983rk,Berends:1981fb}. 
These MC programs, however, are based either on a fixed
NLO calculation (such as BHAGENF and BKQED) or include corrections to all orders in 
perturbation theory, but in the LL approximation only (like BabaYaga v3.5). 
Therefore the precision of these codes can be estimated to lie in the range 
0.5$\div$1\%, depending on the adopted experimental cuts. 

The increasing precision reached on the experimental side during the last years 
led to the development of new dedicated theoretical tools, such as BabaYaga@NLO and
MCGPJ, and the adoption of already well-tested codes, like BHWIDE, the latter
extensively used at the high-energy LEP/SLC colliders for the simulation of the large-angle 
Bhabha process. As already emphasised in Section \ref{NLO-HO}, 
all these three codes include NLO corrections in combination with 
multiple photon contributions and have, therefore, a precision tag of 
$\sim 0.1$\%. As described in the following, the experiments typically use more than 
one generator, to keep the luminosity theoretical error under control through the
comparison of independent predictions.

A list of the MC tools used in the luminosity measurement at meson factories is given 
in Table~\ref{tabmc:1}, which summarises the main ingredients of their formulation for radiative 
corrections and the estimate of their theoretical accuracy.

\begin{table}[t]
\caption{MC generators used for luminosity monitoring at meson factories.}
\label{tabmc:1}
\begin{center}
\begin{tabular}{lll}
%
 \hline
 Generator & Theory & Accuracy
 \\ 
    \hline
BabaYaga  v3.5 &{\small Parton Shower}& $\sim 0.5\div 1\%$ \\ 
    \hline
BabaYaga@NLO & $O (\alpha)+{\rm\small PS}$&$\sim  0.1\%$ \\ 
     \hline
BHAGENF & $O (\alpha)$ & $\sim 1\%$ \\ 
    \hline        
BHWIDE &$O (\alpha)\, {\rm\small YFS}$&$\sim 0.5\% {\tiny({\rm LEP1}})$ \\ 
\hline 
BKQED & $O (\alpha)$& $\sim 1\% $\\ 
\hline
MCGPJ &$O (\alpha)+\mbox{\rm\small SF}$& $< 0.2\%$ \\ 
 \hline
 \end{tabular}
 \end{center}
\end{table}

The basic theoretical and phenomenological features of the different generators are summarised in the following.

\begin{enumerate}

\item BabaYaga v3.5  -- It is a MC generator developed by the Pavia group at the start of the 
DA$\mathrm{\Phi}$NE 
operation using a QED PS approach for the treatment of LL QED 
corrections to luminosity processes and later improved  to account for the interference of radiation emitted by different charged legs in the generation of the momenta 
of the final-state particles. 
The main drawback of  BabaYaga v3.5 is the absence of $O (\alpha)$ non-logarithmic
contributions, resulting in a theoretical precision of $\sim 0.5$\% for large-angle Bhabha scattering and
of about 1\% for $\gamma\gamma$ and $\mu^+ \mu^-$ final states. It is used by the CLEO-c 
collaboration for the study of all the three luminosity processes.

\item BabaYaga@NLO -- It is the presently released version of BabaYaga, based on the 
matching of exact $O (\alpha)$ corrections with QED PS, as described in 
Section \ref{NLO-HO}. The accuracy of the current version is estimated to be at the 0.1\% level for large-angle Bhabha scattering, two-photon and $\mu^+ \mu^-$~\footnote{At present, 
finite mass effects in the virtual corrections to $e^+ e^- \to \mu^+\mu^-$, which should 
be included for precision simulations at the $\mathrm{\phi}$ factories, are not included 
in BabaYaga@NLO.} production. It is presently used by the KLOE and BaBar collaborations, 
and under consideration by the BES-III experiment.
Like  BabaYaga v3.5, BabaYaga@NLO is available at the 
web page of the Pavia phenomenology group\\
 \texttt{www.pv.infn.it/\~\,$\!$hepcomplex/babayaga.html} .

\item BHAGENF/BKQED -- BKQED is the event generator developed by Berends and Kleiss and based
on the classical exact NLO calculations of \cite{Berends:1983rk,Berends:1981fb} for all QED processes. It was 
intensively used at LEP to perform tests of QED through the analysis of
the $e^+ e^- \to \gamma\gamma$ process and is adopted by the BaBar collaboration for
the simulation of the same reaction. BHAGENF is a code 
realised by Drago and  Venanzoni at the beginning of the DA$\mathrm{\Phi}$NE operation to simulate Bhabha
events, adapting the calculations of \cite{Berends:1983rk} to include the contribution of the 
$\mathrm{\phi}$ resonance. 
Both generators lack the effect of HO corrections and, as such, have a precision accuracy of
about 1\%. The BHAGENF code is available at the web address\\
 \texttt{www.lnf.infn.it/\~\,$\!$graziano/bhagenf/bhabha.html}.

\item BHWIDE -- It is a MC code realised in Krakow-Knox\-wille 
 at the time of the LEP/SLC operation and described in~\cite{Jadach:1995nk}. 
In this generator
exact $O (\alpha)$ corrections are matched with the resummation of 
the infrared virtual and real photon contributions through the YFS exclusive exponentiation approach.
According to the authors the precision is 
estimated to be about 0.5\% for c.m. energies around the $Z$ resonance. This accuracy estimate was derived through detailed comparisons of the BHWIDE predictions with those of other LEP tools in the presence 
of the full set of NLO corrections, including purely weak corrections. However, since the latter are phenomenologically unimportant at $e^+e^-$
accelerators of moderately high energies and since the QED theoretical ingredients of BHWIDE are very similar to the formulation of both BabaYaga@NLO and MCGPJ, one can argue that the accuracy of 
BHWIDE for physics at flavour factories is at the level of 0.1\%. It is adopted 
by the KLOE, BaBar and BES collaborations.
The code is available at \\
\texttt{placzek.home.cern.ch/placzek/bhwide/}.

\item MCGPJ -- It is the generator developed by the Dubna-Novosibirsk collaboration and 
used at the VEPP-2M collider. This program includes exact $O (\alpha)$ corrections supplemented
with HO LL contributions related to the emission of collinear photon jets
and taken into account through analytical QED collinear SF, as described 
in Section \ref{NLO-HO}. The theoretical 
precision is estimated to be better than 0.2\%. The generator is available at the web address \\
\texttt{cmd.inp.nsk.su/\~\,$\!$sibid/} .

\end{enumerate}

It is worth noticing that the theoretical uncertainty of the most accurate generators based on the matching of exact NLO with LL resummation starts at the level of 
$O (\alpha^2)$ NNL contributions, as
far as photonic corrections are concerned. Other sources of error affecting their physical precision are discussed in detail in Section \ref{TH}.

\subsection{Numerical results \label{NUMERICS}}

Before showing the results 
which enable us to settle the
technical and theoretical accuracy of the generators, it is worth discussing the impact
of various sources of radiative corrections implemented in the programs used
in the experimental analysis. This allows one to understand which corrections are strictly
necessary to achieve a precision at the per mill level for both the 
calculation of integrated cross sections and the simulation of more exclusive distributions.

\subsubsection{Integrated cross sections \label{xsections}}

The first set of phenomenological results about radiative corrections refer to the Bhabha cross 
section, as obtained by means of the code BabaYaga@NLO, according to different
perturbative and precision levels. 
In Table~\ref{tabnum:1} we show 
the values for the Born cross section $\sigma_0$, the $O (\alpha)$ PS and exact cross section, 
$\sigma_{\alpha}^{\textrm{PS}}$ and $\sigma_{\alpha}^{\textrm{NLO}}$, respectively, as well as the LL PS cross section $\sigma^{\rm PS}$ and the matched cross section $\sigma_{\textrm{matched}}$. 
Furthermore, the cross section in the presence of the vacuum polarisation correction, 
$\sigma_{0}^{\rm VP}$, is also shown. 
The results correspond to the c.m. energies $\sqrt{s} = 1,4,10\ \textrm{GeV}$ and were 
obtained with the selection criteria of Eq.~(\ref{eq:cutsmf}), 
but for an angular acceptance of $55^\circ \leq \theta_{\pm} \leq 125^\circ$ resembling realistic 
data taking at meson factories.
One should keep in mind that the cuts of  Eq.~(\ref{eq:cutsmf}) tend to
single out quasi-elastic Bhabha events and that the energy of final state electron/positron
corresponds to a so-called ``bare'' event selection (i.e. without photon recombination), which 
corresponds to what is done in 
practice at flavour factories. In particular the rather stringent energy and acollinearity 
cuts enhance the impact of soft and collinear radiation with respect to a more inclusive setup.

\begin{table}[t]
\caption{Bhabha cross section (in nb) at meson factories 
according to different precision levels and using the cuts of  
Eq.~(\ref{eq:cutsmf}), 
but with an angular acceptance of 
 $55^\circ \leq \theta_{\pm} \leq 125^\circ$. The numbers in parentheses are 1$\sigma$ 
 MC errors.}
\label{tabnum:1}
\begin{center}
\begin{tabular}{clll}
\hline
        $\sqrt{s} (\textrm{GeV})$        &     1.02 &
      4   &     10 \\
      \hline
      $\sigma_{0}   $      &     $529.4631(2)$  &
      $44.9619(1)$   &     $5.5026(2)$    \\
      \hline
      $\sigma_{0}^{\rm VP}$      &     $542.657(6)$  &
      $46.9659(1)$   &     $5.85526(3)$    \\
      \hline
      $\sigma_{\rm NLO}$    &     $451.523\left (6\right )$  &
      $37.1654\left (6\right )$  &     $4.4256\left (2\right )$    \\
      \hline
      $\sigma^{\rm PS}_\alpha$     &     $454.503\left (6\right )$  &
      $37.4186\left (6\right )$  &     $4.4565\left (1\right )$    \\
      \hline
      $\sigma_{\rm matched}$    &     $455.858\left (5\right )$  &
      $37.6731\left (4\right )$   &     $4.5046\left (3\right )$    \\
      \hline
      $\sigma^{\rm PS}$    &     $458.437\left (4\right )$  &
      $37.8862\left (4\right )$   &     $4.5301\left (2\right )$    \\
      \hline
\end{tabular}
\end{center}
\end{table}

From these cross section values, it is possible to calculate the relative effect of various 
corrections, namely the contribution of vacuum polarisation and 
 exact $O (\alpha)$ QED corrections, 
of non-logarithmic (NLL) terms entering the $O (\alpha)$ cross section, 
of HO corrections in the $O (\alpha)$ matched PS scheme, and finally of 
NNL effects beyond order $\alpha$ largely dominated by $O (\alpha^2 L)$ 
contributions.
The above corrections are shown in Table~\ref{tabnum:2} in per cent and 
can be derived from the cross section results of Table~\ref{tabnum:1} with the following definitions:

\begin{eqnarray*}
   \delta_{\rm VP}       &\equiv& \frac{\sigma_{0}^{\rm VP}-\sigma_{0}}{\sigma_{0}}, ~~~~~~~~~~~~~~
   \delta_{\alpha}    \equiv  \frac{\sigma_{\rm NLO}-\sigma_0}{\sigma_0} , \nonumber\\
\delta_{\alpha}^{\rm NLL} &\equiv&
                              \frac{\sigma_{\rm NLO}-\sigma^{\rm PS}_\alpha}{\sigma_{\rm NLO}} ,~~~~~~~~
   \delta_{\rm HO}       \equiv \frac{\sigma_{\rm matched}-\sigma_{\rm NLO}}{\sigma_{\rm NLO}} ,
   \nonumber\\
   \delta_{\alpha^2L}&\equiv& \frac{\sigma_{\rm matched}-\sigma_{\rm NLO} -\sigma^{\rm PS}+\sigma^{\rm PS}_\alpha}{\sigma_{\rm NLO}} . \nonumber
\label{eq:deltas}                                   
\end{eqnarray*}

From Table~\ref{tabnum:2}  it can be seen that $O (\alpha)$ corrections decrease the Bhabha cross section
by about 15$\div$17\% at the $\mathrm{\phi}$ and $\tau$-charm factories, and 
by about 20\% at the $B$ factories. Within the full set
of $O (\alpha)$ corrections, non-logarithmic terms are of the order of 0.5\%, as expected almost independent of the 
c.m. energy, and with a mild dependence 
on the angular acceptance cuts 
due to box and interference contributions. The effect of HO corrections due to multiple photon
emission is about 1\% at the $\mathrm{\phi}$ and $\tau$-charm factories and reaches about 2\% at the $B$ factories. The 
contribution of (approximate) $O (\alpha^2 L)$ corrections is at the 0.1\% level, 
while vacuum polarisation increases the cross section by about 2\% around 1 GeV, and by 
about 5\% and 6\% at 4~GeV and 10~GeV, respectively. Concerning the latter correction the 
non-perturbative hadronic contribution to the running of $\alpha$ was parameterised in terms of the 
 HADR5N routine \cite{Jegerlehner:1985gq,Burkhardt:1989ky,Jegerlehner:2006ju} included in BabaYaga@NLO both in the 
 LO and NLO diagrams. We have checked that 
the results obtained for the vacuum polarisation correction in terms of the 
parametrisation \cite{rintpl:2008AA} agree at the $10^{-4}$ level with those 
obtained with HADR5N, as shown in detail in Section \ref{TH}. Those routines return a
data driven error, thus affecting the theoretical precision of the calculation of the Bhabha cross 
cross section as will be discussed in Section \ref{CONCLUSIONS}.

Analogous results for the size of radiative 
corrections to the process $e^+ e^- \to \gamma\gamma$ are given in Table~\ref{tabnum:3}
\cite{Balossini:2008xr}. 
They were obtained using BabaYaga@NLO, according 
to the experimental cuts of Eq.~(\ref{eq:cutsmfg}) for the c.m. energies 
$\sqrt{s} = 1, 3, 10$~GeV.

\begin{table}[phtb]
\caption{Relative size of different sources of corrections (in per cent) to the large-angle Bhabha 
cross section for typical selection cuts at $\mathrm{\phi}$, $\tau$-charm and $B$ factories.}
\label{tabnum:2}
\begin{center}
\begin{tabular}{clll}
      \hline
                                      $\sqrt{s} \mathrm{(GeV)}$ &     1.02  &
      4.   &      10.   \\
      \hline
      $\delta_{\alpha}^{}$                          &
      $-14.73$       &     $ -17.32$       &  $-19.57$       \\
      \hline
      $\delta_{\alpha}^{\rm NLL}$                  &
      $-0.66$       &     $-0.68$       &  $-0.70$       \\
      \hline
      $\delta_{\mathrm{HO}}$                      &
      $0.97$        &     $1.35$        &  $1.79$        \\
      \hline
      $\delta_{\alpha^{2}L}$                    &
      $0.09$        &     $0.09$        &  $0.11$        \\
      \hline
      $\delta_{\mathrm{VP}}$                      &
      $2.43$       &     $4.46$        &  $6.03$        \\
      \hline
      \end{tabular}
\end{center}
\end{table}

\begin{table}[!htbp]
\caption{Photon pair production cross sections (in nb) at different 
accuracy levels and relative corrections (in per cent) for the setup 
of Eq.~(\ref{eq:cutsmfg}) and the c.m. energies $\sqrt{s} = 1, 3, 10$~GeV.}
\label{tabnum:3}
\begin{center}
\begin{tabular}{clll}
\hline
$\sqrt{s}\ \left(\textrm{GeV}\right)$ & $1$ & $3$ & $10$ \\
\hline
$\sigma_0$ & $137.53$ & $15.281$ & $1.3753$ \\
\hline
$\sigma_{\rm NLO}$ & $129.45$ & $14.211$ & $1.2620$ \\
\hline
$\sigma_{\alpha}^{\rm PS}$ & $128.55$ & $14.111$ & $1.2529$ \\
\hline
$\sigma_{\rm matched}$ & $129.77$ & $14.263$ & $1.2685$ \\
\hline
$\sigma^{\rm PS}$ & $128.92$ & $14.169$ & $1.2597$ \\
\hline
$\delta_{\alpha}$ & $-5.87$ & $-7.00$ & $-8.24$ \\
\hline
$\delta_{\alpha}^{\rm NLL}$ & $0.70$ & $0.71$ & $0.73$ \\
\hline
$\delta_{\rm HO}$ & $0.24$ & $0.37$ & $0.51$ \\
\hline
\end{tabular}
\end{center}
\end{table}

The numerical errors coming from the MC integration are not shown in Table~\ref{tabnum:2} because
they are beyond the quoted digits. From Table~\ref{tabnum:2} it can be seen that the exact $O(\alpha)$ corrections 
lower the Born cross section by about $5.9\%$ (at the $\mathrm{\phi}$ resonance), $7.0\%$ 
(at $\sqrt{s} = 3$~GeV) and $8.2\%$ (at the $\Upsilon$ resonance).  
The effect due to $O (\alpha^{n}L^{n})$ 
(with $n \geq 2$) terms is quantified by the contribution $\delta_{\rm HO}$, which is 
a positive correction of about $0.2\%$ (at the $\mathrm{\phi}$ resonance), $0.4\%$ ($\tau$-charm factories)
and $0.5\%$ (at the $\Upsilon$ resonance), and therefore important in the light of the per mill accuracy aimed at. 
On the other hand, also next-to-leading $O (\alpha)$ corrections, quantified
by the  contribution $\delta_{\alpha}^{\rm NLL}$, are necessary at the precision level 
of 0.1\%, since their contribution is of about $0.7\%$ almost independent of the c.m. energy. 
To further corroborate
the precision reached in the cross section calculation of 
$e^+ e^- \to \gamma\gamma$, we also evaluated the effect due to the 
most important sub-leading $O (\alpha^2)$ photonic corrections given by 
order $\alpha^2 L$ 
contributions. It turns out that the effect due to  $O (\alpha^2 L)$ corrections 
does not exceed the 0.05\% level. Obviously, the contribution of vacuum polarisation is absent in $\gamma\gamma$ 
production. This is an advantage for particularly precise predictions, 
as the uncertainty associated with the
hadronic part of vacuum polarisation does not affect the cross section calculation.

\subsubsection{Distributions \label{distributions}}

\begin{figure}
\begin{center}
\resizebox{0.475\textwidth}{!}{%
\includegraphics{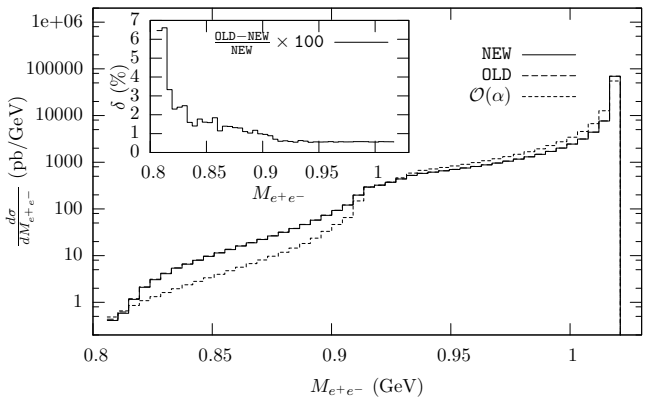}
}
\caption{Invariant mass distribution of the Bhabha process at KLOE, according to BabaYaga v3.5 
(OLD), BabaYaga@NLO (NEW) 
and an exact NLO calculation. The inset shows the relative effect of NLO corrections, given by the difference of BabaYaga v3.5 and BabaYaga@NLO predictions. From 
 \cite{Balossini:2006wc}.} 
\label{fignum:1}
\end{center}
\end{figure}

\begin{figure}
\begin{center}
\resizebox{0.475\textwidth}{!}{%
\includegraphics{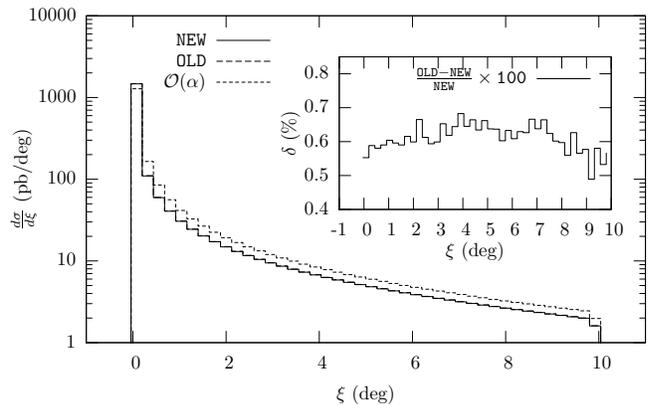}
}
\caption{Acollinearity distribution of the Bhabha process at KLOE, according to BabaYaga v3.5 
(OLD) and BabaYaga@NLO (NEW). The inset shows the relative effect of NLO corrections, given by the difference of BabaYaga v3.5 and BabaYaga@NLO predictions. From  \cite{Balossini:2006wc}.}
\label{fignum:2}
\end{center}
\end{figure}

Besides the integrated cross section, various differential 
cross sections are used by the
experimentalists to monitor the collider luminosity. 
In Figs.~\ref{fignum:1} and \ref{fignum:2} we show two
distributions which are particularly sensitive to the details of
photon radiation, i.e.  the $e^+ e^-$ invariant mass and acollinearity
distribution, in order to quantify the size of NLO and HO corrections. The distributions
are obtained according to the exact $O (\alpha)$ calculation and 
with the two BabaYaga
versions, BabaYaga v3.5 and BabaYaga@NLO. From Figs.~\ref{fignum:1} and \ref{fignum:2} it can be clearly
seen that multiple photon
corrections introduce significant deviations with respect to 
an $O (\alpha)$ simulation, especially in the hard tails of
the distributions where they amount to 
several per cent. To make the 
contribution of exact  $O (\alpha)$ non-logarithmic terms clearly visible, the inset shows the relative 
differences between the predictions of BabaYaga v3.5 (denoted as OLD) and BabaYaga@NLO 
(denoted as NEW). Actually, as discussed in Section \ref{NLO-HO}, these differences mainly come from 
non-logarithmic NLO contributions and to a smaller extent
 from $O (\alpha^2 L)$ terms. Their effect is flat and at 
the level of 0.5\% for the acollinearity distribution, while they reach the several per cent level 
in the hard tail of the invariant mass distribution.

\begin{figure}
\begin{center}
\resizebox{0.475\textwidth}{!}{%
\includegraphics{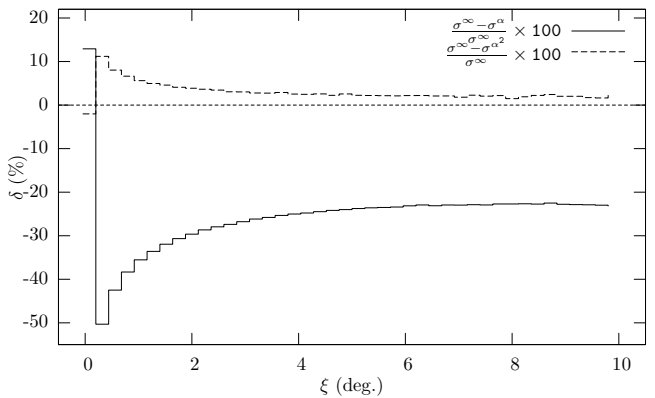}
}
\caption{Relative effect of HO corrections $\alpha^2 L^2$ and
$\alpha^n L^n$ ($n \geq 3$) to the acollinearity distribution
of the Bhabha process at KLOE. From  \cite{Balossini:2006wc}.}
\label{fignum:3}
\end{center}
\end{figure}

It is also worth noticing that LL radiative corrections
beyond $\alpha^2$ can be quite important for accurate 
simulations, at least when considering
differential distributions. This means that even
with a complete NNLO calculation at hand it would be
desirable to match such corrections with the resummation of all the remaining
LL effects. In Fig.~\ref{fignum:3},
the relative effect of  HO corrections beyond $\alpha^2$ dominated by the $\alpha^3$ 
contributions (dashed line) is shown in comparison with that of the $\alpha^2$ 
corrections (solid line) on the acollinearity
distribution for the Bhabha process at DA$\mathrm{\Phi}$NE. As can be seen, 
the $\alpha^3$ effect can be as large as
$10\%$ in the phase space region of soft photon emission, corresponding to
small acollinearity angles  with almost back-to-back final state fermions.

\begin{figure}
\begin{center}
\resizebox{0.475\textwidth}{!}{%
\includegraphics{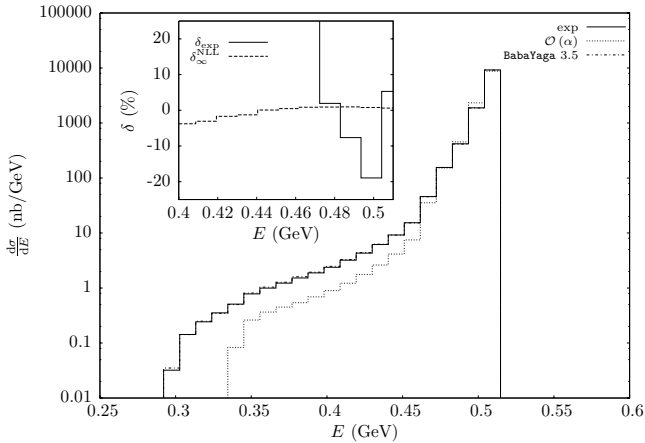}
}
\caption{Energy distribution of the most energetic photon in the process 
$e^+ e^- \to \gamma\gamma$, according to the PS matched 
with $O (\alpha)$ corrections denoted as exp (solid line), the exact $O (\alpha)$ 
calculation (dashed line) and the pure all-order PS as in BabaYaga v3.5 
(dashed-dotted line). lnset: relative effect (in per cent) of multiple photon corrections (solid line) 
and of non-logarithmic contributions of the matched PS algorithm (dashed line). 
From \cite{Balossini:2008xr}.}
\label{fignum:4}
\end{center}
\end{figure}

\begin{figure}
\begin{center}
\resizebox{0.475\textwidth}{!}{%
\includegraphics{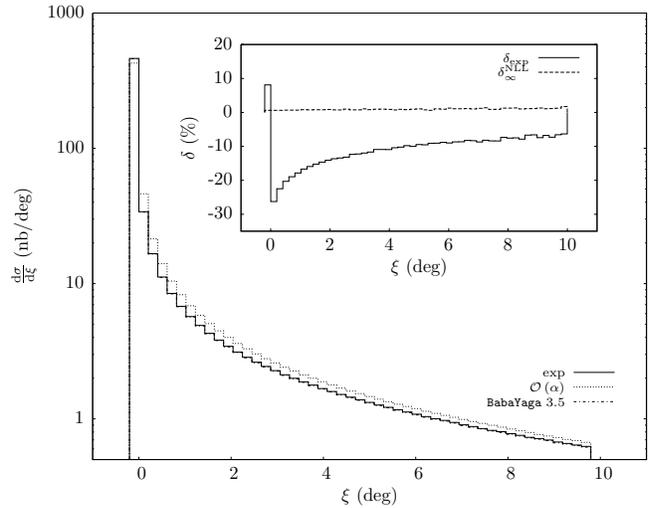}
}
\caption{Acollinearity distribution for the process 
$e^+ e^- \to \gamma\gamma$, according to the PS matched 
with $O (\alpha)$ corrections denoted as exp (solid line), the exact $O (\alpha)$ 
calculation (dashed line) and the pure all-order PS as in BabaYaga v3.5 
(dashed-dotted line). lnset: relative effect (in per cent) of multiple photon corrections (solid line) 
and of non-logarithmic contributions of the matched PS algorithm (dashed line). From \cite{Balossini:2008xr}.}
\label{fignum:5}
\end{center}
\end{figure}

Concerning the process $e^+ e^- \to \gamma\gamma$ 
 we show in  Fig.~\ref{fignum:4}  the energy distribution 
of the most energetic photon,  while the acollinearity distribution of 
the two most energetic photons is represented in Fig.~\ref{fignum:5}. The distributions 
refer to exact  $O (\alpha)$ corrections matched with 
the PS algorithm (solid line), to the exact NLO calculation 
 (dashed line) and to all-order pure PS predictions of BabaYaga v3.5
 (dashed-dotted line). In the inset of each plot, the relative effect due to multiple photon contributions
 ($\delta_{\rm HO}$) and non-logarithmic terms entering the improved PS algorithm
($\delta_{\alpha}^{\rm NLL}$) is also shown, according to the definitions given in 
Eq.~(\ref{eq:deltas}). 

For the energy distribution of the most energetic photon 
particularly pronounced effects due to exponentiation
are present. In the statistically 
dominant region, 
HO corrections reduce the $O (\alpha)$ 
distribution by about 20\%, while they give rise to a significant hard tail close
to the energy threshold of 0.3$\sqrt{s}$ as a consequence of the higher photon multiplicity 
of the resummed calculation with respect to the fixed-order NLO prediction. 
Needless to say, the relative effect of multiple photon corrections below about 0.46 GeV not shown in the inset is finite but huge. This representation with the inset was chosen to make also the contribution of $O (\alpha)$ non-logarithmic terms visible, which otherwise
would be hardly seen in comparison with the multiple photon corrections. Concerning
the acollinearity distribution, the contribution of higher-order corrections is positive and of about
10\% for quasi-back-to-back photon events, whereas it is negative and decreasing
from $\sim -30$\% to $\sim -10$\% for increasing acollinearity values. As far as the contributions of non-logarithmic effects
dominated by next-to-leading  $O (\alpha)$
corrections are concerned, they contribute at the level of several per mill for the acollinearity distribution, while they lie in 
the range of several per cent for the energy distribution.

As a whole, the results of the present Section emphasise that, for a 0.1\% theoretical precision in the calculation of both the cross sections and distributions, both exact $O (\alpha)$ and HO photonic corrections are necessary, as well as the running of $\alpha$.

\subsection{Tuned comparisons \label{TUNED}}

The typical procedure followed in the literature to establish the technical precision of 
the theoretical tools is to perform tuned comparisons between 
the predictions of independent 
programs using the same
set of input parameters and experimental cuts. This strategy was initiated 
in the 90s during the CERN workshops 
for precision physics at LEP and is still in use when considering processes of interest
for physics at hadron colliders
demanding particularly
accurate theoretical calculations. The tuning procedure is a key step in the validation of generators, 
because it allows to check that the different details entering the complex structure of 
the generators, e.g. the implementation of radiative corrections, event selection routines, MC integration and 
event generation, are under control, and to fix possible mistakes. 

The tuned comparisons discussed in the following were performed switching off the
 vacuum polarisation  correction to the Bhabha scattering cross section. 
Actually, the generators implement the non-perturbative hadronic contribution to the 
running of $\alpha$ according to different parameterisations, which differently affect the 
cross section prediction (see Section \ref{sec:4} for discussion). Hence, this simplification is introduced to avoid possible bias in the interpretation of the results and allows to disentangle the effect of pure QED corrections. 
Also, in order to provide useful results for the experiments, the comparisons take into account {\it realistic} event selection cuts. 

The present Section is a merge of results available in the literature 
\cite{Balossini:2006wc} with those of new studies. The results refer to the Bhabha process at the energies of 
$\mathrm{\phi}$, $\tau$-charm and $B$ factories. No tuned comparisons for the two photon
production process have been carried out.

\subsubsection{$\mathrm{\phi}$ and $\tau$-charm factories \label{phic}}

\begin{figure}
\begin{center}
\resizebox{0.475\textwidth}{!}{%
\includegraphics{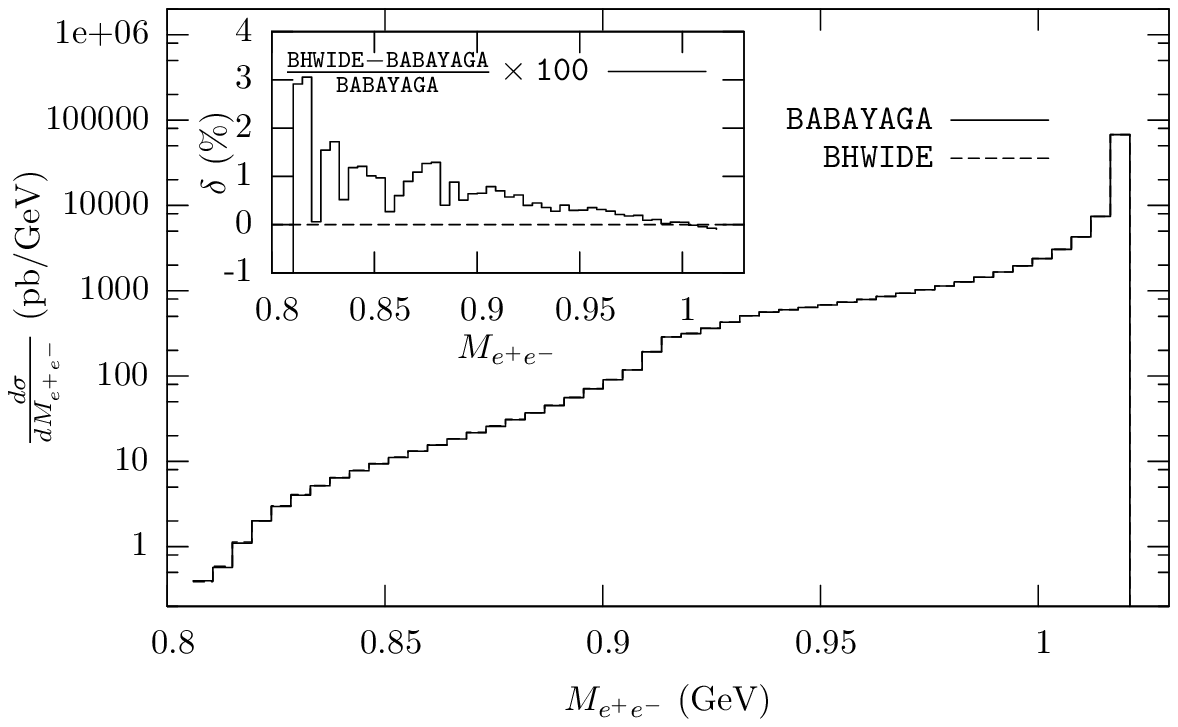}
}
\caption{Invariant mass distribution of the Bhabha process 
according to BHWIDE and BabaYaga@NLO,  for the KLOE experiment at 
DA$\mathrm{\Phi}$NE, and relative differences of the program predictions (inset). From \cite{Balossini:2006wc}.}
\label{figtun:1}
\end{center}
\end{figure}

\begin{figure}
\begin{center}
\resizebox{0.475\textwidth}{!}{%
\includegraphics{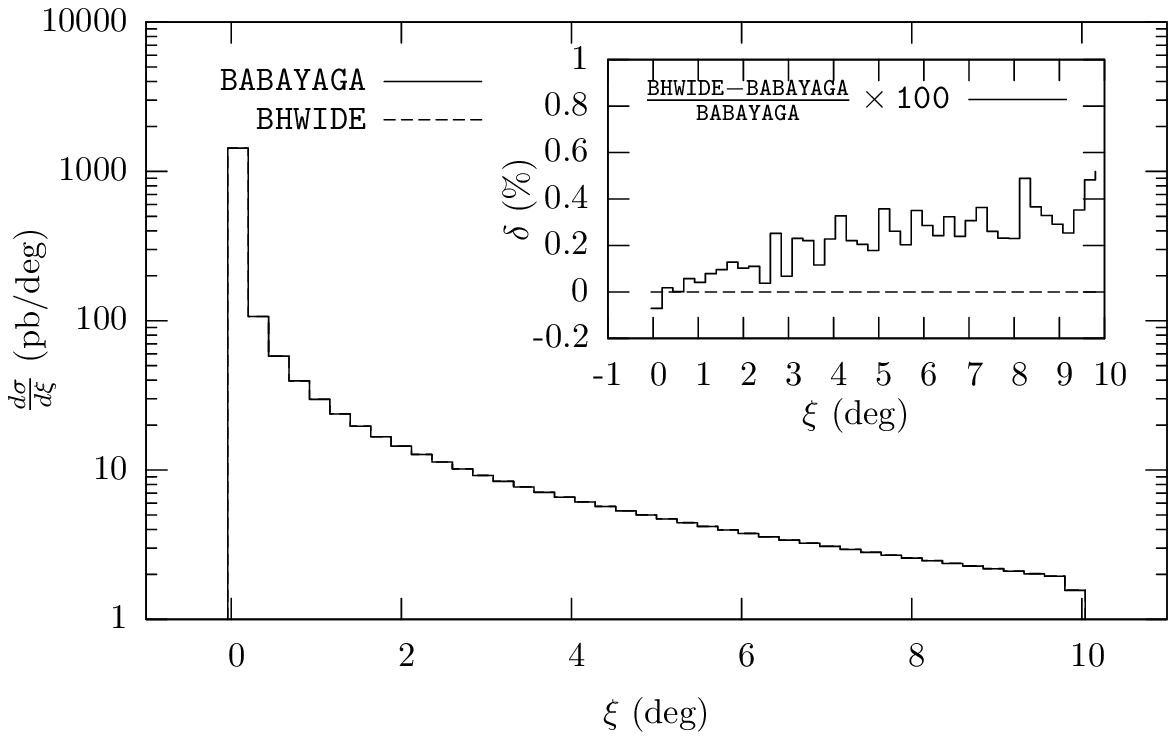}
}
\caption{Acollinearity distribution of the Bhabha process 
according to BHWIDE and BabaYaga@NLO,  for the KLOE experiment at 
DA$\mathrm{\Phi}$NE, and relative differences of the program predictions (inset). From \cite{Balossini:2006wc}.}
\label{figtun:2}
\end{center}
\end{figure}

First we show comparisons between BabaYaga@NLO and BHWIDE according to the 
KLOE selection cuts of Eq.~(\ref{eq:cutsmf}), considering 
also the angular range $20^\circ \leq \vartheta_{\pm} \leq 160^\circ$ 
for cross section results. The predictions of the two codes are reported in 
Table~\ref{tabtun:1} for the two 
acceptance cuts together with their relative deviations. As can be seen the agreement is excellent, 
the relative deviations being well below the 0.1\%. Comparisons between BabaYaga@NLO and 
BHWIDE at the level of differential distributions are
given in Figs.~\ref{figtun:1} and~\ref{figtun:2} where the inset shows the relative deviations
between the predictions of the two codes. As can be seen there is 
very good agreement between the two generators, and the predicted distributions 
appear at a first sight almost indistinguishable. Looking in more detail, there is a 
relative difference of a few per mill for the acollinearity
distribution (Fig.~\ref{figtun:2}) and
of a few per cent for the invariant mass  (Fig.~\ref{figtun:1}), but
only in the very hard tails, where the fluctuations observed are due to limited MC statistics. These
configurations however give a negligible contribution
 to the integrated cross section, a factor $10^{3} \div 10^{4}$ smaller than that around the very dominant peak 
regions. In fact these differences on differential distributions translate into 
agreement on the cross section values well below the one per mill, 
as shown in Table~\ref{tabtun:1}.  

Similar tuned comparisons were performed between the results of 
BabaYaga@NLO, BHWIDE and MCGPJ in the presence of cuts modelling the event selection 
criteria of the CMD-2 experiment at the VEPP-2M collider, for a c.m. energy of $\sqrt{s} = 900$~MeV. 
The cuts used in this case are
\begin{eqnarray}
&& | \theta_- + \theta_+ - \pi | \leq \Delta\theta , \quad 1.1 \leq (\theta_+ - \theta_- + \pi)/2 \leq \pi-1.1, \nonumber\\
&& | |\phi_- + \phi_+|-\pi | \leq 0.15 , \nonumber\\
&& p_- \sin(\theta_-) \geq 90~{\rm MeV} , \qquad \, \qquad p_+ \sin(\theta_+) \geq 90~{\rm MeV},  \nonumber\\
&& (p_- + p_+)/2 \geq 90~{\rm MeV} ,
\label{eq:acmd2}
\end{eqnarray}
where $\theta_-,\theta_+$ are the electron/positron polar angles, respectively, 
$\phi_{\pm}$ their azimuthal angles, and $p_{\pm}$ 
the moduli of their three-momenta. $\Delta\theta$ stands for an acollinearity
cut. 

\begin{table}
\caption{Cross section predictions [nb] of BabaYaga@NLO and BHWIDE for the Bhabha cross section 
corresponding to two different angular acceptances, for 
the KLOE experiment at DA$\mathrm{\Phi}$NE, and their relative differences (in per cent).}
\label{tabtun:1}
\begin{center}
\begin{tabular}{clll}
    \hline
    angular acceptance & BabaYaga@NLO & BHWIDE & $\delta(\%)$ \\ 
    \hline
    $20^\circ \div 160^\circ$ & 6086.6(1) & 6086.3(2)& 0.005\\ 
     \hline 
    $55^\circ \div 125^\circ$ & 455.85(1) & 455.73(1)& 0.030\\ 
    \hline
 \end{tabular}
 \end{center}
\end{table}

\begin{figure}
\begin{center}
\resizebox{0.4\textwidth}{!}{%
\includegraphics{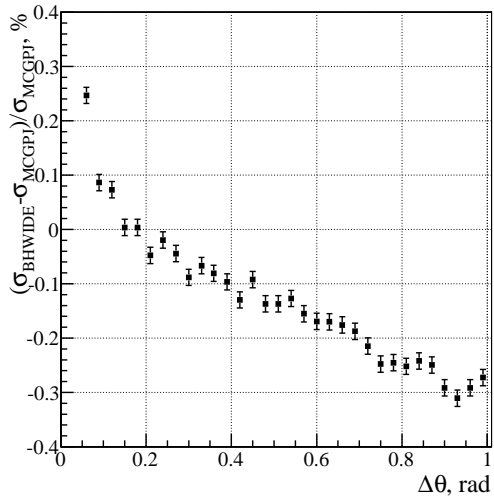}
}
\caption{Relative differences between BHWIDE and MCGPJ Bhabha cross sections as a function 
of the acollinearity cut, for the CMD-2 experiment at VEPP-2M.}
\label{figtun:3}
\end{center}
\end{figure}

\begin{figure}
\begin{center}
\resizebox{0.4\textwidth}{!}{%
\includegraphics{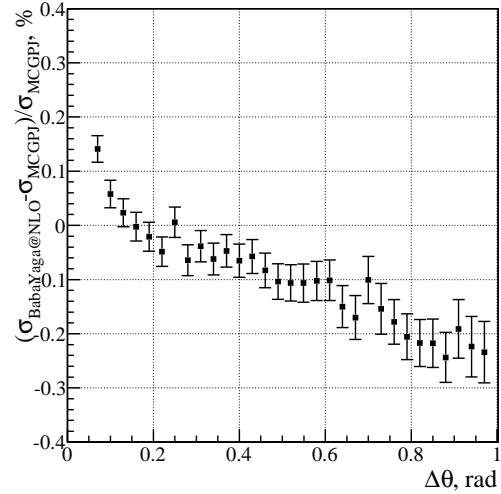}
}
\caption{ Relative differences between BabaYaga@NLO and MCGPJ Bhabha cross sections as a function 
of the acollinearity cut, for the CMD-2 experiment at VEPP-2M.}
\label{figtun:4}
\end{center}
\end{figure}

Figure~\ref{figtun:3} shows the relative differences between the results of BHWIDE and MCGPJ according
to the criteria of Eq.~(\ref{eq:acmd2}), as a function of the acollinearity cut $\Delta\theta$. The
relative deviations between the results of BabaYaga@NLO and MCGPJ for the same cuts
are given in Fig.~\ref{figtun:4}. It can be seen that the predictions of the three generators lie 
within a $0.2\%$ band with differences of $\sim 0.3\%$ for extreme values of the acollinearity cut. 
This agreement can be considered satisfactory since for the acollinearity cut of 
real experimental interest ($\Delta\theta \approx 0.2$~rad) the generators agree 
within one per mill.

A number of comparisons were also performed for a c.m. energy of 3.5~GeV relevant to the experiments 
at $\tau$-charm factories. An example is given in Table~\ref{tabtun:2}
 where the predictions of BabaYaga@NLO and
MCGPJ are compared, using cuts similar to those of Eq.~(\ref{eq:acmd2}) 
and for an acollinearity cut 
of $\Delta\theta = 0.25$~rad. The agreement between the two codes is below
 one per mill. 
Comparisons between the two codes were also done at the level of differential cross 
sections, showing satisfactory agreement in the statistically relevant phase space regions.
Preliminary results \cite{Ping:2009xxxx} for a c.m. energy on top of the $J/\Psi$ resonance show good agreement between BabaYaga@NLO and BHWIDE predictions too.

\begin{table}
\caption{Cross section predictions [nb] of BabaYaga@NLO and MCGPJ for the Bhabha cross section 
at $\tau$-charm factories ($\sqrt{s} = 3.5$~GeV) and their relative difference (in per cent).}
\label{tabtun:2}
\begin{center}
\begin{tabular}{lll}
    \hline
    BabaYaga@NLO & MCGPJ & $\delta(\%)$ \\ 
    \hline
    35.20(2) & 35.181(5)& 0.06 \\ 
     \hline 
 \end{tabular}
 \end{center}
\end{table}

\subsubsection{$B$ factories \label{bfac}}

\begin{figure}
\begin{center}
\resizebox{0.475\textwidth}{!}{%
\includegraphics{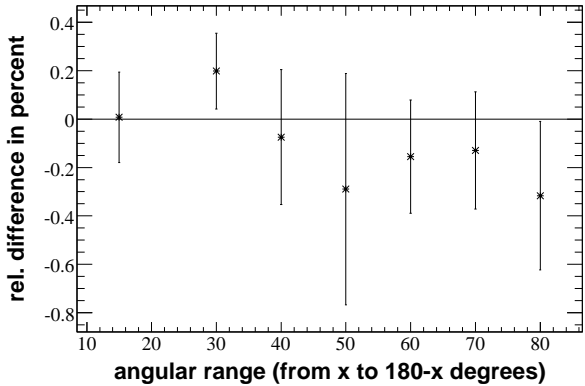}
}
\caption{Relative differences between BabaYaga@NLO and BHWIDE Bhabha cross sections as a function of the angular acceptance cut for the BaBar experiment at PEP-II. From \cite{andreas:2009xxxx}.}
\label{figtun:5}
\end{center}
\end{figure}

Concerning the $B$ factories, a considerable effort was done to establish the level of 
agreement between the generators BabaYaga@NLO and BHWIDE in comparison with BabaYaga v3.5 too.
This study made use of the realistic luminosity cuts 
quoted in Section \ref{4pfinal} for the BaBar experiment. The
cross sections predicted by BabaYaga@NLO and BHWIDE are shown in 
Table~\ref{tabtun:3}, together with
the corresponding relative differences as a function of the considered angular range.  
The latter are also shown in Fig.~\ref{figtun:5},  where the 1$\sigma$ numerical error due to MC statistics 
is also quoted. As can be seen, the two codes agree nicely, the predictions for the 
central value being in general in agreement at the 0.1\% level or statistically compatible whenever a 
two to three per mill difference is present.

\begin{table}
\caption{Cross section predictions [nb] of BabaYaga@NLO and BHWIDE for the Bhabha cross section 
as  a function of the angular selection cuts for the BaBar experiment at PEP-II and 
absolute value of their relative differences (in per cent).}
\label{tabtun:3}
\begin{center}
\begin{tabular}{clll}
    \hline
    angular range (c.m.s.) & BabaYaga@NLO & BHWIDE &$|\delta (\%)|$ \\ 
    \hline
    $15^{\circ}\div 165^{\circ}$ & 119.5(1) & 119.53(8)&0.025\\ 
     \hline 
    $30^{\circ}\div 150^{\circ}$ & 24.17(2) & 24.22(2)&0.207\\
        \hline
    $40^{\circ}\div 140^{\circ}$ & 11.67(3) & 11.660(8)&0.086\\
        \hline
    $50^{\circ}\div 130^{\circ}$ &  6.31(3) & 6.289(4)&0.332\\
        \hline
    $60^{\circ}\div 120^{\circ}$ & 1.928(2) & 1.931(3)&0.141\\
        \hline
    $70^{\circ}\div 110^{\circ}$ & 3.554(6) & 3.549(3)&0.155\\
        \hline
    $80^{\circ}\div 100^{\circ}$ & 0.824(2) & 0.822(1)&0.243\\
    \hline
 \end{tabular}
 \end{center}
\end{table}

\begin{figure}
\begin{center}
\resizebox{0.475\textwidth}{!}{%
\includegraphics{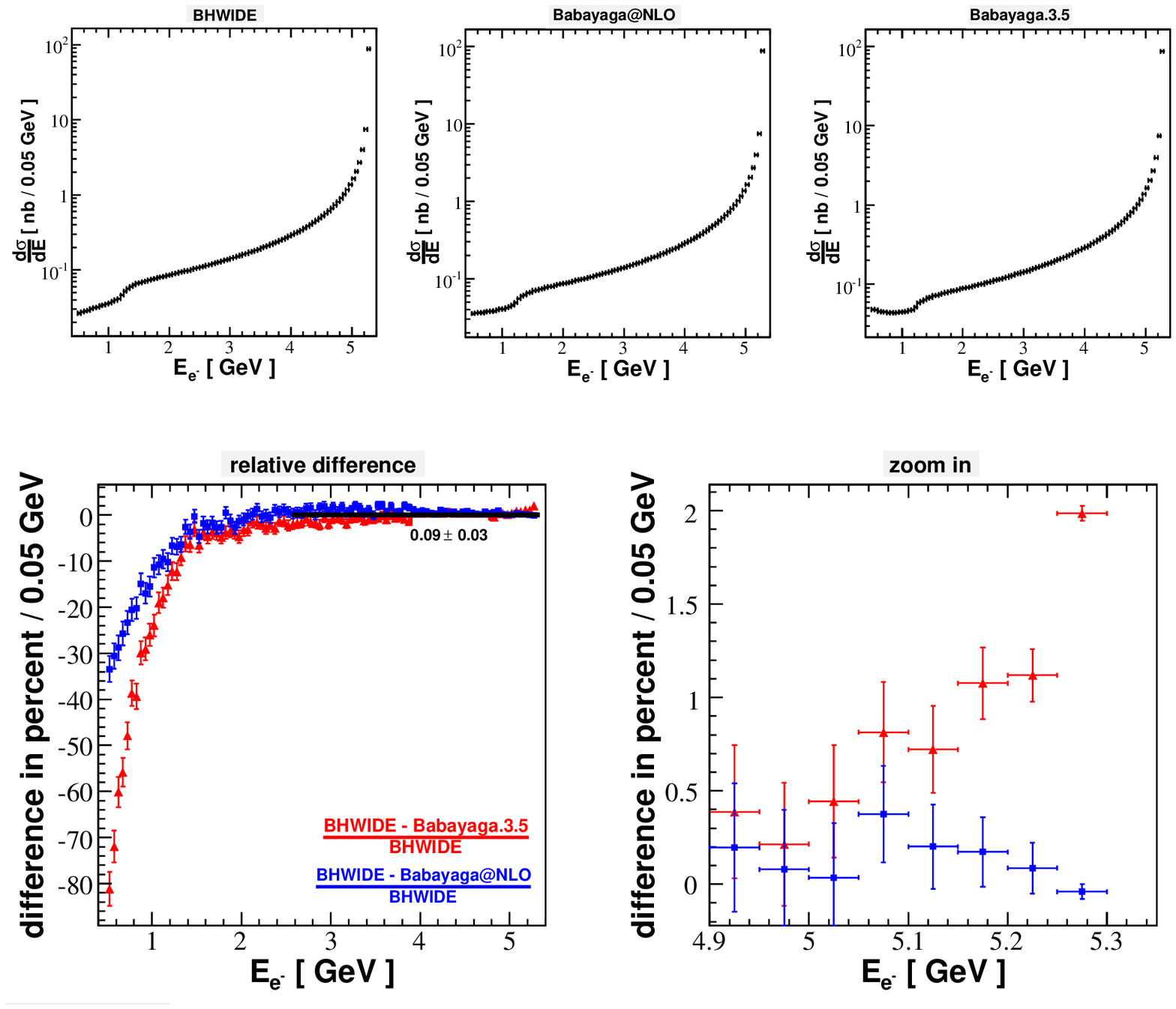}
}
\caption{Electron energy distributions according to BHWIDE, BabaYaga@NLO and BabaYaga v3.5
for the BaBar experiment at PEP-II and relative differences of the 
predictions of the programs. From \cite{andreas:2009xxxx}.}
\label{figtun:6}
\end{center}
\end{figure}

\begin{figure}
\begin{center}
\resizebox{0.475\textwidth}{!}{%
\includegraphics{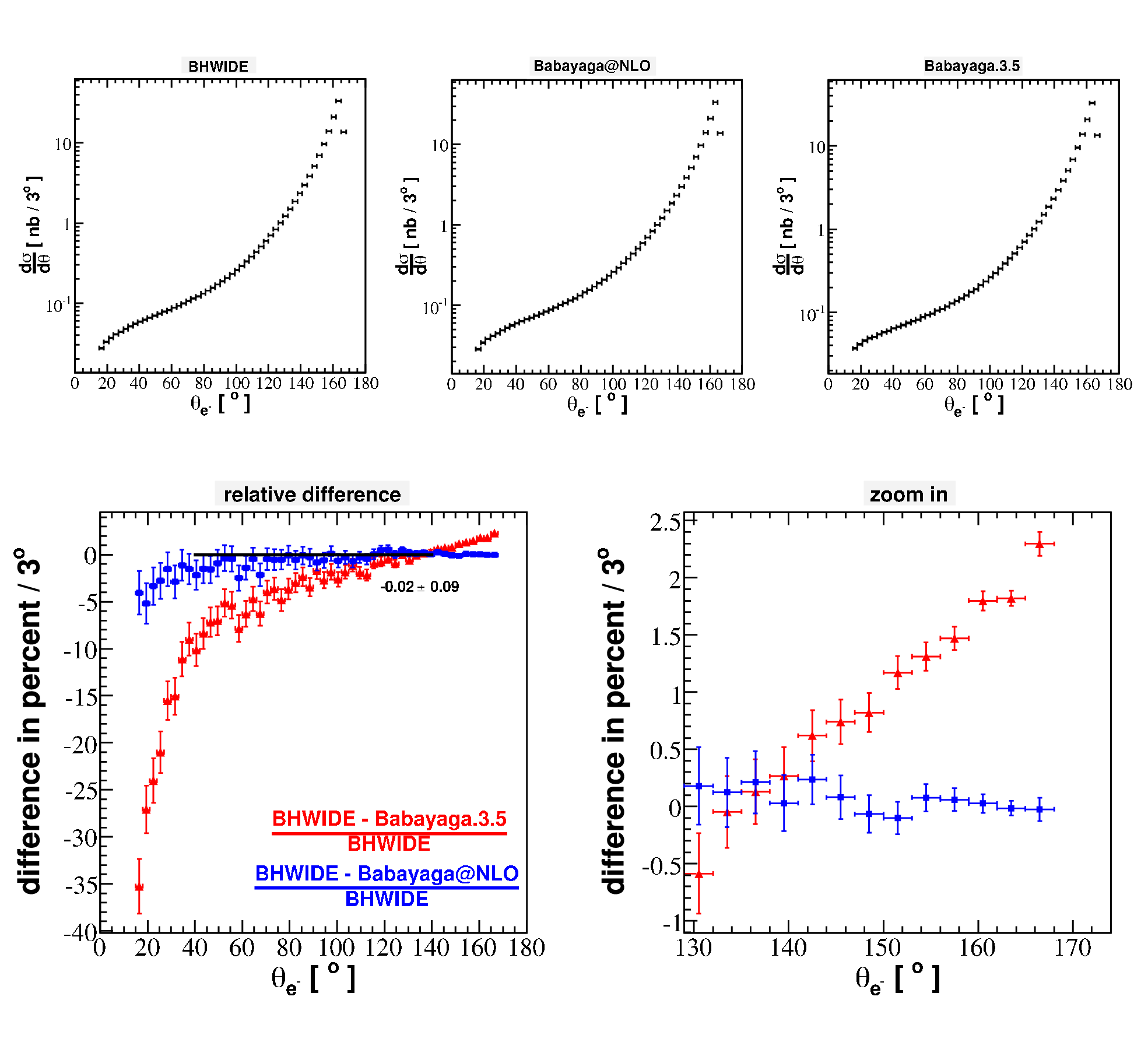}
}
\caption{Electron polar angle distributions according to BHWIDE, BabaYaga@NLO and BabaYaga v3.5
for the BaBar experiment at PEP-II and relative differences of the predictions of the programs. 
From \cite{andreas:2009xxxx}.}
\label{figtun:7}
\end{center}
\end{figure}

\begin{figure}
\begin{center}
\resizebox{0.475\textwidth}{!}{%
\includegraphics{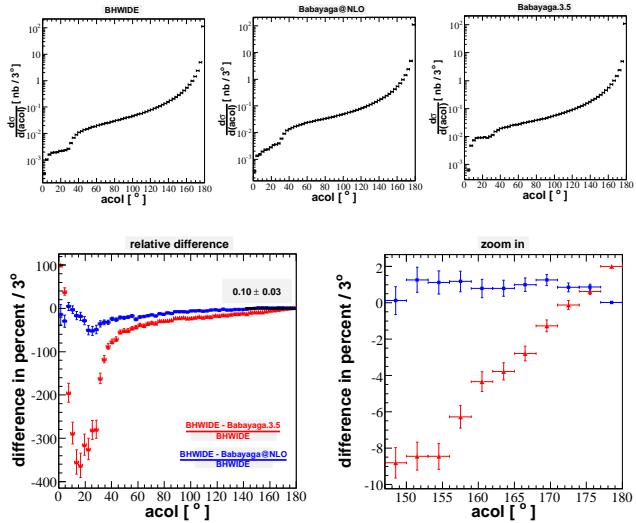}
}
\caption{Acollinearity distributions according to BHWIDE, BabaYaga@NLO and BabaYaga v3.5
for the BaBar experiment at PEP-II and relative differences of the predictions of the programs. 
From \cite{andreas:2009xxxx}.}
\label{figtun:8}
\end{center}
\end{figure}

To further investigate how the two generators compare with each other
 a number of differential cross sections were
studied. The results of this study are shown in Figs.~\ref{figtun:6} and \ref{figtun:7} 
for the distribution of the electron energy and the polar
angle, respectively, and in Fig.~\ref{figtun:8} for the acollinearity. For both the energy and
scattering angle distribution, the two programs agree within the statistical errors showing deviations 
below 0.5\%. For the acollinearity dependence of the cross section, BabaYaga@NLO
and BHWIDE agree within $\sim 1\%$. Therefore, the level of the agreement between the two codes 
around 10~GeV is the same as that observed at 
the $\mathrm{\phi}$ factories.

The main conclusions emerging from the tuned comparisons 
discussed in the present Section can be summarised as follows:

\begin{itemize}

\item The predictions for the Bhabha  cross section of the most precise tools, i.e. BabaYaga@NLO, BHWIDE and MCGPJ, generally 
agree within 0.1\%. If (slightly) lar\-ger differences are present they show up for particularly 
tight cuts
or are due to limited MC statistics. When statistically meaningful discrepancies are observed they 
can be ascribed to the different theoretical recipes for the treatment of radiative corrections and their
technical implementation. For example, as already emphasised, BabaYaga@NLO and 
BHWIDE adopt a fully factorised prescription for the matching of NLO and HO corrections, 
whereas MCGPJ implement some pie\-ces of the 
radiative corrections in additive form. This can give
rise to discrepancies between the programs' predictions, especially in the presence of
tight cuts enhancing the effect of soft radiation. Furthermore, different choices are adopted in the
generators for the scale entering the collinear logarithms in HO corrections 
beyond $O (\alpha)$, 
which are
another possible source of the observed differences. To go beyond the present situation, a 
further nontrivial effort should be done by comparing, for instance, the programs in the 
presence of NLO corrections only (technical test) and by analysing their different 
treatment of the
exponentiation of soft and collinear logarithms. This would certainly shed light on the origin
of the (small) discrepancies still registered at present.

\item Also the distributions predicted by the generators agree well, with relative differences 
below the 1\% level. Slight\-ly larger discrepancies are
only seen in sparsely populated phase space regions corresponding to very hard photon emission 
which do not influence the luminosity measurement noticeably.

\end{itemize}

\subsection{Theoretical accuracy \label{TH}}

As discussed in Section 
\ref{MOTIVATION}, the total luminosity error crucially depends on the theoretical 
accuracy of the MC programs used by the 
experimentalists.  As emphasised in Section 
\ref{MC}, 
some of these generators like BHAGENF, BabaYaga v3.5 and BKQED miss theoretical ingredients 
which are unavoidable for cross section calculations with a precision at the per mill level.
Therefore, they are inadequate for a highly accurate luminosity determination. 
BabaYaga@NLO, BHWIDE and MCGPJ include, however, both NLO and multiple photon corrections, and 
their accuracy aims at a precision tag of 0.1\%. But also these generators are 
affected by uncertainties which must be carefully considered in the light of the very stringent criteria 
of per mill accuracy. The most important components of the theoretical error of 
BabaYaga@NLO, BHWIDE and MCGPJ are mainly due to approximate or partially included pieces of 
radiative corrections and come from the following sources:
\begin{enumerate}

\item The non-perturbative hadronic contributions to the running of $\alpha$. It can be 
reliably evaluated only using the data of the hadron cross section at low energies. Hence, 
the vacuum polarisation correction receives a data driven error which affects in turn the
prediction of the Bhabha cross section, as emphasised in Section \ref{sec:4}.

\item The complete set of $O (\alpha^2)$ QED corrections. In spite of the impressive progress
in this area, as reviewed in Section \ref{NNLO}, an important piece of NNLO corrections, i.e. 
the exact
NLO SV QED corrections to the single hard bremsstrahlung process $e^+ e^- \to e^+ e^-\gamma$,
 is still missing for the full $s+t$ Bhabha process.\footnote{As already remarked and further discussed in the following, the complete calculation of the NLO corrections to hard photon emission in Bhabha scattering was 
 performed during the completion of this report \cite{Actis:2009uq}.} However, partial results 
 obtained for $t$-channel small-angle Bhabha scattering \cite{Jadach:1995hy,Ward:1998ht}
 and large-angle annihilation processes are available \cite{Glosser:2003ux,Jadach:2006fx}.
 
 \item The $O (\alpha^2)$ contribution due to real and virtual (lepton and hadron) pairs. 
 The virtual contributions originate from the NNLO electron, heavy flavour and hadronic loop corrections discussed in Section \ref{NNLO}, while the real corrections are due to the conversion of an 
 external photon into pairs. The latter, as  discussed in Section \ref{4pfinal}, gives rise to a final state with four 
 particles, two of which to be considered as undetected to contribute to the Bhabha signature.

\end{enumerate}

The uncertainty relative to the first point can be estimated by using the routines available in
the literature for the calculation of the non-perturbative hadronic contribution 
$\Delta\alpha^{(5)}_{\rm hadr}(q^2)$ to the running $\alpha$. 
Actually these routines return, in addition to $\Delta\alpha^{(5)}_{\rm hadr}(q^2)$, 
an error $\delta_{\rm hadr}$ on its value. Therefore an estimate of the 
induced error can be simply obtained by computing the Bhabha cross section with
$\Delta\alpha^{(5)}_{\rm hadr}(q^2)\pm\delta_{\rm hadr}$ and taking the
difference as the theoretical uncertainty due to the hadronic
contribution to vacuum polarisation. In Table~\ref{tabth:1}, the Bhabha cross sections, as obtained 
in the presence of the vacuum polarisation correction according to the parameterisations
of \cite{Jegerlehner:1985gq,Burkhardt:1989ky,Jegerlehner:2006ju} (denoted as J) 
and of \cite{rintpl:2008AA} (denoted as HMNT), respectively, are shown for 
$\mathrm{\phi}$, $\tau$-charm and $B$ factories. The applied angular cuts 
refer to the typically adopted acceptance $55^\circ \leq \theta_{\pm} \leq 125^\circ$.

\begin{table}[h]
\caption{Bhabha scattering cross section in the presence of the vacuum polarisation 
correction, according to \cite{Jegerlehner:1985gq,Burkhardt:1989ky,Jegerlehner:2006ju} (J) and 
\cite{rintpl:2008AA} (HMNT), at meson factories. The 
notation J$_-$/HMNT$_-$, J/HMNT and J$_+$/HMNT$_+$ indicates minimum, central and maximum 
value of the two parametrisations.}
\label{tabth:1}
\begin{center}
\begin{tabular}{clll}
\hline
Parametrisation & $\mathrm{\phi}$ & $\tau$-charm& $B$ \\
\hline
J$_-$ & 542.662(4) & 46.9600(1) & 5.85364(2) \\
\hline
$\!\!\!\!\!$ J & 542.662(4) & 46.9658(1) & 5.85529(2) \\
\hline
J$_+$ & 542.662(4) & 46.9715(1) & 5.85693(2) \\
\hline
HMNT$_-$  & 542.500(5) & 46.9580(1) & 5.85496(1) \\
\hline
$\!\!\!\!$HMNT          & 542.391(5) & 46.9638(1) & 5.85621(1) \\
\hline
HMNT$_+$ & 542.283(5) & 46.9697(1) & 5.85746(2) \\
\hline
\end{tabular}
\end{center}
\end{table}

From Table~\ref{tabth:1} it can be seen that the two treatments of $\Delta\alpha^{(5)}_{\rm hadr}(q^2)$ 
induce effects on the Bhabha cross section in very good agreement, the relative differences 
between the central values being 0.05\% ($\mathrm{\phi}$ factories), 0.005\% ($\tau$-charm 
factories) and 
0.02\% ($B$ factories). This can be understood in terms of the dominance of $t$-channel 
exchange for large-angle Bhabha scattering at meson factories. Indeed, the two routines provide 
results in excellent agreement for space-like momenta, as we explicitly checked, whereas 
differences in the predictions show up for time-like momenta which, however, contribute only marginally to the
Bhabha cross section. Also the spread between the minimum/maximum values and
the central one as returned by the two routines agrees rather well, 
also a consequence of the dominance of 
$t$-channel exchange. This spread amounts to a few units in $10^{-4}$ and is presented in detail
in Table~\ref{tabcon:1} in the next Section.

\begin{figure}
\begin{center}
\resizebox{0.475\textwidth}{!}{%
\includegraphics{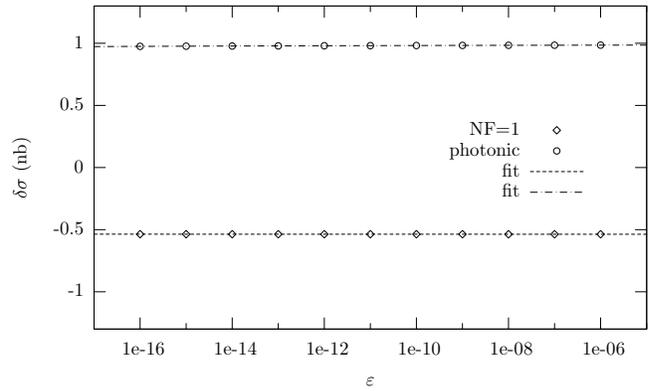}
}
\caption{Absolute differences (in nb) between the $\sigma^{\alpha^2}_{\rm SV}$ 
prediction of BabaYaga@NLO and 
the NNLO calculations of the photonic corrections
\cite{Penin:2005kf,Penin:2005eh} (photonic) and of the electron loop corrections
\cite{Bonciani:2005im,Bonciani:2003te,Bonciani:2003cj0,Bonciani:2004gi,Bonciani:2004qt} ($N_F = 1$)
 as a function of the infrared regulator $\epsilon$ for typical KLOE cuts. From 
 \cite{Balossini:2006wc}.}
\label{figth:1}
\end{center}
\end{figure}

\begin{figure}
\begin{center}
\resizebox{0.475\textwidth}{!}{%
\includegraphics{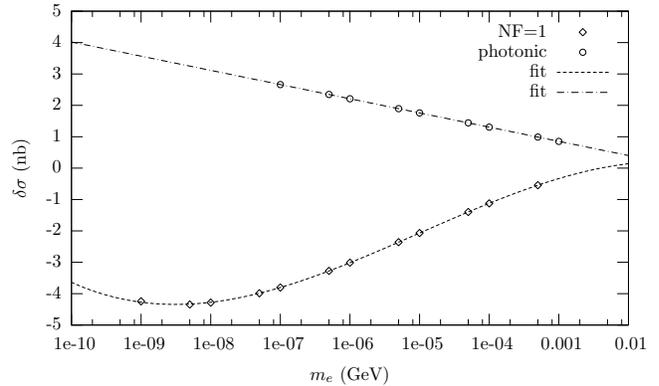}
}
\caption{Absolute differences (in nb) between the $\sigma^{\alpha^2}_{\rm SV}$ 
prediction of BabaYaga@NLO and 
the NNLO calculations of the photonic corrections 
\cite{Penin:2005kf,Penin:2005eh} (photonic) and of the electron loop corrections 
\cite{Bonciani:2005im,Bonciani:2003te,Bonciani:2003cj0,Bonciani:2004gi,Bonciani:2004qt} ($N_F = 1$) as a function of a fictitious electron mass for typical KLOE cuts. From \cite{Balossini:2006wc}.}
\label{figth:2}
\end{center}
\end{figure}

Concerning the second point a general strategy to evaluate the size of missing 
NNLO corrections consists in deriving a cross section expansion up to 
 $O (\alpha^2)$ from the theoretical
formulation implemented in the generator of interest. It can be cast in
general into the following form
\begin{eqnarray}
\sigma^{\alpha^2} \, = \, \sigma^{\alpha^2}_{\rm SV}
+ \sigma^{\alpha^2}_{\rm SV,H} + \sigma^{\alpha^2}_{\rm HH} ,
\label{eq:a2}
\end{eqnarray}
where in principle each of the $O (\alpha^2)$ contributions is affected by an uncertainty 
to be properly estimated. In Eq.~(\ref{eq:a2}) the first contribution is the cross section 
including $O (\alpha^2)$ SV corrections, whose uncertainty can be evaluated 
through a comparison with some of the available NNLO calculations 
reviewed in Section \ref{NNLO}. In particular, in 
\cite{Balossini:2006wc} the $\sigma^{\alpha^2}_{\rm SV}$ of 
the BabaYaga@NLO generator was compared with
the calculation of photonic corrections by Penin \cite{Penin:2005kf,Penin:2005eh} 
and the calculations by Bonciani 
{\it et al.}~\cite{Bonciani:2005im,Bonciani:2003te,Bonciani:2003cj0,Bonciani:2004gi,Bonciani:2004qt} 
who
computed two-loop fermionic corrections (in the one-family approximation $N_F = 1$) with finite mass
terms and the addition of soft bremsstrahlung and real pair contributions.\footnote{To provide 
meaningful results, the contribution of the vacuum polarisation was switched off in BabaYaga@NLO
to compare with the calculation by Penin consistently. For the same reason the real soft and 
some pieces of virtual electron pair corrections
were neglected in the 
comparison with the calculation by Bonciani {\it et al.}} The 
results of such comparisons are shown in Figs.~\ref{figth:1} 
and \ref{figth:2} for realistic cuts at the $\mathrm{\phi}$ factories.
In  Fig.~\ref{figth:1} $\delta\sigma$ is the difference between $\sigma^{\alpha^2}_{\rm SV}$ of BabaYaga@NLO and the cross sections of the two $O (\alpha^2)$ calculations, denoted
as photonic (Penin) and $N_F = 1$ (Bonciani {\it et al.}), as a function of the logarithm of the
infrared regulator $\epsilon$. It can be seen that the differences are given by flat functions,
demonstrating that such differences are infrared-safe, as expected, a consequence of
the universality and factorisation properties of the infrared divergences. In Fig.~\ref{figth:2},  
$\delta\sigma$ is shown as a function of the logarithm of a fictitious electron mass and for a
fixed value of $\epsilon = 10^{-5}$. Since the difference with the calculation by Penin is given by
a straight line, this indicates that the soft plus virtual two-loop photonic corrections missing in BabaYaga@NLO are $O (\alpha^2 L)$ contributions, as 
already remarked. On the other hand, the difference with the
calculation by Bonciani {\it et al.} is fitted by a quadratic function, showing that the 
electron two-loop
effects missing in BabaYaga@NLO are of the order of $\alpha^2 L^2$. 
However, it is important to emphasise
that, as shown in detail in~\cite{Balossini:2006wc}, the sum of the 
relative differences with the two $O (\alpha^2)$ calculations 
does not exceed the $2 \times 10^{-4}$ level for experiments at 
$\mathrm{\phi}$ and $B$ factories. 

The second term in Eq.~(\ref{eq:a2}) is the cross section containing the one-loop corrections to single
hard photon emission, and its uncertainty can be estimated by relying on partial results 
existing in the literature. Actually the exact perturbative expression of $\sigma^{\alpha^2}_{\rm SV,H}$
is not yet available for full $s+t$ Bhabha scattering, but using the results valid for 
small-angle Bhabha scattering~\cite{Jadach:1995hy,Ward:1998ht} and large-angle 
annihilation processes \cite{Glosser:2003ux,Jadach:2006fx} the
relative uncertainty of the theoretical tools in the calculation of $\sigma^{\alpha^2}_{\rm SV,H}$ can be 
conservatively estimated to be at the level of 0.05\%. Indeed the papers 
~\cite{Jadach:1995hy,Ward:1998ht,Glosser:2003ux,Jadach:2006fx} show that a YFS 
matching of NLO and HO corrections gives SV one-loop results for the $t$-channel process $e^+ e^- 
\to e^+ e^- \gamma$ and $s$-channel annihilation $e^+ e^- \to f \bar{f} \gamma$ ($f$ = fermion) 
differing from the exact perturbative calculations by a few units in $10^{-4}$ at most. This
conclusion also holds when photon energy cuts are varied. It is worth noting that during the completion of the present work a complete calculation of the NLO QED corrections to hard bremsstrahlung emission in full 
$s+t$ Bhabha scattering appeared in the literature \cite{Actis:2009uq}, along the lines described in Section \ref{Penta}. Explicit comparisons between the results of such an exact calculation with the predictions of the most accurate MC tools according to the typical luminosity cuts used at meson factories would be worthwhile to 
make the present error estimate related to the calculation of $\sigma^{\alpha^2}_{\rm SV,H}$ 
more robust.

The third contribution in Eq.~(\ref{eq:a2}) is the double hard 
bremsstrahlung cross section whose uncertainty can be directly evaluated by 
explicit comparison with the 
exact $e^+ e^- \to e^+ e^- \gamma\gamma$ cross section. It was shown in~\cite{Balossini:2006wc} that the differences between $\sigma^{\alpha^2}_{\rm HH}$ as in BabaYaga@NLO 
and the matrix element calculation, which exactly describes the 
contribution of two hard photons, are really negligible, being at the $10^{-5}$ level.

The relative effect due to lepton ($e,\mu,\tau$) and hadron ($\pi$) pairs has been numerically analysed in 
Section \ref{4pfinal}, in the presence of realistic selection cuts. 
This evaluation makes use of the complete NNLO virtual corrections combined with an 
exact matrix element calculation of the four-particle production processes. It 
supersedes previous approximate estimates which
underestimated the impact of those corrections. According to this new evaluation, the pair contribution, 
dominated by the electron pair correction, amounts to about 0.3\% for KLOE and 0.1\% 
for BaBar. These contributions are partially included in the BabaYaga@NLO code, as well as in other generators, through the insertion of the vacuum polarisation correction in the NLO diagrams, and detailed
comparisons between the exact calculation and the BabaYaga@NLO predictions are in progress 
\cite{pairs}.

\subsection{Conclusions and open issues \label{CONCLUSIONS}}

During the last few years a remarkable progress occurred in reducing the error 
of the luminosity measurements at flavour factories. 

Dedicated event generators like
BabaYaga@NLO and MCGPJ were developed in 2006 to provide predictions for the cross section 
of the large-angle Bhabha process, as well as for other QED reactions of interest, with a
theoretical accuracy at the level of  0.1\%. In parallel codes well-known since 
the time of the LEP/SLC operation such as BHWIDE were extensively used by the experimentalists in
data analyses. All these MC programs include, albeit according to different formulations, 
exact $O (\alpha)$ QED corrections matched with LL contributions describing
multiple photon emission. Such ingredients, together with the vacuum polarisation 
correction, are strictly necessary to achieve a physical 
precision down to the per mill level. Indeed, when considering typical selection, cuts
the NLO photonic corrections amount to about 15$\div$20\%, vacuum polarisation contributes 
at the several per cent level and HO effects lie between 1$\div$2\%.

The generators mentioned are, however, affected by 
an uncertainty due to HO effects neglected in their formulation, such as light pair
corrections or exact perturbative contributions present in NNLO calculations. 
From this point of view  the great progress in the calculation of two-loop corrections to the Bhabha scattering cross section was essential to establish the 
theoretical accuracy of the existing generators and 
will be crucial if an improvement 
of the precision below the one per mill level will be required.

A particular effort was done to compare the predictions of the generators 
consistently, in order to assess the technical precision obtained by the implementation of radiative corrections 
and related computational details. These comparisons were performed in the presence of realistic 
event selection criteria and at different c.m. energies. For the KLOE and CMD-2 experiments around the
$\mathrm{\phi}$-resonance, where the statistics of Bhabha events is the highest and the experimental luminosity
error at a few per mill level, the cross section results of BabaYaga@NLO, BHWIDE and MCGPJ agree 
within $\sim 0.1$\%. If (slightly) larger discrepancies are observed, they show up only for particularly
tight cuts or exclusive distributions in specific phase space regions which do not influence the 
luminosity determination. Very similar results were obtained for $\tau$-charm and $B$ factories.
The main conclusion of the work on tuned comparisons is that the technical precision of MC programs
is well under control, the discrepancies being 
due to different details in the treatment of the same sources of radiative corrections and their
technical implementation. For example, BabaYaga@NLO and 
BHWIDE adopt a fully factorised prescription for the matching of NLO and HO corrections, 
whereas MCGPJ implement some radiative corrections pieces in additive form. This can give
rise to some discrepancies between their predictions, especially in the presence of
tight cuts enhancing the effect of soft radiation. Furthermore, different choices are adopted in the
generators for the energy scale in the treatment of HO corrections beyond 
$O (\alpha)$, which are
another possible source of the observed differences. To go beyond the present situation, a 
further, nontrivial effort should be done by comparing, for instance, the programs in the 
presence of NLO corrections only (technical test) and for the specific effect due to the 
 exponentiation of soft and collinear logarithms. This would certainly shed light on the origin
of the (minor) discrepancies still registered at present.

On the theoretical side, a new exact evaluation of lepton and hadron pair corrections to the Bhabha scattering cross section was carried out, taking into account realistic cuts. This calculation provides results in 
substantial agreement with estimates based on singlet SF but supersedes previous evaluations in the soft-photon approximation. The results of the new exact calculation were preliminarily compared with the predictions of BabaYaga@NLO, which includes the bulk of such corrections (due to reducible contributions) through the insertion of the vacuum polarisation correction in the NLO diagrams, but neglects the effect of real pair radiation and two-loop form factors. It turns out that the error induced by the approximate treatment of pair corrections amounts to a few units in $10^{-4}$, both at KLOE and BaBar. Further work is in progress to arrive at a 
more solid and 
quantitative error estimate for these corrections when considering other 
selection criteria and c.m. energies too \cite{pairs}.
Also, the contribution induced by the uncertainty 
related to the non-perturbative contribution to the running of $\alpha$ was revisited,
making use of and comparing the two independent parameterisations 
derived in \cite{Jegerlehner:1985gq,Burkhardt:1989ky,Jegerlehner:2006ju} and \cite{rintpl:2008AA}.

A summary of the different sources of theoretical error and their relative impact on the Bhabha 
cross section is given in Table~\ref{tabcon:1}.  In Table~\ref{tabcon:1}, $|\delta^{\rm err}_{\rm VP}|$ 
is the error induced by the hadronic component of the vacuum polarisation, 
$|\delta^{\rm err}_{\rm pairs}|$ the error due to missing pair corrections, 
$|\delta^{\rm err}_{\rm SV}|$ the uncertainty coming from SV NNLO corrections, 
 $|\delta^{\rm err}_{\rm HH}|$ the uncertainty in the calculation of the double 
 hard bremsstrahlung 
 process 
 and $|\delta^{\rm err}_{\rm SV,H}|$ the error 
 estimate for one-loop corrections to single hard bremsstrahlung. As can be seen,
 pair corrections and exact NLO corrections to $e^+ e^- \to e^+ e^-\gamma$ are the
 dominant sources of error. 

\begin{table}[thb]
\caption{Summary of different
sources of theoretical uncertainty for the most precise generators used
for luminosity measurements and the corresponding total
theoretical errors for the calculation of the large-angle Bhabha cross section 
at meson factories.}
\label{tabcon:1}
\begin{center}
\begin{tabular}{llll}
\hline
Source of error (\%) & $\mathrm{\phi}$  & $\tau$-charm & $B$ \\
\hline
 $|\delta^{\rm err}_{\rm VP}|$~\cite{Jegerlehner:1985gq,Burkhardt:1989ky,Jegerlehner:2006ju}        & 0.00 & 0.01 & 0.03 \\
\hline
 $|\delta^{\rm err}_{\rm VP}|$~\cite{rintpl:2008AA}      & 0.02 & 0.01 & 0.02 \\ 
  \hline
 $|\delta^{\rm err}_{\rm SV}|$ & 0.02 & 0.02 & 0.02\\
 \hline
 $|\delta^{\rm err}_{\rm HH}|$       & 0.00 & 0.00 & 0.00\\
 \hline
 $|\delta^{\rm err}_{\rm SV,H}|$      & 0.05 & 0.05 & 0.05 \\
 \hline
 $|\delta^{\rm err}_{\rm pairs}|$     & 0.05 & 0.1 
 & 0.02 \\
\hline
 $|\delta^{\rm err}_{\rm total}|$     & 0.12$\div$0.14 & 0.18 &  0.11$\div$0.12\\
\hline
\end{tabular}
\end{center}
\end{table}

The total theoretical uncertainty as
obtained by summing the different contributions linearly is 0.12$\div$0.14\% 
at the $\mathrm{\phi}$ factories, 
0.18\% at the $\tau$-charm factories and $0.11\div 0.12$\% at the $B$ factories.  As can be seen from Table~\ref{tabcon:1}, the slightly larger uncertainty at the $\tau$-charm factories is mainly due to the pair contribution error, which is presently based on a very preliminary evaluation and for which a deeper analysis is ongoing \cite{pairs}. The
total uncertainty is slightly affected by the particular choice of the 
routine for the calculation of  $\Delta\alpha^{(5)}_{\rm hadr}(q^2)$, 
 since the two parameterisations considered here give rise to 
 similar errors, with the exception of the $\mathrm{\phi}$ factories 
 for which the two recipes return uncertainties differing by $2 \times 10^{-4}$. However 
 the ``parametric'' error induced by the hadronic contribution to the vacuum polarisation may become 
 a relevant source of uncertainty  when considering predictions for a c.m. energy on top of and closely 
 around very narrow resonances. For such a specific situation of interest, for instance for the BES experiment, 
 the appropriate treatment of the running $\alpha$ in the calculation of the Bhabha cross section should be scrutinised deeper because of the differences observed between the predictions for $\Delta\alpha^{(5)}_{\rm hadr}(q^2)$ obtained by means of the different parametrisation routines available (see Section \ref{sec:4} for a more detailed discussion).
 
Although the theoretical uncertainty quoted in Table~\ref{tabcon:1} could be put on firmer 
ground thanks to further studies in progress, it appears to be quite robust and sufficient for 
present and planned precision luminosity measurements at meson factories, where the experimental 
error currently is about a factor of two or three larger. 
Adopting the strategy followed during the LEP/SLC operation
one could arrive at a more aggressive error estimate by summing the relative
contributions in quadrature. However, for the time
being, this does not seem to be necessary in the light of the current experimental errors.

In conclusion, the precision presently reached by large-angle Bhabha programs used in the luminosity measurement at meson factories is comparable 
with that achieved about ten years ago for luminosity monitoring 
through small-angle Bhabha scattering at LEP/SLC. 

Some issues are still left open.
In the context of tuned comparisons, no 
effort was done to compare the available codes for the process of photon pair production. Since it contributes relevantly to the luminosity determination and as precise predictions for its 
cross section can be obtained by means of the codes BabaYaga@NLO and MCGPJ, this work should 
be definitely carried out. This would lead to a 
better understanding of the luminosity on the experimental side. In the framework of new theoretical advances, 
an evaluation of 
NNLO contributions to the process $e^+ e^- \to \gamma\gamma$ would be 
worthwhile to better assess the
precision of the generators which, for the time being, do not include such corrections 
exactly. 
More importantly, the exact one-loop corrections to the radiative process $e^+ e^- \to e^+ e^- \gamma$ should be calculated going beyond the partial results scattered in the literature 
(and referring to selection criteria valid for high-energy $e^+ e^-$ colliders)  
or limited to the soft-photon approximation.\footnote{As already remarked in Section 
\ref{TH}, during the completion of the present work a complete calculation of the NLO QED corrections to hard bremsstrahlung emission in full $s+t$ Bhabha scattering was performed in \cite{Actis:2009uq}. However, explicit comparisons between the predictions of this new calculation and the corresponding results of the most precise luminosity tools  are still missing and would be needed to better assess the theoretical error induced by such 
contributions in the calculation of the luminosity cross section.} Furthermore, to get a better control of the theoretical uncertainty in the sector of NNLO corrections to Bhabha scattering, the radiative Bhabha process at one-loop should be evaluated taking into account the typical experimental cuts used at meson factories. Incidentally this calculation would be also
of interest for other studies at $e^+ e^-$ 
colliders of moderately high energy, such as the search for new physics 
phenomena (e.g. dark matter candidates), for which radiative Bhabha scattering is a very important background. \\



\section{$R$ measurement from energy scan}
\label{sec:2}
%

\def\gsim{\mathrel{\raise.3ex\hbox{$>$\kern-.75em\lower1ex\hbox{$\sim$}}}}
\def\lsim{\mathrel{\raise.3ex\hbox{$<$\kern-.75em\lower1ex\hbox{$\sim$}}}}



In this section we will consider some theoretical and experimental aspects 
of the direct $R$ measurement and related quantities in experiments with energy scan. 
As discussed in the Introduction, the cross section of $e^+e^- $ annihilation 
into hadrons is involved in evaluations of various problems of particle physics and, 
in particular, in the definition of the hadronic contribution to vacuum 
polarisation, which is crucial for the precision tests of the Standard Model 
and searches for new physics.

The ratio of the radiation-corrected hadronic cross sections to the cross
section for muon pair production, calculated in the lowest order, 
is usually denoted as (see Eq.~(\ref{Rhad}))
\begin{equation} \label{r_def}
R \equiv R(s) = \frac{\sigma^0_{\rm had}(s)}
{4\pi \alpha^{2} /(3s)}.
\end{equation}
In the numerator of Eq.~(\ref{r_def})
one has to use the so called {\em undressed} hadronic cross section which does not
include vacuum polarisation corrections. 

The value of  $R$ has been measured in many experiments 
in different energy regions 
from the pion pair production threshold up to the $Z$ mass.
Practically all electron-positron colliders contributed to the global data
set on the hadronic annihilation cross section~\cite{Amsler:2008zzb}.
The value of $R$ extracted from the experimental data is then
widely used for various QCD tests as well as for the calculation
of dispersion integrals.
At high energies and away from resonances, 
the experimentally determined values of $R$ are in good agreement with
predictions of perturbative QCD, confirming, in particular, 
the hypothesis of three colour degrees of freedom for quarks.
On the other hand, for the low energy range the direct $R$ 
measurement~\cite{Amsler:2008zzb,Akhmetshin:2001ig}   
at experiments with energy scan is necessary.\footnote{Lattice QCD computations (see, e.g., Ref.~\cite{Shintani2009xxx}) 
of the hadronic vacuum polarisation 
are in progress, but they are not yet able to provide 
the required precision.} 
Matching between the two regions is performed at energies of a few GeV, where
both approaches for the determination of $R$ are in fair agreement.

For the best possible compilation of $R(s)$, 
data from different channels and different experiments have to be combined. 
For $\sqrt{s} \leq$ 1.4 GeV, the total hadronic cross section 
is a sum of about 25 
exclusive final states. At the present level of precision, 
a careful treatment of the radiative corrections is required. 
As mentioned above, it is mandatory to remove VP effects from the 
observed cross sections, 
but the final state radiation off hadrons should be kept. 
The major contribution to the uncertainty comes from the 
systematic error of the $R(s)$ measurement at low energies
($s<$ 2 GeV$^2$), which, in turn, is dominated by the 
systematic error of the measured cross section $e^+e^-\to\pi^+\pi^-$. 

\subsection{Leading-order annihilation cross sections}

Here we present the lowest-order expressions for the processes of
electron-positron annihilation into pairs of muons, pions and kaons.

For the muon production channel
\begin{eqnarray} \label{eemumu}
e^-(p_-) + e^+(p_+)\to \mu^-(p_-') + \mu^+(p_+')
\end{eqnarray}
within the Standard Model at Born level we have
\begin{eqnarray}
&& \frac{\dd{\sigma}^{\mu\mu}_0}{\dd\Omega_-} 
= \frac{\alpha^2\beta_\mu}{4s}
\left(2-\beta_\mu^2(1-c^2)\right)(1+K_{W}^{\mu\mu}),
\\ \nonumber 
&& s = (p_-+p_+)^2=4\eps^2, \quad c = \cos\theta_-, \quad
\theta_- = \widehat{\vec{p}_-\vec{p}}'_-, 
\end{eqnarray}
where $\beta_\mu=\sqrt{1-m_{\mu}^2/\eps^2}$ is the muon velocity. Small
terms suppressed by the factor $m_e^2/s$ are omitted.
Here $K_W^{\mu\mu}$ represents contributions due to $Z$-boson intermediate states
including $Z-\gamma$ interference, see, e.g., 
Refs.~\cite{Berends:1980yz,Arbuzov:1991pr}:
\begin{eqnarray}
&& K_W^{\mu\mu} =
\frac{s^2(2-\beta_\mu^2(1-c^2))^{-1}}
{(s-M_Z^2)^2+M_Z^2\Gamma_Z^2}\biggl\{(2-\beta_\mu^2(1-c^2))
\nonumber \\ && \quad \times
\left(c_v^2\biggl(3-2\frac{M_Z^2}{s}\biggr)+c_a^2\right) 
- \frac{1-\beta_\mu^2}{2}(c_a^2+c_v^2)
\nonumber \\ && \quad 
+c\beta_\mu\left[4\biggl(1-\frac{M_Z^2}{s}\biggr)c_a^2+8c_a^2c_v^2\right]\biggr\}, 
\nonumber \\ && 
 c_a=-\frac{1}{2\sin 2\theta_W},\quad
c_v=c_a(1-4\sin^2\theta_W),
\end{eqnarray}
where $\theta_W$ is the weak mixing angle.

The contribution of $Z$ boson exchange is suppressed, in the energy range under
consideration, by a factor $s/M_Z^2$ which reaches per mill level only
at $B$ factories. 

In the Born approximation the differential cross section
of the process
\begin{equation} \label{eepipi}
e^+(p_+)\ +\ e^-(p_-)\ \to\ \pi^+(q_+)\ +\ \pi^-(q_-)
\end{equation}
has the form
\begin{eqnarray} \label{seepipi}
&& \frac{\dd\sigma_0^{\pi\pi}}{\dd\Omega}(s)=\frac{\alpha^2\beta_\pi^3}{8s}\sin^2\theta\;
|F_{\pi}(s)|^2, \\ \nonumber 
&& \beta_\pi=\sqrt{1-m_{\pi}^2/\eps^2},\quad 
\theta=\widehat{\vec{p}_-\vec{q}}_- .
\end{eqnarray}
The pion form factor $F_{\pi}(s)$ takes into account
non-pertur\-ba\-tive virtual vertex corrections due to strong 
interactions~\cite{Bramon:1997un,Drago:1997px}.
We would like to emphasise that in the approach under discussion 
the final state QED corrections are not included into $F_{\pi}(s)$. 
The form factor is extracted from the experimental data on the 
same process as discussed below. 

The annihilation process with three pions in the final state was 
considered in Refs.~\cite{Ahmedov:2002tg,Kuraev:1995hc} including 
radiative corrections relevant to the energy region close to the threshold.
A stand-alone Monte Carlo event generator for this channel is 
available~\cite{Ahmedov:2002tg}. The channel was also included in the
MCGPJ generator~\cite{Arbuzov:2005pt} on the same footing as other 
processes under consideration in this report.

In the case of $K_LK_S$ meson pair production the differential
cross section in the Born approximation reads
\begin{equation}
\frac{\dd\sigma_{0}(s)^{K_LK_S}}{\dd\Omega_L}=\frac{\alpha^2\beta_{K}^3}{4s}
\sin^2\theta\; |F_{LS}(s)|^2.
\end{equation}
Here, as well as in the case of pion production, we assume that the
form factor $F_{LS}$ also includes the vacuum polarisation operator
of the virtual photon.
The quantity $\beta_{K}=\sqrt{1-4m_K^2/s}$ is the $K$ meson c.m.s.
velocity, and $\theta$ is the angle between the directions of motion of
the long living kaon and the initial electron.

In the case of $K^+K^-$ meson production near threshold, the Sakharov-Sommerfeld
factor for the Coulomb final state interaction 
should additionally be taken into account:
\begin{eqnarray} \label{bornk}
&& \frac{\dd\sigma_{0}(s)^{K^+K^-}}{\dd\Omega_-}=\frac{\alpha^2\beta_{K}^3}{4s}
\sin^2\theta |F_{K}(s)|^2\frac{Z}{1-\exp(-Z)}, 
\nonumber \\ 
&& Z=\frac{2\pi\alpha}{v}, \ \ 
v = 2\sqrt{\frac{s-4m_K^2}{s}}\left(1+\frac{s-4m_K^2}{s}\right)^{-1},
\end{eqnarray}
where $v$ is the relative velocity of the kaons~\cite{Arbuzov:1993qc} which has
the proper non-relativistic and ultra-relativistic limits. 
When $s=m_{\phi}^2$, we have $v\approx 0.5$ and the final state interaction
correction gives about 5\%  enhancement in the cross section.

\subsection{QED radiative corrections}

One-loop radiative corrections (RC) for the 
processes~(\ref{eemumu},\ref{eepipi})  
can be separated into two natural parts according to the parity 
with respect to the substitution $c\to -c$.

The $c$-even part of the one-loop soft and virtual contribution to the muon 
pair creation channel can be combined from the well known Dirac and Pauli
form factors and the soft photon contributions. It reads
\begin{eqnarray} \label{beven_mumu}
&& \frac{\dd\sigma^{B+S+V}_{\mu\mu-\mathrm{even}}}{\dd\Omega} =
\frac{\dd\sigma_0^{\mu\mu}}{\dd\Omega}\,\frac{1}{|1-\Pi(s)|^2}\Biggl\{1
+ \frac{2\alpha}{\pi}\Biggl[\biggl[L - 2
\nonumber \\ && \quad 
+ \frac{1+\beta_\mu^2}{2\beta_\mu}l_\beta \biggr]
\ln\frac{\Delta\eps}{\eps} 
+ \frac{3}{4}(L-1)
+ K^{\mu\mu}_{\mathrm{even}} \Biggr]\Biggr\}, \\ \nonumber\label{js}
&& K_{\mathrm{even}}^{\mu\mu} = \frac{\pi^2}{6} - \frac{5}{4}
+ \rho\biggl(\frac{1+\beta_\mu^2}{2\beta_\mu} - \frac{1}{2}
+ \frac{1}{4\beta_\mu}\biggr) 
\nonumber \\ && \quad 
+ \ln\frac{1+\beta_\mu}{2}\left(\frac{1}{2\beta_\mu}
+ \frac{1+\beta_\mu^2}{\beta_\mu}\right) 
\\ \nonumber && \quad
- \frac{1-\beta_\mu^2}{2\beta_\mu}\frac{l_\beta}{2-\beta_\mu^2(1-c^2)}
\\ \nonumber && \quad
+ \frac{1+\beta_\mu^2}{2\beta_\mu}\biggl[ \frac{\pi^2}{6}
+ 2{\mathrm{Li}}_2\left(\frac{1-\beta_\mu}{1+\beta_\mu}\right)
\\ \nonumber && \quad
+ \rho\ln\frac{1+\beta_\mu}{2\beta_\mu^2} 
+ 2\ln\frac{1+\beta_\mu}{2}\ln\frac{1+\beta_\mu}{2\beta_\mu^2} \biggr], \\ 
\nonumber
l_{\beta} &=& \ln\frac{1+\beta_\mu}{1-\beta_\mu}\, , \quad
\rho=\ln\frac{s}{m_{\mu}^2}\quad L=\ln\frac{s}{m_{e}^2},
\end{eqnarray}
where ${\rm Li}_2(z)=-\int_0^z dt \ln(1-t)/t$ is the dilogarithm and 
$\Delta\eps\ll\eps$ is the maximum energy of soft photons
in the centre--of--mass (c.m.) system.
$\Pi(s)$ is the vacuum polarisation operator. Here we again see
that the terms with the large logarithm $L$ dominate numerically.

The $c$-odd part of the one--loop correction comes from the interference
of Born and box Feynman diagrams and from the interference part
of the soft photon emission contribution.
It causes the charge asymmetry of the process:
\begin{equation}
\eta={\dd\sigma(c)-\dd\sigma(-c)\over\dd\sigma(c)+\dd\sigma(-c)}\neq 0.
\end{equation}
The $c$-odd part of the differential cross section has the following 
form~\cite{Arbuzov:1997pj}:
\begin{eqnarray} \label{eqll}
\frac{\dd\sigma^{S+V}_{\mathrm{odd}}}{\dd\Omega} &=&
\frac{\dd\sigma_0^{\mu\mu}}{\dd\Omega}\,\frac{2\alpha}{\pi}\Biggl[
2\ln\frac{\Delta\eps}{\eps}\ln\frac{1-\beta c}{1+\beta c}
+ K_{\mathrm{odd}}^{\mu\mu} \biggr]. 
\end{eqnarray}
The expression for the $c$-odd form factor can be found 
in Ref.~\cite{Arbuzov:1997pj}.
Note that in most cases the experiments have a symmetric angular 
acceptance, so that the odd part of the cross section does not 
contribute to the measured quantities.

Consider now the process of hard photon emission
\begin{eqnarray}
e^+(p_+) + e^-(p_-) \rightarrow \mu^+(q_+) + \mu^-(q_-) + \gamma(k).
\end{eqnarray}
It was studied in detail in Refs.~\cite{Arbuzov:1997pj,Eidelman:1979gc}.
The photon energy is assumed to be larger than $\Delta\eps$.
The differential cross section has the form
\begin{eqnarray} \label{rmu}
&& \dd\sigma^{\mu\mu\gamma}=\frac{\alpha^3}{2\pi^2 s^2}R \dd\Gamma, 
\\ \nonumber 
&& \dd\Gamma = \frac{\dd^3q_-\dd^3q_+\dd^3k}
{q_-^0q_+^0k^0}\delta^{(4)}(p_++p_--q_--q_+-k), 
\\ \nonumber
&& R=\frac{s}{16(4\pi\alpha)^3}\sum\limits_{spins}^{}
|M|^2=R_e + R_{\mu} + R_{e\mu}.
\end{eqnarray}
The quantities $R_i$ are found directly from the matrix elements 
and read
\begin{eqnarray*}
&& R_e = \frac{s}{\chi_-\chi_+}B
- \frac{m_e^2}{2\chi_-^2}\;\frac{(t_1^2+u_1^2+2m_{\mu}^2s_1)}{s_1^2}
\\ && \quad
- \frac{m_e^2}{2\chi_+^2}\;\frac{(t^2+u^2+2m_{\mu}^2s_1)}{s_1^2}
+ \frac{m_{\mu}^2}{s_1^2} \Delta_{s_1s_1},
\nonumber \\
&& R_{e\mu} = B\biggl( \frac{u}{\chi_-\chi_+'} + \frac{u_1}{\chi_+\chi_-'}
- \frac{t}{\chi_-\chi_-'} 
\\ && \quad
- \frac{t_1}{\chi_+\chi_+'}\biggr)
+ \frac{m_{\mu}^2}{ss_1} \Delta_{ss_1},
\\ \nonumber
&& R_{\mu} = \frac{s_1}{\chi_-'\chi_+'}B + \frac{m_{\mu}^2}{s^2} \Delta_{ss}\, ,
\\ \nonumber
&& B = \frac{u^2+u_1^2+t^2+t_1^2}{4ss_1}\, ,
\\ \nonumber
&& \Delta_{s_1s_1} = - \frac{(t+u)^2+(t_1+u_1)^2}{2\chi_-\chi_+}\, ,
\\ \nonumber
&& \Delta_{ss} = - \frac{u^2+t_1^2+2sm_{\mu}^2}{2(\chi_-')^2}
- \frac{u_1^2+t^2+2sm_{\mu}^2}{2(\chi_+')^2}
\\ && \quad
+ \frac{1}{\chi_-'\chi_+'}\bigl( ss_1 - s^2 + tu +t_1u_1 - 2sm_{\mu}^2\bigr),
\\ \nonumber
&& \Delta_{ss_1} = \frac{s+s_1}{2}\biggl( \frac{u}{\chi_-\chi_+'}
+ \frac{u_1}{\chi_+\chi_-'} - \frac{t}{\chi_-\chi_-'}
\\ && \quad
- \frac{t_1}{\chi_+\chi_+'} \biggr)
+ \frac{2(u-t_1)}{\chi_-'} + \frac{2(u_1-t)}{\chi_+'},
\end{eqnarray*}
where
\begin{eqnarray*}
&& s_1=(q_++q_-)^2, \ \ 
t=-2p_-q_-, \ \ 
t_1=-2p_+q_+,
\\ 
&& u=-2p_-q_+,\ \ 
u_1=-2p_+q_-, \ \
\chi_{\pm}=p_{\pm}k, \ \ 
\chi_{\pm}'=q_{\pm}k.
\end{eqnarray*}

The bulk of the hard photon radiation comes from ISR in collinear regions.
If we consider photon emission inside two narrow cones along the beam axis 
with restrictions
\begin{eqnarray}
\widehat{\vec{p}_{\pm}\vec{k}}=\theta\leq\theta_0\ll 1,\quad
\theta_0\gg\frac{m_e}{\eps},
\end{eqnarray}
we see that the corresponding contribution takes the factorised form
\begin{eqnarray} \label{eqcd}
&& \left(\frac{\dd\sigma^{\mu\mu}}{\dd\Omega_-}\right)_{\mathrm{coll}}=
C_e + D_e, 
\\ \nonumber 
&& C_e=\frac{\alpha}{2\pi}\biggl(\ln\frac{s}{m_e^2}-1\biggr)
\int\limits_{\Delta}^{1}\!\!\dd x \frac{1+(1-x)^2}{x} A_0,
\\ \nonumber
&& D_e=\frac{\alpha}{2\pi}\int\limits_{\Delta}^{1}\!\!\dd x \biggl\{x
+ \frac{1+(1-x)^2}{x}\ln\frac{\theta_0^2}{4}\biggr\} A_0, 
\\ \nonumber
&& A_0 = \biggl[\frac{\dd\tilde{\sigma}_0(1-x,1)}{\dd\Omega_-}
+ \frac{\dd\tilde{\sigma}_0(1,1-x)}{\dd\Omega_-}\biggr],
\end{eqnarray}
where the {\em shifted\/} 
Born differential cross section describes the process
$e^+(p_+(1-x_2)) + e^-(p_-(1-x_1)) \to \mu^+(q_+) + \mu^-(q_-)$,
\begin{eqnarray}
&& \frac{\dd\tilde{\sigma}_0^{\mu\mu}(z_1,z_2)}{\dd\Omega_-}
= \frac{\alpha^2}{4s}
\nonumber \\ && \quad \times
\frac{y_1[z_1^2(Y_1-y_1c)^2
+ z_2^2(Y_1+y_1c)^2+8z_1z_2m_{\mu}^2/s]}
{z_1^3z_2^3[z_1+z_2-(z_1-z_2)cY_1/y_1]}, 
\nonumber \\
&& y_{1,2}^2=Y_{1,2}^2-\frac{4m_{\mu}^2}{s},\quad
Y_{1,2}=\frac{q_{-,+}^0}{\eps}, \quad z_{1,2}=1-x_{1,2},
\nonumber \\
&& Y_1= \frac{4m_{\mu}^2}{s}(z_2-z_1)c \biggl[
2z_1z_2
\nonumber \\ && \quad 
+\sqrt{4z_1^2z_2^2-4(m_{\mu}^2/s)((z_1+z_2)^2
-(z_1-z_2)^2c^2)}\biggr]^{-1} 
\nonumber \\ && \quad
+ \frac{2z_1z_2}{z_1+z_2-c(z_1-z_2)}.
\end{eqnarray}
One can recognise that the large logarithms related to the collinear
photon emission appear in $C_e$ in agreement with the structure 
function approach
discussed in the Luminosity Section. In analogy to the definition of
the QCD structure functions, one can move the factorised logarithmic 
corrections $C_e$
into the QED electron structure function. 
Adding the higher-order radiative corrections in the leading logarithmic
approximation to the complete one-loop result,
the resulting cross section can be written as
\begin{eqnarray} \label{eemmg}
&& \frac{\dd\sigma^{e^+e^-\to\mu^+\mu^-(\gamma)}}{\dd \Omega_-}
=\int\limits_{z_{\mathrm{min}}}^{1}
\int\limits_{z_{\mathrm{min}}}^{1}\dd z_1\dd z_2
\frac{{\cal D}(z_1,s){\cal D}(z_2,s)
}{|1-\Pi(sz_1z_2)|^2} 
\nonumber \\ && \quad \times
\frac{\dd\tilde{\sigma}_0^{\mu\mu}(z_1,z_2)}{\dd \Omega_-}
\biggl(1+\frac{\alpha}{\pi}K_{\mathrm{odd}}^{\mu\mu}
+\frac{\alpha}{\pi}K_{\mathrm{even}}^{\mu\mu}\biggr)
\nonumber \\ \nonumber && \quad
+ \biggl\{
\frac{\alpha^3}{2\pi^2s^2} \int\limits_{\stackrel{k^0>\Delta\eps}
{\widehat{k p}_{\pm}>\theta_0}}
\frac{R_e|_{m_e=0}}{|1-\Pi(s_1)|^2}\frac{\dd\Gamma}{\dd \Omega_-}
 + \frac{D_e}{|1-\Pi(s_1)|^2}\biggr\}
\\  && \quad
+ \biggl\{\frac{\alpha^3}{2\pi^2s^2}\int\limits_{k^0>\Delta\eps}\!\!\!
\biggl(\mbox{Re}\, \frac{R_{e\mu}}{(1-\Pi(s_1))(1-\Pi(s))^{*}}
\nonumber \\ && \quad 
+ \frac{R_{\mu}}{|1-\Pi(s)|^2}\biggr)\frac{\dd\Gamma}{\dd \Omega_-} 
+ \mbox{Re}\,\frac{C_{e\mu}}{(1-\Pi(s_1))(1-\Pi(s))^{*}}
\nonumber \\  && \quad
+ \frac{C_{\mu}}{|1-\Pi(s)|^2} \biggr\},                     
\\ && \nonumber
C_{\mu} = \frac{2\alpha}{\pi}\frac{\dd\sigma_0^{\mu\mu}}{\dd\Omega_-}
\ln\frac{\Delta\eps}{\eps}
\biggl(\frac{1+\beta^2}{2\beta}\ln\frac{1+\beta}{1-\beta}-1\biggr),
\\ \nonumber &&
C_{e\mu} = \frac{4\alpha}{\pi}\frac{\dd\sigma_0^{\mu\mu}}{\dd\Omega_-}
\ln\frac{\Delta\eps}{\eps}\ln\frac{1-\beta c}{1+\beta c},
\end{eqnarray}
where $D_e$, $C_{e\mu}$ and $C_{\mu}$ are compensating terms,
which provide the cancellation of the auxiliary parameters $\Delta$ and $\theta_0$
inside the curly brackets. In the first term, containing ${\cal D}$ functions,
we collect all the leading logarithmic terms. 
A part of non-leading terms proportional
to the Born cross section is written as the $K$-factor. The rest of the 
non-leading contributions are written as two additional terms. The compensating
term $D_e$ (see Eq.~(\ref{eqcd})) comes from the integration in the collinear
region of hard photon emission. The quantities $C_{\mu}$ and $C_{e\mu}$
come from the even and odd parts of the differential cross section
(arising from soft and virtual corrections), respectively.
Here we consider the phase space of two
$(\dd\Omega_-)$ and three $(\dd\Gamma)$ final particles as
those that already include all required experimental cuts.
Using specific experimental conditions one can determine the lower
limits of the integration over $z_1$ and $z_2$ instead of the kinematical
limit $z_{\mathrm{min}} = 2m_{\mu}/(2\eps-m_{\mu})$.

\begin{figure}[htb]
\centering\includegraphics[width=0.49\textwidth]{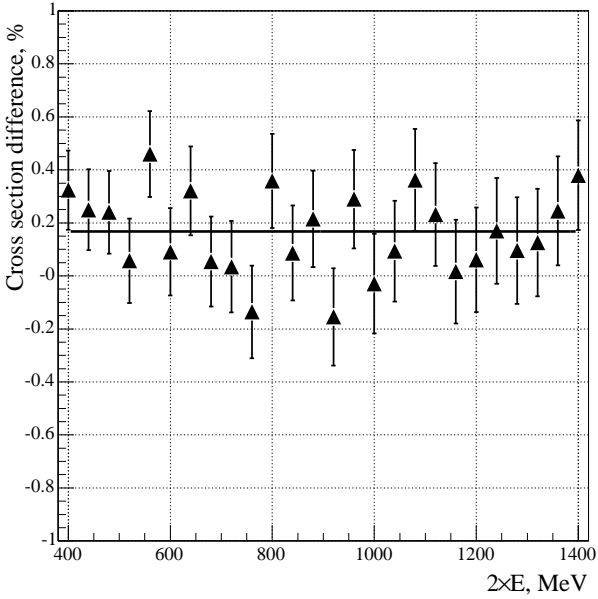}
\caption{Comparison of the $e^+e^-\to\mu^+\mu^-$ total cross sections
computed by the MCGPJ and KKMC generators versus the c.m. energy.}
\label{mumu_kkmc} 
\end{figure}

Matching of the complete $O (\alpha)$ RC with higher-order leading
logarithmic corrections can be performed in different schemes. The above approach
is implemented in the MCGPJ event generator~\cite{Arbuzov:2005pt}. The solution of the
QED evolution equations in the form of parton showers (see the Luminosity Section), 
matched again with the first order corrections, is implemented in the BabaYaga@NLO 
generator~\cite{CarloniCalame:2003yt}. Another possibility is realised in the 
KKMC code~\cite{Jadach:1999vf,Jadach:2000ir} with the Yennie-Frautschi-Suura exponentiated 
representation of the photonic higher-order corrections.
A good agreement was obtained in~\cite{Arbuzov:2005pt} for various differential 
distributions for the $\mu^+\mu^-$ channel between MCGPJ, BabaYaga@NLO and KKMC,
see Fig.~\ref{mumu_kkmc} for an example.

Since the radiative corrections to the initial $e^+e^-$ state 
are the same for annihilation into hadrons and muons
as well as that into pions, they cancel out in part in the ratio~(\ref{piform}). 
However, the experimental conditions and systematic are different 
for the muon and hadron channels.
Therefore, a separate treatment of the processes has to be performed and the 
corrections to the initial states have to be included in the analysis.

For the $\pi^+\pi^-$ channel the complete set of $O (\alpha)$ 
corrections matched with the leading logarithmic electron structure functions
can be found in Ref.~\cite{Arbuzov:1997je}. There the RC calculation 
was performed within scalar QED. 

Taking into account only final state corrections calculated within scalar QED,
it is convenient to introduce the {\em bare} 
$e^+e^-\to\pi^+\pi^-(\gamma)$ cross
section as 
\begin{equation}
\label{bare}
\sigma^0_{\pi\pi(\gamma)} = 
\frac{\pi\alpha^2}{3s}\beta^3_\pi \left| F_\pi(s)\right|^2 |1-\Pi(s)|^2 
\left( 1+\frac{\alpha}{\pi}\Lambda(s) \right),
\end{equation}
where the  factor $|1-\Pi(s)|^2$ with the polarisation operator 
$\Pi(s)$ gives the effect of leptonic and hadronic vacuum 
polarisation. The final state radiation (FSR) correction is
denoted by $\Lambda(s)$. For an inclusive measurement without cuts
it reads~\cite{Schwinger:1989ix,Drees:1990te,Melnikov:2001uw,Hoefer:2001mx}
\begin{eqnarray}
&& \Lambda(s)  =  \frac {1+\beta_{\pi}^2} {\beta_{\pi}} 
\biggl\{ 4\mathrm{Li}_2 (\frac {1-\beta_{\pi}} {1+\beta_{\pi}})+
2\mathrm{Li}_2 (-\frac {1-\beta_{\pi}} {1+\beta_{\pi}})
\nonumber \\ && \quad
- \biggl[3\ln(\frac{2}{1+\beta_{\pi}})
+ 2\ln\beta_{\pi} \biggr]\ln\frac{1+\beta_{\pi}}{1-\beta_{\pi}}\biggr\} 
-  3 \ln  (\frac {4}{1-\beta_{\pi}^2})
\nonumber \\ && \quad
-4\ln\beta_{\pi} 
+  \frac {1} {\beta_{\pi}^3}
 \left[\frac{5}{4}(1+\beta_{\pi}^2)^2-2\right]\ln\frac{1+\beta_{\pi}}{1-\beta_{\pi}}
\nonumber \\ && \quad
 + \frac {3} {2} \frac{1+\beta_{\pi}^2}{\beta_{\pi}^2}.
\end{eqnarray}

For the neutral kaon channel the corrected cross section has the form
\begin{eqnarray} \nonumber
\frac{\dd\sigma^{e^+e^-\to K_LK_S}(s)}{\dd\Omega_L}=
\int\limits_{0}^{\Delta}\dd x\;
\frac{\dd\sigma_0^{e^+e^-\to K_LK_S}(s(1-x))}{\dd\Omega_L}
F(x,s).
\end{eqnarray}
The radiation factor $F$ takes into account radiative
corrections to the initial state within the leading logarithmic 
approximation with exponentiation of the numerically important
contribution of soft photon radiation, see Ref.~\cite{Kuraev:1985hb}:
\begin{eqnarray}
&& F(x,s) = bx^{b-1}\biggl[ 1 + \frac{3}{4}b
+ \frac{\alpha}{\pi}\biggl(\frac{\pi^2}{3}-\frac{1}{2}\biggr)
- \frac{b^2}{24}\biggl(\frac{1}{3}L  - 2\pi^2 
\nonumber \\ \nonumber && \quad 
- \frac{37}{4}\biggr)\biggr] 
- b\biggl(1-\frac{x}{2}\biggr)
+ \frac{1}{8}b^2\biggl[4(2-x)\ln\frac{1}{x}
\\ \nonumber && \quad
+ \frac{1}{x}(1+3(1-x)^2)\ln\frac{1}{1-x} - 6 + x\biggr] 
\\ \nonumber && \quad
+ \biggl(\frac{\alpha}{\pi}\biggr)^2\biggl\{\frac{1}{6x}
\biggl(x-\frac{2m_e}{\eps}\biggr)^{b}\biggl[ (2-2x+x^2)
\biggl(\ln\frac{sx^2}{m_e^2}-\frac{5}{3}\biggr)^2
\\ \nonumber && \quad
+ \frac{b}{3}\biggl(\ln\frac{sx^2}{m_e^2}-\frac{5}{3}\biggr)^3\biggr]
+ \frac{1}{2}L^2\biggl[\frac{2}{3}\;\frac{1-(1-x)^3}{1-x}
\\ \nonumber && \quad
+ (2-x)\ln(1-x) + \frac{x}{2}\biggr]\biggr\} \Theta(x-\frac{2m_e}{\eps}).
\end{eqnarray}

Radiative corrections to the $K^+K^-$ channel in the point-like particle
approximation are the same as for the case of charged pion pair 
(with the substitution $m_\pi\to m_K$).
Usually, for the kaon channel we deal with the energy range
close to $\phi$ mass. There one may choose the maximal energy of a radiated
photon as
\begin{equation}
\omega\leq\Delta E=m_{\phi}-2m_K\ll m_K,\quad
\Delta\equiv\frac{\Delta E}{m_K}\approx\frac{1}{25}.
\end{equation}
For these photons one can use the soft photon approximation. 

\subsection{Experimental treatment of hadronic cross sections and $R$}

For older low energy data sets obtained at various $e^+e^-$ colliders, 
the correct treatment of radiative corrections is difficult and
sometimes ambiguous. 
So, to avoid uncontrolled possible
systematic errors, it may be reasonable not to include all previous 
results except the recent data from CMD-2 and SND. 
Both experiments at the VEPP-2M collider in
Novosibirsk have delivered independent new measurements. 
The covered energy range is crucial for
($g_{\mu}$-2)/2 of muon and for running $\alpha$. 
As for the two-pion channel $\pi^+\pi^-$, which gives more than
70\% of the total hadronic contribution, both experiments have very good
agreement over the whole energy range. The relative deviation ``SND -
CMD-2'' is (-0.3 $\pm$ 1.6)\% only, well within the quoted errors.       

The CMD-2 and SND detectors were located in the opposite straight sections of
VEPP-2M and were taking data in parallel until the year 2000 when the   
collider was shut down to prepare for the construction of the new collider 
VEPP-2000. Some important features of the CMD-2
detector allowed one to select a sample of the ``clean'' 
collinear back-to-back 
events. The drift chamber (DC) was used to 
separate $e^+e^-$, $\mu^+\mu^-$, $\pi^+\pi^-$ and $K^+K^-$ events from other
particles. The Z-chamber allowed one to significantly improve 
the determination 
of the polar angle of charged particle tracks in the DC that, in turn, 
provided the detector acceptance with 0.2\% precision. 
The barrel electromagnetic calorimeter based on CsI crystals 
helped to separate the Bhabha from other collinear events. 

The SND detector consisted of three spherical layers of the
electromagnetic calorimeter with 1620 crystals (NaI) and a total
weight of 3.6 tons.  The solid angle of the calorimeter is about 90\% of
$4\pi$ steradians, which makes the detector practically hermetic for photons
coming from the interaction point. The angular and energy resolution 
for photons was found to be $1.5^{\circ}$ and 
$\sigma(E)/E = 4.2\%/E$(GeV)$^{1/4}$, respectively. 
More detail about CMD-2 and SND can be found 
elsewhere~\cite{Anashkin1988:xxx,Achasov:1999ju}.\\

\subsubsection{Data taking and analysis of the $\pi^+\pi^-$ channel}
The detailed data on the pion form factor are crucial for a number of problems
in hadronic physics and they are used to extract $\rho(770)$ meson
parameters and its radial excitations. Besides, the detailed data allow to 
extrapolate the pion form factor to the point $s=0$ and determine 
the value of the pion electromagnetic radius.

From the experimental point of view the form factor
can be defined as~\cite{Akhmetshin:2001ig}
\begin{eqnarray} \label{piform}
|F_\pi|^2 &=& \frac{N_{\pi\pi}}{N_{ee}+N_{\mu\mu}}
\frac{
\sigma_{ee}(1+\delta_{ee})\varepsilon_{ee} +
\sigma_{\mu\mu}(1+\delta_{\mu\mu})\varepsilon_{\mu\mu}
}{\sigma_{\pi\pi}(1+\delta_{\pi\pi})
(1+\Delta_N)(1+\Delta_D)\varepsilon_{\pi\pi}}
\nonumber \\
&-& \Delta_{3\pi},
\end{eqnarray}
where the ratio $N_{\pi\pi}/(N_{ee}+N_{\mu\mu})$ is derived from the
observed numbers of events,
$\sigma$ are the corresponding Born cross sections,
$\delta$ are the radiative corrections (see below),
$\epsilon$ are the detection efficiencies,
$\Delta_D$ and $\Delta_N$ are the corrections for the pion losses
caused by decays in flight and nuclear interactions respectively,
and $\Delta_{3\pi}$  is the correction for misidentification of
$\omega\to\pi^+\pi^-\pi^0$ events as $e^+e^-\to\pi^+\pi^-$. In the case of the
latter process, $\sigma_{\pi\pi}$ corresponds to point-like pions. \\

\begin{figure}[htb]
\centering\includegraphics[width=0.49\textwidth]{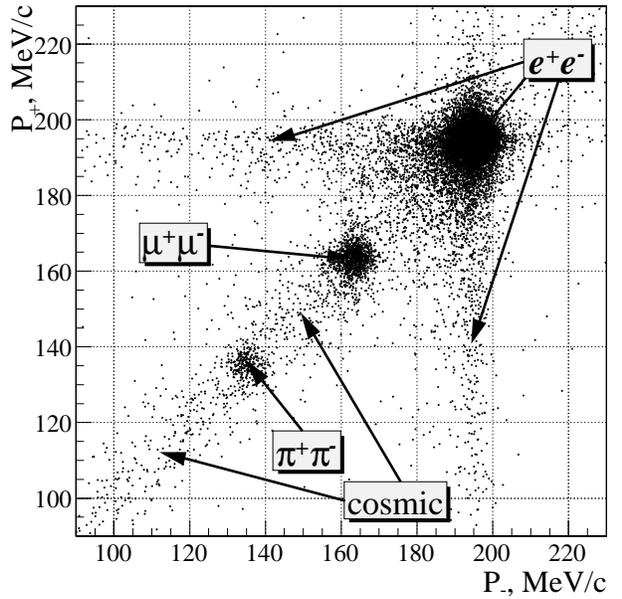}
\caption{Two-dimensional plot of the $e/\mu/\pi$ events. 
Cosmic events are distributed predominantly along a corridor which
extends from the right upper to the left bottom corner. 
Points in this plot correspond to the momenta of particles 
for the beam energy of 195~MeV. }
\label{fig:p1-vs-p2} 
\end{figure}

\begin{figure}[htb]
\centering\includegraphics[width=0.49\textwidth]{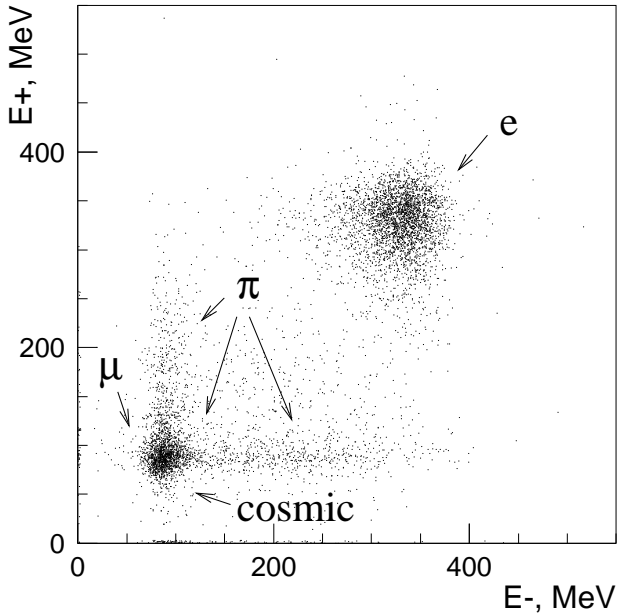}
\caption{ Energy deposition of collinear events 
for the beam energy of 460~MeV.}
\label{fig:e1-vs-e2}
\end{figure}

The data were collected in the whole energy range of VEPP-2M and 
the integrated luminosity of about 60~pb$^{-1}$ was recorded by both
detectors. The beam energy was controlled and measured with a relative 
accuracy not worse than $\sim 10^{-4}$ by using the method of resonance depolarisation. A sample of the $e^+e^-$, $\mu^+\mu^-$ and $\pi^+\pi^-$  
events was selected for analysis. 
As for CMD-2, the procedure of the
$e/\mu/\pi$ separation for energies 2E $\leq$ 600 MeV was based on the momentum
measurement in the DC. For these energies the average difference 
between the momenta of $e/\mu/\pi$ is large enough with respect to the momentum 
resolution (Fig.~\ref{fig:p1-vs-p2}). On the contrary, for energies 
2E $\geq$ 600~MeV,
the energy deposition of the particles in the calorimeter is quite different 
and allows one to separate electrons from muons and pions 
(Fig.~\ref{fig:e1-vs-e2}).
At the same time, muons and pions cannot be separated by their energy
depositions in the calorimeter. So, the ratio 
$N(\mu^+ \mu^-)/N(e^+e^-)$ was fixed
according to QED calculations taking into account the detector acceptance and
the radiative corrections. Since the selection criteria were the same for  
all collinear events, many effects of the detector
imperfections were partly cancelled out. It allowed one to measure the cross
section of the process $e^+e^- \to \pi^+\pi^-$ with better precision than
that of the luminosity.  

Separation of $e^+e^-$, $\mu^+\mu^-$ and $\pi^+\pi^-$
events was based on the minimisation of the unbinned likelihood
function. This method is described in detail 
elsewhere~\cite{Akhmetshin:1999uj}. 
To simplify the error calculation of the pion form
factor, the likelihood function had the global fit parameters $(N_{ee}
+ N_{\mu\mu})$ and $N_{\pi\pi}/(N_{ee} + N_{\mu\mu})$, 
through $|F_{\pi}(s)|^2$ 
given by Eq.~(\ref{piform}). 
The pion form factor measured by CMD-2 has a systematic 
error of about 0.6-0.8\% for $\sqrt{s} \leq $ 1 GeV. 
For energies above 1 GeV it varies from 1.2\% to 4.2\%. 

Since at low energies all three final states could be separated independently, 
the cross section of the process $e^+e^- \to \mu^+\mu^-$ was also measured,
providing an additional consistency test. The experimental value
$\sigma^{\rm exp}_{\mu\mu} / \sigma^{\rm QED}_{\mu\mu}$ = (0.980 $\pm$ 0.013 $\pm$
0.007) is in good agreement with the expected value of 1 within 1.4
statistical deviations.  

Another method  to discriminate electrons and pions from other
particles was used in SND. The event separation was 
based on the difference in longitudinal energy deposition profiles 
(energy deposition in three calorimeter layers) 
for these particles. To use the correlations between energy 
depositions in calorimeter layers in the most complete way, 
the separation parameter was
based on the neural network approach~\cite{Achasov:2005rg,Achasov:2006vp}. 
The network had an input layer consisting of 7 neurons, two hidden layers with
20 neurons each, and the output layer with one neuron. As input data, the
network used the energy depositions of the particles in calorimeter layers
and the polar angle of one of the particles. The output signal $R_{e/\pi}$ is
a discrimination parameter between different particles. The network was
tuned by using simulated events and was checked with experimental 3$\pi$ and
$e^+e^-$ events. The misidentification ratio between electrons and pions was
found to be 0.5 - 1\%. SND measured the
$e^+e^- \to \pi^+\pi^-$ cross section in the energy range 0.36 - 0.87
GeV with a systematic error of 1.3\%.   

\begin{figure}[htb]
  \centering\includegraphics[width=0.49\textwidth]{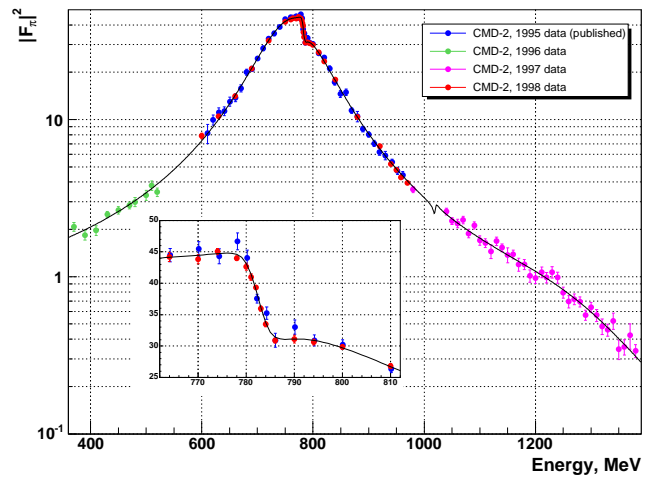}
  \caption{ Pion form factor data from CMD-2 and GS
    fit. The energy range around the $\omega$ meson is scaled up and
    presented in the inset.}
  \label{fig:fpicmd2}
\end{figure}

The  Gounaris-Sakurai (GS) parametrisation was used to fit the pion form factor.    
Results of the fit are shown in Fig.~\ref{fig:fpicmd2}.
The 
$\chi^2$
 was found to be $\chi^2_{\rm min}/\mathrm{n.d.f.} = 122.9/111$ that corresponds to the
 probability P$(\chi^2_{\rm min}/\mathrm{n.d.f.})$ = 0.21.
 The average deviation between SND \cite{Achasov:2005rg,Achasov:2006vp} 
and CMD-2 \cite{Akhmetshin:2003zn} data
 is: $\Delta$(SND -- CMD-2) $\sim (1.3 \pm 3.6)\%$ for the energy range
 $\sqrt{s}$ $\leq$ 0.55 GeV and $\Delta$(SND -- CMD-2)
 $\sim (-0.53 \pm 0.34)\%$ for the energy range $\sqrt{s}$ $\geq$ 0.55 GeV.
The obtained $\rho$ meson parameters are:\\ 
CMD-2 -- $M_{\rho} = 775.97 \pm 0.46 \pm 0.70 $ MeV,\\
$\Gamma_{\rho} = 145.98 \pm 0.75 \pm 0.50 $ MeV,\\
$\Gamma_{ee} = 7.048 \pm 0.057 \pm 0.050 $ keV,\\
$\mathrm{Br}(\omega \to \pi^+\pi^-) = (1.46 \pm 0.12 \pm 0.02) \%$;\\
SND -- $M_{\rho} = 774.6 \pm 0.4 \pm 0.5$ MeV,\\
$\Gamma_{\rho} = 146.1 \pm 0.8 \pm 1.5  $ MeV,\\
$\Gamma_{ee} = 7.12 \pm 0.02 \pm 0.11 $ keV,\\
$\mathrm{Br}(\omega \to \pi^+\pi^-) = (1.72 \pm 0.10 \pm 0.07) \%$.\\ 
The systematic errors were carefully studied and are listed in 
Table~\ref{tsyst-err}.   

\begin{table*}
\begin{center}
\begin{tabular}{|l|c|c|c|}
    \hline
    Sources of errors & CMD-2             & SND & CMD-2 \\
    & $\sqrt {s} < 1 $ GeV &     & $1.4 > \sqrt {s} > 1$ GeV \\
    \hline
    Event separation method & 0.2 - 0.4\% & 0.5\% & $0.2 - 1.5\%$ \\
    Fiducial volume &  0.2\%  & 0.8\% & $0.2 - 0.5\%$ \\
    Detection efficiency & 0.2 - 0.5\% & 0.6\% & $0.5 - 2\%$ \\
    Corrections for pion losses & 0.2\% & 0.2\% & 0.2\% \\
    Radiative corrections & 0.3 - 0.4\% & 0.2\% & $0.5 - 2\%$ \\
    Beam energy determination & 0.1 - 0.3\% & 0.3\% & $0.7 - 1.1\%$  \\
    Other corrections & 0.2\% & 0.5\% & $0.6 - 2.2\%$ \\
    \hline
    The total systematic error & 0.6 - 0.8\% & 1.3 \% & $1.2 - 4.2\%$ \\
    \hline
\end{tabular}
\caption{ The main sources of the systematic errors for different energy regions.}
\label{tsyst-err}
\end{center}
\end{table*}

\begin{figure}[htb]
  \centering\includegraphics[width=0.49\textwidth]{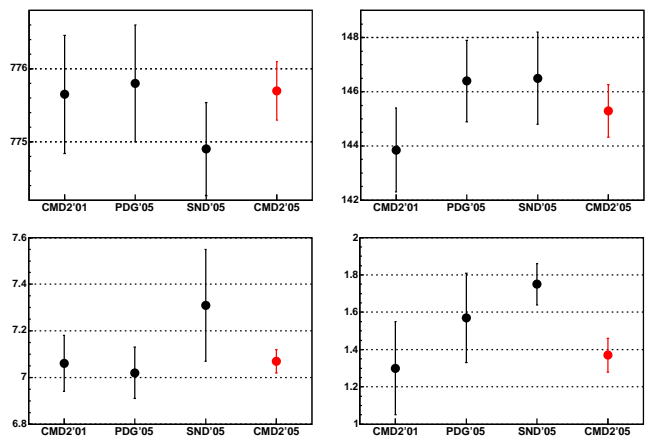}
  \caption{Comparison of $\rho$ meson parameters from CMD-2 and SND
    with corresponding PDG values. The panels (top-left to
    bottom-right) refer to the mass (MeV), width (MeV), leptonic width
    (keV) and the branching fraction of the decay
    $\omega \to \pi^+\pi^-$ (\%).} 
  \label{fig:rho-par} 
\end{figure}

The comparison of the $\rho$ meson parameters determined by CMD-2 and
SND with the values from the PDG is presented in
Fig.~\ref{fig:rho-par}. Good agreement is observed for all parameters, 
except for the branching fraction of $\omega$ decaying to $\pi^+\pi^-$,
where a difference $\sim$ 1.6 standard deviations is observed. 

\subsubsection{Cross section of the process $e^+e^- \to \pi^+\pi^-\pi^{0}$}

This channel was studied
by SND in the energy range $\sqrt{s}$ from 0.6 to 1.4 
GeV~\cite{Achasov:2002ud,Achasov:2003ir}, 
while CMD-2 has reported results of the measurements in
vicinity of the $\omega$~\cite{Akhmetshin:2003zn} and 
$\phi$ meson peaks~\cite{Akhmetshin:2006sc}.   
For both the $\omega$ and $\phi$ resonances CMD-2 and SND obtain consistent 
results for the product of the resonance branching fractions into
$e^+e^-$ and $\pi^+\pi^-\pi^0$, for which they have the world's best accuracy
(SND for the $\omega$ and CMD-2 for the $\phi$ resonance).

CMD-2 has also performed a detailed Dalitz plot analysis of the 
dynamics of $\phi$ decaying to $\pi^+\pi^-\pi^{0}$. 
Two models of $3\pi$ production were used: a $\rho\pi$ mechanism 
and a contact amplitude.  
The result obtained for the ratio of the contact and $\rho\pi$
amplitudes is in good agreement with that of KLOE~\cite{Aloisio:2003ur}. 
 
\begin{figure}[htb]

\centering\includegraphics[width=0.49\textwidth]{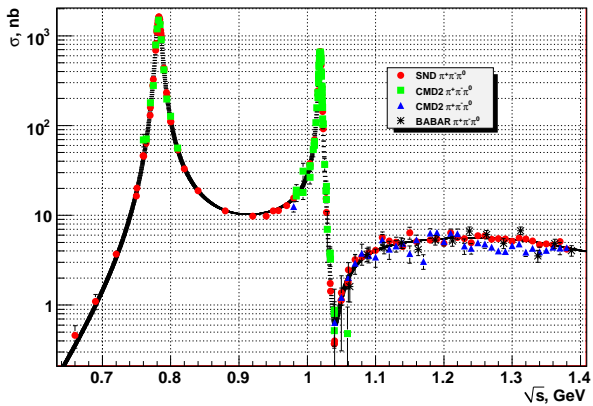}
  \caption{Cross section of the process 
  $e^+e^- \to \pi^+\pi^-\pi^{0}$.}
  \label{fig:3pi-crsect} 
\end{figure}

The systematic accuracy of the measurements is about 1.3\% around  the $\omega$
meson energy region, 2.5\% in the $\phi$ region, and about 5.6\%  for  
higher energies. The results of different experiments are collected in
Fig.~\ref{fig:3pi-crsect}. The curve is the fit which takes into account
the $\rho,\ \omega,\ \phi,\ \omega' $ and $\omega''$ mesons.  

\subsubsection{Cross section of the process $e^+e^- \to 4\pi$}

This cross section becomes 
important for energies above the $\phi$ meson region. CMD-2  
showed that the $a_1(1260)\pi$ mechanism
is dominant for the process  $e^+e^- \to \pi^+\pi^-\pi^+\pi^-$, whereas for
the channel $e^+e^- \to \pi^+\pi^-\pi^{0}\pi^{0}$ in addition the
intermediate state $\omega\pi$ is required to describe the energy
dependence of the cross section~\cite{Akhmetshin:1998df}.
The SND analysis confirmed these conclusions~\cite{Achasov:2003bv}. 
The knowledge of the dynamics of $4\pi$ production allowed to determine
the detector acceptance and efficiencies with better precision
compared to the previous measurements. 

The cross section of the process  $e^+e^- \to \pi^+\pi^-\pi^+\pi^-$ was
measured with a total systematic error of 15\% for CMD-2 and
7\% for SND. For the channel $e^+e^- \to \pi^+\pi^-\pi^{0}\pi^{0}$
the systematic uncertainty was 15 and 8\%, respectively. 
The CMD-2 reanalysis of the process
$e^+e^- \to \pi^+\pi^-\pi^+\pi^-$, 
with a better procedure for the efficiency determination, reduced the
systematic error to (5-7)\%~\cite{Akhmetshin:2004dy}, and these new
results are now in remarkable agreement with other experiments. \\ 

\subsubsection{Other modes}

CMD-2 and SND have also measured the cross sections of the processes 
$e^+e^- \to K_{S} K_{L}$ and $e^+e^- \to K^+K^-$ from threshold and 
up to 1.38 GeV with much better accuracy than before~\cite
{Akhmetshin:2002vj,Achasov:2006bv,Achasov:2007kg}. 
These cross sections were studied 
thoroughly in the vicinity of the
$\phi$ meson, and 
their systematic errors were determined 
with a precision of about 1.7\% (SND) and 4\% (CMD-2), 
respectively. The analyses were based
on two decay modes of the $K_{S}$: 
$K_{S} \to \pi^{0}\pi^{0}$ and $\pi^+\pi^-$.
As for the process $e^+e^- \to K^+ K^-$, 
the systematic uncertainty was studied in detail 
and found to be 2.2\% (CMD-2) and 7\% (SND). 

At energies $\sqrt{s}$ above 1.04 GeV the cross 
sections of the processes $e^+e^- \to K_{S} K_{L}, 
K^+ K^-$ were measured with a statistical 
accuracy of about 4\% and systematic errors of about 
4-6\% and 3\%, respectively, and  are in good agreement
with other experiments. 

\begin{figure}[htb]
\centering\includegraphics[width=0.49\textwidth]{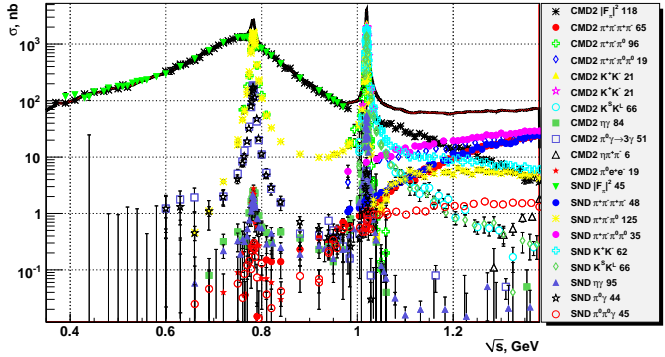}
  \caption{Hadronic cross sections measured by
  CMD-2 and SND in the whole energy range of VEPP-2M. The curve represents
  a smooth spline of the sum of all data.}
  \label{fig:crscmd2snd-all} 
  \end{figure}

To summarise, the experiments performed in 1995--2000 with 
the CMD-2 and SND detectors at VEPP-2M allowed one to measure
the exclusive cross sections of $e^+e^-$ annihilation into
hadrons in the energy range $\sqrt{s}$ = 0.36 - 1.38 GeV with
larger statistics and smaller systematic errors compared to the previous
experiments. 
Figure~\ref{fig:crscmd2snd-all} summarises the cross
section measurements from CMD-2 ans SND.
The results of these experiments determine the current accuracy of the
  calculation of the muon anomaly, and they are one of the main
  sources of information about physics of vector mesons at low energies. 

\subsubsection{$R$ measurement at CLEO}

Two important measurements of the $R$ ratio have been recently
reported by the CLEO Collaboration~\cite{:2007qwa,CroninHennessy:2008yi}.

\begin{figure}
\centering\includegraphics[width=0.49\textwidth]{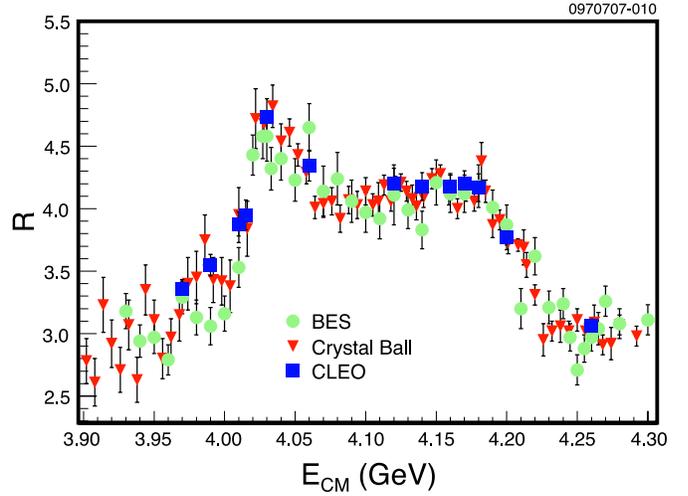}
\caption{Comparison of the $R$ values from CLEO
(the inclusive determination) with those from Crystal Ball and BES.}
\label{fig:low}
\end{figure}

In the energy range just above the open charm threshold,
they collected statistics at thirteen c.m. energy points
from 3.97 to 4.26~GeV~\cite{CroninHennessy:2008yi}.
Hadronic cross sections 
in this region exhibit a
rich structure, reflecting the production of $c\bar{c}$ resonances.
Two independent measurements 
have been performed. In one of them they determined a sum of the 
exclusive cross sections for final
states consisting of two charm mesons ($D\bar{D}$, $D^*\bar{D}$,
 $D^*\bar{D}^*$, $D^+_sD^-_s$,  $D^{*+}_sD^-_s$, and
$D^{*+}_sD^{*-}_s$)
and of processes in which the charm-meson pair is accompanied by a
pion. In the second one they measured the inclusive cross section
with a systematic uncertainty between 5.2 and 6.1\%.
The results of  both measurements are in excellent agreement, which
leads to the 
important conclusion that in this energy range the sum of the two- and 
three-body cross sections
saturates the total cross section for charm production. In 
Fig.~\ref{fig:low} the inclusive cross section measured by CLEO 
is compared with the previous measurements by
Crystal Ball~\cite{Osterheld:1986hw} and BES~\cite{Bai:2001ct}. 
Good agreement is observed between the data.

\begin{figure}
\centering\includegraphics[width=0.49\textwidth]{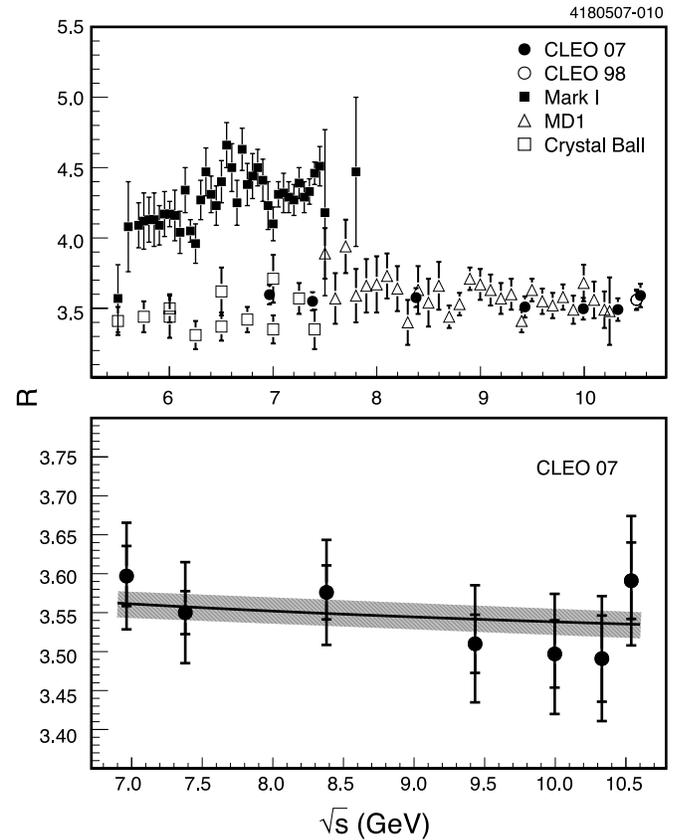}
\caption{Top plot: comparison of the $R$ values from CLEO
with those from MARK I, Crystal Ball and MD-1; bottom plot: comparison
of the new CLEO results with the QCD prediction at $\Lambda=$0.31~GeV.}
\label{fig:high}
\end{figure}

CLEO has also performed a new measurement of $R$ at higher energy. 
They collected statistics at seven c.m.
energy points from 6.964 to 10.538~GeV~\cite{:2007qwa} and reached
a very small systematic uncertainty of 2\% only. Results of their 
scan are presented in Fig.~\ref{fig:high} and are in good 
agreement with those of Crystal Ball~\cite{Osterheld:1986hw}, 
MD-1~\cite{Blinov:1993fw} and the previous 
measurement of CLEO~\cite{Ammar:1997sk}. However, they are obviously  
inconsistent with those of the old MARK I 
measurement~\cite{Siegrist:1981zp}. 
    
\subsubsection{$R$ measurement at BES}

Above 2 GeV the number of final states becomes too large for completely
exclusive measurements, so that  the values of $R$ are measured 
inclusively.

In 1998, as a feasibility test of $R$ measurements, 
BES took data at six c.m. energy
points between 2.6 and 5.0 GeV~\cite{Bai:1999pk}. The integrated
luminosity collected at each energy point changed from 85 to 292~nb$^{-1}$.
The statistical error was  around 3\% per point 
and the systematic error ranged from 7 to 10\%.

Later, in 1999, BES performed a systematic fine scan over the
c.m. energy range from 2 to 4.8~GeV~\cite{Bai:2001ct}. 
Data were taken at 85 energy
points, with an integrated luminosity varying from 
9.2 to 135~nb$^{-1}$ per point. In this experiment, besides 
the continuum region below the 
charmonium threshold, the high charmonium states from 3.77 to
4.50~GeV were studied~\cite{Ablikim:2007gd} in detail. The statistical error 
was between 2 to 3\%,
while the systematic error ranged from 
5 to 8\%, due to improvement 
on hadronic event selection and Monte Carlo simulation of hadronisation
processes. The uncertainty due to the luminosity
determination varied from 2 to 5.8\%.

More recently, in 2003 and 2004, before BES-II
was shut down for the upgrade to BES-III, a high-statistics 
data sample was taken at 2.6, 3.07 and 3.65~GeV, with an integrated luminosity
of 1222, 2291 and 6485~nb$^{-1}$, respectively~\cite{:2009js}. The systematic
error, which exceeded the statistical error, was reduced to
3.5\%  due to further refinement on
hadronic event selection and Monte Carlo simulation.

For BES-III, the main goal of the $R$ measurement is to perform a fine scan
over the whole energy region which BEPC-II can cover. For a continuum region
(below 3.73~GeV), the step size should not exceed 100~MeV, and for
the resonance region (above 3.73~GeV), the step size should be 10 to
20~MeV. Since the luminosity of BEPC-II is two orders of magnitude
higher than at BEPC, the scan of the resonance region will provide
precise information on the $1^{--}$  charmonium states up to
4.6~GeV.

\subsection{Estimate of the theoretical accuracy}

Let us discuss the accuracy of the description of the
processes under consideration. This accuracy can be subdivided into
two major parts: theoretical and technical one.
The first one is related to the precision in the actual computer codes.  
It usually does not take into account all known contributions
in the {\em best} approximation. The technical precision can be verified
by special tests within a given code (e.g., by looking at the
numerical cancellation of the dependence on auxiliary parameters)
and tuned comparisons of different codes. 

The pure theoretical precision consists of unknown high\-er-order 
corrections, of uncertainties in the treatment of photon radiation off
hadrons, and of errors in the phenomenological definition of such
quantities as the hadronic vacuum polarisation and the pion form factor.

Many of the codes used at meson factories do not include contributions
from weak interactions even at Born level. As discussed above, these
contributions are suppressed at least by a factor  of $s/M_Z^2$ and do not
spoil the precision up to the energies of $B$ factories.

Matching the complete one-loop QED corrections with the higher-order 
corrections in the leading logarithmic approximation, 
certain parts of the second-order next-to-leading corrections 
are taken into account~\cite{Balossini:2006wc}. For the case of Bhabha
scattering, where, e.g., soft and virtual photonic corrections 
in $O (\alpha^2L)$ are known analytically,
one can see that their contribution in the relevant kinematic 
region does not exceed 0.1\%.\footnote{The proper choice of the factorisation 
scale~\cite{Arbuzov:2006mu} is important here.}

The uncertainty coming from the 
the hadronic vacuum polarisation has been estimated~\cite{Eidelman:1995ny}
to be of order 0.04\%. For measurements performed with the c.m.
energy at a narrow resonance (like at the $\phi$-meson factories), 
a systematic error in the determination of the resonance contribution to vacuum
polarisation is to be added.

The next point concerns non-leading terms of order $(\alpha/\pi)^2L$.
There are several sources of them. One is the emission of two extra
hard photons, one in the collinear region and one
at large angles. Others are related to virtual and soft-photon
radiative corrections to single hard photon emission and Born processes.
Most of these contributions were not considered up to now.
Nevertheless we can estimate the coefficient in front of the quantity
$(\alpha/\pi)^2L\approx 1\cdot 10^{-4}$ to be of order one.
This was indirectly confirmed by our complete calculations of
these terms for the case of small--angle Bhabha scattering.

Considering all sources of uncertainties mentioned above as
independent, we conclude that the systematic error of our formulae
is about 0.2\% or better, both for muons and pions.
For the former it is a rather safe estimate.
Comparisons between different codes which treat higher-order QED
corrections in different ways typically show agreement at the 
0.1\% level. Such comparisons test the technical and partially  
the theoretical uncertainties.
As for the $\pi^+\pi^-$ and two kaon channels, the uncertainty is 
enhanced due to the presence of form factors and due to the application
of the point-like approximation for the final state hadrons. 




\section{Radiative return }
\label{sec:3}
  \newcommand{\ta}[1]{#1\hspace{-.42em}/\hspace{-.07em}} 
\newcommand{\beq}{\begin{equation}}
\newcommand{\eeq}{\end{equation}}
\newcommand{\non}{\nonumber}
\newcommand{\nn}{\nonumber}
\newcommand{\eq}[1]{eq. (\ref{#1})}         
\newcommand{\Eq}[1]{Eq. (\ref{#1})}
\newcommand{\e}{\varepsilon} 

\def\Li{\hbox{Li}}
\def\DAF{DA\char8NE}

\subsection{History and evolution of radiative return in precision physics}
\label{radret:hist}

The idea to use \emph{Initial State Radiation} to measure
hadronic cross sections from the threshold of a reaction up to the
centre-of-mass (c.m.) energy of colliders with fixed energies $\sqrt{s}$,
to reveal reaction mechanisms and to search for new mesonic states
consists in exploiting the process $e^+e^-\rightarrow hadrons + n
\gamma$, thus reducing the c.m. energy of the colliding
electrons and positrons and consequently the mass squared
$M^2_{\rm had}= s - 2 \sqrt{s} \: \: E_{\gamma}$ of the hadronic
system in the final state by emission of one or more photons. The
method is particularly well suited for modern meson factories
like DA$\mathrm{\Phi}$NE (detector KLOE), running at the
$\mathrm{\phi}$-resonance, BEPC-II (detector BES-III),
commissioned in 2008 and running at the $J/\psi$ and ${\psi}(2S)$-resonances,
PEP-II (detector BaBar) and KEKB (detector
Belle) at the $\Upsilon(4S)$-resonance. Their high
luminosities compensate for the $\alpha / \pi$ suppression
of the photon emission. DA$\mathrm{\Phi}$NE, BEPC-II,
PEP-II and KEKB cover the regions in $M_{\rm had}$ up to
1.02, 3.8 (maximally 4.6) and 10.6 GeV,
respectively (for the latter actually restricted to 4--5 GeV
if hard photons are detected). A big advantage of ISR
is the low point-to-point systematic errors of the
hadronic energy spectra. 
This is because 
the luminosity, the energy of the
electrons and positrons and many other contributions to
the detection efficiencies are determined once for the whole
spectrum. As a consequence, the overall normalisation error is the
same for all energies of the hadronic system. The term
\emph{Radiative Return} alternately used for ISR refers to
the appearance of pronounced resonances (e.g. $\rho,\ \omega,\ \phi,\ 
J/\psi,\ Z$) with energies below the collider energy. Reviews and
updated results can be found in the Proceedings of the
International Workshops in Pisa (2003) \cite{Pisa}, Nara (2004)
\cite{tau04}, Novosibirsk (2006) \cite{Budker}, Pisa (2006)
\cite{tau06}, Frascati (2008) \cite{Frascati08}, and Novosibirsk
(2008) \cite{tau08}.

\begin{figure}[htb]
\includegraphics[width=8cm,height=7cm]{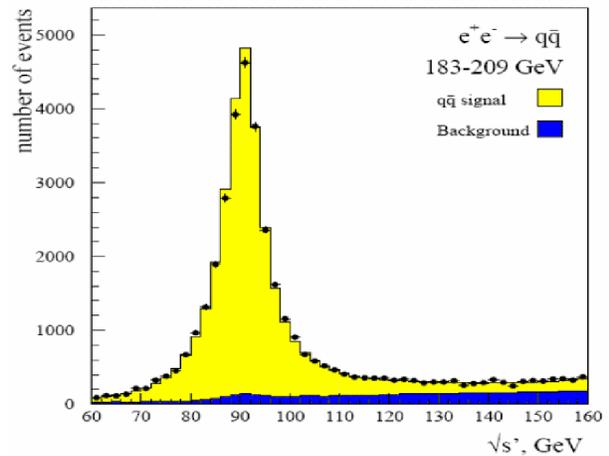}
\caption{
The reconstructed distribution of $e^+ e^-   \rightarrow q \bar q$ events as a function of the invariant mass of the quark-antiquark system. The data has been taken for a collider energy range of 182 - 209 GeV. The prominent peak around 90 GeV represents the Z-resonance, populated after emission of photons in the initial state \cite{Abdallah:2005ph}.}
\label{fig:Rsigma.eps}
\end{figure}

Calculations of ISR date back to the sixties to seventies of
the $20^{th}$ century. For example, photon emission for muon pair
production in electron-positron collisions has been calculated in
Ref. \cite{Baier:1965bg}, for the $2 \pi $-final state in Refs.\ 
\cite{Baier:1965jz,Pancheri:1969yx}; the resonances $\rho,\ \omega$ and $\phi$ have been
implemented in Ref. \cite{Pancheri:1969yx}, the excitations 
$\psi(3100)$ and $\psi^{\prime}(3700)$ in Ref. \cite{Greco:1975rm}, and
the possibility to determine the pion form factor was discussed in
Ref. \cite{Chen:1974wv}. The application of ISR to the new high
luminosity meson factories, originally aimed at the determination
of the hadronic contribution to the vacuum polarisation, more
specifically the pion form factor, has materialised in the late
nineties. Early calculations of ISR for the colliders
DA$\mathrm{\Phi}$NE, PEP-II and KEKB can be found in
\cite{Spagnolo:1998mt,Khoze:2000fs,Benayoun:1999hm,Arbuzov:1998te}. In Ref. \cite{Arbuzov:1997je}
calculations of radiative corrections for pion and kaon production
below energies of 2 GeV have been reported. An impressive example
of ISR is the \emph{Radiative Return} to the region of the
\emph{Z}-resonance at LEP-2 with collider energies around
200 GeV \cite{Abbiendi:2003dh,Abdallah:2005ph,Achard:2005nb,Schael:2006wu} (see Fig.~\ref{fig:Rsigma.eps}).

\vglue 1.0 cm

ISR became a powerful tool for the analysis of experiments
at low and intermediate energies with the development of
EVA-PHOKHARA, a Monte Carlo 
generator which is user
friendly, flexible and easy to implement into the software of the
existing detectors
\cite{Binner:1999bt,Czyz:2000wh,Rodrigo:2001jr,Kuhn:2002xg,Rodrigo:2001kf,Czyz:2002np,Czyz:2003ue,Czyz:2004rj,Czyz:2004nq,Czyz:2005as,Czyz:2008kw,Czyz:2004ua,Czyz:2007wi,Czyz:2007ue,Czyz:2006xf,Czyz:2008zz,Grzelinska:2008eb}.

EVA and its successor PHOKHARA  allow to simulate the
process $e^+e^- \to hadrons + \gamma$ for a variety of exclusive
final states.
 As a starting point EVA was constructed
\cite{Binner:1999bt} to simulate leading order ISR and FSR
for the $\pi^+ \pi^-$ channel, and additional soft and collinear ISR
was included on the basis of structure functions taken from
\cite{Caffo:1994dm}. Subsequently EVA was extended to
include the four-pion state \cite{Czyz:2000wh}, albeit without
FSR. Neglecting FSR and radiative corrections, i.e. including one-photon emission from the initial state only, the
cross section for the radiative return can be cast into the
product of a radiator function $H(M^2_{\rm had},s)$ and the cross
section $\sigma(M^2_{\rm had})$ for the reaction $e^+e^- \to hadrons$:

\noindent
$s \:  {{\rm d}\sigma(e^{+} e^{-} \rightarrow hadrons \: \gamma)}/{{\rm d}
M_{\rm had}^2}=\sigma (M_{\rm had}^2) \: H(M_{\rm had}^2,s)$.

However, for a precise evaluation of $\sigma(M^2_{\rm had})$, the
leading logarithmic approximation inherent in EVA is
insufficient. Therefore, in the next step, the exact one-loop
correction to the ISR process was evaluated analytically,
first for large angle photon emission \cite{Rodrigo:2001jr}, then for
arbitrary angles, including collinear configurations \cite{Kuhn:2002xg}. This
was and is one of the key ingredients of the generator called
PHOKHARA \cite{Rodrigo:2001kf,Czyz:2002np}, which also includes
soft and hard real radiation, evaluated using exact matrix
elements formulated within the framework of helicity amplitudes
\cite{Rodrigo:2001kf}. FSR in NLO approximation was
addressed in \cite{Czyz:2003ue} and incorporated in
\cite{Czyz:2004rj,Czyz:2004nq}. The importance of the charge
asymmetry, a consequence of interference between ISR and
FSR amplitudes, for a test of the (model dependent)
description of FSR has been emphasised already in
Ref. \cite{Binner:1999bt} and was further studied in  \cite{Czyz:2004nq}.


Subsequently the generator was extended to allow for the generation of
many more channels with mesons, like $K^+K^-$, $K^0\bar K^0$,
$\pi^+\pi^-\pi^0$, for an improved description of the $4\pi$ modes
\cite{Czyz:2005as,Czyz:2008kw} and for improvements in the description of
FSR for the $\mu^+\mu^-$ channel
\cite{Czyz:2004rj,Czyz:2004nq}. Also the nucleon channels $p \bar{p}$
and $n \bar{n}$ were implemented \cite{Czyz:2004ua}, and it was
demonstrated that the separation of electric and magnetic proton
form factors is feasible for a wide energy range. In fact, for the
case of $\Lambda \bar{\Lambda}$ and including the 
polarisation-sensitive weak decay of $\Lambda$ into the simulation, it was
shown that even the relative phase between the two independent
form factors could be disentangled \cite{Czyz:2007wi}.

Starting already with \cite{Melnikov:2000gs}, various improvements
were made to include the direct decay $\phi \to \pi^+ \pi^-
\gamma$ as a specific aspect of FSR into the generator, a
contribution of specific importance for data taken on top of the
$\mathrm{\phi}$ resonance.

This was further pursued in the event generators
FEVA and FASTERD based on EVA-PHOKHARA. FEVA
includes the effects of the direct decay $\phi
\rightarrow \pi^{-} \pi^{+} \gamma$ and the decay via the
$\rho$-resonance $\phi \rightarrow \rho^{\pm} \pi^{\mp}
\rightarrow \pi^{-} \pi^{+} \gamma$
\cite{Dubinsky:2004xv,Pancheri:2006cp,Pancheri:2007xt}. The code
FASTERD takes into account \emph{Final State Radiation} in
the framework of both Resonance Perturbation Theory and sQED,
\emph{Initial State Radiation}, their interference and also the
direct decays $ e^{+} e^{-} \rightarrow \phi \rightarrow (f_{o};
f_{o}+\sigma) \gamma \rightarrow \pi^{+} \pi^{-} \gamma$, $e^{+}
e^{-} \rightarrow \phi \rightarrow \rho^{\pm} \pi^{\mp}
\rightarrow \pi^{+} \pi^{-} \gamma$ and $e^{+} e^{-} \rightarrow
\rho \rightarrow \omega \pi^{o} \rightarrow \pi^{o} \pi^{o}
\gamma$ \cite{Shekhovtsova:2009yn}, with the possibility to include
additional models.

EVA-PHOKHARA was applied for the first time to an
experiment to determine the cross section $e^+e^-\rightarrow
\pi^{+} \pi^{-}$ from the reaction threshold up to the maximum
energy of the collider with the detector KLOE at DA$\mathrm{\Phi}$NE \cite{Cataldi:1999aa,Denig:2001ra,Aloisio:2001xq,Denig:2002ps,Valeriani:2002yk,Venanzoni:2002jb,Muller:2004mb,Aloisio:2003dw,Denig:2003jn,Valeriani:2004mp,Kluge:2004mc,Denig:2005eb,Kluge:2005ac,Aloisio:2005tm,Denig:2006kj,Muller:2006bk,Leone:2006bm,Venanzoni:2007zz,Nguyen:2008rv,Muller:2007zzd,Ambrosino:2007vj,Aloisio:2004bu,:2008en,Kluge:2008fb,Venanzoni:2009aq} (Section
\ref{rr:kloe}). The motivation was
the determination of the $2 \pi$ final state contribution to the hadronic
vacuum polarisation.

The determination of the hadronic contribution to the vacuum
polarisation, which arises from the coupling of virtual photons to
quark-antiquark pairs, $ \gamma ^ \star \rightarrow q \bar {q}
\rightarrow \gamma ^\star $, is possible by measuring the cross
section of electron-positron annihilation into hadrons, $e^+e^-
\rightarrow \gamma^* \rightarrow q \bar {q} \rightarrow hadrons$, and
applying the optical theorem. It is of great importance for the
interpretation of the precision measurement of the anomalous
magnetic moment of the muon $a_\mu$ in Brook\-ha\-ven (E821) \cite{Bennett:2002jb,Bennett:2004pv,Bennett:2006fi,Hertzog:2008zz} and for the determination of the
value of the running QED coupling at the $Z^o$
resonance, $\alpha (m_{Z}^{2})$, which contributes to precision tests of
the \emph{Standard Model} of particle physics, for details see e.g. \emph{Jegerlehner} \cite{Jegerlehner:2008zza}, also \emph{Davie} and
\emph{Marciano} \cite{Davier:2004gb}, or \emph{Teubner et al.}
\cite{Teubner:2008zz,Jegerlehner:2008zz,Stockinger:2008zz}. 
The hadronic contribution to $a_\mu$ below about 2 GeV is
dominated by the $2 \pi$ final state, which contributes about 70\% 
due to the dominance of the $\rho-$resonance. Other major
contributions come from the three- and four-pion final states. These
hadronic final states constitute at present the largest error to
the \emph{Standard Model} values of $a_\mu$ and $\alpha
(m_{Z}^{2})$ and can be determined only experimentally. This is because
calculations within perturbative QCD are unrealistic,
calculations on the lattice are not yet available with the necessary
accuracy, and calculations in the framework of chiral perturbation
theory are restricted to values close to the reaction thresholds.
At energies above about 2 to 2.5 GeV, perturbative QCD
calculations start to become possible and reliable, see e.g. Refs.
\cite{Kuhn:1998ze,Eidelman:1998vc}, and also
\cite{Chetyrkin:1996ia}.

The Novosibirsk groups CMD-2 \cite{Budker,Akhmetshin:2001ig,Akhmetshin:2002vj,Akhmetshin:2003ag,Akhmetshin:2003zn,Aulchenko:2006na,Akhmetshin:2004gw,Akhmetshin:2006wh,Ignatov:2008zz,Akhmetshin:2006bx}
and SND \cite{Achasov:2003ir,Achasov:2005rg,Achasov:2006xc,Achasov:2006vp,Achasov:2007kg,Achasov:2006bv} measured hadronic cross sections below
1.4 GeV by changing the collider energy (\emph{energy scan}, see
the preceding Section \ref{sec:2}). The \emph{Initial State Radiation}
method used by KLOE represents an alternative, independent
and complementary way to determine hadronic cross sections with
different systematic errors. KLOE has determined the cross
section for the reaction $e^{+} e^{-}\rightarrow \pi^{+} \pi^{-}$
in the energy region between 0.63 and 0.958 GeV by measuring the
reaction $e^{+} e^{-}\rightarrow \pi^{+} \pi^{-} \gamma$ and
applying a radiator function based on PHOKHARA. 
For the hadronic contribution to the anomalous magnetic moment of
the muon due to the $2 \pi$ final state it obtained $a_{\mu}^{\pi\pi} = (356.7
\pm 3.1_{{\rm stat}+{\rm syst}}) \cdot 10^{-10}$ \cite{:2008en}. This value
is in good agreement with those from SND \cite{Achasov:2006bv} and
CMD-2 \cite{Akhmetshin:2006bx}, $a_{\mu}^{\pi\pi} = (361.0 \pm
5.1_{{\rm stat}+{\rm syst}}) \cdot 10^{-10}$ and $a_{\mu}^{\pi\pi} = (361.5
\pm 3.4_{stat+syst})\cdot 10^{-10}$, respectively, leading to an evaluation of 
$a_{\mu}$~\cite{Jegerlehner:2008zza,Davier:2004gb,Teubner:2008zz,Jegerlehner:2008zz,Stockinger:2008zz,Davier:2009ag} which differs
by about three standard deviations from the BNL experiment~\cite{Bennett:2006fi}.
A different evaluation using $\tau$ decays into
two pions results in a reduced discrepancy~\cite{Davier:2004gb,Davier:2009ag}.
The difference between $e^{+}e^{-}$ and $\tau$ based analyses
 is at present not understood.  
But one has to be aware that the evaluation with $\tau$ data needs more theoretical input.


Soon after the application of EVA-PHOKHARA to KLOE
\cite{Cataldi:1999aa}, the BaBar collaboration also started the
measurement of hadronic cross sections exploiting ISR
\cite{Solodov:2002xu} and using PHOKHARA (Section
\ref{rr:babar}). In recent years a
plethora of final states has been studied, starting with the
reaction $e^+e^- \rightarrow J/\psi \: \gamma \rightarrow \mu^{+}
\mu^{-}\: \gamma$ \cite{Aubert:2003sv}. While detecting a hard
photon, the upper energy for the hadron cross sections is limited
to roughly 4.5 GeV. Final states with 3, 4, 5, 6 charged and
neutral pions, 2 pions and 2 kaons, 4 kaons, 4 pions and 2 kaons,
with a $\phi$ and an $f_{o}(980)$, $J/\psi$ and 2 pions or 2 kaons,
pions and $\eta$, kaons and $\eta$, but also baryonic final states
with protons and antiprotons, $\Lambda^{o}$ and $\bar
\Lambda^{o}$, $\Lambda^{o}$ and $\bar {\Sigma^{o}}$, $\Sigma^{o}$
and $\bar {\Sigma^{o}}$, $D \bar D$, $D \bar D^*$, and $D^* \bar
D^*$ mesons, etc. have been investigated \cite{Aubert:2004kj,Aubert:2005eg,Aubert:2005cb,Aubert:2006jq,Aubert:2006bu,Aubert:2007ur,Aubert:2007ym,Aubert:2007ef,Aubert:2007uf,Aubert:2006mi,:2008ic,:2009xs,Denig:2008zz}.
In preparation are final states with 2 pions \cite{Davier:2009aa}
and 2 kaons. Particularly important final states are those with 4
pions (including $\omega \pi^o$). They contribute significantly to
the muon anomalous magnetic moment and were poorly known
before the ISR measurements. In many of these channels
additional insights into isospin symmetry breaking are expected from
the comparison between $e^+ e^-$ annihilation and $\tau$ decays.

More recently also Belle joined the ISR programme
with emphasis on final states containing mesons with hidden and
open charm: $J/\psi$ and $\psi(2S)$, 
$D^{(*)}$ and $\bar D^{(*)}$,
$\Lambda_c{^+} \Lambda_c{^-}$ 
\cite{:2007sj,:2007ea,:2007bt,Abe:2006fj,Pakhlova:2008zza,Pakhlova:2007fq,Pakhlova:2008vn,Pakhlova:2009jv} 
(Section \ref{rr:belle}).

A major surprise in recent years was the opening of a totally new
field of hadron spectroscopy by applying ISR. Several new, 
relatively narrow highly excited states with $J^{PC}= 1^{--}$, the
quantum numbers of the photon, have been discovered (preliminarily
denoted as \emph{X, Y, Z}) at the $B$ factories PEP-II and
KEKB with the detectors BaBar and Belle,
respectively. The first of them was found by BaBar in the
reaction $e^+e^-\rightarrow Y(4260)\: \gamma \rightarrow J/\psi \:
\pi^{+} \pi^{-} \gamma$ \cite{Aubert:2005rm},
 a state around 4260 MeV with
a width of 90 MeV, later confirmed by Belle via ISR
\cite{Abe:2006hf,:2007sj} and by CLEO in an direct energy scan
\cite{Coan:2006rv} and a radiative return \cite{He:2006kg}. Another state was detected at 2175 MeV by
BaBar in the reaction $e^+e^-\rightarrow Y(2175)\: \gamma
\rightarrow \phi f_{o}(980)\gamma$ \cite{Aubert:2006bu}. Belle
found new states at 4050, 4360, 4660 MeV in the reactions
$e^+e^-\rightarrow Y\: \gamma \rightarrow J/\psi \: \pi^{+}
\pi^{-}\gamma$ and $e^+e^-\rightarrow Y\: \gamma \rightarrow
\psi(2S)\: \pi^{+} \pi^{-}\gamma$ 
\cite{:2007sj,:2007ea}. The structure of
basically all of these new states (if they will survive) is
unknown so far. Four-quark states, e.g. a $[cs][\bar{c} \bar{s}]$
state for $Y(4260)$, a $[ss][\bar{s}\bar{s}]$ state for $Y(2175)$,
hybrid and molecular structures are discussed, see also
\cite{Kalashnikova:2008zz}.

Detailed analyses allow, in addition, also the identification of
intermediate states, and consequently a study of reaction
mechanisms. For instance, in the case of the final state with 2
charged and 2 neutral pions $(e^+ e^- \rightarrow \pi^{+} \pi^{-}
\pi^{o} \pi^{o} \gamma)$, the dominating intermediate states are
$\omega \pi^{o}$ and $a_{1}(1260) \pi $, while $\rho^{+} \rho^{-}$
and $\rho^{o} f_{o}(980)$ contribute significantly less.

Many more highly excited states with quantum numbers different
from those of the photon have been found in decay chains of the
primarily produced heavy mesons at the $B$ factories
PEP-II and KEKB. These analyses without ISR
have clearly been triggered and encouraged by the unexpected
discovery of highly excited states with $J^{PC}= 1^{--}$ found
with ISR.

Also baryonic final states with protons and antiprotons,
$\Lambda^{o}$ and $\bar \Lambda^{o}$, $\Lambda^{o}$ and $\bar
{\Sigma^{o}}$, $\Sigma^{o}$ and $\bar {\Sigma^{o}}$ have been
investigated using ISR. The effective proton form factor (see Section~\ref{rr:babar}) shows a strong increase down to the $p\bar{p}$ threshold and
nontrivial structures at invariant $p\bar{p}$ masses of 2.25 and
3.0 GeV, so far unexplained \cite{Aubert:2005cb,Maas:2008zza,Salme:2008an,Dmitriev:2008zz,Baldini:2007qg}. Furthermore, it
should be possible to disentangle electric and magnetic form
factors and thus shed light on discrepancies between different
measurements of these quantities in the space-like region
\cite{Arrington:2003df}.

Prospects for the \emph{Radiative Return} at the
Novosibirsk collider VEPP-2000 and BEPC-II are
discussed in Sections \ref{rr:vepp2000} and \ref{rr:besiii}.

\subsection{Radiative return: a theoretical overview}
\label{radret:theo}

\subsubsection{Radiative return at leading order}
\label{sec:LO}

\begin{figure}[h]
\begin{center}
\epsfig{file=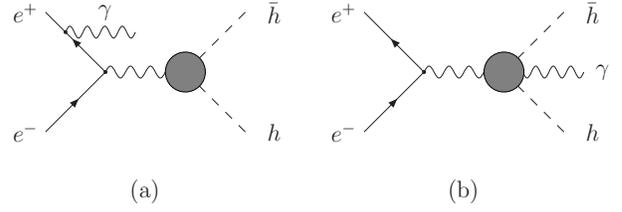,width=8.5cm} 
\caption{Leading order contributions to the reaction
$e^+e^-\to h \, \bar h + \gamma$ from ISR (a) and FSR (b).
Final state particles are pions or muons, or any other
multi-hadron state. The blob represents the hadronic 
form factor.}
\label{fig:born}
\end{center}
\end{figure}

We consider the $e^+ e^-$ annihilation process
\beq
e^+(p_1) + e^-(p_2) \rightarrow {\rm hadrons} + \gamma(k_1)~,
\label{eq:LO}
\eeq
where the real photon is emitted either from the initial 
(Fig.~\ref{fig:born}a) or the final state (Fig.~\ref{fig:born}b).
The former process is denoted initial state radiation (ISR), 
while the latter is called final state radiation (FSR).

The differential rate for the ISR process 
can be cast into the product of a leptonic $L^{\mu\nu}$ 
and a hadronic $H^{\mu\nu}$ tensor and the corresponding 
factorised phase space
\bea
{\rm d}\sigma_{\rm ISR} &=& \frac{1}{2s} L^{\mu \nu}_{\rm ISR} H_{\mu \nu}
\nn \\ && \times {\rm d} \Phi_2(p_1,p_2;Q,k_1) {\rm d} \Phi_n(Q;q_1,\cdot,q_n) 
\frac{{\rm d}Q^2}{2\pi}~,
\eea
where ${\rm d} \Phi_n(Q;q_1,\cdot,q_n)$ denotes the 
hadronic $n$-body phase-space with all the statistical factors 
coming from the hadro\-nic final state included, 
$Q = \sum q_i$ and $s=(p_1+p_2)^2$.

For an arbitrary hadronic final state, the matrix element for 
the diagrams in Fig.~\ref{fig:born}a is given by  
\begin{align}
{\cal A}_{\rm ISR}^{(0)} 
&= M^{(0)}_{\rm ISR} \cdot J^{(0)} =  \non \\
&= - \frac{e^2}{Q^2}  \bar{v}(p_1) \bigg(   
\frac{\ta{\varepsilon}^*(k_1)[\ta{k}_1-\ta{p}_1+m_e]
\gamma^{\mu}}{2 k_1 \cdot p_1} \non \\ 
& \qquad + \frac{\gamma^{\mu}[\ta{p}_2-\ta{k}_1+m_e]\ta{\varepsilon}^*(k_1)}
{2 k_1 \cdot p_2} 
\bigg) u(p_2) \; J_{\mu}^{(0)}~, 
\label{AISR}
\end{align}
where $J_{\mu}$ is the hadronic current.
The superscript $(0)$ indicates that the scattering amplitude 
is evaluated at tree-level. 
Summing over the polarisations of the final real photon,
averaging over the polarisations of the initial $e^+ e^-$ state,
and using current conservation, $Q \cdot J^{(0)} = 0$, 
the leptonic tensor
\begin{equation*}
L_{\rm ISR}^{(0), \mu \nu} = \overline{M_{\rm ISR}^{(0), \, \mu} 
( M_{\rm ISR}^{(0), \, \nu} )^\dagger}
\end{equation*}
can be written in the form
\begin{align}
L_{\rm ISR}^{(0),\, \mu \nu} &= 
\frac{(4 \pi \alpha)^2}{Q^4} \; \bigg[ \left( 
\frac{2 m^2 q^2(1-q^2)^2}{y_1^2 y_2^2}
- \frac{2 q^2+y_1^2+y_2^2}{y_1 y_2} \right) g^{\mu \nu} \non \\ & 
+ \left(\frac{8 m^2}{y_2^2} - \frac{4q^2}{y_1 y_2} \right) 
\frac{p_1^{\mu} p_1^{\nu}}{s} 
+ \left(\frac{8 m^2}{y_1^2} - \frac{4q^2}{y_1 y_2} \right) 
\frac{p_2^{\mu} p_2^{\nu}}{s} \non \\
& - \left( \frac{8 m^2}{y_1 y_2} \right) 
\frac{p_1^{\mu} p_2^{\nu} + p_1^{\nu} p_2^{\mu}}{s} \bigg]~, 
\label{Lmunu0}
\end{align}
with 
\begin{equation}
y_i = \frac{2 k_1 \cdot p_i}{s}~, 
\qquad m^2=\frac{m_e^2}{s}~, \qquad q^2=\frac{Q^2}{s}~.
\label{dimensionless}
\end{equation}
The leptonic tensor is symmetric under the exchange of the electron and 
the positron momenta. Expressing the bilinear products $y_i$ 
by the photon emission angle in the c.m. frame,
\begin{equation*}
y_{1,2} = \frac{1-q^2}{2}(1 \mp \beta \cos \theta)~, 
\qquad \beta = \sqrt{1-4m^2}~,
\end{equation*}
and rewriting the two-body phase space as 
\begin{equation}
{\rm d} \Phi_2(p_1,p_2;Q,k_1) = \frac{1-q^2}{32 \pi^2} {\rm d} \Omega~,
\end{equation}
it is evident that expression (\ref{Lmunu0}) contains several 
singularities: soft singularities for $q^2\rightarrow 1$ and 
collinear singularities for $\cos \theta \rightarrow \pm 1$.
The former are avoided by requiring a minimal photon energy.
The latter are regulated by the electron mass.
For $s \gg m_e^2$ the expression (\ref{Lmunu0}) can  
nevertheless be safely taken in the limit $m_e\rightarrow 0$ if the 
emitted real photon lies far from the collinear region.
In general, however, one encounters spurious singularities in the 
phase space integrations if powers of $m^2=m_e^2/s$ are 
neglected prematurely.

Physics of the hadronic system, whose description is model dependent, 
enters through the hadronic tensor 
\begin{equation}
H_{\mu \nu} = J^{(0)}_{\mu} (J^{(0)}_{\nu})^\dagger~,
\end{equation}
where the hadronic current has to be parametrised through form factors.
For two charged pions in the final state, the current 
\begin{equation}
J^{(0), \, \mu}_{\pi^+\pi^-} = i e F_{2\pi}(Q^2) \; (q_1-q_2)^{\mu}~,
\end{equation}
where $q_1$ and $q_2$ are the momenta of the $\pi^+$ and
$\pi^-$, respectively, is determined by only one function,
the pion form factor $F_{2\pi}$.
The current for the \(\mu^+\mu^-\) final state is 
obviously defined by QED:
\begin{eqnarray}
 J^{(0), \, \mu}_{\mu^+\mu^-} = i e \, \bar u(q_2)\gamma^\mu  v(q_1)~.
\end{eqnarray}

Integrating the hadronic tensor over the hadronic 
phase space, one gets
\begin{equation}
\int H^{\mu \nu} {\rm d}\Phi_n(Q;q_1, \cdot, q_n) 
= \frac{e^2}{6\pi} (Q^{\mu}Q^{\nu} - g^{\mu \nu} Q^2) R(Q^2)~,
\end{equation}
where $R(Q^2) = \sigma(e^+ e^- \rightarrow {\rm hadrons})/
\sigma_0(e^+ e^- \rightarrow \mu^+ \mu^-)$, with 
\beq
\sigma_0(e^+ e^- \rightarrow \mu^+ \mu^-) = \frac{4\pi \, \alpha^2}{3 Q^2}
\eeq
the tree-level muonic cross section in the limit $Q^2\gg 4m_{\mu}^2$.
After the additional integration over the photon angles,  
the differential distribution 
\begin{align}
Q^2 \frac{{\rm d}\sigma_{\rm ISR}}{{\rm d}Q^2} = \frac{4\alpha^3}{3 s} R(Q^2)
\left\{ \frac{s^2+Q^4}{s(s-Q^2)} \left( L - 1 \right) \right\}~,
\label{diff1n}
\end{align}
with $L=\log(s/m_e^2)$ is obtained. 
If instead the photon polar angle is restricted to be 
in the range $\theta_{\rm min}< \theta < \pi-\theta_{\rm min}$,
this differential distribution is given by 
\begin{align}
Q^2 \frac{{\rm d}\sigma_{\rm ISR}}{{\rm d}Q^2} &= \frac{4\alpha^3}{3 s} R(Q^2)
\bigg\{ \frac{s^2+Q^4}{s(s-Q^2)} 
\log \frac{1+\cos \theta_{\rm min}}{1-\cos \theta_{\rm min}} \non \\
& - \frac{s-Q^2}{s} \cos \theta_{\rm min} \bigg\}~.
\label{diff2n}
\end{align}

\begin{figure}[th]
\begin{center}
\epsfig{file=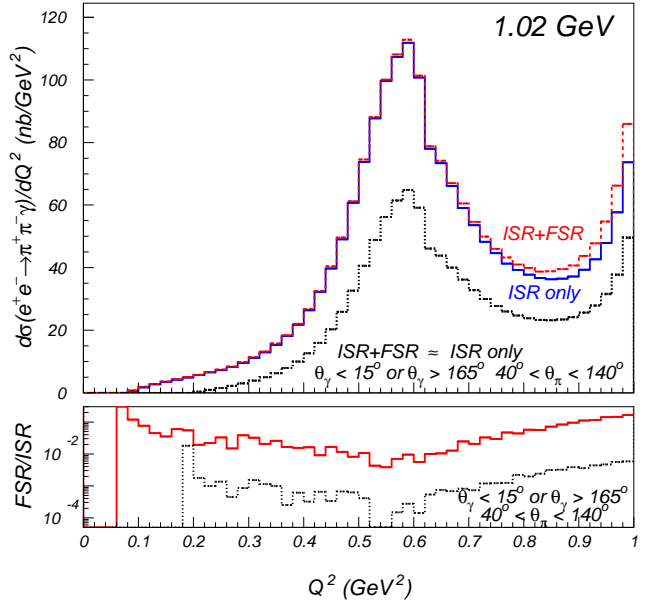,width=8.5cm}
\caption{Suppression of the FSR contributions
to the cross section by a suitable choice of angular cuts; results from the PHOKHARA generator; no cuts (upper curves) and suitable cuts applied (lower curves). }
\label{fig:isrtofsr}
\end{center}
\end{figure}

In the latter case, the electron mass can be taken equal to zero
before integration, since the collinear region is excluded
by the angular cut. The contribution of the two-pion exclusive
channel can be calculated from \Eq{diff1n} and \Eq{diff2n} with 
\beq
R_{\pi^+\pi^-}(Q^2) = \frac14 \left(1-\frac{4m_{\pi}^2}{Q^2}\right)^{3/2} 
|F_{2\pi}(Q^2)|^2~,
\eeq 
and the corresponding muonic contribution with 
\beq
R_{\mu^+\mu^-}(Q^2) = \sqrt{1-\frac{4m_{\mu}^2}{Q^2}} 
\left(1+\frac{2m_\mu^2}{Q^2}\right)~.
\eeq

A potential complication for the measurement of the hadronic cross section 
from the radiative return may arise from the interplay between photons 
from ISR and FSR \cite{Binner:1999bt}.
Their relative strength is strongly dependent
on the photon angle relative to the beam and to the direction of the final 
state particles, the c.m. energy of the reaction and the invariant mass 
of the hadronic system. While ISR is independent of the hadronic final state,
FSR is not. Moreover, it cannot be predicted from first principles and 
thus has to be modelled.

\begin{figure*}[ht]
\begin{center}
\epsfig{file=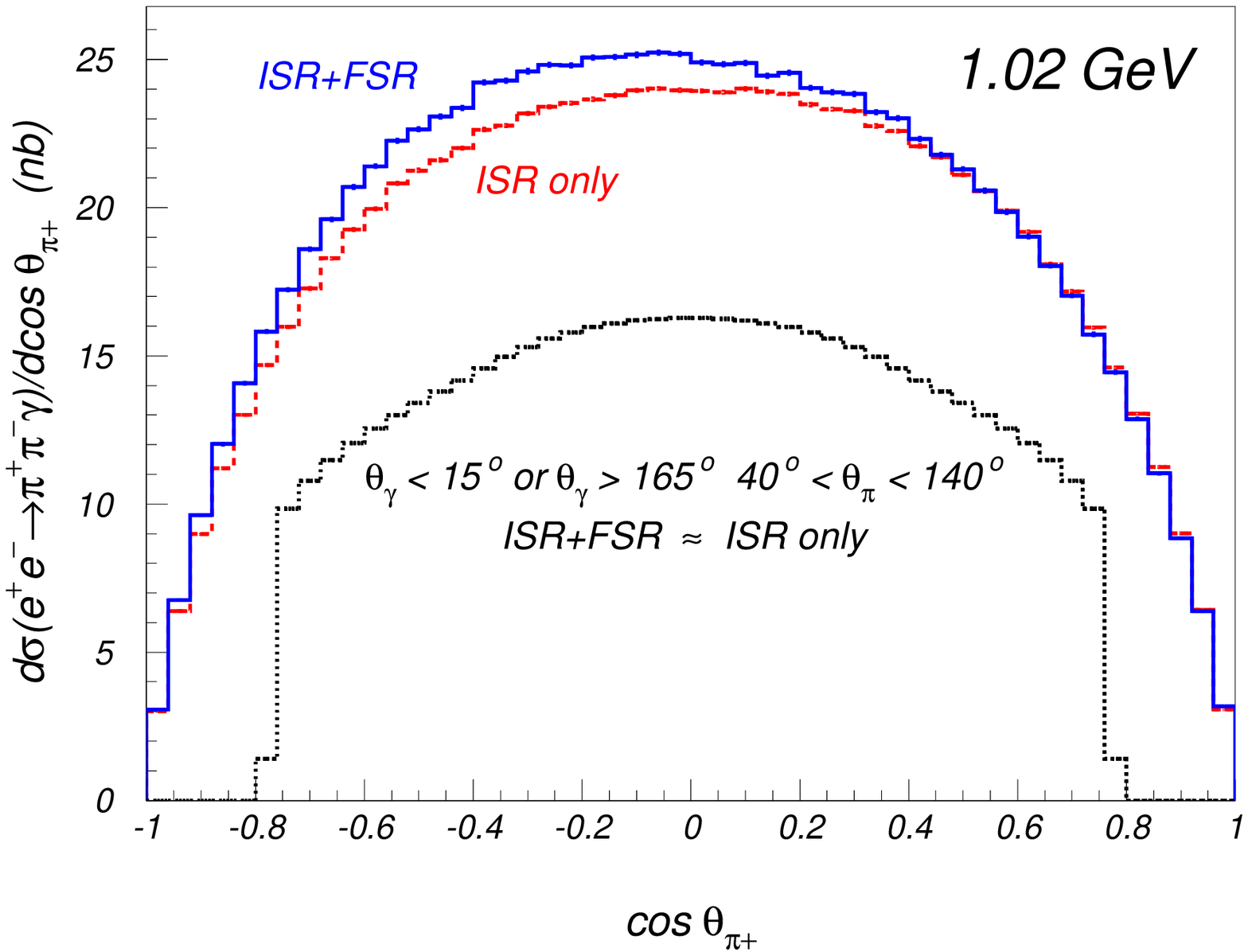,width=8cm,height=6cm} 
\epsfig{file=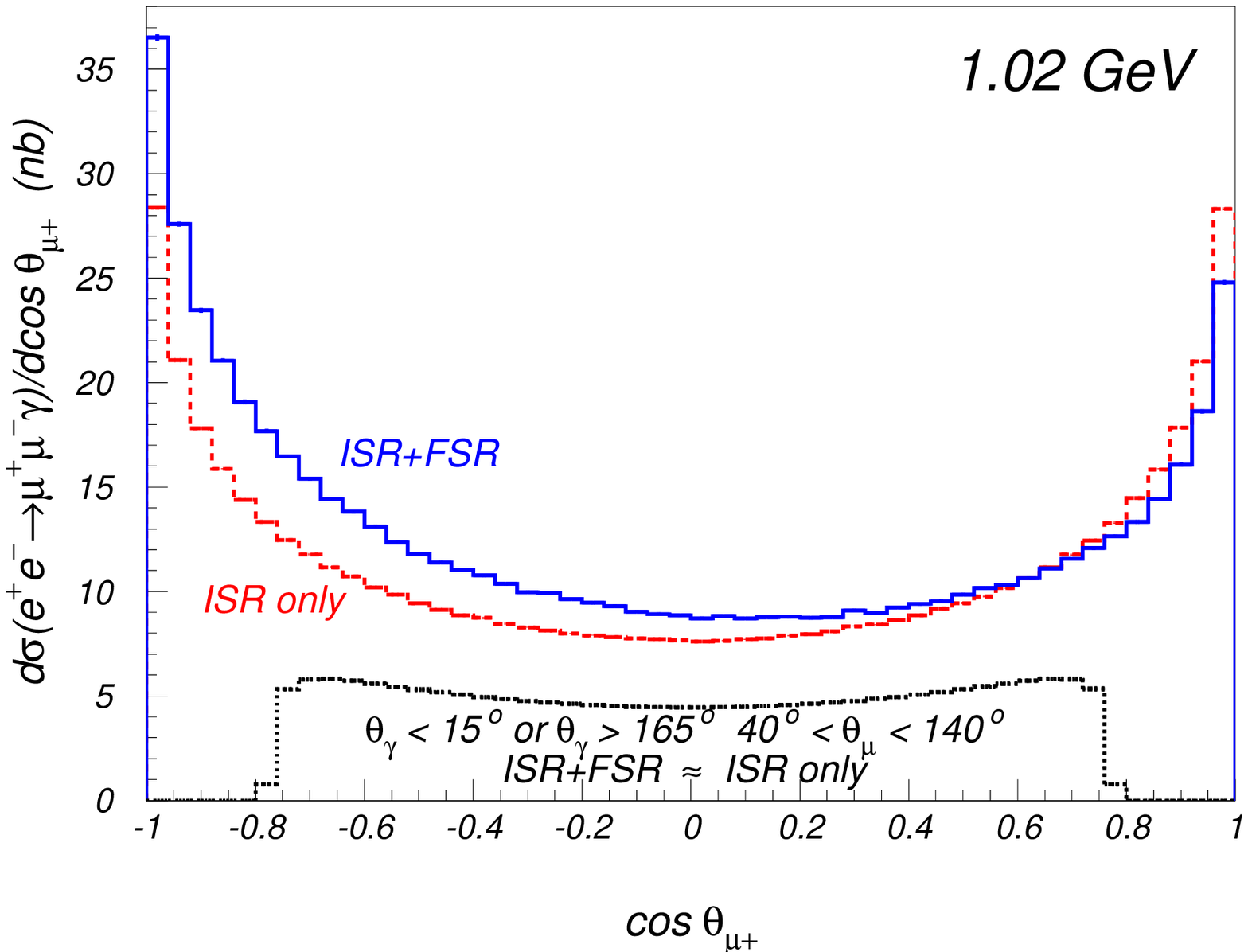,width=8cm,height=6cm} 
\end{center}
\caption{Angular distributions of $\pi^+$ and $\mu^+$ at 
$\sqrt{s}=1.02$~GeV with and without FSR 
for different angular cuts.}
\label{fig:angular}
\end{figure*}

\begin{figure*}[ht]
\begin{center}
\epsfig{file=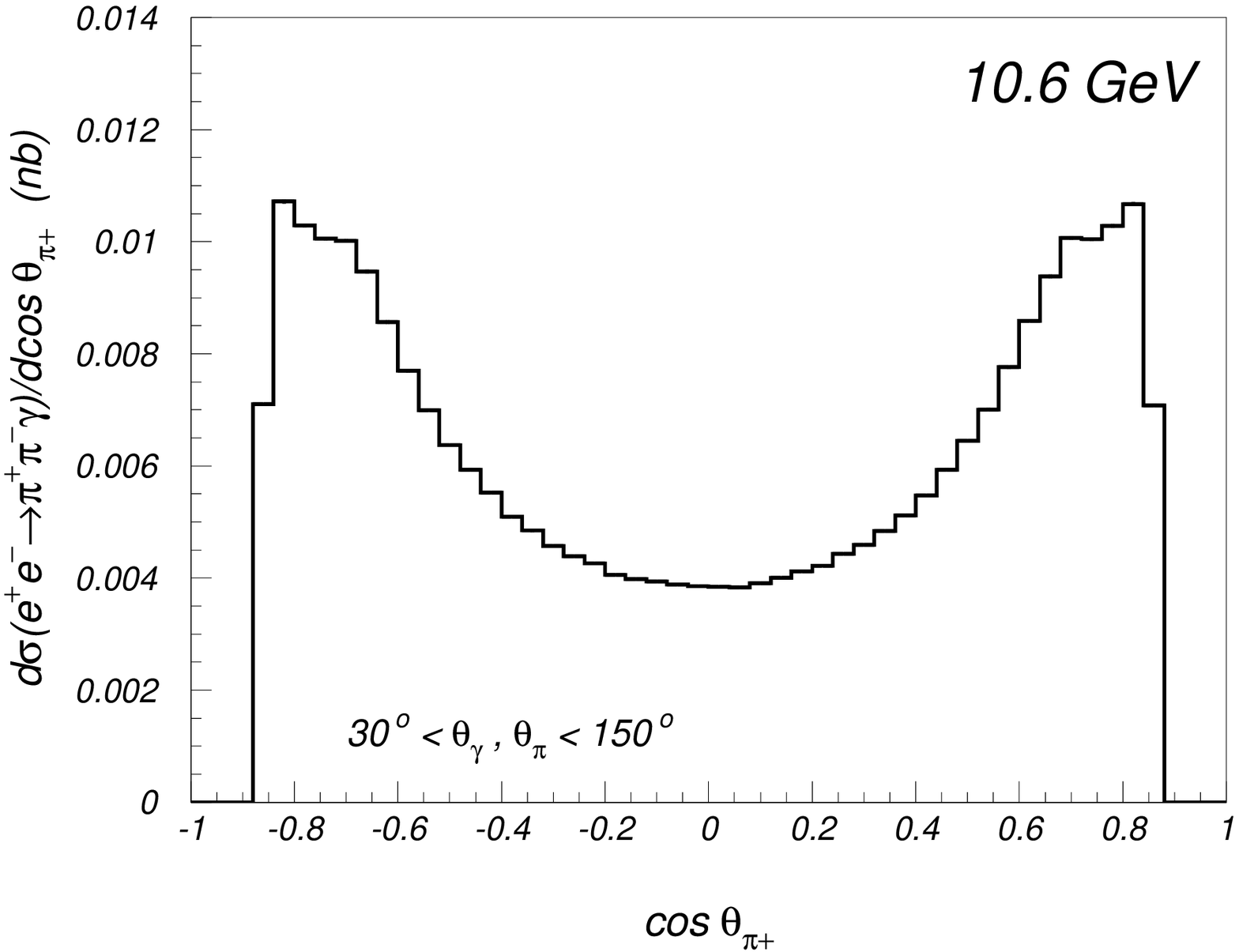,width=8.5cm,height=6cm} 
\epsfig{file=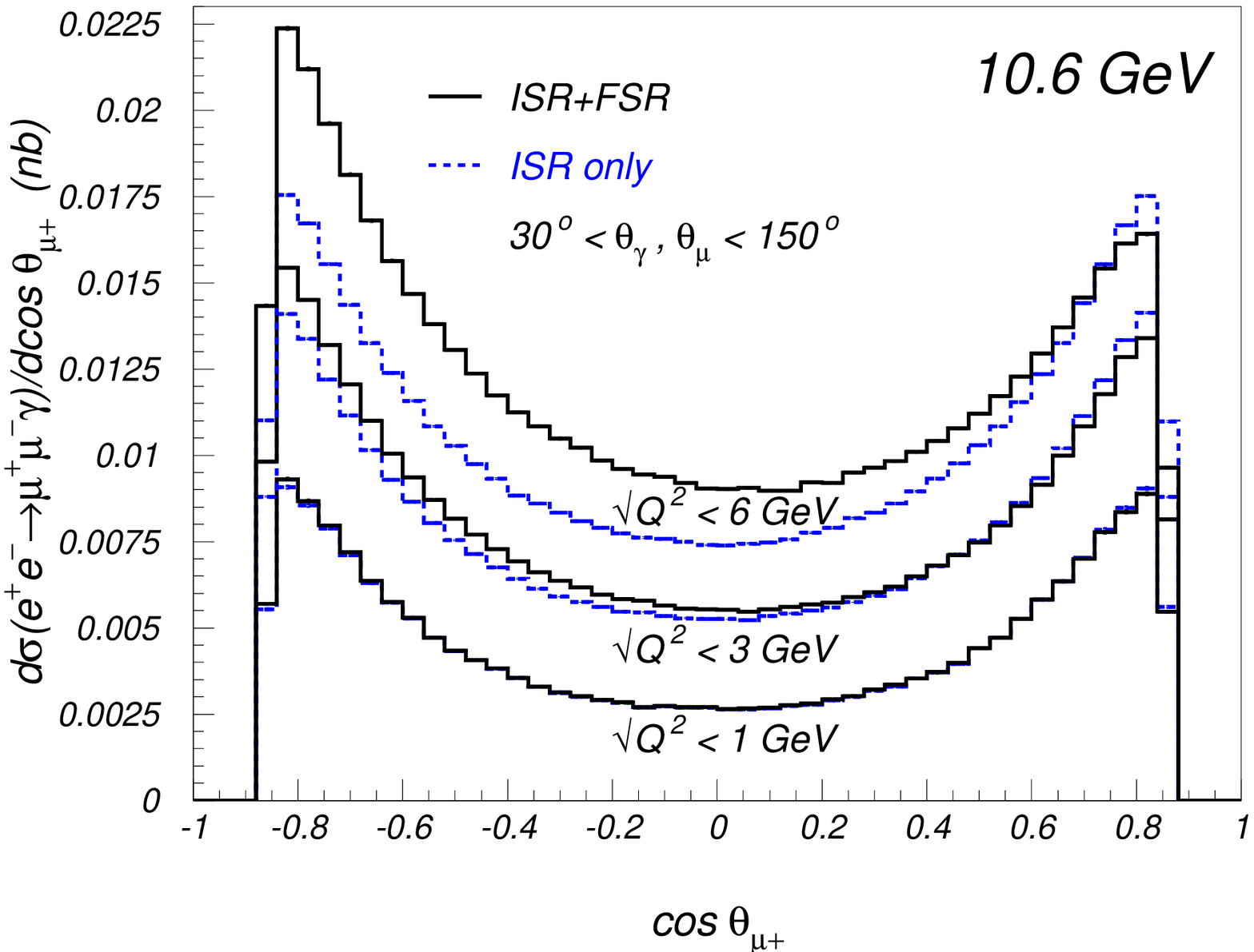,width=8.5cm,height=6cm} 
\end{center}
\caption{Angular distributions of $\pi^+$ 
(ISR \(\simeq\) FSR+ISR) and $\mu^+$ at $\sqrt{s}$=10.6~GeV for various $Q^2$ cuts.}
\label{fig:angular10GeV}
\end{figure*}

The amplitude for FSR (Fig.~\ref{fig:born}b) factorises as well as
\beq
{\cal A}_{\rm FSR}^{(0)} = M^{(0)} \cdot J_{\rm FSR}^{(0)}~,
\label{AFSR}
\eeq
where
\beq
M^{(0)}_\mu = \frac{e}{s} \, \bar{v}(p_1) \gamma_{\mu} u(p_2)~.
\eeq
Assuming that pions are point-like, the FSR current for 
two pions in scalar QED (sQED) reads  
\bea
J_{\rm FSR}^{(0), \, \mu} 
&=& - i \, e^2 \, F_{2\pi}(s) \nn \\ &\times&  
\left[ -2 g^{\mu\sigma} 
+ (q_1+k_1-q_2)^{\mu} \, \frac{(2q_1+k_1)^\sigma }{2k_1 \cdot q_1}
\right. \nn \\ && \left.
- (q_1-k_1-q_2)^{\mu} \, \frac{(2q_2+k_1)^\sigma }{2k_1 \cdot q_2}
\right] \epsilon_\sigma^*(k_1)~.  
\eea
Due to momentum conservation, $p_1+p_2=q_1+q_2+k_1$, and 
current conservation, this expression can be simplified further to 
\bea
J_{\rm FSR}^{(0), \, \mu} &=& 2 i \, e^2 \, F_{2\pi}(s) \,
\left[ g^{\mu\sigma} 
+ \frac{q_2^{\mu}\, q_1^\sigma}{k_1 \cdot q_1}
+ \frac{q_1^{\mu}\, q_2^\sigma}{k_1 \cdot q_2}
\right] \epsilon_\sigma^*(k_1)~. \nn \\
\eea
This is the basic model adopted in EVA~\cite{Binner:1999bt} 
and in PHO\-KHARA \cite{Rodrigo:2001jr,Kuhn:2002xg,Rodrigo:2001kf,Czyz:2002np,Czyz:2003ue,Czyz:2004rj,Czyz:2004nq,Czyz:2005as,Czyz:2007wi,Rodrigo:2001cc} 
to simulate FSR off charged pions.
The corresponding FSR current for muons is 
given by QED.

The fully differential cross section describing photon emission 
at leading order can be split into three pieces
\begin{equation}
{\rm d}\sigma^{(0)} = {\rm d}\sigma_\mathrm{ISR}^{(0)} 
+ {\rm d}\sigma_\mathrm{FSR}^{(0)} + {\rm d}\sigma_\mathrm{INT}^{(0)}~,
\label{LOxsection}
\end{equation}
which originate from the squared ISR and FSR amplitudes and the interference term, respectively. 
The ISR--FSR interference is odd under charge conjugation, 
\beq
{\rm d}\sigma_{\rm INT}^{(0)}(q_1,q_2) = - {\rm d}\sigma_{\rm INT}^{(0)}(q_2,q_1)~, 
\eeq
and its contribution vanishes after angular integration.
It gives rise, however, to a relatively large charge asymmetry and, 
correspondingly, to a forward--backward asymmetry
\begin{equation}
A(\theta) = \frac{N_{h}(\theta)-N_{h}(\pi-\theta)}
{N_{h}(\theta)+N_{h}(\pi-\theta)}~.
\end{equation}
The asymmetry can be used for the calibration of the FSR amplitude,
and fits to the angular distribution $A(\theta)$ can test
details of its model dependence~\cite{Binner:1999bt}. 

The second option to disentangle ISR from FSR exploits
the markedly different angular distribution of the photon 
from the two processes. This observation is completely general
and does not rely on any model like sQED for FSR.
FSR is dominated by photons collinear to the final state 
particles, while ISR is dominated by photons collinear to 
the beam direction. This suggests that we should consider 
only events with photons well separated from the charged 
final state particles and preferentially close to the 
beam~\cite{Binner:1999bt,Rodrigo:2001kf,Czyz:2002np}.

\begin{figure}[ht]
\begin{center}
\epsfig{file=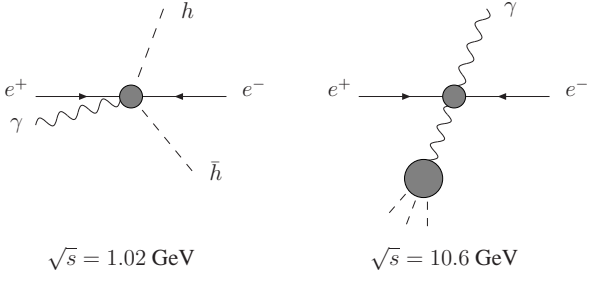,width=8.5cm} 
\end{center}
\caption{Typical kinematic configuration of the radiative return at low 
and high energies. }
\label{fig:kinematics}
\end{figure}

This is illustrated in Fig.~\ref{fig:isrtofsr}, 
which has been generated running PHOKHARA at leading order (LO). 
After introducing suitable angular cuts, the contamination of 
events with FSR is easily reduced to less than a few per mill.
The price to pay, however, is a suppression of the threshold 
region too. To have access to that region, 
photons at large angles need to be tagged and a better 
control of FSR is required. In Fig.~\ref{fig:angular}
the angular distribution of $\pi^+$ and $\mu^+$ at 
DA$\mathrm{\Phi}$NE energies, $\sqrt{s}=1.02$~GeV, are shown 
for different angular cuts. The angles are defined with respect 
to the incoming positron. If no angular cut is applied, 
the angular distribution in both cases is highly asymmetric as 
a consequence of the ISR--FSR interference contribution. If cuts 
suitable to suppress FSR, and therefore the ISR--FSR interference,
are applied, the distributions become symmetric. 

Two complementary analyses are therefore possible (for details see 
Section~\ref{rr:kloe}): 
The {\it small photon angle} analysis, where the photon is untagged 
and FSR can be suppressed below some reasonable limit. 
This analysis is suitable 
for intermediate values of the invariant mass of the hadronic 
system. And the {\it large photon angle} analysis, giving access to 
the threshold region, where FSR is more pronounced and 
the charge asymmetry is a useful tool to probe its model
dependence. 

These considerations apply, however, only to low beam energies,
around $1$~GeV. 
At high energies, e.g. at  $B$ facto\-ries, very hard
tagged photons are needed to access the region with low 
hadronic invariant masses, and the hadronic system is mainly 
produced back-to-back to the hard photon. The suppression of FSR 
is naturally accomplished and no special angular cuts are 
needed. This kinematical situation is illustrated 
in Fig.~\ref{fig:kinematics}.  The suppression of FSR contributions 
to $\pi^+\pi^-\gamma$ events is also a consequence of the rapid 
decrease of the form factor above $1$~GeV.
The relative size of FSR is of the order of a few per mill
(see Fig.~\ref{fig:angular10GeV}). 
For $\mu^+\mu^-$ in the final state, the amount of FSR 
depends on the invariant mass of the muons. 
For $\sqrt{Q^2} < 1$~GeV FSR is still tiny, 
but becomes more relevant for larger values of $Q^2$
(see Fig.~\ref{fig:angular10GeV}).

\subsubsection{Structure functions}
The original and default version of EVA~\cite{Binner:1999bt}, simulating 
the process $e^+ e^- \rightarrow \pi^+ \pi^- \gamma$ at LO, allowed for 
additional initial state radiation of soft and collinear photons by the 
structure function (SF) method~\cite{Caffo:1997yy,Caffo:1994dm}.

In the leading logarithmic approximation (LL), the multiple 
emission of collinear photons off an electron is described 
by the convolution integral 
\beq
\sigma(e^-X\to Y+ n\gamma) = \int_0^1 {\rm d}x \, f_e(x,Q^2) \, 
\sigma(e^- X\to Y)~,
\eeq
where $f_e(x,Q^2)$ is the probability distribution of the electron 
with longitudinal momentum fraction $x$, and $Q$ is the transverse 
momentum of the collinear photons. The function $f_e(x,Q^2)$ fulfils 
the evolution equation 
\bea
&& \!\!\!\!\!\!\!\!\!\!\!\!\!\!\!\!\!\!\!\!
\frac{{\rm d}}{{\rm d} \log Q} f_e(x,Q^2) = \int_x^1
\frac{{\rm d} z}{z} \nn \\ && \frac{\alpha}{\pi}
\left( \frac{1+z^2}{(1-z)_+} + \frac32 \delta(1-z)\right)
f_e(\frac{x}{z},Q^2)
\label{evoleq}
\eea
with initial conditions
\beq
\left. f_e(x,Q^2)\right|_{Q^2=m_e^2} = \delta(1-x)~,
\eeq
and the $+$ prescription defined as 
\beq
\int_0^1 {\rm d}x \, \frac{f(x)}{(1-x)_+}
= \int_0^1 {\rm d}x \, \frac{f(x)-f(1)}{(1-x)}~.
\eeq

The analytic solution to \Eq{evoleq} provided in 
Refs.~\cite{Caffo:1997yy,Caffo:1994dm} allows to resum soft 
photons to all orders in perturbation theory, accounting for  
large logarithms of collinear origin, $L=\log(s/m_e^2)$, up to two loops.
The resummed cross section, 
\beq
\sigma_{\rm SF} =
\int_0^1 {\rm d}x_1 \int_0^1  {\rm d}x_2 \, D(x_1) \, D(x_2) 
\, \sigma_{e^+e^-\to {\rm had.}+\gamma}(x_1 x_2 s)~,
\eeq
is thus obtained by convoluting 
the Born cross section of the hard photon emission 
process $e^+e^- \to {\rm hadrons} +\gamma$ 
with the SF distribution~\cite{Caffo:1997yy,Caffo:1994dm} 
\bea
D(x) &=& \left[1+\delta_N  \right]^{1/2} \, 
\frac{\beta_e}{2} (1-x)^{\frac{\beta_e}{2}-1} \nn \\
&& \times\bigg\{ \frac12 (1+x^2) + 
\frac12 \frac{(1-x)^2}{L-1} \nn \\ 
&& + \frac{\beta_e}{8}\left( -\frac{1}{2}(1+3x^2)
\log x - (1-x)^2 \right) \bigg\}~,
\eea
with 
\beq
\beta_e = 2 \, \frac{\alpha}{\pi} \, (L-1)
\eeq
and 
\bea
\delta_N &=& \frac{\alpha}{\pi} \left( \frac{3}{2}L+\frac{\pi^2}{3}-2\right)
\nn \\ &&
+ \beta_e^2 \frac{\pi^2}{8} + \left(\frac{\alpha}{\pi}\right)^2
\left( \frac{11}{8}-\frac{2\pi^2}{3}\right)L^2~. 
\eea

In the SF approach, the additional emission of collinear 
photons reduces the effective c.m. energy of the collision
to $\sqrt{x_1 x_2 s}$.
Momentum conservation is not accomplished because the 
extra radiation is integrated out. 
In order to reduce the kinematic distortion of the events,
a minimal invariant mass of the observed particles, 
hadrons plus the tagged photon, was required in~\cite{Binner:1999bt},
introducing in turn a cut dependence.
Therefore the SF predictions are not accurate enough for 
a high precision measurement of the hadronic cross section 
from radiative return, and a next-to-leading order (NLO) calculation  
is in order. The NLO prediction contains the large logarithms  
$L=\log(s/m_e^2)$ at order $\alpha^3$ and additional sub-leading terms, 
which are not taken into account within the SF method. Furthermore, 
it allows for a better control of the kinematical configurations
because momentum conservation is fulfilled. 
A comparison between SF and NLO predictions can be found 
in~\cite{Rodrigo:2001kf}.

\begin{figure}[ht]
\epsfig{file=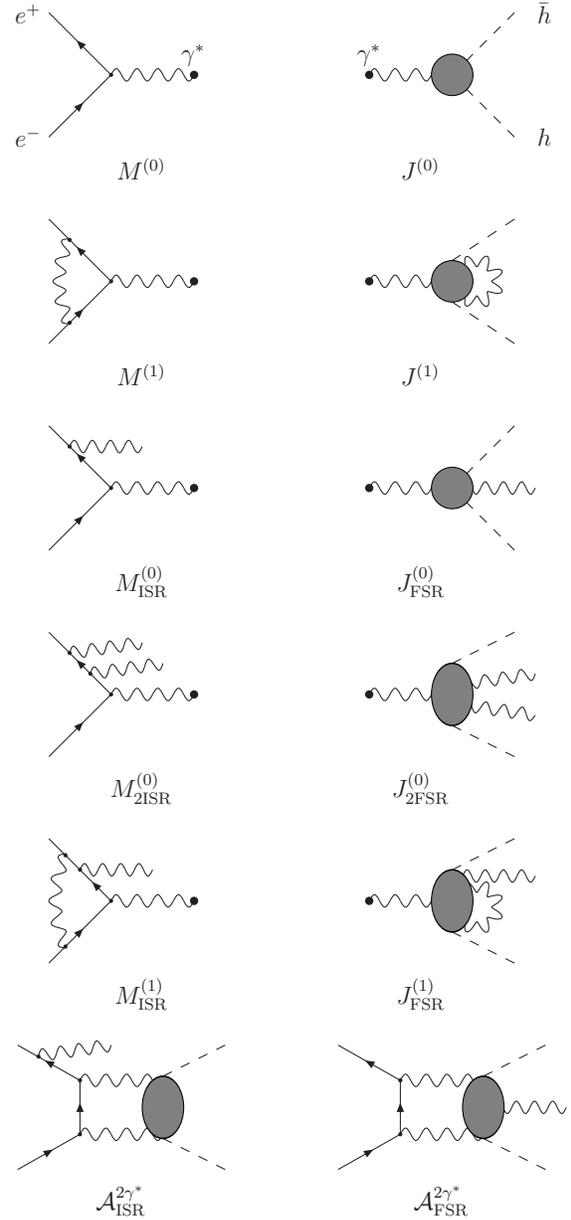,width=8cm} 
\caption{Typical sub-amplitudes describing virtual and real 
corrections to the reaction $e^+e^-\to h \bar h + \gamma (\gamma)$, 
where $h = \pi^-$, $\mu^-$. The superscripts $(0)$ and $(1)$ denote
tree-level and one-loop quantities, respectively. 
ISR and FSR indicate that real photons are emitted from the initial 
or final state. The last two diagrams, with exchange of 
two virtual photons, are non-factorisable. 
Permutations are omitted.}
\label{fig:subamplitudes}
\end{figure}

\subsubsection{Radiative return at NLO}
\label{rratnlo}

At NLO, the $e^+ e^-$ annihilation process in~\Eq{eq:LO}
receives contributions from one-loop corrections 
and from the emission of a second real photon 
(see Fig.~\ref{fig:subamplitudes}). 
After renormalisation, the one-loop matrix elements still contain  
infrared divergences. These are cancelled by adding the two-photon contributions to the one-loop corrections. 
There are several well established methods to perform this cancellation. 
The slicing method, where amplitudes are evaluated in dimensional 
regularisation and the two photon contribution is integrated 
analytically in phase space for one of the photon energies
up to an energy cutoff $E_{\gamma}<w\sqrt{s}$ far below $\sqrt{s}$,
was used in \cite{Rodrigo:2001jr,Kuhn:2002xg} 
to calculate the NLO corrections to ISR. 
Here the sum of the virtual and soft contributions is finite, but it depends on the soft photon cutoff. 
The contribution from the emission of the second photon with 
energy $E_{\gamma}>w\sqrt{s}$, which is evaluated numerically, 
completes the calculation and cancels this dependence. 

The size and sign of the NLO corrections do depend on the 
particular choice of the experimental cuts. Hence, only using a 
Monte Carlo event generator one can realistically 
compare theoretical predictions with experiment.
This is the main motivation behind PHOKHARA~\cite{Rodrigo:2001jr,Kuhn:2002xg,Rodrigo:2001kf,Czyz:2002np,Czyz:2003ue,Czyz:2004rj,Czyz:2004nq,Czyz:2005as,Czyz:2007wi,Rodrigo:2001cc}. 

The full set of scattering amplitudes at tree-level and one-loop 
can be constructed from the sub-amplitudes 
depicted in Fig.~\ref{fig:subamplitudes}.
The one-loop amplitude with emission of a single photon is given by 
\bea
 {\cal A}^{(1)}_{1\gamma} &=& 
  {\cal A}^{(1)}_{\rm ISR} + {\cal A}^{(1)}_{\rm FSR} \nn \\ &+&
  M^{(1)}\cdot J^{(0)}_{\rm FSR} 
+ M^{(0)}_{\rm ISR}\cdot J^{(1)}   \nn \\ &+&
  {\cal A}^{2\gamma^*}_{\rm ISR} + {\cal A}^{2\gamma^*}_{\rm FSR}~,
\eea
where
\beq
{\cal A}^{(1)}_{\rm ISR} = M^{(1)}_{\rm ISR}\cdot J^{(0)}~, \qquad
{\cal A}^{(1)}_{\rm FSR} = M^{(0)}\cdot J^{(1)}_{\rm FSR}~,
\eeq
while the amplitude with emission of two real photons reads 
\bea
{\cal A}^{(0)}_{2\gamma} &=&  
  {\cal A}^{(0)}_{\rm 2ISR}
+ {\cal A}^{(0)}_{\rm 2FSR} \nn \\ &+&
  \left( M^{(0)}_{\rm ISR}(k_1)\cdot J^{(0)}_{\rm FSR}(k_2)  
+ (k_1 \leftrightarrow k_2) \right)~,   
\eea
where
\beq
{\cal A}^{(0)}_{\rm 2ISR} = 
M^{(0)}_{\rm 2ISR}\cdot J^{(0)}~, \qquad
{\cal A}^{(0)}_{\rm 2FSR} = 
M^{(0)}\cdot J^{(0)}_{\rm 2FSR}~.
\eeq

PHOKHARA includes the full LO amplitudes and 
the most relevant C-even NLO contributions: 
\beq
{\rm d}\sigma = {\rm d}\sigma^{(0)} + {\rm d}\sigma^{(1)}_{\rm ISR} 
+ {\rm d}\sigma^{(1)}_{\rm IFS}~,
\eeq
where ${\rm d}\sigma^{(0)}$ is the LO differential cross section 
(\Eq{LOxsection}), 
\bea
{\rm d}\sigma^{(1)}_{\rm ISR} &=& \frac{1}{2s} \bigg[
2 {\rm Re} \left\{   
{\cal A}^{(1)}_{\rm ISR}
\left({\cal A}^{(0)}_{\rm ISR}\right)^\dagger \right\} \, 
{\rm d}\Phi_3(p_1,p_2;q_1,q_2,k_1) \nn \\
&& + \left|{\cal A}^{(0)}_{\rm 2ISR}\right|^2 \,
{\rm d}\Phi_4(p_1,p_2;q_1,q_2,k_1,k_2) \bigg]
\eea
is the second order radiative correction to ISR, and 
\bea
{\rm d}\sigma^{(1)}_{\rm IFS} &=& \frac{1}{2s} \bigg[
2 {\rm Re} \bigg\{   
M^{(0)}_{\rm ISR}\cdot J^{(1)} 
\left({\cal A}^{(0)}_{\rm ISR}\right)^\dagger
\nn \\ &&
+ M^{(1)}\cdot J^{(0)}_{\rm FSR} \, 
\left({\cal A}^{(0)}_{\rm FSR}\right)^\dagger
\bigg\} \, {\rm d}\Phi_3(p_1,p_2;q_1,q_2,k_1) \nn \\
&& +
\left( \left|M^{(0)}_{\rm ISR}(k_1)\cdot J^{(0)}_{\rm FSR}(k_2)\right|^2 
+ (k_1 \leftrightarrow k_2) \right) \nn \\ && \times
\, {\rm d}\Phi_4(p_1,p_2;q_1,q_2,k_1,k_2) \bigg]
\label{sigmaIFS}
\eea
is the contribution of events with simultaneous emission of 
one photon from the initial state and another one from the 
final state, together with ISR amplitudes with final state
one-loop vertex corrections, and FSR amplitudes with 
initial state one-loop vertex corrections. We denote these
 corrections as IFS.

Vacuum polarisation corrections are included in the hadronic 
currents multiplicatively: 
\bea
 J^{(i)} &\to& C_{\rm VP}(Q^2) \, J^{(i)}~, \nn \\ 
 J^{(i)}_{\rm FSR}(k_j) &\to& 
C_{\rm VP}((Q+k_j)^2) \, J^{(i)}_{\rm FSR}(k_j)~, \nn \\ 
 J^{(0)}_{\rm 2FSR} &\to& 
C_{\rm VP}(s) \, J^{(0)}_{\rm 2FSR} ~. 
\eea
The virtual photon propagator is by definition included 
in the leptonic sub-amplitudes $M^{(i)}$, $M^{(i)}_{\rm ISR}$ 
and $M^{(0)}_{\rm 2ISR}$:
\bea
&& M^{(i)} \sim \frac{1}{s}~, \nn \\ 
&& M^{(i)}_{\rm ISR}(k_j) \sim \frac{1}{(p_1+p_2+k_j)^2}~, \nn \\ 
&& M^{(0)}_{\rm 2ISR} \sim \frac{1}{Q^2}~. 
\eea

Neither diagrams where two photons are emitted from the final state, nor
final-state vertex corrections with associated real radiation from
the final state are included.
These constitute radiative corrections to FSR and will give 
non-negligible contributions only for those cases where at least one 
photon is collinear with one of the final state particles.
Box diagrams with associated real radiation from the initial- or the 
final-state leptons, as well as pentagon diagrams, are also neglected.
As long as one considers charge symmetric
observables only, their contribution is divergent neither in the soft nor
the collinear limit and is thus of order $\alpha/\pi$ without any enhancement
factor. One should stress that PHOKHARA includes only C-even gauge 
invariant sets of diagrams at NLO. 
The missing contributions are either small or do not
contribute for charge symmetric cuts. Nevertheless their implementation 
is underway.

The calculation of the NLO corrections to ISR,  
${\rm d}\sigma^{(1)}_{\rm ISR}$, is independent of the final 
state.  These corrections are included by default
for all the final state channels implemented in PHOKHARA, 
and can be easily added for any other new channel, 
with the sole substitution of the tree-level final 
state current. The radiative corrections of the IFS process 
depend on the final state. The latest version of PHOKHARA 
(version 6.0~\cite{Czyz:2007wi})
includes these corrections for two charged pions, kaons 
and muons.

{$\\$}
{\noindent \it Virtual and soft corrections to ISR\\}

The virtual and soft QED corrections to ISR in $e^+ e^-$ annihilation
were originally implemented in PHO\-KHA\-RA through the leptonic tensor. 
For future applications, however, it will be more convenient to 
implement those corrections directly at the amplitude level (in preparation).
In terms of sub-amplitudes, the leptonic tensor is given by  
\bea
 L^{\mu \nu}_{\rm ISR} &=& L^{(0),\mu \nu}_{\rm ISR} 
+ M^{(1),\, \mu}_{\rm ISR} \left(M^{(0),\, \nu}_{\rm ISR}\right)^\dagger 
+ M^{(0),\, \mu}_{\rm ISR} \left(M^{(1),\, \nu}_{\rm ISR}\right)^\dagger \nn \\
&& + \frac{1}{2(2\pi)^{d-1}} \int_0^{w\sqrt{s}} E^{d-3}\, {\rm d}E \, {\rm d}\Omega 
\, M^{(0),\, \mu}_{2\rm ISR} 
\left( M^{(0),\, \nu}_{2\rm ISR}\right)^\dagger~, \nn \\ &&
\eea
where $E$ and $\Omega$ are the energy and the solid angle of the soft photon,
respectively, and $d=4-2\epsilon$ is the number of dimensions in dimensional 
regularisation. 
The leptonic tensor has the general form 
\begin{align}
L^{\mu \nu}_{\rm ISR} &=
\frac{(4 \pi \alpha)^2}{Q^4} \;
\bigg[ a_{00} \; g^{\mu \nu} + a_{11} \; \frac{p_1^{\mu} p_1^{\nu}}{s}
 + a_{22} \; \frac{p_2^{\mu} p_2^{\nu}}{s} \non \\
&+ a_{12} \; \frac{p_1^{\mu} p_2^{\nu} + p_2^{\mu} p_1^{\nu}}{s}
+ i \pi \; a_{-1} \; 
\frac{p_1^{\mu} p_2^{\nu} - p_2^{\mu} p_1^{\nu}}{s} \bigg]~,
\label{generaltensor}
\end{align}
where the scalar coefficients $a_{ij}$ and $a_{-1}$ allow 
the following expansion: 
\begin{equation}
a_{ij} = a_{ij}^{(0)} + \frac{\alpha}{\pi} \; a_{ij}^{(1)}~, \qquad
a_{-1} = \frac{\alpha}{\pi} \; a_{-1}^{(1)}~.
\end{equation}
The imaginary antisymmetric piece, which is proportional to $a_{-1}$,
appears for the first time at second order and is particularly relevant for 
those cases where the hadronic current receives contributions from 
different amplitudes with nontrivial relative phases. This is 
possible, e.g., for final states with three or more mesons, or for 
$p\bar{p}$ production.

The LO coefficients $a_{ij}^{(0)}$ can be read directly from~\Eq{Lmunu0}
\begin{align}
a_{00}^{(0)} &= \frac{2 m^2 q^2(1-q^2)^2}{y_1^2 y_2^2}
- \frac{2 q^2+y_1^2+y_2^2}{y_1 y_2}~, \non \\
a_{11}^{(0)} &= \frac{8 m^2}{y_2^2} - \frac{4q^2}{y_1 y_2}~, \qquad
a_{22}^{(0)} = a_{11}^{(0)} (y_1 \leftrightarrow y_2)~, \non \\
a_{12}^{(0)} &= - \frac{8 m^2}{y_1 y_2} ~. 
\end{align}

The NLO coefficients $a_{ij}^{(1)}$ and $a_{-1}^{(1)}$ are 
obtained by combining the one-loop and the soft contributions. 
It is convenient to split the coefficients $a^{(1)}_{ij}$ 
into a part that contributes at large photon angles and a part 
proportional to $m_e^2$ and $m_e^4$ which is relevant only in the 
collinear regions. These coefficients are denoted by $a^{(1,0)}_{ij}$ 
and $a^{(1,m)}_{ij}$, respectively:
\begin{align}
a^{(1)}_{ij} &= a_{ij}^{(0)} \bigg[ -\log(4w^2) [1+\log(m^2)] \non \\ & 
-\frac{3}{2} \log(\frac{m^2}{q^2}) 
- 2 + \frac{\pi^2}{3} \bigg] 
+ a^{(1,0)}_{ij}+a^{(1,m)}_{ij}~.
\end{align}
The factor proportional to the LO coefficients $a^{(0)}_{ij}$ 
contains the usual soft and collinear logarithms.
The quantity $w$ denotes the dimensionless 
value of the soft photon energy cutoff, $E_{\gamma}<w\sqrt{s}$.
It is enough to present four out of the five coefficients 
because exchanging the positron with the electron momenta
leads to the symmetry relation
\begin{align}
a_{22}^{(1)} = a_{11}^{(1)} (y_1 \leftrightarrow y_2)~.
\label{p2p2}
\end{align}

The large-angle contributions have been calculated 
in Ref.~\cite{Rodrigo:2001jr}.
The coefficient proportional to $g^{\mu \nu}$ reads
\begin{align}
a_{00}^{(1,0)} &= \frac{1}{y_1\, y_2} \bigg[
-\frac{q^2(1-q^2)}{2}  - y_1 y_2
 - \bigg[ q^2 + \frac{2y_1 y_2}{1-q^2} \bigg] \log(q^2) \non \\
&+ \bigg\{ \frac{y_1}{2} \bigg[ 4-y_1-\frac{3(1+q^2)}{1-y_2} \bigg] 
\log(\frac{y_1}{q^2}) \non \\
& - \bigg[ 1 + (1-y_2)^2 + \frac{y_1 q^2}{y_2} \bigg] L(y_1) + 
(y_1 \leftrightarrow y_2)\bigg\}
\bigg]~,
\label{gmunu}
\end{align}
where the function $L$ is defined as
\begin{align}
L(y_i) &= \Li_2(-\frac{y_i}{q^2})-\Li_2(1-\frac{1}{q^2}) \non \\ & + 
\log(q^2+y_i) \log(\frac{y_i}{q^2})~, 
\end{align}
with $\Li_2$ the Spence (or dilogarithmic) function defined below
Eq.~(\ref{js}). 
The coefficient in front of the tensor structure 
$p_1^{\mu}p_1^{\nu}$ is given by 
\begin{align}
\label{p1p1}
a_{11}^{(1,0)} &= 
\frac{1}{y_1\, y_2} \bigg[
(1+q^2)^2 \left(\frac{1}{1-y_1}-\frac{1}{1-q^2} \right) -
\frac{4(1-y_2)y_1}{1-q^2} \non \\
&- \frac{2 q^2}{1-q^2} \bigg[(1-y_2)\left(\frac{1}{y_2}+\frac{q^2}{y_1}
+ \frac{2 y_1}{1-q^2} \right) \non \\ & +\frac{2q^2}{1-q^2}
\bigg] \log(q^2) 
- q^2 \bigg[ 1 + \frac{2}{y_2} \bigg] \log(\frac{y_1}{q^2}) \non \\ 
&- q^2 \bigg[ \frac{(2-3y_1)(1-y_2)^2}{y_1(1-y_1)^2} \bigg] 
 \log(\frac{y_2}{q^2}) \non \\
& - 2 q^2 \bigg[ 1+\frac{1}{y_2^2} \bigg] L(y_1)
- 2 q^2 \bigg[ 3+\frac{2q^2}{y_1}+\frac{q^4}{y_1^2} \bigg] L(y_2)
\bigg]~.
\end{align}
For the symmetric tensor structure 
($p_1^{\mu}p_2^{\nu}+p_2^{\mu}p_1^{\nu}$) one gets 
\begin{align}
\label{p1p2}
a_{12}^{(1,0)} &= \frac{1}{y_1\, y_2} \bigg[
- \frac{4q^2+(y_1-y_2)^2}{1-q^2} \non \\
&-2 q^2 \bigg[ \frac{q^2}{y_1 y_2}+\frac{1+q^2-2 y_1 y_2}{(1-q^2)^2} \bigg] 
\log(q^2) + \bigg\{ \frac{q^2}{1-y_1} \non \\
& - \frac{2 q^2}{1-y_2} \bigg[1-y_1+\frac{q^2}{y_2}
-\frac{q^2}{2(1-y_2)} \bigg] \log(\frac{y_1}{q^2}) \non \\  
& - 2 q^2 \bigg[ 1+\frac{q^2}{y_2}+\frac{q^2}{y_2^2} \bigg] L(y_1) 
+ (y_1 \leftrightarrow y_2) \bigg\}
\bigg]~. 
\end{align}
Finally, the antisymmetric coefficient $a_{-1}$ 
accompanying ($p_1^{\mu}p_2^{\nu}-p_2^{\mu}p_1^{\nu}$) reads 
\begin{align}
a_{-1}^{(1,0)} &= \frac{q^2}{y_1\, y_2} 
\bigg[ \frac{2\log(1-y_1)}{y_1} 
+ \frac{1-q^2}{1-y_1} + \frac{q^2}{(1-y_1)^2} \bigg] \non \\ &
 - (y_1\leftrightarrow y_2)~.
\label{asym}
\end{align}

The mass-suppressed coefficients $a_{ij}^{(1,m)}$ 
are given by~\cite{Kuhn:2002xg}
\begin{align}
a_{00}^{(1,m)} &=  
\frac{m^2 q^2}{y_1^2} \bigg[
\log(q^2) \log(\frac{y_1^4}{m^4 q^2}) + 4 \Li_2(1-q^2) \non \\ &
+ \Li_2(1-\frac{y_1}{m^2}) - \frac{\pi^2}{6}  \bigg] 
- \frac{m^2(1-q^2)}{y_1^2} \bigg[ 1 - \log(\frac{y_1}{m^2}) \non \\ &
+ \frac{m^2}{y_1} \bigg( \Li_2(1-\frac{y_1}{m^2})-\frac{\pi^2}{6} \bigg) 
\bigg] + \frac{q^2}{2} n(y_1,\frac{1-3q^2}{q^2}) \non \\ &
+ (y_1 \leftrightarrow y_2)~,
\label{apiupiu}
\end{align}
whereas 
\begin{align}
a_{11}^{(1,m)} &= \frac{q^2}{1-q^2} \bigg\{
\frac{4m^2}{y_1^2} \bigg[1-\log(\frac{y_1}{m^2}) \non \\ &
+ \frac{m^2}{y_1} \bigg( \Li_2(1-\frac{y_1}{m^2})-\frac{\pi^2}{6} \bigg) \bigg]
- n(y_1,1) \non \\ &
+\frac{2m^2 q^2}{y_1(m^2(1-q^2)-y_1)} \bigg[
\frac{1}{q^2} \log(\frac{y_1}{m^2}) + \frac{\log(q^2)}{1-q^2} \non \\ &
+ \bigg(1+\frac{m^2}{m^2(1-q^2)-y_1}\bigg) N(y_1) \bigg] \bigg\} + \non \\ &
+ \frac{1}{1-q^2} \bigg\{
 \frac{4m^2(1-q^2)}{y_2^2} \bigg[ \log(q^2) \log(\frac{y_2^4}{m^4 q^2}) 
\non \\ & + 4 \Li_2(1-q^2) 
+ 2 \bigg( \Li_2(1-\frac{y_2}{m^2})-\frac{\pi^2}{6} \bigg) \bigg] \non \\ & 
+ \frac{4m^2 q^2}{y_2^2} \bigg[1-\log(\frac{y_2}{m^2}) 
+ \bigg(1+\frac{m^2}{y_2}\bigg)
 \bigg( \Li_2(1-\frac{y_2}{m^2}) \non \\ & 
 -\frac{\pi^2}{6} \bigg) \bigg] 
- \frac{1-2q^4}{q^2} n(y_2,\frac{3-8q^2+6q^4}{1-2q^4}) \non \\ &
+\frac{2m^2}{y_2(m^2(1-q^2)-y_2)} \bigg[
\frac{1}{q^2} \log(\frac{y_2}{m^2}) + \frac{\log(q^2)}{1-q^2}  \non \\ &
+ \bigg(3+\frac{m^2}{m^2(1-q^2)-y_2}\bigg) N(y_2) \bigg] \bigg\}~,
\end{align}
and
\begin{align}
a_{12}^{(1,m)} &= \frac{q^2}{1-q^2} \bigg\{
 \frac{4m^2}{y_1^2} \bigg[1-\log(\frac{y_1}{m^2}) \non \\ &
 + \bigg( \frac{1}{2} + \frac{m^2}{y_1} \bigg)
 \bigg( \Li_2(1-\frac{y_1}{m^2})-\frac{\pi^2}{6} \bigg) \bigg] \non \\ &
- \frac{1-q^2}{q^2} n(y_1,\frac{1}{1-q^2})
+ \frac{2m^2}{y_1(m^2(1-q^2)-y_1)} \non \\ & \times \bigg[
\frac{1}{q^2}\log(\frac{y_1}{m^2}) + \frac{\log(q^2)}{1-q^2} \non \\ &
+ \bigg(2+\frac{m^2}{m^2(1-q^2)-y_1}\bigg) N(y_1)  \bigg] \bigg\} 
 + (y_1 \leftrightarrow y_2)~.
\end{align}
The asymmetric coefficient does not get mass corrections,
\begin{align}
a_{-1}^{(1,m)} &= 0~.
\end{align}
The functions $n(y_i,z)$ and $N(y_i)$ are defined through
\begin{align}
n(y_i,z) & = \frac{m^2}{y_i(m^2-y_i)} \bigg[ 1 + z \; \log(\frac{y_i}{m^2}) 
\bigg] \non \\ &
+ \frac{m^2}{(m^2-y_i)^2} \log(\frac{y_i}{m^2})~,
\end{align}
and
\begin{align}
N(y_i) &= \log(q^2) \log(\frac{y_i}{m^2})+\Li_2(1-q^2) \non \\ & +
\Li_2(1-\frac{y_i}{m^2})-\frac{\pi^2}{6}~.
\end{align}
The apparent singularity of the function 
$n(y_i,z)$ inside the phase space limits is 
compensated by the zero in the numerator. 
In the region $y_i$ close to $m^2$ it behaves as 
\begin{align}
n(y_i,z) \big|_{y_i \to m^2}&=
\frac{1}{y_i} \bigg[ 1 + z \; \log(\frac{y_i}{m^2}) \bigg] \non \\ &
- \frac{1}{m^2} \sum_{n=0} \bigg(\frac{1}{n+2}+\frac{z}{n+1}\bigg)
\left( 1-\frac{y_i}{m^2}\right)^n~.
\end{align}
Similarly, the function $N(y_i)$ guarantees that the coefficients 
$a^{(1)}_{ij}$ are finite in the limit $y_i \rightarrow m^2 (1-q^2)$:
\begin{equation}
\frac{m^2 N(y_i)}{m^2(1-q^2)-y_i} \bigg|_{y_i \to m^2(1-q^2)}  = 
-\frac{\log(1-q^2)}{q^2}-\frac{\log(q^2)}{1-q^2}~.
\end{equation}

{$\\$}
{\noindent \it Virtual and soft corrections to IFS\\}

The virtual plus soft photon corrections of 
the initial-state and final-state 
vertex (see \Eq{sigmaIFS}) to FSR and ISR, respectively,  
can be written as~\cite{Berends:1973tz,Berends:1987ab}
\bea
{\rm d}\sigma^{\rm V+S}_{\rm IFS} &=& \frac{\alpha}{\pi} \, 
\left[ \delta^{\rm V+S}(w) \, 
{\rm d}\sigma^{(0)}_{\rm FSR}(s) \right. \nn \\ && + \left. 
\eta^\mathrm{V+S}(s',w) \, 
{\rm d}\sigma^{(0)}_{\mathrm{ISR}}(s') \right]~,
\label{sigvspi}
\eea
where ${\rm d}\sigma^{(0)}_{\mathrm{FSR}}$ and ${\rm d}\sigma^{(0)}_{\mathrm{ISR}}$ 
are the leading order FSR and ISR differential cross sections, 
respectively, $w=E_\gamma^{\rm cut}/\sqrt{s}$ with $E_\gamma^{\rm cut}$
the maximal energy of the soft photon in the $e^+e^-$ 
c.m. rest frame, and $s'$ corresponds to the squared mass of the 
$h \bar h \gamma$ system. The function $\delta^{\mathrm{V+S}}(w)$
is independent of the final state. In the limit $m_e^2\ll s$,
\begin{align}
\delta^{\mathrm{V+S}}(w) = 2 \left[
(L-1)\log{(2w)}+\frac{3}{4}L-1+\frac{\pi^2}{6}\right]~,
\label{delvs}
\end{align}
where $L=\log(s/m_e^2)$.
For two pions in the final state, 
the function $\eta^{\rm V+S}(s',w)$ is given by 
\begin{align}
& \eta^{\rm V+S}(s',w)  = 
- 2 \left[ \frac{1+\beta_\pi^2}
{2\beta_\pi} \log(t_\pi) + 1 \right] \nn \\ & \times
\left[ \log(2w) + 1 + \frac{s'}{s'-s} 
\log \left(\frac{s}{s'} \right) \right]
+ \log \left(\frac{m_\pi^2}{s'}\right) \nn \\ & 
- \frac{1+\beta_\pi^2}{\beta_\pi} \left[  
2 \Li_2 (1-t_\pi) + \log(t_{\pi}) \log (1+t_\pi) 
- \frac{\pi^2}{2} \right] \nn \\ &
- \frac{2+\beta_\pi^2}{\beta_\pi} \log(t_\pi) - 2~, 
\label{etavspi}
\end{align}
where 
\bea
\beta_\pi = \sqrt{1 - \frac{4m_\pi^2}{s'}}~, 
\qquad t_\pi = \frac{1-\beta_\pi}{1+\beta_\pi}~.
\label{betapi}
\eea

The function $\eta^\mathrm{V+S}(s',w)$ is equivalent to the familiar
correction factor derived in~\cite{Schwinger:1989ix,Drees:1990te} for the reaction 
$e^+e^- \to \pi^+ \pi^- \gamma$ in the framework of sQED 
(see also \cite{Hoefer:2001mx}) in the limit $s \to s'$:
\beq
\left. \log(2w) + 1 + \frac{s'}{s'-s} 
\log \left(\frac{s}{s'} \right) \right|_{s \to s'} = \log(2w')
\label{changeframe}
\eeq
with $w' = E_\gamma^{\rm cut}/\sqrt{s'}$.
The factor on the right hand side of \Eq{changeframe}
for $s\ne s'$ arises from defining the soft photon cutoff 
in the $e^+e^-$ laboratory frame. 

Correspondingly, the function $\eta^{\rm V+S}(s',w)$
for two muons in the final state reads 
\begin{align}
&\eta^{\rm V+S}(s',w) 
= -2 \left[\frac{1+\beta_{\mu}^2}
{2\beta_{\mu}} \log(t_\mu) + 1 \right] \nn \\ & \times
\left[ \log(2w) + 1 + \frac{s'}{s'-s} 
\log \left(\frac{s}{s'} \right) \right]
+ \log \left(\frac{m_\mu^2}{s'}\right)  \nn \\ &
-\frac{1+\beta_{\mu}^2}{2\beta_{\mu}} \left[
4 \Li_2 ( 1-t_\mu) - 2 \log(t_\mu) \log\left(\frac{1+\beta_{\mu}}{2}\right)
- \pi^2 \right] \nn \\ & 
- \frac{1}{\beta_\mu} 
\left[\frac{3}{3-\beta_\mu^2} + \beta_\mu^2 \right] \log(t_\mu) - 2~, 
\label{etavsmu}
\end{align}
where
\bea
\beta_\mu = \sqrt{1 - \frac{4m_\mu^2}{s'}}~, 
\qquad t_\mu = \frac{1-\beta_{\mu}}{1+\beta_{\mu}}~. 
\label{betamu}
\eea

{$\\$}
{\noindent \it Real corrections \\}

Matrix elements for the emission of two real photons, 
\begin{align}
e^+(p_1) + e^-(p_2) \rightarrow  {\rm hadrons}\, (Q) 
+ \gamma(k_1) + \gamma(k_2)~,
\end{align}
are calculated in PHOKHARA following 
the helicity amplitude method with the conventions 
introduced in~\cite{Jegerlehner:1999wu,Kolodziej:1991pk}.
The Weyl representation for fermions is used where the Dirac matrices 
\begin{align}
\gamma^{\mu}= \left( 
\begin{array}{cc}
0 & \sigma^{\mu}_+ \\
\sigma^{\mu}_- & 0
\end{array} \right)~, \qquad \mu=0,1,2,3~,
\end{align}
are given in terms of the unit $2 \times 2$ matrix $I$ and the Pauli 
matrices $\sigma_i,\ i=1,2,3$, with $\sigma^{\mu}_{\pm}=(I,\pm\sigma_i)$.
The contraction of any four-vector $a^{\mu}$ with the $\gamma^{\mu}$ 
matrices has the form
\begin{align}
\ta{a} = a_{\mu} \gamma^{\mu} = \left( 
\begin{array}{cc}
0   & a^+ \\
a^- & 0
\end{array} \right)~,
\end{align}
where the $2 \times 2$ matrices $a^{\pm}$ are given by 
\begin{align}
a^{\pm} =  a^{\mu} \sigma_{\mu}^{\pm} &= \left( \begin{array}{cc}
a^0 \mp a^3      & \mp (a^1 - i a^2) \\
\mp (a^1 + i a^2) & a^0 \pm a^3
\end{array} \right)~.
\end{align}

The helicity spinors $u$ and $v$ for a particle and an antiparticle 
of four-momentum $p=(E, {\bf p})$ and helicity $\lambda=\pm 1/2$ 
are given by
\begin{align}
u(p,\lambda=\pm 1/2) &= \left( \begin{array}{c}
\sqrt{E \mp |{\bf p}|} \; \chi({\bf p},\pm) \\
\sqrt{E \pm |{\bf p}|} \; \chi({\bf p},\pm)
\end{array} \right) \equiv \left( \begin{array}{c}
u_I \\ u_{II} \end{array} \right)~, \non \\ 
v(p,\lambda=\pm 1/2) &= \left( \begin{array}{c}
\mp \sqrt{E \pm |{\bf p}|}\; \chi({\bf p},\mp) \\
\pm \sqrt{E \mp |{\bf p}|}\; \chi({\bf p},\mp)
\end{array} \right) \equiv \left( \begin{array}{c}
v_I \\ v_{II} \end{array} \right)~.
\end{align}
The helicity eigenstates $\chi({\bf p},\lambda)$ can be expressed
in terms of the polar and azimuthal angles of the momentum
vector ${\bf p}$ as
\begin{align}
\chi({\bf p},+) &= \left( \begin{array}{r}
\cos{(\theta/2)} \\ e^{i\phi} \sin{(\theta/2)}
\end{array} \right)~, \non \\ 
\chi({\bf p},-) &= \left( \begin{array}{r}
-e^{-i\phi}\sin{(\theta/2)} \\ \cos{(\theta/2)}
\end{array} \right)~.
\end{align}
Finally, complex polarisation vectors in the helicity basis are defined
for the real photons: 
\begin{eqnarray}
\varepsilon^{\mu}(k_i,\lambda_i=\pm) &=& \frac{1}{\sqrt{2}} \big(  0,
\mp \cos \theta_i \cos \phi_i + i \sin \phi_i, \non \\  
&\mp& \cos \theta_i \sin \phi_i - i \cos \phi_i,
\pm \sin{\theta_i} \big)\,,
\end{eqnarray}  
with $i=1,2$.

{$\\$}
{\noindent \it Phase space \\}

One of the key ingredients of any Monte Carlo simulation is an 
efficient generation of the phase space. 
The generation of the multi-particle phase space in PHOKHARA
is based on the Lorentz-invariant representation 
\begin{align} 
& {\rm d} \Phi_{m+n}(p_1,p_2;k_1,\cdot,k_m,q_1,\cdot,q_n)
= \non \\ & \qquad {\rm d} \Phi_m(p_1,p_2;Q,k_1,\cdot,k_m) {\rm d} \Phi_n(Q;q_1,\cdot,q_n) 
\frac{{\rm d}Q^2}{2\pi}~,
\end{align}
where $p_1$ and $p_2$ are the four-momenta of the initial particles,
$k_1 \ldots k_m$ are the four momenta of the emitted photons and 
$q_1 \ldots q_n$, with $Q = \sum q_i$, label the four-momenta of the 
final state hadrons. 

When two particles of the same mass are produced in the final state, 
$q_i^2=M^2$, their phase space is given by
\begin{equation}
{\rm d} \Phi_2(Q;q_1,q_2) = \frac{\sqrt{1-\frac{4 M^2}{Q^2}}}{32 \pi^2} 
{\rm d} \Omega~,
\end{equation}
where ${\rm d} \Omega$ is the solid angle of one of the final state 
particles at, for instance, the $Q^2$ rest frame.
 
Single photon emission is described by the corresponding 
leptonic part of the phase space, 
\begin{equation}
{\rm d} \Phi_2(p_1,p_2;Q,k_1) = \frac{1-q^2}{32 \pi^2} {\rm d} \Omega_1~,
\end{equation}
with $q^2=Q^2/s$ and $d \Omega_1$ the solid angle of the emitted 
photon at the $e^+ e^-$ rest frame. The polar angle $\theta_1$
is defined with respect to the positron momentum $p_1$.
In order to make the Monte Carlo generation more efficient,
the following substitution is performed:
\begin{align}
\cos \theta_1 = \frac{1}{\beta} \tanh(\beta \; t_1) ~, \quad
t_1 = \frac{1}{2\beta} \log \frac{1+\beta \cos \theta_1}{1-\beta \cos \theta_1}~,
\label{t1}
\end{align}
with $\beta = \sqrt{1-4m_e^2/s}$,
which accounts for the collinear emission peaks  
\begin{equation}
\frac{{\rm d} \cos \theta_1}{1-\beta^2 \cos^2 \theta_1} = {\rm d}t_1~.
\end{equation}
With this the azimuthal angle and the new variable $t_1$ are generated flat. 

Considering the emission of two real photons
in the c.m. of the initial particles, the four-momenta 
of the po\-si\-tron, the electron and the two emitted photons are given by 
\begin{align}
p_1 &= \frac{\sqrt{s}}{2}(1,0,0,\beta)~, \qquad 
p_2 = \frac{\sqrt{s}}{2}(1,0,0,-\beta)~, \non \\
k_1 &= w_1 \sqrt{s} (1,\sin \theta_1 \cos \phi_1,\sin \theta_1 \sin \phi_1,
\cos \theta_1)~, \non \\
k_2 &= w_2 \sqrt{s} (1,\sin \theta_2 \cos \phi_2,\sin \theta_2 \sin \phi_2,
\cos \theta_2)~,
\end{align}
respectively. The polar angles $\theta_1$ and $\theta_2$ are
again defined with respect to the positron momentum $p_1$.
Both photons are generated with energies larger than 
the soft photon cutoff: $w_i>w$ with $i=1,2$. At least 
one of these exceeds the minimal detection energy: 
$w_1 > E_{\gamma}^{\rm min}/\sqrt{s}$ or 
$w_2 > E_{\gamma}^{\rm min}/\sqrt{s}$.
In terms of the solid angles ${\rm d} \Omega_1$ and ${\rm d} \Omega_2$ of the 
two photons and the normalised energy of one of them, e.g. $w_1$,
the leptonic part of the phase space reads 
\begin{align}
{\rm d} \Phi_3(p_1,p_2;Q,k_1,k_2) &= \frac{1}{2!} \;\frac{s}{4(2\pi)^5} 
\non \\ \times &  \frac{w_1 w_2^2}{1-q^2-2w_1}
\; {\rm d}w_1 \; {\rm d}\Omega_1 \; {\rm d}\Omega_2~,
\end{align}
where the limits of the phase space are determined from the constraint 
\begin{equation}
q^2 = 1-2(w_1+w_2)+2w_1 w_2(1-\cos \chi_{12})~,
\label{dpslimits}
\end{equation}
with $\chi_{12}$ being the angle between the two photons
\begin{equation}
\cos \chi_{12} = \sin \theta_1 \sin \theta_2 \cos (\phi_1 - \phi_2) 
+ \cos \theta_1 \cos \theta_2~.
\end{equation}

Again, the matrix element squared contains several peaks, soft and 
collinear, which should be softened by choosing suitable substitutions
in order to achieve an efficient Monte Carlo generator.
The leading behaviour of the matrix element squared is given 
by $1/(y_{11} \; y_{12} \; y_{21} \; y_{22})$, where  
\begin{equation}
y_{ij} = \frac{2 k_i \cdot p_j}{s} = w_i (1 \mp \beta \cos \theta_i)~.
\end{equation}
In combination with the leptonic part of the phase space, we have
\begin{align}
& \frac{{\rm d}\Phi_3(p_1,p_2;Q,k_1,k_2)}{y_{11} \; y_{12} \; y_{21} \; y_{22}}
\sim
\frac{{\rm d}w_1}{w_1(1-q^2-2w_1)} \; \non \\ & \qquad \times 
\frac{{\rm d}\Omega_1}{1-\beta^2 \cos^2{\theta_1}}
\; \frac{{\rm d}\Omega_2}{1-\beta^2 \cos^2{\theta_2}}~.
\end{align}
The collinear peaks are then flattened with the help of \Eq{t1}, with 
one change of variables for each photon polar angle. The remaining soft 
peak, $w_1 \rightarrow w$, is reabsorbed with the following substitution
\begin{align}
w_1 = \frac{1-q^2}{2+e^{-u_1}}~, \quad u_1 = \log \frac{w_1}{1-q^2-2w_1}~,
\end{align}
or
\begin{align}
\frac{{\rm d} w_1}{w_1(1-q^2-2w_1)} = \frac{{\rm d} u_1}{1-q^2}~,
\end{align}
where the new variable $u_1$ is generated flat. 
Multi-channe\-ling is used to absorb simultaneously the soft and 
collinear peaks, and the peaks of the form factors.

{$\\$}
{\noindent \it NLO cross section and theoretical uncertainty \\}

\begin{figure}
\begin{center}
\epsfig{file=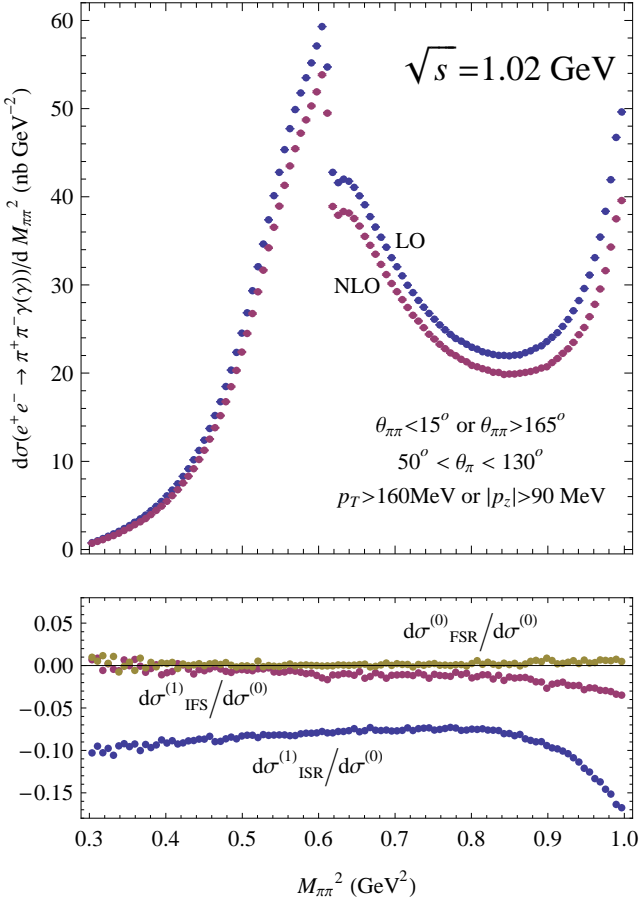,width=9cm} 
\end{center}
\caption{Differential cross section for the process 
$e^+ e^- \rightarrow \pi^+ \pi^- \gamma$ at LO and NLO for 
$\sqrt{s}=1.02$~GeV. The cuts are the same as in the 
small angle analysis of KLOE, including the cut on the 
track mass. The lower plot shows the relative size of FSR 
at LO, ISR at NLO and IFS contributions with respect to 
the full LO prediction.}
\label{fig:nlo1gev}
\end{figure}

The LO and NLO predictions for the differential cross section of the 
process $e^+ e^- \rightarrow \pi^+ \pi^- \gamma (\gamma)$ at DA$\mathrm{\Phi}$NE 
energies, $\sqrt{s} = 1.02$~GeV, are presented in 
Fig.~\ref{fig:nlo1gev} as a function of 
the invariant mass of the hadronic system $M_{\pi\pi}$.
We choose the same kinematical cuts as in the small angle analysis 
of KLOE~\cite{:2008en}; pions are restricted to be in the central region, 
$50^o < \theta_{\pi}< 130^o$, with $|p_T|>160$~MeV or $|p_z|>90$~MeV, 
the hard photon is not tagged and the sum of the momenta of the 
two pions, which flows in the opposite direction to the photon's momenta,  
is close to the beam ($\theta_{\pi\pi}<15^o$ or $\theta_{\pi\pi}>165^o$). 
The track mass, which is calculated from the equation 
\bea
&& \!\!\!\!\!\!\!\!\!\!
\left(\sqrt{s}-\sqrt{|\vec{p}_{\pi^+}|^2+M_{\rm trk}^2} 
- \sqrt{|\vec{p}_{\pi^-}|^2+M_{\rm trk}^2} \right)^2 \nn \\ &&
- (\vec{p}_{\pi^+}+\vec{p}_{\pi^-})^2 = 0~,
\label{eq:mtrkdef}
\eea
lies within the limits $130$~MeV$<M_{\rm trk}<220$~MeV 
and $M_{\rm trk}< (250 - 105 \sqrt{1-(M_{\pi\pi}^2/0.85)^2})$~MeV,
with $M_{\pi\pi}$ in GeV,
in order to reject $\mu^+\mu^-$ and $\pi^+\pi^-\pi^0$ events. 
The cut on the track mass, however, does not have any effect for single 
photon emission, as obviously $M_{\rm trk}=m_\pi$ for such 
events. 

The lower plot in Fig.~\ref{fig:nlo1gev} shows the relative 
size, with respect to the LO prediction, of FSR at LO, ISR 
corrections at NLO, and IFS contributions. The NLO ISR radiative 
corrections are almost flat and of the order of $-8$\%, FSR is 
clearly below $1$\%, while IFS corrections are also small 
although they become of the order of a few per cent at high 
values of $M_{\pi\pi}$.

To estimate the systematic uncertainty of the NLO prediction, we observe that
leading logarithmic two-loop $O (\alpha^2)$ corrections and the 
associate real emission are not included. 
For samples with untagged photons the process 
$e^+e^- \to e^+e^-\pi^+\pi^-$ might also become a sizable background. 
This process, however, can be simulated with the Monte Carlo event generator
EKHARA~\cite{Czyz:2006dm,Czyz:2005ab}. Its contribution depends on the pion
pair invariant mass, ranges 
from $0.1-0.8$\% for the KLOE event selection,
 and has been taken into account in the 
KLOE analysis~\cite{:2008en}.

From na\"{\i}ve exponentiation one expects that LL corrections 
at next-to-next-to-leading order (NNLO) are of 
the order of $\frac{1}{2} ( \frac{3}{2} (\alpha/\pi) \log(s/m_e^2) )^2 
\approx 0.1$--$0.2\%$ for inclusive observables. 
For less inclusive distributions, a larger error is expected. 
The conservative estimate of the accuracy of PHOKHARA 
from ISR is $0.5$\%. 
This has been confirmed by comparisons with 
KKMC~\cite{Jadach:2006fx,Glosser:2003ux}, where
the biggest observed difference is about
$0.3$\% in the invariant mass regions which are not
close to the nominal energies of the experiments.
Improving the accuracy of PHOKHARA below $0.5$\%,
however, will be required to meet the growing experimental 
requirements in the near future.

\subsubsection{FSR beyond sQED$*$VMD model}
\label{sQEDVMD}

  The model for FSR from pions described in details in
 Sections \ref{sec:LO} and  \ref{rratnlo} will be called for short
 the sQED$*$VMD model.
  The question arises how well
it
  can reflect the data. 
As shown in \cite{Baier:1965jz}, the first two terms in the
 expansion of the FSR amplitude as a function of $k^0/\sqrt{ Q^2}$
(i.e. the divergence and the constant) are fully given by the pion
 form factor. Thus one could expect that going
  beyond this approximation is necessary only for a hard photon emission. Moreover,
   the pion form factor is extremely big in the $\rho$ resonance
  region, and thus the validity of this approximation is further extended.
  In the kinematical regions where resonance contributions are 
  not contained in  the pion form factor, and also
   near the $\pi^+\pi^-$ threshold, where the emitted photon is hard
  and the pion form factor is relatively small, it is necessary to go
  beyond the sQED$*$VMD model and one needs a more general description
  of the  amplitude  $ M(\gamma ^{\ast
}( Q)\rightarrow \gamma (k)+\pi ^{+}(q_1)+\pi ^{-}(q_2))$.

In the general case the amplitude of the reaction

\noindent
 $\gamma ^{\ast
}( Q)\rightarrow \gamma (k)+\pi ^{+}(q_1)+\pi ^{-}(q_2)$
depends on  three 4-momenta, which can be chosen as $Q$, $k$ and
$l\equiv q_1-q_2$. The second-rank Lorentz
tensor $M^{\mu \nu }( Q,k,l)$ that describes the FSR amplitude 
can be decomposed through ten independent tensors
\cite{Tarrach:1975tu,Drechsel:1996ag}. Taking into account the charge
conjugation symmetry of the S-matrix element

 \noindent
($\langle \gamma
(k),\pi ^{+}(q_1)\pi ^{-}(q_2)|S|\gamma ^{\ast }(Q)\rangle
=$ 

\hskip 3 cm $\langle \gamma (k),\pi ^{-}(q_1)\pi ^{+}(q_2)|S|\gamma ^{\ast
}(Q)\rangle $),

 \noindent
 the photon crossing symmetry ($Q\leftrightarrow
-k$ and $\mu \leftrightarrow \nu $) and the gauge invariance
conditions $Q_{\mu }M^{\mu \nu }(Q,k,l)=0$ and $M_F^{\mu \nu
}(Q,k,l)k_{\nu }=0$, the number of independent tensors
decreases to five. For  a
final  real photon, i.e. $k^{2}=0$ and $k^{\nu }\epsilon _{\nu
}=0$ ($\epsilon _{\nu }$ being the polarisation
vector of the final photon) and  the initial virtual photon produced in
$e^{+}e^{-}$ annihilation ( $Q^{2}\geq
4m_{\pi}^{2}$), the FSR tensor can be rewritten in 
terms of  three gauge invariant tensors \cite{Tarrach:1975tu,Drechsel:1996ag}
\begin{equation}
M^{\mu \nu }(Q,k,l)=\tau _{1}^{\mu \nu }f_{1}+\tau _{2}^{\mu \nu
}f_{2}+\tau _{3}^{\mu \nu }f_{3}, \label{eq:A3}
\end{equation}
where the gauge invariant tensors $\tau_i^{\mu\nu}$ read
\begin{eqnarray}
\tau _{1}^{\mu \nu } &=&k^{\mu }Q^{\nu }-g^{\mu \nu }k\cdot Q,  \\
\tau _{2}^{\mu \nu } &=&k\cdot l(l^{\mu }Q^{\nu }-g^{\mu \nu
}k\cdot
l)+l^{\nu }(k^{\mu }k \cdot l-l^{\mu }k \cdot Q),  \nonumber \\
\tau _{3}^{\mu \nu } &=&Q^{2}(g^{\mu \nu }k\cdot l-k^{\mu }l^{\nu
})+Q^{\mu }(l^{\nu }k\cdot Q-Q^{\nu }k\cdot l).\nonumber  \label{eq:A4}
\end{eqnarray}

It thus follows that the evaluation of the FSR tensor amounts to
the calculation of the scalar functions

 $f_{i}(Q^2,Q\cdot k,k\cdot l) $ \ ($i=1,2,3$).

  As is clear from the above discussion, the extraction of 
  the pion form factor from radiative return experiments is a demanding
  task. The main problem is that in the same experiment one has
  to test the models describing the pion-photon interactions (see 
   Section \ref{expvsth}) and to extract the pion form factor needed
  for the evaluation of the muon anomalous magnetic moment. Fortunately,
  there are event selections, which naturally suppress the FSR contributions,
  independently of their nature. These were already discussed in Section
   \ref{sec:LO} in the context of the sQED$*$VMD model.

  Extensive theoretical studies of the role of the FSR emission
  beyond the sQED$*$VMD model were performed 
  \cite{Czyz:2004nq,Melnikov:2000gs,Pancheri:2006cp,Shekhovtsova:2009yn,Pancheri:2007xt}. They
  are important  mainly for the KLOE measurements at
  DA$\mathrm{\Phi}$NE, 
  as at $B$ factories FSR is naturally suppressed and the accuracy
 needed in its modelling is by far less demanding than that for KLOE purposes.

For DA$\mathrm{\Phi}$NE, running on or near the $\mathrm{\phi}$ resonance,
the following mechanisms of the
$\pi^+\pi^-$ final state photon emission have to be considered:

 -- bremsstrahlung process
\begin{eqnarray}
e^++e^-&\to&\pi^++\pi^-+\gamma \ ,
  \label{brem}
\end{eqnarray}

\noindent
which is modelled by sQED$*$VMD;

-- direct $\mathrm{\phi}$ decay
\begin{eqnarray}
e^++e^-&\to&\phi\to (f_0;f_0+\sigma)\gamma\to\pi^++\pi^-+\gamma \ ,
\label{phi_direct} 
\end{eqnarray}

\noindent
and 

-- double resonance process
\begin{eqnarray}
e^++e^-&\to&(\phi;\omega')\to \rho\pi\to\pi^++\pi^-+\gamma \ .
\label{phi_vmd}
\end{eqnarray}

\noindent
The resonance chiral theory
(R$\chi$T)~\cite{Ecker:1989yg,Ecker:1988te}
  was used in \cite{Pancheri:2006cp,Pancheri:2007xt}
 to estimate the contributions
beyond sQED$*$VMD. They were implemented at leading order
into the event generator FASTERD \cite{Shekhovtsova:2009yn}.
Having in mind that at present these models still await accurate
 experimental tests, 
 other models \cite{Achasov:1996xz,Achasov:2005hm} were also
 implemented in the event generator FASTERD.
To include both next-to-leading-order radiative corrections
 and the mechanisms discussed 
for FSR, a part of the FASTERD code, based on 
the models \cite{Achasov:1996xz,Achasov:2005hm}, was implemented
by O.~Shekhovtsova in PHOKHARA v6.0 (PHOKHARA v6.1 \cite{OlgaS}) and the studies
presented below are based on this code. The model
used there, even if far from an ideal,
 is the best tested model available in literature.

 We briefly 
describe main features of the models 
 used to describe processes contributing to FSR photon emission
listed above.
 For a more detailed description and the calculation of the
function $f_i$ we refer the reader to 
 \cite{Czyz:2004nq,Melnikov:2000gs,Shekhovtsova:2009yn} (see also references therein).

The sQED$*$VMD part gives contributions to $f_1$  and $f_2$.

The direct $\mathrm{\phi}$ decay is assumed to proceed through the 
intermediate scalar meson state:
 $\phi\to (f_0+\sigma)\gamma\to\pi\pi\gamma$.
 Various models  are proposed to
describe the $\phi$-scalar-$\gamma$ vertex: either
it is the direct decay $\phi\to \text(scalar)\gamma$, or the
vertex is generated dynamically through a loop of the charged
kaons.  As shown in~\cite{Melnikov:2000gs}, in the framework of any model, the
direct $\mathrm{\phi}$ decay affects only the form factor $f_1$ of
Eq.~({\ref{eq:A3}}). 

The double resonance contribution consists of the off-shell $\mathrm{\phi}$ meson
decay into ($\rho^\pm \pi^\mp$)  and subsequent decay
$\rho \to\pi\gamma$. In the energy region around $1$ GeV the
tail  of the excited $\omega$ meson  can  also play a role, and 
$\gamma^*\to\omega'\to\rho\pi$ has to be 
considered. The double resonance mechanism affects all three
form factors $f_i$ of Eq.~({\ref{eq:A3}}).

 Assuming isospin symmetry, this part can be deduced from the 
 measurement of the neutral pion pair production. Various models
 \cite{Achasov:1996xz,Achasov:2005hm}
 were confronted with data by KLOE \cite{Ambrosino:2006hb}
  for the neutral mode. The model that was 
 reproducing the data in the best way was adopted to be used for the
 charged pion pair production relying on the isospin symmetry \cite{OlgaS}.

   In \cite{Czyz:2004nq} it was shown that an important tool 
  for testing the various models of FSR is
  the charge asymmetry. At leading order it originates from the fact that 
  the pion pair couples to an even (odd) number of photons if the final state
  photon is emitted  from the final (initial) state. The interference diagrams
  do not give any contribution to the integrated cross section for C--even event
  selections, but produce an asymmetry in the angular distribution.
  The definitions and experimental studies based on the charge asymmetry
  are presented in Section \ref{radret:KLOEasymFSR}.
%
 
\begin{figure}[ht]
 \vspace{-0.6 cm}
\includegraphics[width=8cm,height=7cm]{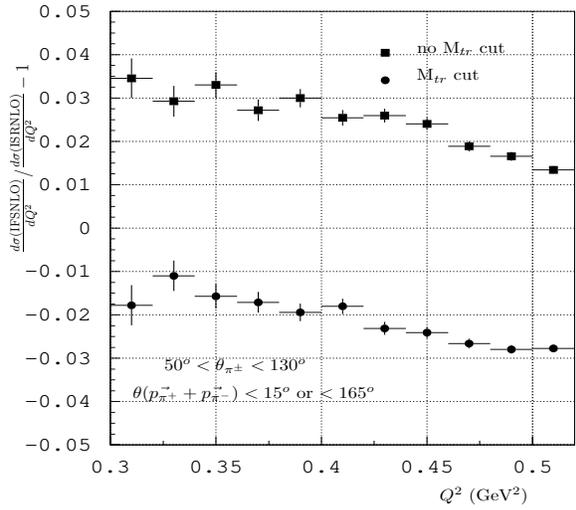}
\caption{Relative contribution of the FSR to the differential cross section
 of the reaction $e^+e^-\to \pi^+\pi^-\gamma(\gamma)$ for $\sqrt{s}=m_\phi$
and low invariant masses of pion pairs.
 KLOE small angle event selection \cite{:2008en} was used, 
and for this event selection
 the relative contribution of the FSR is almost identical also
 for the off peak cross section. The effect of a trackmass cut
 (see Section  \ref{rr:kloe}) is  shown. ISRNLO refers to initial state
 corrections at next-to-leading order (NLO). The IFSNLO cross section contains
 the final state emissions at NLO. 
}
 \vspace{-0.5 cm}
\label{f2}
\end{figure}

  Few strategies can be adopted to profit in the best way
 from the KLOE data taken on and off peak.
  The 'easiest' part is to look for the event selections where the FSR
 contributions are negligible. This was performed by KLOE \cite{:2008en}
  (see Section \ref{rr:kloe}),
 giving important information on the pion form factor relevant for 
  the prediction of the hadronic contributions to the muon anomalous
   magnetic moment $a_\mu$. Typical contributions of the FSR (1 -- 4\%)
  to the differential cross section (Figs.~\ref{fig:nlo1gev} and \ref{f2})
  allow for excellent control of the accuracy of these corrections.
 One disadvantage of using this event selection is that it does not allow
  to perform measurements near the pion production threshold. 

\begin{figure}[tbp]
\includegraphics[width=8cm,height=8cm]{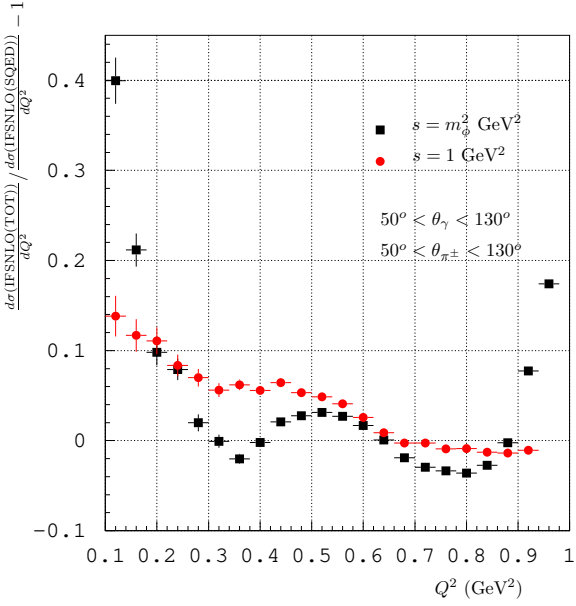}
\caption{The contributions of FSR beyond the sQED$*$VMD approximation (see 
Eqs.~(\ref{phi_direct}) and~(\ref{phi_vmd}))
 for KLOE large angle event selection \cite{Beltrame:2009zz,Leone:2007zz}
 for $\sqrt{s}=m_\phi$
 and for $\sqrt{s}= 1\ {\rm GeV}$. }
\label{fig3s}
\end{figure}

\begin{figure}[ht]
 \vspace{-0.65 cm}
\includegraphics[width=8cm,height=7cm]{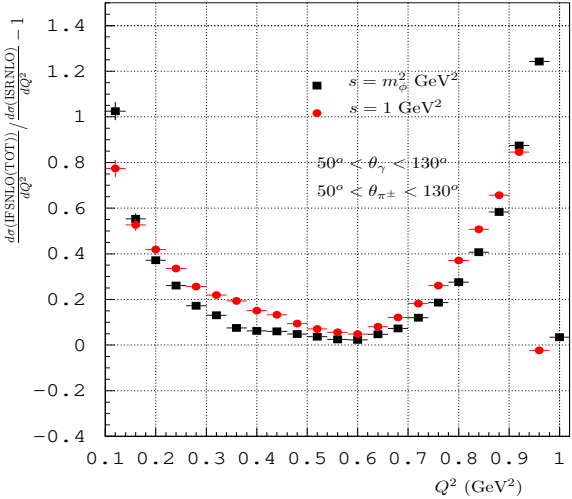}
\caption{Relative contribution of FSR to the differential cross section
 of the reaction $e^+e^-\to \pi^+\pi^-\gamma(\gamma)$ for $\sqrt{s}=m_\phi$
 and for $\sqrt{s}= 1\ {\rm GeV}$.
 KLOE large angle event selection \cite{Beltrame:2009zz,Leone:2007zz} was used. 
}
 \vspace{-0.6 cm}
\label{f:fsrla}
\end{figure}

The next step, partly discussed in Section \ref{radret:KLOEasymFSR},
   is to confront the models based on isospin symmetry and the
  neutral channel data with charged pion data taken off-peak, where the
  contributions from models beyond the sQED$*$VMD approximation 
  are relatively small
  (Fig.~\ref{fig3s}).
For the off-peak data \cite{Beltrame:2009zz}
 the region below $Q^2=0.3\ {\rm GeV}^2$ can be
   covered experimentally. However, the small statistics in this region
makes it difficult to perform high-pre\-ci\-sion tests of the models.
 For this analysis an accurate knowledge of the
   pion form factor at the nominal energy of the experiment is
   important, 
  as it defines the sQED$*$VMD predictions and as the FSR corrections 
  (Fig.~\ref{f:fsrla})  are sizeable.

  The last step, which  allows for the most accurate FSR model testing
   and profits from the knowledge of the pion form factor from previous
  analysis, is the on-peak large angle measurement. The large 
  FSR corrections coming from sources  beyond the sQED$*$VMD approximation
 (Figs.~\ref{fig3s} and \ref{f:fsrla})
  make these data \cite{Leone:2007zz} the most valuable source of information
  on these models. In this case, the accumulated data set is much larger than
  the off-peak data set and one is able to cover also the region
  below $Q^2=0.3\ {\rm GeV}^2$.

\subsection{Experiment confronting theory } 
\label{expvsth} 

\subsubsection{Study of the process $e^+e^- \to \pi^+\pi^-\gamma$ with FSR with
the CMD-2 detector at VEPP-2M}

\newcommand{\eeee}{\ensuremath{e^+e^-\to e^+e^-}\xspace}
\newcommand{\eemm}{\ensuremath{e^+e^-\to\mu^+\mu^-}\xspace}
\newcommand{\eepp}{\ensuremath{e^+e^-\to\pi^+\pi^-}\xspace}
\newcommand{\mm}{\ensuremath{\mu^+\mu^-}\xspace}
\newcommand{\pp}{\ensuremath{\pi^+\pi^-}\xspace}

The process $e^+e^- \to \pi^+\pi^-\gamma$ with final state radiation can be used to
answer the question whether one can treat pions as point-like particles and 
apply scalar QED to calculate the radiative corrections to the cross 
section. In particular, one can compare the photon spectra obtained using scalar 
QED with those found in data. 

The radiative corrections due to photon emission in the final state (FSR) 
contribute about 1\% to the cross section. The hadronic contribution 
of the process $e^+e^- \to \pi^+\pi^-$ to the value $a^{\rm had}_{\mu}$ amounts to 
$\sim$50 ppm, while the anomalous magnetic moment of the muon was measured 
in the E821 experiment at BNL with an accuracy of 0.5 ppm~\cite{Bennett:2006fi}. 
Therefore the theoretical precision of the 
cross section calculation for
this process should be several times smaller than 1\%. In this case we can  
neglect the error of this contribution to the value $a^{\rm had}_{\mu}$ 
compared to 0.5 ppm. These facts are the main motivation to study this process.  

\vspace{0.2cm}
{\noindent \it Event selection\\}

\begin{figure}[!htb]
\begin{center}
\subfigure[]{\includegraphics[width=0.45\textwidth]{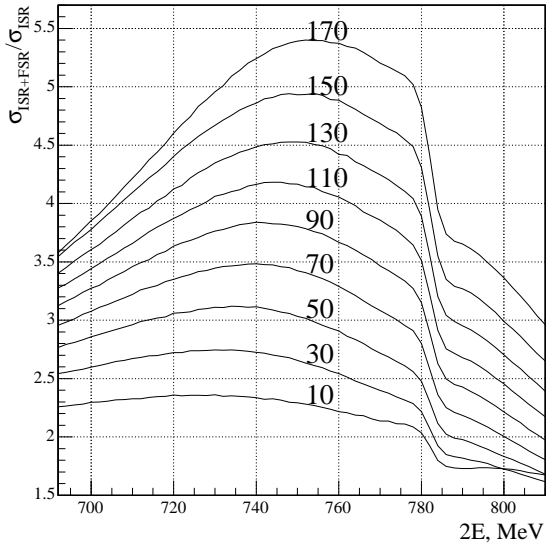}}
\subfigure[]{\includegraphics[width=0.5 \textwidth]{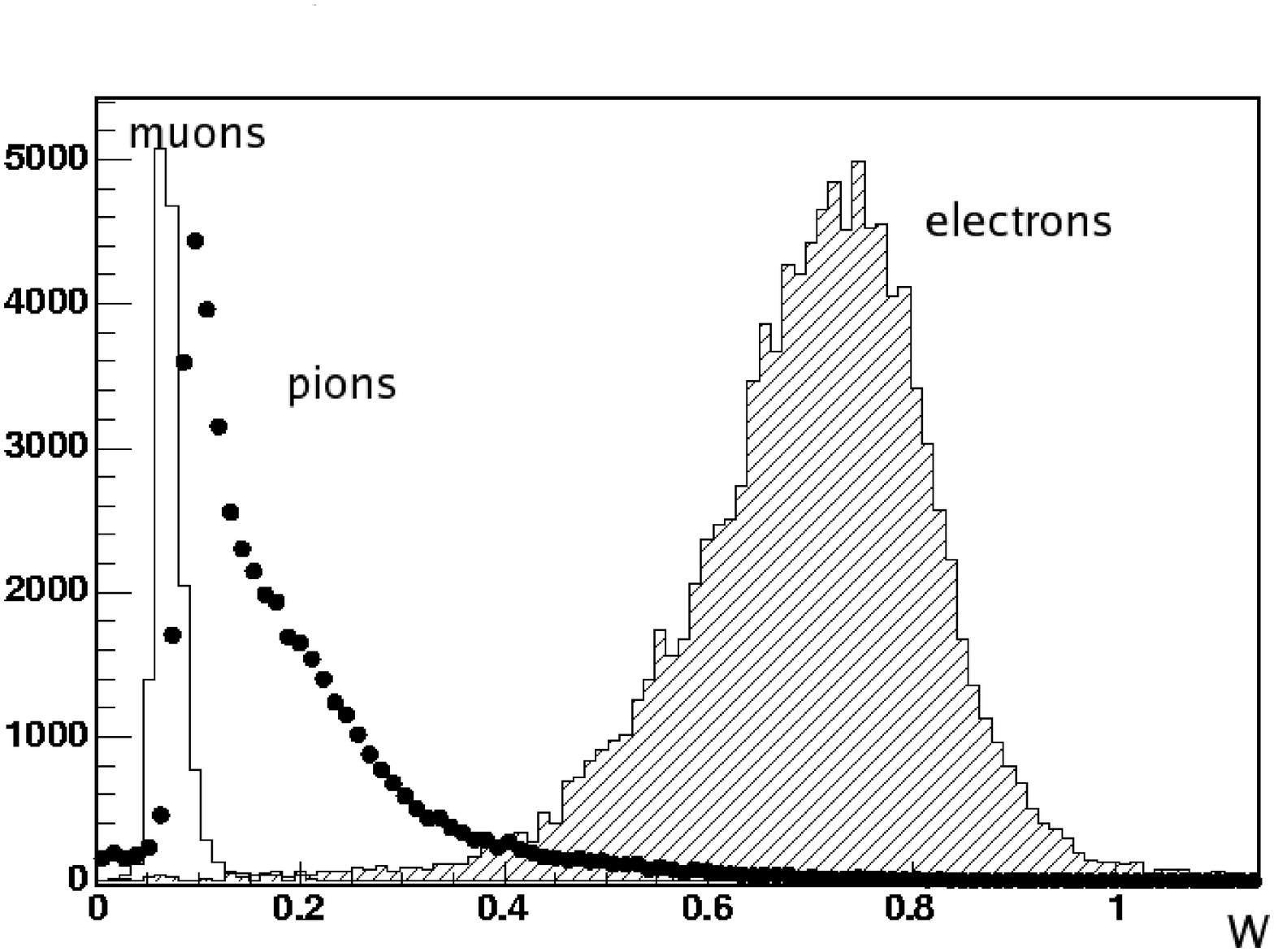}}
\caption{\label{isrfsr} (a) Ratio
$\sigma_{{\rm ISR} + {\rm FSR}}/\sigma_{\rm ISR}$ vs the c.m. energy. The set of curves indicates
how this ratio depends on the threshold energy for the detected photons. 
The threshold energy in MeV is stated over the curves.
(b) Distributions of the parameter $W$ for events of 
the processes
$e^+e^- \to \pi^+\pi^-\gamma$,
$e^+e^- \to \mu^+\mu^-\gamma$ and
$e^+e^- \to e^+e^-\gamma$, for a c.m. energy of 780 MeV.}
\end{center}
\end{figure}

For the analysis, data were taken in a c.m. energy range from 720 
to 780 MeV, with one photon detected in the CsI calorimeter. 
Events from the processes $e^+e^- \to e^+ e^-\gamma$ and $e^+e^- \to \mu^+\mu^-\gamma$ have a
very similar topology in the detector, compared to $e^+e^- \to \pi^+\pi^-\gamma$ events. 
In addition, the cross section of the process $e^+e^- \to \pi^+\pi^-\gamma$ 
with FSR is more than ten times smaller than the one for the similar process with ISR. 
On the other hand, the cross section of the process 
$e^+e^- \to \pi^+\pi^-\gamma$ has a strong energy dependence 
due to the presence of the $\rho$-resonance. This fact allows to 
significantly enrich the fraction of the events $e^+e^- \to \pi^+\pi^-\gamma$ with FSR 
for energies below the $\rho$-peak. Indeed, ISR shifts the c.m. energy to smaller   
values and, as a result, the cross section falls down dramatically, 
whereas the process with FSR is almost energy-independent. 
Several curves describing the ratio 
$\sigma^{{\rm FSR} + {\rm ISR}}_{\pi^+\pi^-\gamma}/\sigma^{\rm ISR}_{\pi^+\pi^-\gamma}$
plotted against the c.m. energy, are presented in Fig.~\ref{isrfsr} (a) for different energy
thresholds for photons detected in the calorimeter. It is clearly visible that
the optimal energy range to be used in this study goes from 720 MeV up to 780 MeV. 

It is also seen that this ratio increases with the threshold energy 
for photons to be detected. 
This means that the fraction of the  
$\pi^+\pi^-\gamma$ events with FSR (with respect to events without FSR) 
grows with increasing photon energy. 
It allows to enrich the number of $\pi^+\pi^-\gamma$ events with FSR.
Let us recollect that the shape of the distribution of $\pi^+\pi^-\gamma$ 
events, at photon energies of the same order as the pion mass or larger, 
is of special interest. First of all, namely in that part of 
the photon spectrum we can meet a discrepancy with the sQED prediction. 

A typical $\pi^+\pi^-\gamma$ event in the CMD-2 detector has two tracks in the 
drift chamber with two associated clusters in the CsI calorimeter
and a third cluster representing the radiated photon. 
To suppress multi-photon events and significantly
cut off collinear $\pi^+\pi^-$ events the following
requirements were applied: the angle between the direction of photon momentum and  
missing momentum must be larger than 1 rad and the angle between one of the two   
tracks and the photon direction must be smaller than 0.2 rad.

\begin{figure}[!htb]
\begin{center}
\subfigure[]{\includegraphics[width=0.45 \textwidth]{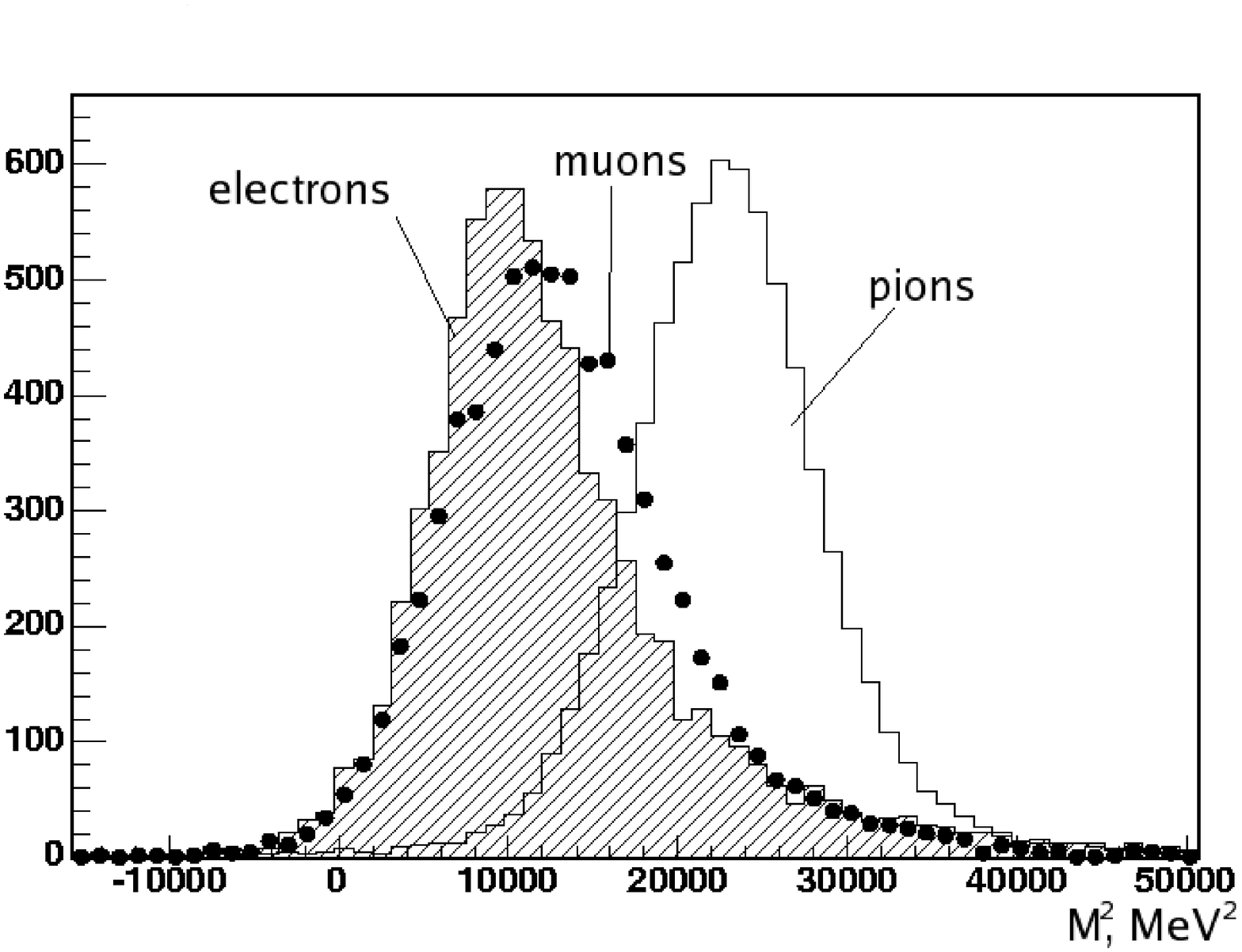}}
\subfigure[]{\includegraphics[width=0.45 \textwidth]{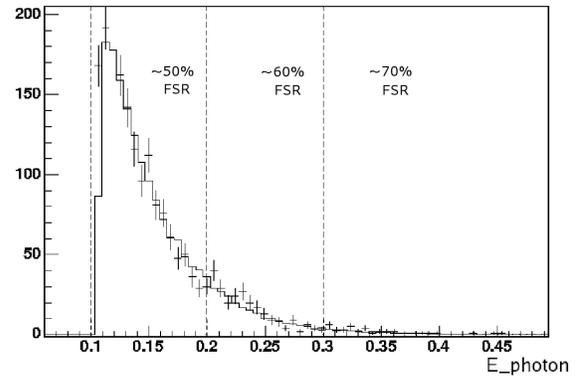}}
\caption{\label{M2sim} (a) 
Distributions of the parameter $M^2$ for events of the processes
$e^+e^- \to \pi^+\pi^-\gamma$,
$e^+e^- \to \mu^+\mu^-\gamma$ and
$e^+e^- \to e^+e^-\gamma$ for a c.m. energy of 780 MeV.
(b) Distribution of the $\pi^+\pi^-\gamma$ events against 
the photon energy in relative units. 
Also stated is the fraction of $\pi^+\pi^-\gamma$ events with FSR for each region as indicated by the vertical lines.}
\end{center}
\end{figure}

To suppress $e^+e^-\gamma$ events, a parameter $W = p/E$
was used, in which the particle momentum $p$ (measured in the drift chamber) 
is divided by the energy $E$
(measured in the CsI calorimeter). Simulation results are presented in 
Fig.~\ref{isrfsr} (b). The condition $W < 0.4$ reduces the electron contribution to 
the level of $\sim$ 1\%. The square of the invariant mass for electrons, muons and 
pions is plotted in Fig.~\ref{M2sim} ~(a). The condition $M^2 > 10 000$~MeV$^2$  
further rejects the number of electrons and muons by a factor of 1.5. 
About 1\% of the pion events are lost with these cuts.   

{\noindent \it Preliminary results of the analysis\\}

The histogram of the number $\pi^+\pi^-\gamma$ events against
the photon energy in relative units is presented in Fig.~\ref{M2sim} (b).
The histogram represents the simulation, while the points with error bars 
show the experimental data. Vertical
dotted lines divide the plot area into three zones. The inscription inside each
zone indicates the fraction of $\pi^+\pi^-\gamma$ events with FSR  with respect 
to others. The number of the simulated events was normalised to the
experimental one. The average deviation between the two distributions was found
to be $(-2.1 \pm 2.3)\%$. Therefore, one can conclude that there is no 
evidence that photon radiation by pions needs to be described beyond 
the framework of scalar QED. In other words, pions can be treated as 
point-like objects, and the application of  scalar QED is found to be valid 
within the stated accuracy. Unfortunately, the lack of statistics in the 
energy range under study 
does not allow us to check this assumption with better accuracy. Forthcoming
experiments at VEPP-2000 will significantly improve the 
statistical error.

 \subsubsection{Study of the process $e^+e^- \to \pi^+\pi^-\gamma$ with FSR with
 KLOE detector}
 \label{radret:KLOEasymFSR}

As has been explained in Section~\ref{radret:theo}, the forward-backward asymmetry 
\begin{eqnarray}
 {\mathcal A}_{FB}(Q^2) = 
 \frac{N(\theta_{\pi^+}>90^\circ)- N(\theta_{\pi^+}<90^\circ) }
 {N(\theta_{\pi^+}>90^\circ)+ N(\theta_{\pi^+}<90^\circ)}\left(Q^2\right)
\label{KLOEasymfb}
\end{eqnarray}
can be used to test the validity of the description of the 
various mechanisms of the $\pi^+\pi^-$ final state photon emission, by 
confronting the output of the Monte Carlo generator with data. In the 
following studies, the Monte Carlo generator PHOKHARA v6.1~\cite{OlgaS} was used.
The parameters for the pion form factor were taken from~\cite{Bruch:2004py}, based 
on the parametrisation of K\"uhn and Santamaria~\cite{Kuhn:1990ad}. The 
parameters for the description of the direct $\phi$ decay and the double 
resonance contribution were taken from the KLOE analysis of the neutral 
mode~\cite{Ambrosino:2006hb}.

To suppress higher order effects, for which the interference and thus the 
asymmetry is not implemented in the Monte Carlo generator, a rather tight 
cut on the track mass variable (see Section~\ref{rr:kloe} and 
Fig.~\ref{fig:radret_kloe45}) of $|M_\mathrm{trk} - M_{\pi^\pm}| < 10$ MeV has 
been applied in the data, in addition to the {\it large angle} selection cuts
described in Section~\ref{rr:kloe}. This should reduce events with more than one 
hard photon emitted and enhance the contribution of the final state radiation 
processes under study over the dominant ISR process.

The datasets used in the analysis were taken in two different periods:
\begin{itemize}
\item[$\bullet$]{The data taken in 2002 were collected with DA$\mathrm{\Phi}$NE operating at the $\phi$-peak, at $\sqrt{s}=M_\phi$ (240 pb$^{-1}$).} 
\item[$\bullet$]{The data taken in 2006 were collected with DA$\mathrm{\Phi}$NE operating 20 MeV 
{\it below} the $\phi$-peak, at $\sqrt{s}=1000$ MeV (230 pb$^{-1}$).}
\end{itemize}

\begin{figure}
\begin{center}
\subfigure[]{\includegraphics[width=80mm]{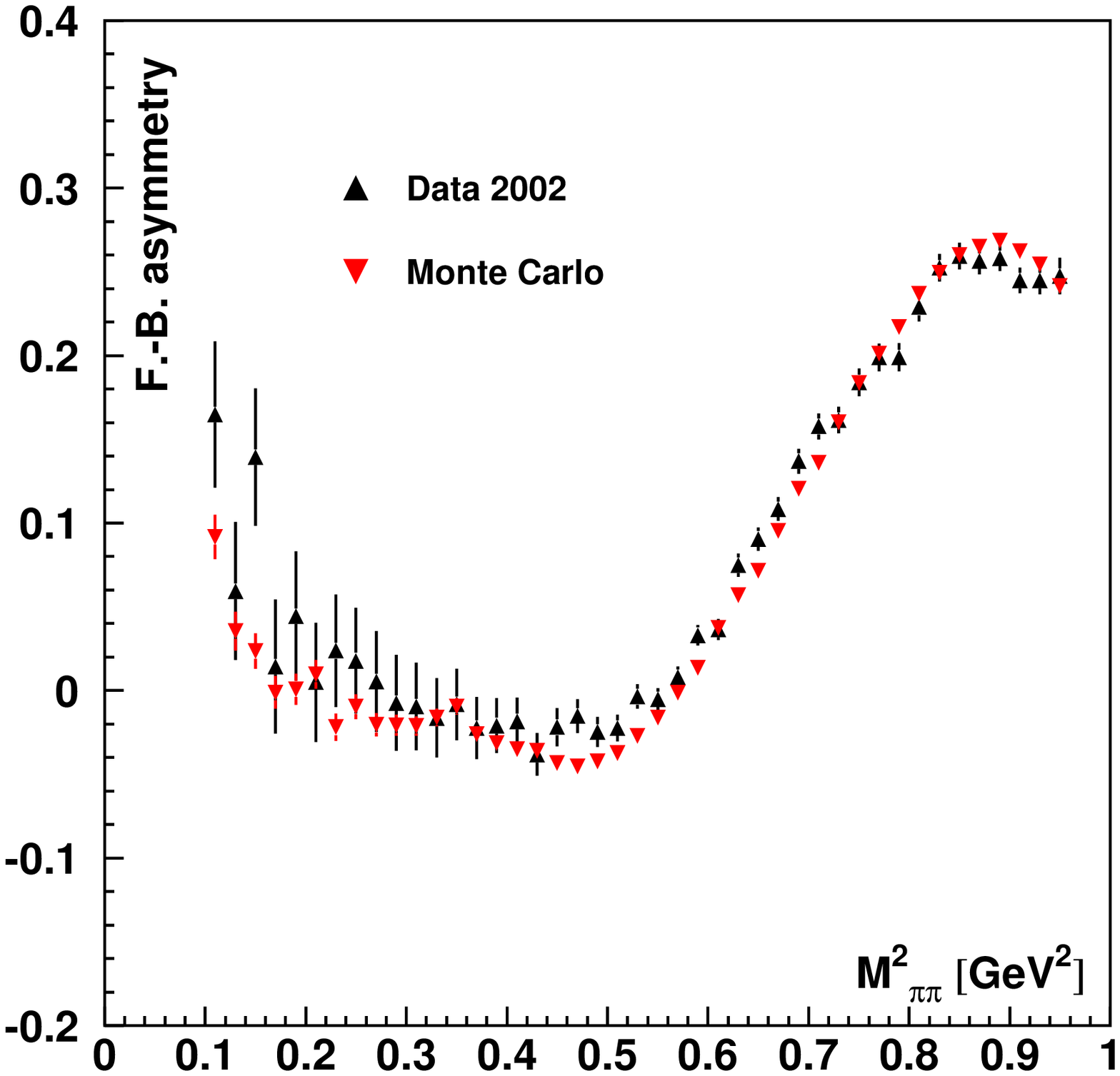}}
\hspace*{0.cm}
\subfigure[]{\includegraphics[width=80mm]{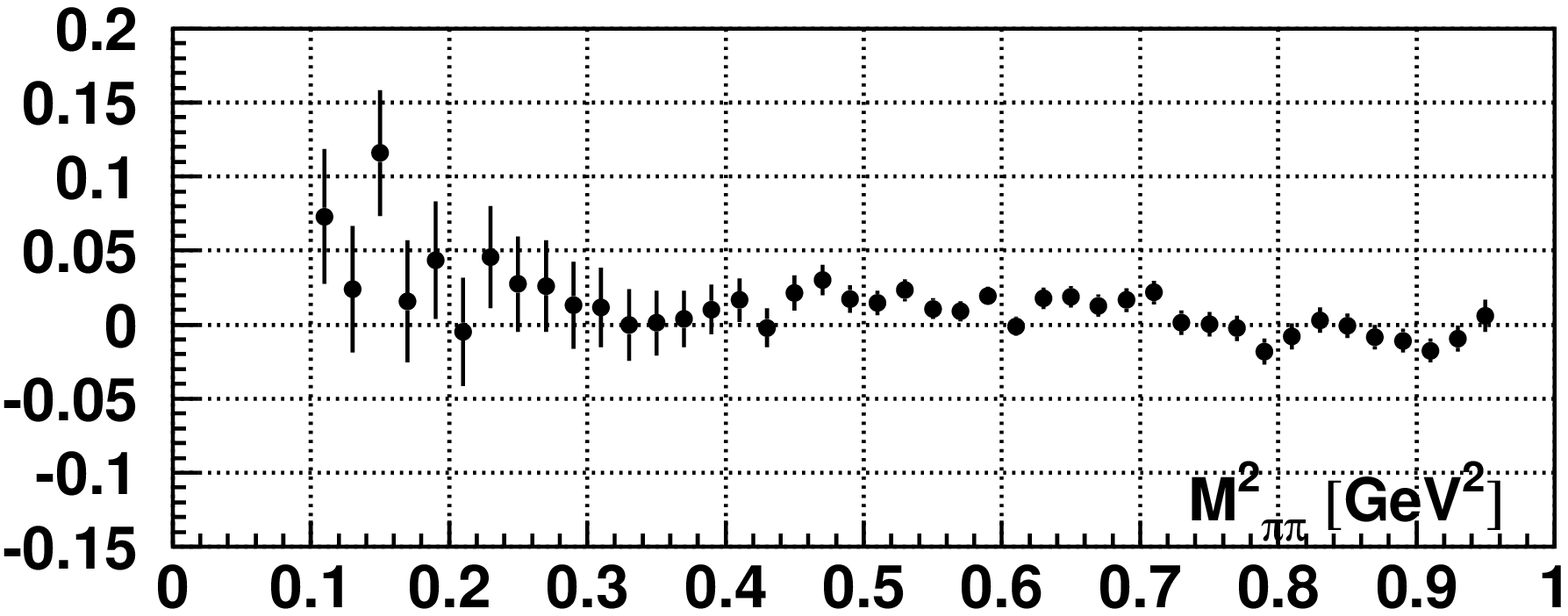}}
\caption{(a) Preliminary Forward--Backward asymmetry for data taken at $\sqrt{s}=M_\phi$  in 2002, and the corresponding Monte Carlo prediction using the PHOKHARA v6.1 generator. (b) Absolute difference between the asymmetries from data and Monte Carlo prediction. Used with permission 
of the KLOE collaboration.} 
\label{fig:radret_kloeasy1}
\end{center}
\end{figure}

\begin{figure}
\begin{center}
\subfigure[]{\includegraphics[width=80mm]{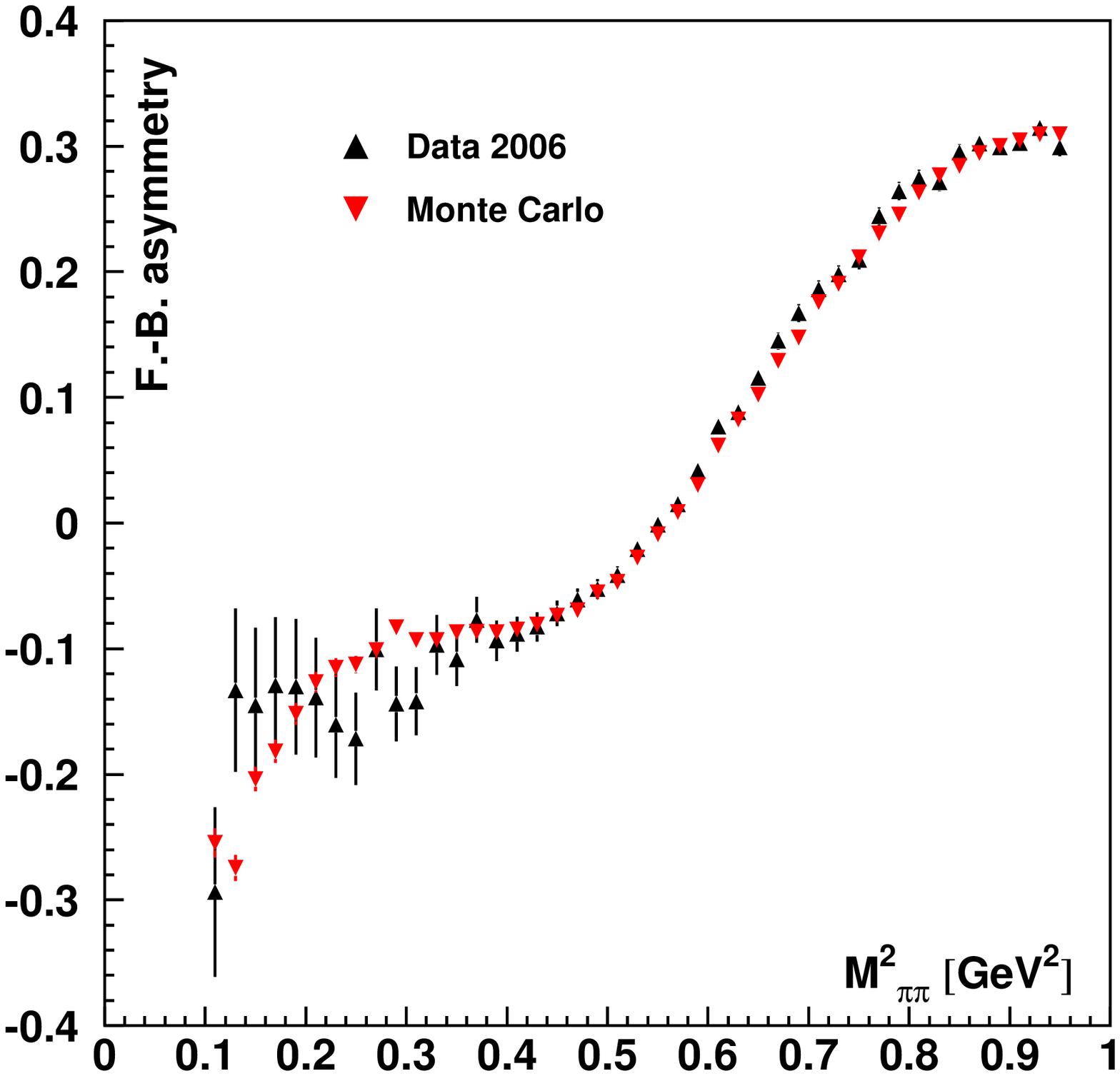}}
\hspace*{0.cm}
\subfigure[]{\includegraphics[width=80mm]{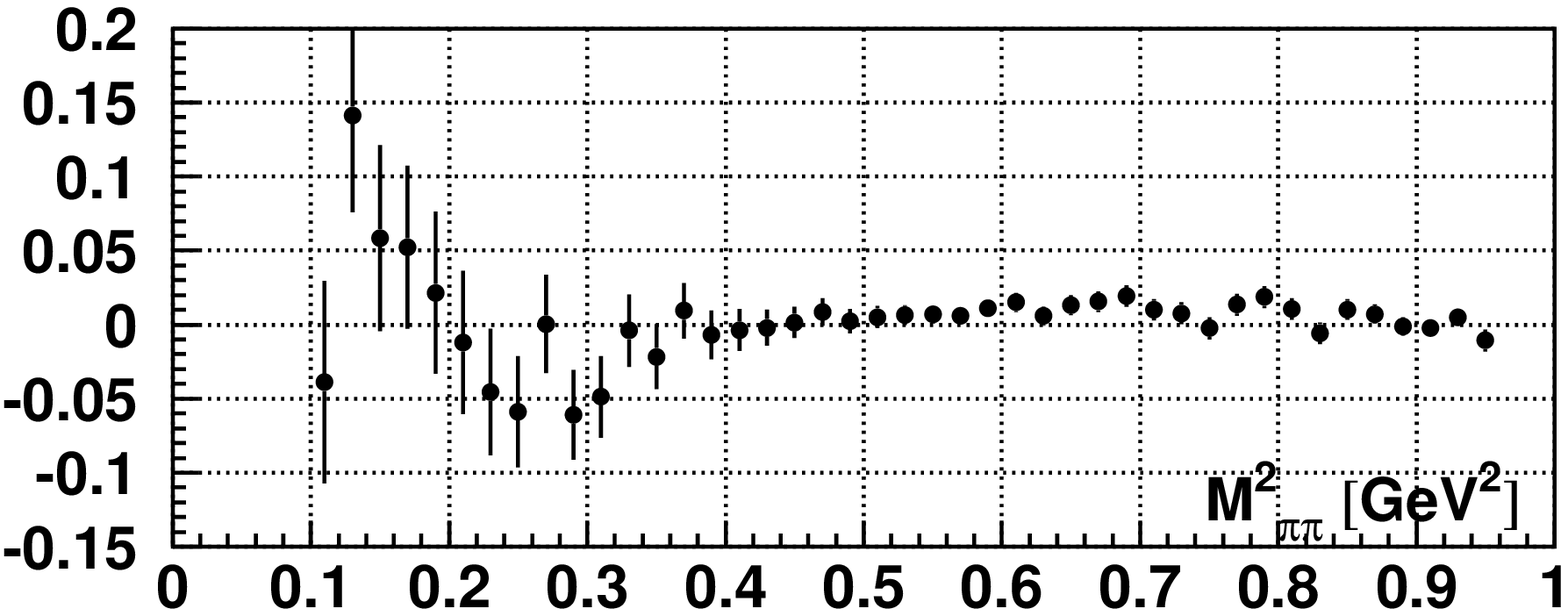}}
\caption{(a) Preliminary Forward--Backward asymmetry for data taken at $\sqrt{s} \simeq 1000$ MeV 
in 2006, and the corresponding Monte Carlo prediction using the PHOKHARA v6.1 generator. (b) Absolute difference between the asymmetries from data and Monte Carlo prediction. Used with permission of the KLOE collaboration.}
\label{fig:radret_kloeasy2}
\end{center}
\end{figure}

Since the 2006 data were taken more than 4$\Gamma_\phi$ below the resonant peak 
($\Gamma_\phi=4.26$ MeV), one expects the contributions from the direct $\phi$ 
decay and the double resonance contribution to be suppressed compared to the 
data taken on-peak in 2002 (see Fig.~\ref{fig3s}). 
In fact one observes a very different shape 
of the forward-backward asymmetry for the two different datasets, as can be 
seen in Figs.~\ref{fig:radret_kloeasy1} and \ref{fig:radret_kloeasy2}. 
Especially in the region below 0.4 GeV$^2$ and in the vicinity of the 
$f_0(980)$ at 0.96 GeV$^2$, one observes different trends in the asymmetries 
for the two datasets.

One can also see that, qualitatively, the theoretical description used to  
model the different FSR contributions agrees well with the data, although, 
especially at 
low $M_{\pi\pi}^2$, the data statistics becomes poor and the data points for the asymmetry have large errors. In particular, the {\it off-peak} data in 
Fig.~\ref{fig:radret_kloeasy2} show very good agreement above 0.35 GeV$^2$. In this case, the asymmetry is dominated fully by the bremsstrahlung-process, 
as the other processes do not contribute outside the $\phi$-resonance. The 
assumption of point-like pions (sQED) used to describe the bremsstrahlung in 
the Monte Carlo generator seems to be valid above 0.35 GeV$^2$, while below it 
is difficult to make a statement due to the large statistical errors of the 
data points.

However, to obtain a solid quantitative statement on the validity of the models, as needed, e.g., in the radiative return analyses at the KLOE experiment, 
one needs to understand how a discrepancy between theory and data in the 
forward-backward asymmetry affects the cross section, as it is the 
cross section one wants to measure. This requires further work, which 
at the moment is still in progress.

It should also be mentioned that the KLOE experiment has taken almost ten times 
more data in the years 2004--2005 than what is shown in 
Fig.~\ref{fig:radret_kloeasy1}, with DA$\mathrm{\Phi}$NE operating at the $\phi$-peak energy. This is unfortunately not the case for the {\it off-peak} data, which is restricted to the dataset shown in Fig.~\ref{fig:radret_kloeasy2}. 
In the future, the larger dataset from 2004--2005 may be used,
together with the results from the neutral channel and the assumption
of isospin symmetry, to determine the parameters of the direct $\phi$
decay and the double resonance contribution with high precision. 

 \subsection{The use of radiative return as an experimental tool}

\subsubsection{Radiative return at KLOE}
\label{rr:kloe}

The KLOE experiment, in operation at the DA$\mathrm{\Phi}$NE $e^+e^-$ collider in 
Frascati between 1999 and 2006, utilises radiative return to obtain precise 
measurements of hadronic cross sections in the energy range below 1 GeV. As
the DA$\mathrm{\Phi}$NE machine was designed to operate as a meson factory with
collision energy equal to the mass of the $\phi$-meson ($m_\phi =$
1.01946 GeV), with limited possibility to change the energy of the
colliding beams while maintaining stable running conditions, the
use of events with initial state radiation of hard photons from the
$e^+$ or the $e^-$ is the only way to access energies below
DA$\mathrm{\Phi}$NE's nominal collision energy. These low-energy cross sections are 
important for the theoretical evaluation of the muon magnetic moment
anomaly $a_\mu=(g_\mu-2)/2$~\cite{Eidelman:1995ny}, and high
precision is needed since the uncertainty on the cross section data
enters the uncertainty of the theoretical prediction. The channel
$e^+e^- \to \pi^+\pi^-$ gives the largest contribution to the hadronic
part $a_\mu^{\rm had}$ of the anomaly. Therefore, so far KLOE efforts have
concentrated on the derivation of the pion pair-production cross
section $\sigma_{\pi\pi}$ from measurements of the differential cross
section $\frac{{\rm d}\sigma_{\pi\pi\gamma(\gamma)}}{{\rm d}M^2_{\pi\pi}}$, in
which $M^2_{\pi\pi}$ is the invariant mass squared of the di-pion system in
the final state.

\begin{figure}[htb]
\begin{center}
\includegraphics[width=70mm] {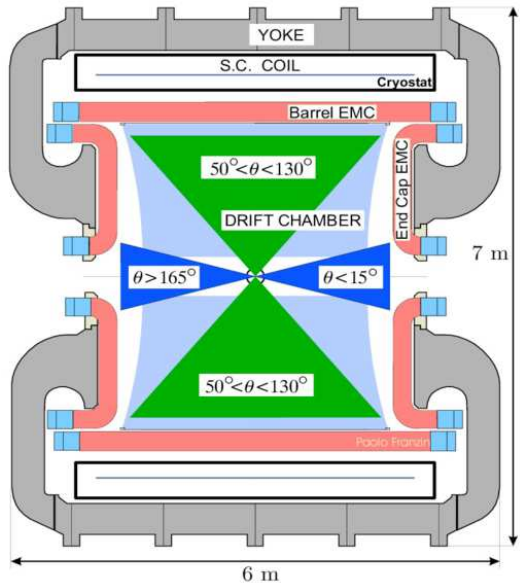}
\caption{KLOE detector with the selection regions for small
  angle photons (narrow cones)
and for pion tracks and large angle photons (wide cones). Used with permission 
of the KLOE collaboration.}
\label{fig:radret_kloe1}
\end{center}
\end{figure}

The KLOE detector (shown in Fig.~\ref{fig:radret_kloe1}), which
consists of a high
resolution drift chamber ($\sigma_{p} / p \leq 0.4\%$) and an 
electromagnetic calorimeter with excellent time ($\sigma_t\sim
54 ~\mathrm{ps}/\sqrt{E~[\mathrm{GeV}]}$ $\oplus100~\mathrm{ps}$) and
good energy 
($\sigma_E/E\sim 5.7\%/\sqrt
{E~[\mathrm{GeV}]}$) resolution, is optimally suited for this kind of
analyses.\\ 
 
{\noindent \it The KLOE $\pi\pi\gamma$ analyses\\}

The KLOE analyses for $\sigma_{\pi\pi}$ use two
different sets of acceptance cuts: 
\begin{itemize}
\item[$\bullet$] In the {\it small angle}
analysis, photons are emitted
within a cone of $\theta_\gamma<15^\circ$ around the 
beamline (narrow cones in Fig.~\ref{fig:radret_kloe1}), and the two
charged pion tracks 
 have $50^\circ<\theta_\pi<130^\circ$. The photon is
not explicitly detected; its direction
is reconstructed from the track momenta
by closing the kinematics:
$\vec{p}_\gamma\simeq\vec{p}_{miss}= -(\vec{p}_{\pi^+}
+\vec{p}_{\pi^-})$. 
In this analysis, the separation of pion- and photon selection
  regions greatly reduces the 
contamination from the resonant process $e^+e^-\to
  \phi\to\pi^+\pi^-\pi^0$ in which the $\pi^0$ mimics the missing
  momentum of the photon(s) and from the final state 
radiation process $e^+e^-\to \pi^+\pi^-\gamma_{\rm FSR}$.
Since ISR-photons are mostly collinear with the beam line, a high
  statistics for the ISR signal events remains. On the other hand, a high 
energy 
photon emitted at angles close to the incoming beams forces the pions also to 
have a small angle with respect to the beamline (and thus outside the selection cuts), resulting in a 
kinematical suppression of events with $M^2_{\pi\pi}< 0.35$
  GeV$^2$.  
\item[$\bullet$] The {\it large angle} analysis requires both photons and pions
  to be emitted at $50^\circ<\theta_{\pi,\gamma}<130^\circ$ (wide
  cones in Fig.~\ref{fig:radret_kloe1}), allowing for a detection of
  the photons in the barrel of the calorimeter. This analysis allows to reach
  the 2$\pi$ threshold region, at the price of higher background
  contributions from the $\pi^+\pi^-\pi^0$ final state and events with
  final state radiation. In addition, events from the
  decays $\phi \to f_0\gamma \to \pi^+\pi^-\gamma$ and $\phi\to\pi^\pm\rho^\mp
  \to \pi^\pm\pi^\mp\gamma$, which need to be described by model-dependent 
parameterisations, contribute to the spectrum of the selected events 
(running at the $\phi$ peak).
\end{itemize}

Two analyses based on the {\it small angle} acceptance cuts have been
carried out. The first one, using 140 pb$^{-1}$ of data taken in the
year 2001, was published in 2005~\cite{Aloisio:2004bu}. The second one,
based on 240 pb$^{-1}$ of data taken in 2002, was published in
2008~\cite{Ambrosino:2008en}.

\begin{figure}[htb]
\begin{center}
\includegraphics[width=70mm] {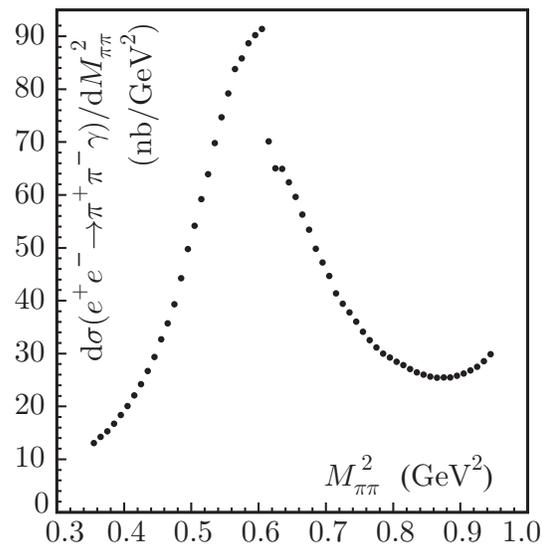}
\caption{Differential radiative cross section ${d\sigma_{\pi\pi\gamma(\gamma)}}/{dM^2_{\pi\pi}}$, inclusive in $\theta_\pi$ and with
$0^o<\theta_{\gamma}< 15^o$ or $165^o<\theta_{\gamma}< 180^o$ measured by the KLOE experiment~\cite{Ambrosino:2008en}. Used with permission 
of the KLOE collaboration.}
\label{fig:radret_kloe2}
\end{center}
\end{figure}

The differential cross section is obtained from the spectrum of
selected events $N^{\mathrm{sel}}$ subtracting the residual background
(mostly $\mu\mu\gamma(\gamma)$, $\pi\pi\pi$ and radiative Bhabha
events) and dividing by the selection efficiencies and the integrated
luminosity:
\begin{equation}
\frac{{\rm d}\sigma_{\pi\pi\gamma(\gamma)}}{{\rm d}M^2_{\pi\pi}} =
\frac{N^{\mathrm{sel}}-N^{\mathrm{bkg}}}{\Delta
  M^2_{\pi\pi}}\cdot\frac{1}{\varepsilon_{\mathrm{sel}}}\cdot\frac{1}{\int L{\rm d}t}~. 
\label{eq:radret_kloe1}
\end{equation}
 $\Delta M^2_{\pi\pi}$ is the bin width used in the analysis
  (typically 0.01 GeV$^2$), and $\int L{\rm d}t$ is the integrated
  luminosity obtained from Bhabha events detected at large angles
  ($55^\circ<\theta_e<125^\circ$) and 
 the reference cross
  section from the BabaYaga
  generator~\cite{CarloniCalame:2000pz,Balossini:2006wc} 
  (discussed in Section~\ref{sec:1}). 
The total cross section is then obtained from the formula
\begin{equation}
\sigma_{\pi\pi}(M^2_{\pi\pi})= s\cdot
\frac{{\rm d}\sigma_{\pi\pi\gamma(\gamma)}}{{\rm d}M^2_{\pi\pi}}
\frac{1}{H(s,M^2_{\pi\pi})}~.
\label{eq:radret_kloe2}
\end{equation}
In Eq.~(\ref{eq:radret_kloe2}), $s$ is the squared energy at which the DA$\mathrm{\Phi}$NE
collider is operated during data taking, and $H(s,M^2_{\pi\pi})$ is
  the radiator function describing the emission of photons from the
  $e^+$ or the $e^-$ in the initial state. Note that
  Eq.~(\ref{eq:radret_kloe2}) does not contain the effects from  final
  state radiation from pions. These effects complicate the analysis, since the KLOE
  detector can not distinguish whether photons in an event were emitted in the
  initial or the final state. The PHOKHARA Monte Carlo 
  generator~\cite{Czyz:2003ue}, which includes final state radiation
  at next-to-leading order and in the pointlike-pion approximation, is used
  to properly take into account final state radiation in the
  analyses. This is important because the {\it bare} cross section used
  to evaluate $a_\mu^{\rm had}$ via an appropriate dispersion integral should be
  inclusive with respect to final state radiation, and also needs to
  be undressed from vacuum polarisation effects present in the virtual
  photon produced in the $e^+e^-$ annihilation. For the latter, we
  use a function provided by
  F.~Jegerlehner~\cite{Jegerlehner:alphaWEB} (see Section~\ref{sec:4}), and correct the cross
  section via
 \begin{equation}
\sigma_{\pi\pi}^{\mathrm{bare}}(M^2_{\pi\pi})= \sigma_{\pi\pi}^{\mathrm{dressed}}(M^2_{\pi\pi}) \left( \frac{\alpha(0)}{\alpha(M^2_{\pi\pi})}
\right)^2\,.
\label{eq:radret_kloe3}
\end{equation}
Here $\alpha(0)$ is the fine structure constant in the limit $q=0$,
and $\alpha(M^2_{\pi\pi})$ represents the value of the effective
coupling at the scale of the invariant mass of the di-pion system. 
Since the hadronic contributions to $\alpha(M^2_{\pi\pi})$ are
calculated via a dispersion integral which includes the hadronic cross
section itself in the integrand (see Section~\ref{sec:4}), the correct
procedure has to be iterative and should include the same data that
must be corrected. However, since 
the correction is at the few percent level, the inclusion of the new
KLOE data will not change $\alpha(M^2_{\pi\pi})$ at a level which would 
significantly affect the analyses. We therefore have used the values
for $\alpha(M^2_{\pi\pi})$ derived from the existing hadronic cross
section database. As an example, Fig.~\ref{fig:radret_kloe2} shows the KLOE result
for ${{\rm d}\sigma_{\pi\pi\gamma(\gamma)}}/{{\rm d}M^2_{\pi\pi}}$ obtained from data taken in the year 
2002~\cite{Ambrosino:2008en}. Inserting this differential cross section into Eq.~(\ref{eq:radret_kloe2}) and the result 
into Eq.~(\ref{eq:radret_kloe3}), 
one derives $\sigma_{\pi\pi}^{\mathrm{bare}}$. Using the {\it bare} cross section to get the $\pi\pi$-contribution to 
$a_\mu^{\rm had}$ between 0.35 and 0.95 GeV$^2$ then gives the value (in units of $10^{-10}$)
\begin{displaymath}
a_\mu^{\pi\pi}(0.35 -0.95 {\rm GeV}^2) =  (387.2\pm0.5_{\rm stat}\pm2.4_{\rm exp}\pm2.3_{\rm th})\,.
\end{displaymath}
Table~\ref{tab:radret_kloe1} shows the contributions to the systematic
errors on $a_\mu^{\pi\pi}(0.35 -0.95$ GeV$^2)$.\\

\begin{table}
\begin{center}
\begin{tabular}{||l|c||}
\hline
Reconstruction Filter & negligible\\
Background subtraction & 0.3 \% \\
Trackmass & 0.2 \% \\
Particle ID & negligible\\
Tracking & 0.3 \% \\
Trigger & 0.1 \% \\
Unfolding & negligible \\
Acceptance ($\theta_{\pi\pi}$) & 0.2 \% \\
Acceptance ($\theta_\pi$) & negligible \\
Software Trigger (L3) & 0.1 \% \\
Luminosity ($0.1_{th}\oplus 0.3_{exp}$)\% & 0.3 \% \\
$\sqrt{s}$ dep. of $H$ & 0.2 \%\\
\hline
Total exp systematics & 0.6 \% \\
\hline
\hline
Vacuum Polarisation &  0.1 \% \\
FSR resummation & 0.3 \% \\
Rad. function $H$   & 0.5 \% \\
\hline
Total theory systematics & 0.6 \% \\
\hline
\end{tabular}
\caption{List of systematic errors on the $\pi\pi$-contribution to 
$a_\mu^{\rm had}$ between 0.35 and 0.95 GeV$^2$ when using the
  $\sigma_{\pi\pi}$ cross section measured
  by the KLOE experiment in the corresponding
  dispersion integral~\cite{Ambrosino:2008en}.}
\label{tab:radret_kloe1}
\end{center}
\end{table}

{\noindent \it Radiative corrections and Monte Carlo tools\\}

The radiator function is a crucial ingredient in this kind of
radiative return analyses, and is obtained using the
relation
 \begin{equation}
H(s,M^2_{\pi\pi}) = s\cdot\frac{3
  M^2_{\pi\pi}}{\pi\alpha^2\beta_\pi^3}\cdot\left.\frac{{\rm
  d}\sigma^{\rm ISR}_{\pi\pi\gamma(\gamma)}}{{\rm d}M^2_{\pi\pi}}\right|_{|F_{2\pi}|^2=1},
\label{eq:radret_kloe4}
\end{equation}
in which $\left.\frac{{\rm d}\sigma^{\rm
ISR}_{\pi\pi\gamma(\gamma)}}{{\rm d}M^2_{\pi\pi}}\right|_{|F_{2\pi}|^2=1}$
is evaluated using the PHOKHARA Monte Carlo generator in
next-to-leading order ISR-only
configuration, with the squared pion form factor $|F_{2\pi}|^2$ set to $1$. 
$\beta_\pi=\sqrt{1-\frac{4m_\pi^2}{M^2_{\pi\pi}}}$ is the pion
velocity. While Eq.~(\ref{eq:radret_kloe4}) provides a convenient mechanism to
extract the dimensionless quantity $H(s,M^2_{\pi\pi})$ also for
specific angular 
regions of pions and photons by applying the relevant cuts to
$\left.\frac{{\rm d}\sigma^{\rm ISR}_{\pi\pi\gamma(\gamma)}}{{\rm d}M^2_{\pi\pi}}\right|_{|F_{2\pi}|^2=1}$,
in the published KLOE analyses. $H(s,M^2_{\pi\pi})$ is evaluated
fully inclusive for pion and photon angles in the range 
$0^\circ < \theta_{\pi,\gamma} < 180^\circ$. 
\begin{figure}[tb]
\begin{center}
\includegraphics[width=70mm] {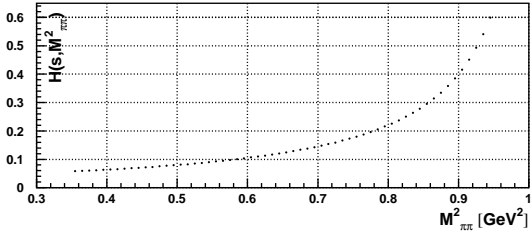}
\caption{The dimensionless radiator function $H(s,M^2_{\pi\pi})$,
  inclusive in $\theta_{\pi,\gamma}$. The value used for $s$ in the
  Monte Carlo production was 
$s = M_\phi^2=(1.019456$ GeV$)^2$.}
\label{fig:radret_kloe3}
\end{center}
\end{figure}
Figure~\ref{fig:radret_kloe3} shows the radiator function in the range
of $0.35 < M^2_{\pi\pi} < 0.95$ GeV$^2$. As can be seen from 
Table~\ref{tab:radret_kloe1}, the  0.5\%  uncertainty of the radiator function
quoted by the authors of PHOKHARA translates into an
uncertainty of 0.5\% in the $\pi\pi$-contribution to 
$a_\mu^{\rm had}$ between 0.35 and 0.95 GeV$^2$, giving the
largest individual contribution and dominating the theoretical
systematic error.  \\

The presence of events with final state radiation in the data sample
affects the analyses in several ways:
\begin{itemize}
\item[$\bullet$] {Passing from $M^2_{\pi\pi}$ to $(M^0_{\pi\pi})^2$.}
The presence of final state radiation 
shifts the observed value of $M^2_{\pi\pi}$ (evaluated from the momenta of the
two charged pion tracks in the events) away from the value of the
invariant mass squared of the virtual photon
produced in the collision of the electron and the 
positron, $(M^0_{\pi\pi})^2$. 
The transition from $M^2_{\pi\pi}$ to $(M^0_{\pi\pi})^2$ is performed 
using a modified version of the PHOKHARA Monte Carlo 
generator, which allows to (approximately) determine whether a generated 
photon comes from
the initial or the final
state~\cite{Czyz:PHOKHARA_omega}. Figure~\ref{fig:radret_kloe4} shows
the probability matrix relating $M^2_{\pi\pi}$ to
$(M^0_{\pi\pi})^2$ by giving the probability for an event in a bin of $M^2_{\pi\pi}$ to end up in a bin of $(M^0_{\pi\pi})^2$. 
It can be seen that the shift is only in one
direction, $(M^0_{\pi\pi})^2 \ge M^2_{\pi\pi}$, so events with one
photon from initial state radiation and one photon from final state
radiation move to a higher value of $(M^0_{\pi\pi})^2$. The entries lining up
above $(M^0_{\pi\pi})^2 \simeq 1.03$ GeV$^2$ represent events with two
pions and only one photon, emitted in the final
state. Events of this type have $(M^0_{\pi\pi})^2= s$, there is no
hard photon from initial state radiation present. Since in the KLOE analyses,
the maximum value of $(M^0_{\pi\pi})^2$ for which the cross sections
are measured is $0.95$ GeV$^2$ and sufficiently smaller than $s\simeq M^2_\phi$ of the
DA$\mathrm{\Phi}$NE collider, these {\it leading-order} final state radiation
events need to be taken out in the analysis. By moving these events to
$(M^0_{\pi\pi})^2 = s$, the passage from $M^2_{\pi\pi}$ to
$(M^0_{\pi\pi})^2$ automatically performs this
task. Figure~\ref{fig:radret_kloe5} shows the fraction of events from
{\it leading-order} final state radiation contributing to the total
number of events, evaluated with the PHOKHARA event generator. Since
in the {\it small angle} analysis the angular regions for pions and
photons are separated, final state radiation, for which the photons
are emitted preferably along the direction of the pions, is suppressed to
less than 0.5\%. Using {\it large angle} acceptance cuts, the effect
is much bigger, especially above and below the $\varrho$-resonance,
where it can reach 20-30\%. 
\begin{figure}[tb]
\begin{center}
\includegraphics[width=55mm] {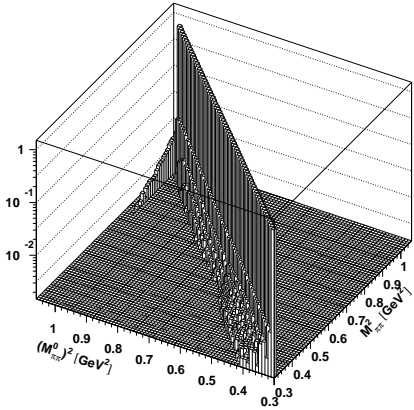}
\caption{Probability matrix relating the measured quantity
  $M^2_{\pi\pi}$ to $(M^0_{\pi\pi})^2$. To produce this plot, a
  private version of the 
PHOKHARA Monte Carlo 
generator was used~\cite{Czyz:PHOKHARA_omega}. The photon angle is
restricted to $\theta_\gamma <15^\circ$ ($\theta_\gamma >165^\circ$).}
\label{fig:radret_kloe4}
\end{center}
\end{figure}
The correction of the shift in $ M^2_{\pi\pi}$ depends on the implementation of final state
radiation in the Monte Carlo generator in terms of model dependence
and missing contributions. It also relies on the correct assignment of
photons coming from the initial or the final state; however, in case
of symmetrical cuts in $\theta_\gamma$, interference effects between
the two states vanish and the separation of initial and final state
amplitudes is feasible.  
\item[$\bullet$] The acceptance in $\theta_\gamma$. Since the direction of the photons
  emitted in the final state is peaked along the direction
  of the pions, and the photons are emitted in the initial state along
  the $e^+$/$e^-$ direction, the choice of the acceptance cuts affects
  the amount of final state radiation in the
  analyses. Using the {\it small angle} analysis cuts, a large part of
  final state radiation is suppressed by the separation of the pion
  and photon acceptance regions, and consequently needs to be
  reintroduced using corrections obtained from Monte Carlo
  simulations to arrive at a result which is inclusive with respect to
  final state radiation (as needed in the dispersion integral for
  $a_\mu^{\pi\pi}$). Even if in the {\it large angle} analysis the fraction
  of events with final state radiation surviving the selection is larger, again the missing
  part has to be added using Monte Carlo simulations. The acceptance
  correction for the cut in $\theta_\gamma$ is evaluated for initial and
  final state radiation using the PHOKHARA generator, and the
  small differences found in the comparison of data and Monte Carlo
  distributions contribute to the systematic uncertainty of the
  measurement (see Table~\ref{tab:radret_kloe1} and~\cite{KLOE_Note221}).

\item[$\bullet$] The distributions of kinematical variables. Cuts on
  the kinematical {\it trackmass} variable
 $M_\mathrm{trk}$ (see Eq.~(\ref{eq:mtrkdef})),
  introduced in the analyses to remove background from the process
  $e^+e^- \to \phi \to \pi^+\pi^-\pi^0$, take out also a fraction of
  the events with final state radiation, necessitating a correction to
  obtain an inclusive result. Figure~\ref{fig:radret_kloe45} shows the
  effect final state radiation has on the distribution of the
  trackmass variable. The radiative tail of multi-photon events to the
  right of the peak at the $\pi^\pm$ mass increases because the additional
  radiation moves events from the peak to higher values in
  $M_\mathrm{trk}$. The width of the peak at $M_{\pi^\pm}$ is due to the
  detector resolution; the plot was produced using the PHOKHARA event
  generator interfaced with the KLOE detector
  simulation~\cite{Ambrosino:2004qx}. Between 150 and 200 MeV, an
  $M_{\pi\pi}^2$-dependent cut is used in the event selection to
  reject the $\pi^+\pi^-\pi^0$ events which have a value of
  $M_\mathrm{trk} > M_{\pi^\pm}$. In this region, the cut also acts on
  the signal events. Missing terms concerning final state radiation in
  the Monte Carlo simulation or the non-validity of the pointlike-pion
  approximation used in PHOKHARA may affect the shape of the radiative
  tail in the trackmass variable. To overcome this, in the KLOE
  analyses, small corrections are applied to the momenta and the
  angles of the charged particles in the event in the simulation, and
  good agreement in the shape of $M_\mathrm{trk}$ is obtained between
  Monte Carlo simulation and data~\cite{KLOE_Note221}. 

 \begin{figure}
\begin{center}
\subfigure[]{\includegraphics[width=80mm]{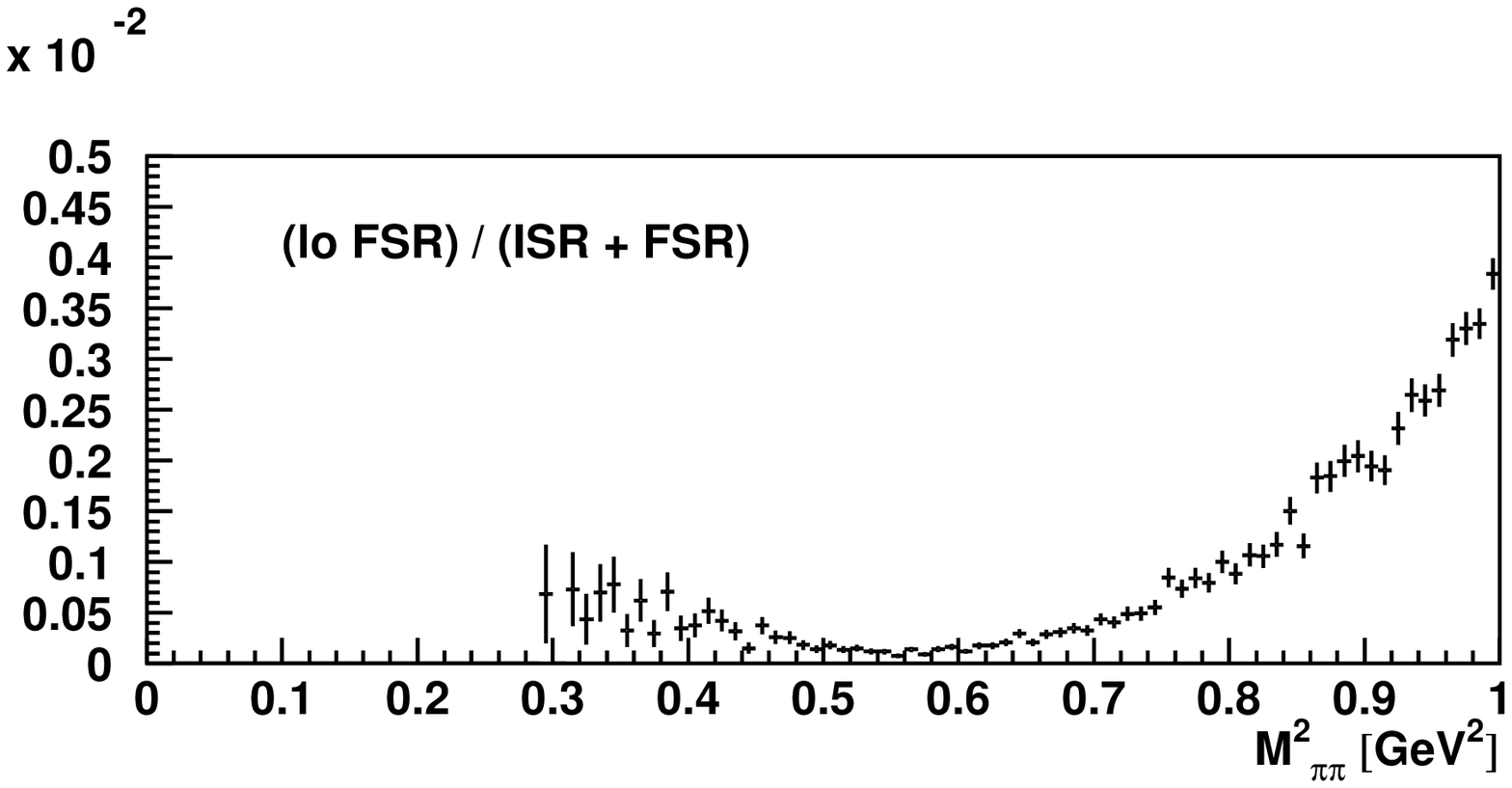}}
\hspace*{0.cm}
\subfigure[]{\includegraphics[width=80mm]{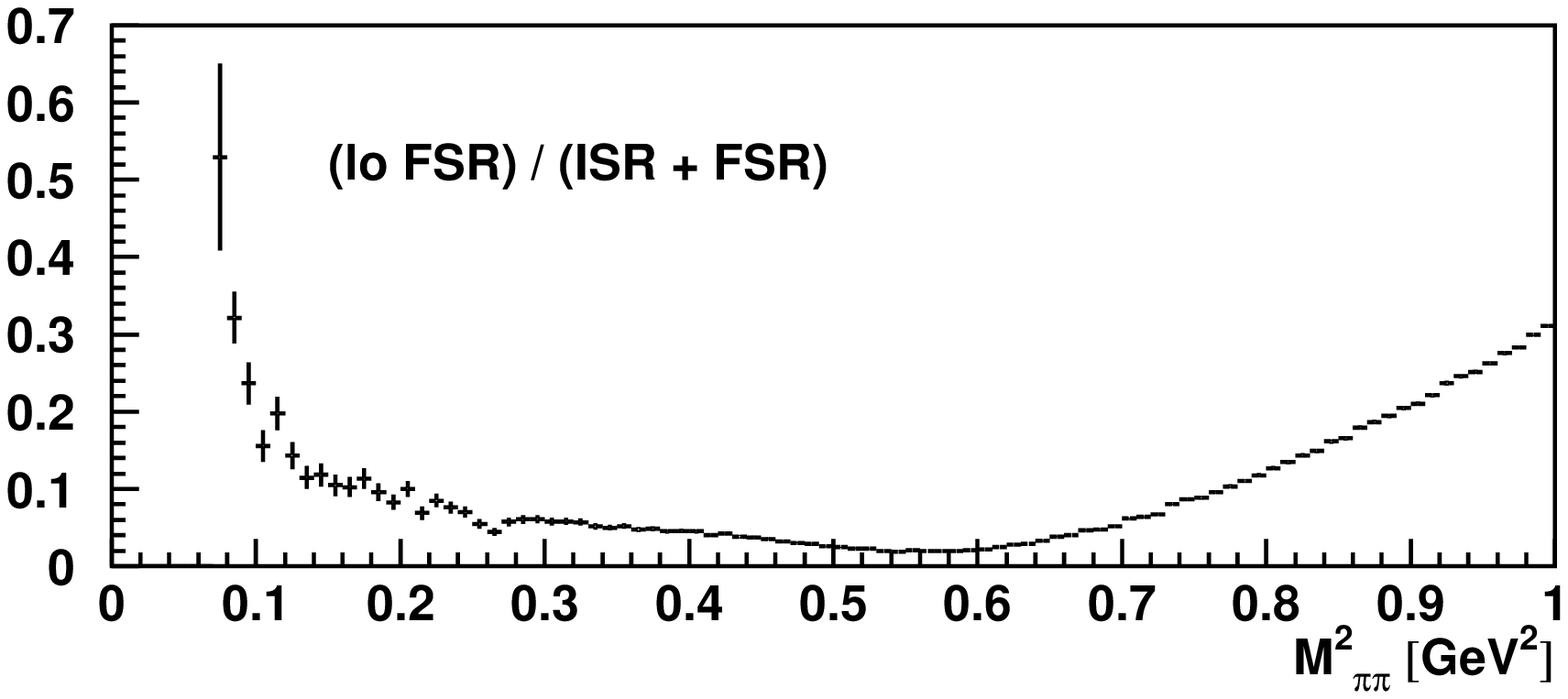}}
\caption{(a) Fraction of events with leading order final state radiation in the
  {\it small angle} selection: $50^\circ<\theta_\pi<130^\circ$ and
  $\theta_\gamma<15^\circ$ ($\theta_\gamma>165^\circ$). (b) Fraction of events with leading
  order final state radiation in the
  {\it large angle} selection: $50^\circ<\theta_\pi<130^\circ$ and
  $50^\circ<\theta_\gamma<130^\circ$. The PHOKHARA generator was used to produce the plots.}
\label{fig:radret_kloe5}
\end{center}
\end{figure}

 \begin{figure}
\begin{center}
\includegraphics[width=90mm]{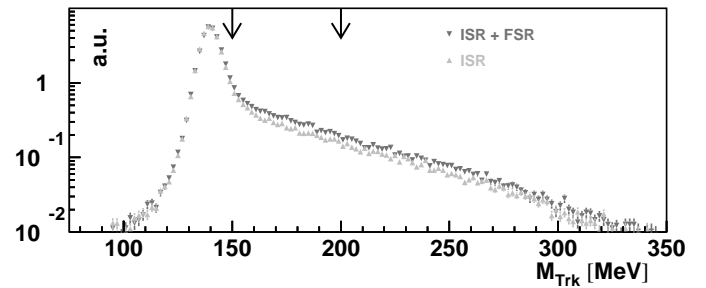}
\caption{Modification of the distribution of the trackmass variable due
to the presence of final state radiation (dark grey triangles) compared to 
the one
with initial state radiation only (light grey triangles). The arrows indicate
the region in which the $M_{\pi\pi}^2$-dependent cut is applied in the
analysis. The plot was created with the PHOKHARA generator interfaced
to the KLOE detector simulation~\cite{Ambrosino:2004qx}.}
\label{fig:radret_kloe45}
\end{center}
\end{figure}

\item[$\bullet$] The division by the radiator function $H(s,M^2_{\pi\pi})$. In 
  this case,
  one assumes perfect
  factorisation between the ISR and the FSR process. This has been
  tested by performing the analysis in an inclusive and  exclusive
  approach with respect to final state
  radiation. 
  The assumption was found to be valid within $0.2\%$~\cite{Aloisio:2004bu,KLOE_Note192}. 
\end{itemize}

It has been argued that contributions from events with two hard photons in the final 
state, which are not included in the PHOKHARA generator, may have an effect on the 
analyses~\cite{Jegerlehner:2008zza}.\\     

\begin{figure}[tb]
\centering
\subfigure[]{\includegraphics[width=80mm]{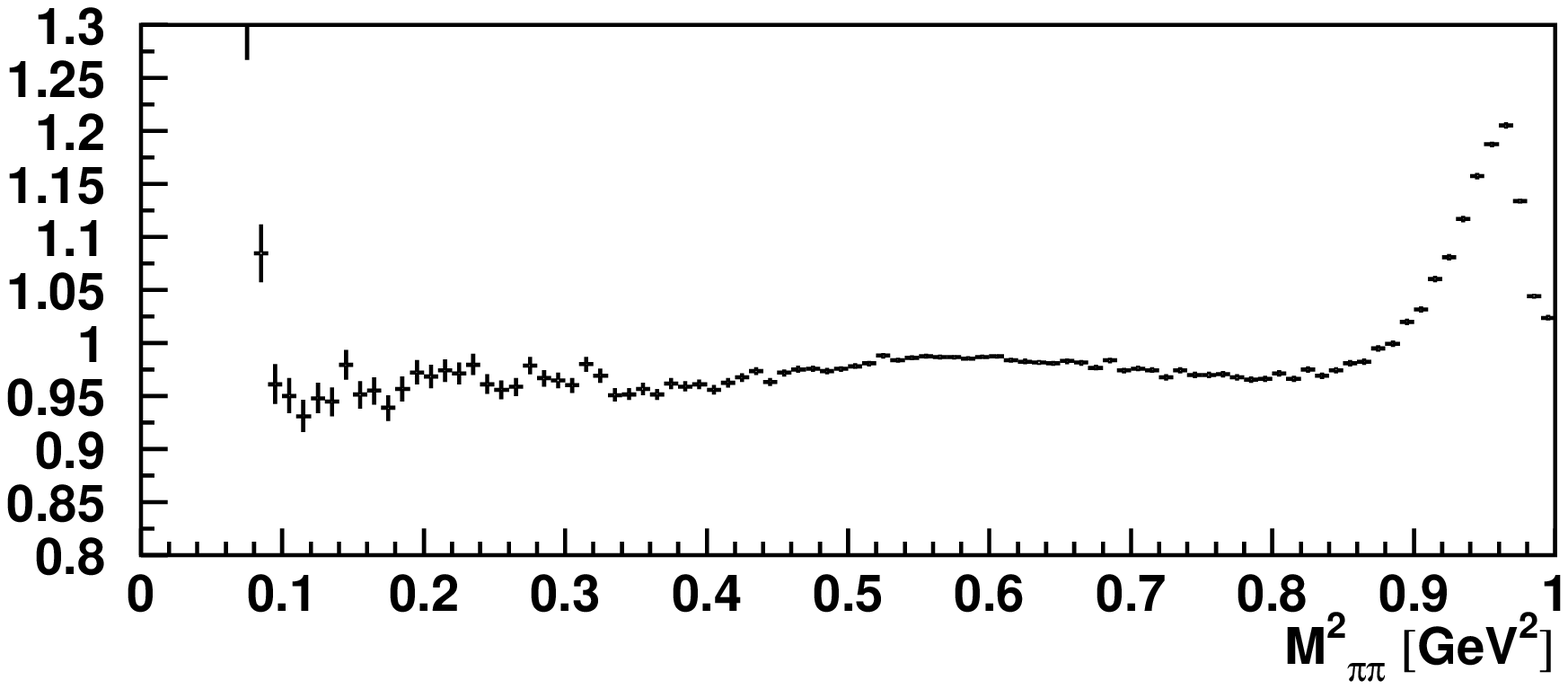}}
\subfigure[]{\includegraphics[width=80mm]{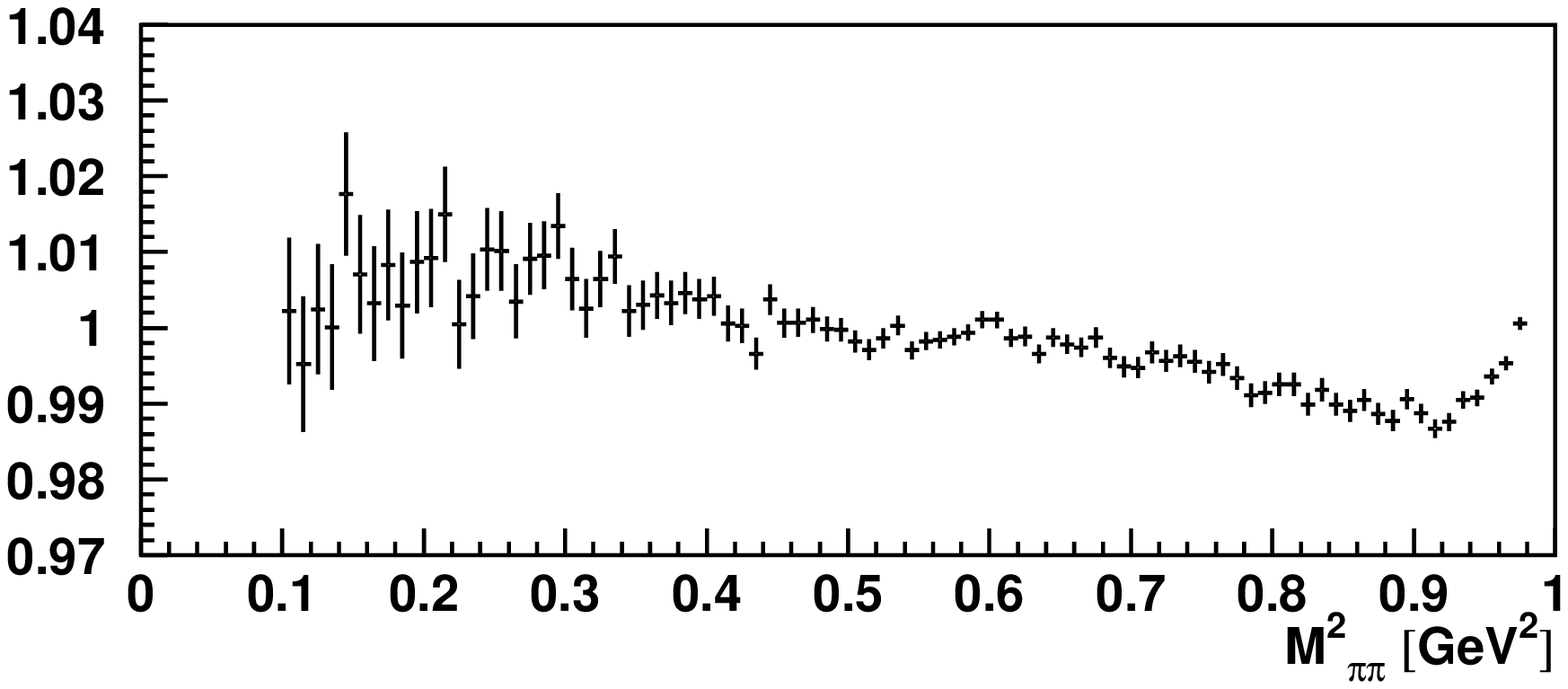}}
\caption{(a): ${{\rm d}\sigma_{\pi\pi\gamma}^{({\rm ISR}+{\rm FSR}+f_0+\varrho\pi)}}/{{\rm d}\sigma_{\pi\pi\gamma}^{({\rm ISR}+{\rm FSR})}}$ for $\sqrt{s}=1.019$ GeV. (b): 
${{\rm d}\sigma_{\pi\pi\gamma}^{({\rm ISR}+{\rm
FSR}+f_0+\varrho\pi})}/{{\rm d}\sigma_{\pi\pi\gamma}^{({\rm ISR}+{\rm FSR})}}$
for $\sqrt{s}=1$ GeV. Both plots were produced with the PHOKHARA v6.1 event
generator using {\it large
    angle} acceptance regions for pions and photons, with model
parameters for the $f_0$ and $\varrho\pi$ contributions from~\cite{Achasov:2005hm,Ambrosino:2006hb}.}
\label{fig:radret_kloe6}
\end{figure}

The effect of the direct decay $\phi\to \pi^+\pi^-\gamma$ on the
radiative return analysis has been addressed already 
in~\cite{Melnikov:2000gs}. Running at $\sqrt{s}\simeq1.02$ GeV, the
amplitude for the processes  $\phi\to (f_0(980)+f_0(600))\gamma \to
\pi^+\pi^-\gamma$ interferes with the amplitude for the final state
radiation process. Due to the yet unclear nature of the scalar states
$f_0(980)$ and $f_0(600)$, the effect on the $\pi^+\pi^-\gamma(\gamma)$
cross section depends on the model used to describe the scalar
mesons. The possibility to simulate $\phi$ decays together with the
processes for initial and final state radiation has been implemented
in the PHOKHARA event generator in~\cite{Czyz:2004nq}, using two
characteristic models for the $\phi$ decays: the ``no structure''
model of~\cite{Bramon:1992ew} and the $K^+K^-$ loop model
of~\cite{LucioMartinez:1994yu}. A refined version of the $K^+K^-$ loop
model~\cite{Achasov:2005hm} and the double vector resonance
$\phi\to\pi^\pm\varrho^\mp(\to \pi^\mp\gamma)$ have been included as
described in~\cite{Pancheri:2007xt}. Using parameter values for the
different $\phi$ decays found in the analysis of the neutral channel 
$\phi\to (f_0(980)+f_0(600))\gamma \to
\pi^0\pi^0\gamma$~\cite{Achasov:2005hm,Ambrosino:2006hb}, one can
estimate the effect on the different analyses. While in the {\it small
  angle} analysis there is no significant effect due to the choice of
the acceptance cuts, in the {\it large angle} selection the effect is
of the order of several percent and can reach up to 20\% in the
vicinity of the $f_0(980)$, see 
Fig.~\ref{fig:radret_kloe6} (a). While this allows to study the
different models for the direct decays of $\phi$-mesons (see also
Section~\ref{radret:KLOEasymFSR}), it prevents a precise measurement of
$\sigma_{\pi\pi}$ until the model and the parameters are understood
with better accuracy. An obvious way out is to use
data taken at a value of $\sqrt{s}$ outside the narrow peak of the
$\phi$ resonance ($\Gamma_\phi = 4.26\pm0.04$
MeV~\cite{Amsler:2008zzb}). 
In 2006, the KLOE experiment has taken $\sim$ 250
pb$^{-1}$ of data at $\sqrt{s}=1$ GeV, 20 MeV below $M_\phi$. As can
be seen in Fig.~\ref{fig:radret_kloe6} (b), this reduces the effect
due to contributions from $f_0\gamma$ and $\varrho\pi$ decays of the
$\phi$-meson to be within $\pm 1\%$.\\ 

{\noindent \it Normalisation with muon events\\}

An alternative method to extract the pion form factor is to normalise
the differential cross section
${{\rm d}\sigma_{\pi\pi\gamma(\gamma)}}/{{\rm d}M^2_{\pi\pi}}$ directly to
the process $e^+e^- \to \mu^+\mu^-\gamma(\gamma)$, ${{\rm
d}\sigma_{\mu\mu\gamma(\gamma)}}/{{\rm d}M^2_{\mu\mu}}$, in
each bin of $\Delta M^2_{\pi\pi}= \Delta M^2_{\mu\mu}$. Radiative
corrections like the effect of vacuum polarisation, the radiator
function and also the integrated luminosity $\int L{\rm d}t$ cancel out in
the ratio of pions over muons, and only the effects from final state
radiation (which is different for pions and muons) need to be taken
into account consistently. An approach currently under way at KLOE
uses the following equation to obtain $|F_{2\pi}|^2$: 
\begin{equation}
|F_{2\pi}(s')|^2\cdot(1+\eta(s'))=\frac{4(1+2m_\mu^2/s')\beta_\mu}{\beta^3_\pi}
\cdot\frac{(\frac{{\rm d}\sigma_{\pi\pi\gamma(\gamma)}}
{{\rm d}M^2_{\pi\pi}})^{{\rm ISR}+{\rm FSR}}}{(\frac{{\rm d}\sigma_{\mu\mu\gamma(\gamma)}}{{\rm d}M^2_{\mu\mu}})^{{\rm ISR}}}
\label{}
\end{equation}
In this formula, the measured differential cross section
${{\rm d}\sigma_{\pi\pi\gamma(\gamma)}}/{{\rm d}M^2_{\pi\pi}}$ should be
inclusive with respect to pionic final state radiation, while the
measured  cross section
${{\rm d}\sigma_{\mu\mu\gamma(\gamma)}}/{{\rm d}M^2_{\mu\mu}}$ should be
exclusive for muonic final state radiation. 
$s'=M^2_{\pi\pi}=M^2_{\mu\mu}$ is the squared invariant mass of the
di-pion or the di-muon system after the respective corrections 
for final state radiation. Using this approach, one gets on the
left-hand side the pion form factor times the factor $(1+\eta(s'))$,
which describes the effect of the pionic final state radiation. This
{\it bare} form factor is the quantity needed in the dispersion
integral for the $\pi\pi$-contribution to $a_\mu^{\rm had}$. While
the measurement of ${{\rm d}\sigma_{\pi\pi\gamma(\gamma)}}/{{\rm d}M^2_{\pi\pi}}$
and its corrections for pionic final state radiation 
are very similar to the one using the normalisation with
Bhabha events already performed at KLOE, the corrections needed to
subtract the muonic final state radiation from the 
${{\rm d}\sigma_{\mu\mu\gamma(\gamma)}}/{{\rm d}M^2_{\mu\mu}}$ cross section are
pure QED and can be obtained from the PHOKHARA generator, which
includes final state radiation for muon pair production at
next-to-leading order~\cite{Czyz:2004rj}. Due to the fact that 
the KLOE detector does
not provide particle IDs, pions and muons have to be separated and
identified using kinematical variables (e.g. the aforementioned
trackmass variable)~\cite{Muller:2006bk}. The analysis is in progress and a systematic precision similar to the one obtained in the absolute measurement 
is expected.

\subsubsection{Radiative return at BaBar}
\label{rr:babar}

The BaBar radiative return program aims at the study of all significant hadronic processes in 
electron-positron annihilation, $e^+e^-\to hadrons$, for energies  
from threshold up to about 4.5 GeV. Moreover, hadron spectroscopy of the initial 
J$^{PC} = 1^{--}$ states, which are produced in $e^+e^-$ collisions, and of their 
decay products is performed. In this chapter BaBar results for
processes with 3, 4, 5 and 6 hadrons in the final state, as well as measurements of 
baryon form factors in the time-like region are reported. 
A precision analysis of the pion form factor,
i.e. of the cross section $e^+e^-\to \pi^+\pi^-$, which is essential for an improved
determination of the hadronic contribution to the anomalous magnetic moment of the muon, 
appeared most recently~\cite{Aubert:2009fg}. 
The results presented in this chapter are based on a total integrated luminosity
of 230 fb$^{-1}$, except for the $3\pi$ and 4 hadron channels of Ref.~\cite{Aubert:2005eg}, which 
were analysed using a data sample of 90 fb$^{-1}$. The total BaBar data sample 
collected between the years 1999 to 2008 amounts to 530 fb$^{-1}$.
A typical feature common to all radiative return analyses at BaBar is
a wide coverage of the entire mass range of interest
in one single experiment, with reduced point-by-point uncertainties compared
to previous experiments. 
\\
\\
{\bf \it $e^+e^-\to 3$ pions}
\\ 
The $\pi^+\pi^-\pi^0$ mass spectrum has been measured from $1.05$ GeV up to the $J/\psi$ mass region 
with a systematic error of $\sim 5\%$ below $2.5$ GeV, and up to $\sim 20\%$ at 
higher masses~\cite{Aubert:2004kj}. 
The spectrum is dominated by the $\omega$, $\phi$ and $J/\psi$ resonances. 
The BaBar measurement was able to significantly improve the world knowledge on the 
excited $\omega$ states. The spectrum has been fitted 
up to $1.8$ GeV and the following results for the masses and widths of the
$\omega^\prime$ and $\omega^{\prime\prime}$ states have been found:
$M(\omega^\prime)=(1350\pm20\pm20)$ MeV, $\Gamma(\omega^\prime)=(450\pm70\pm70)$ MeV,
$M(\omega^{\prime\prime})=(1660\pm10\pm2)$ MeV, 
$\Gamma(\omega^{\prime\prime})=(230\pm30\pm20)$ MeV.
Note that below $1.4$ GeV the results from BaBar are in good agreement
with those from SND~\cite{Achasov:2002ud}, while above this
energy the cross sections measured by BaBar are much higher than those
from DM2~\cite{Antonelli:1992jx}.
\begin{figure}[t]
\begin{center}
\epsfig{file=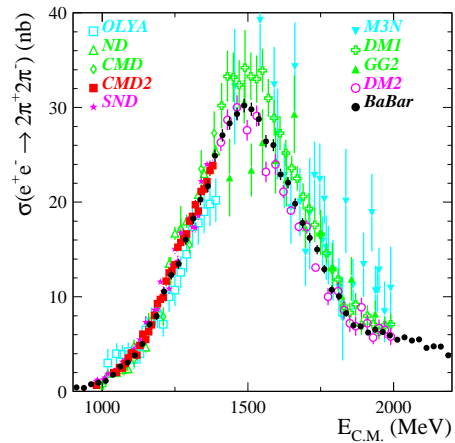, width=6.cm}
\caption{\small{BaBar measurement of the energy dependence of the 
$e^+e^-\to \pi^+\pi^-\pi^+\pi^-$ 
cross section obtained  
by radiative return in comparison with the world data set.}}
\label{fig.4pi}
\end{center}
\end{figure}
\\
\\
{\bf \it $e^+e^-\to 4$ hadrons}
\\
The $\pi^+\pi^-\pi^+\pi^-$, $K^+K^-\pi^+\pi^-$ and $K^+K^-K^+K^-$ exclusive final states 
have been measured from threshold up to $4.5$ GeV with systematic errors of $5\%$, $15\%$ and
$25\%$, respectively~\cite{Aubert:2005eg}. 
The $K^+K^-K^+K^-$ measurement is the first measurement of this process at all.
Figure~\ref{fig.4pi} shows the mass distribution of the $\pi^+\pi^-\pi^+\pi^-$ channel. 
We identify an impressive improvement with respect to previous experiments.
Background is relatively low for all channels under study 
(e.g. a few percent at $1.5$ GeV for $\pi^+\pi^-\pi^+\pi^-$) 
and is dominated by ISR-events of higher multiplicities and by continuum non-ISR 
events at higher masses. 
The $\pi^+\pi^-\pi^+\pi^-$ final state is dominated by the two-body intermediate state $a_1(1260)\pi$;
the $K^+K^-\pi^+\pi^-$ final state shows no significant two-body states, but a rich three-body
structure, including $K^*(890)K\pi$, $\phi\pi\pi$, $\rho K K$ and $K_2^*(1430)K\pi$. 
\\
Figure~\ref{fig2n} shows BaBar preliminary results for the process
$e^+e^-\to\pi^+\pi^-\pi^0\pi^0$. The current systematic error of the measurement varies from 8\% around
the peak of the cross section to 14\% at 4.5 GeV.
BaBar results are in agreement with 
SND~\cite{AchasovSND:2001zz} in the energy range below 1.4 GeV and show a significant  improvement for higher 
energies ($>$ 1.4 GeV). In the energy range above 2.5 GeV this is the first measurement at all.
\begin{figure}
\begin{center}
\epsfig{file=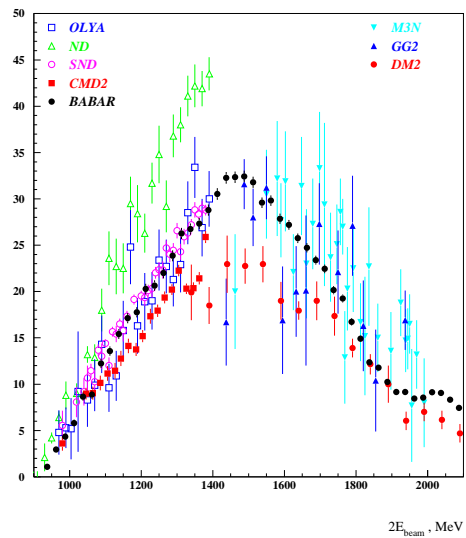, width=6.cm}
\caption{\label{fig2n} Preliminary BaBar data for the $e^+e^-\to\pi^+\pi^-\pi^0\pi^0$ cross section 
in comparison with previous experiments.}
\end{center}
\end{figure}
The $e^+e^-\to\pi^+\pi^-\pi^0\pi^0$ final state is dominated by the
$\omega\pi^0$, $a_1(1260)\pi$ and $\rho^+\rho^-$ intermediate channels, where the 
latter channel has been observed for the first time.
\\
A specific analysis was devoted to the intermediate structures in the 
$e^+e^-\to K^+K^-\pi^+\pi^-$ and $e^+e^-\to K^+K^-\pi^0\pi^0$ channels~\cite{Aubert:2007ur}.
Of special interest is 
the intermediate state $\phi f_0(980)$, where the decays $f_0(980)\to\pi^+\pi^-$ and 
$f_0(980)\to\pi^0\pi^0$ have been looked at. 
A peak is observed in the $\phi f_0(980)$ channel at a mass $M = 2175\pm 18$ MeV and a width $\Gamma = 58\pm 2$ MeV.
The new state is usually denoted as Y(2175) and is also clearly visible in the $K^+K^-f_0$ spectrum. 
\\
\\
{\it \bf $e^+e^-\to 2(\pi^+\pi^-)\pi^0$, $2(\pi^+\pi^-)\eta$}
\\
The $e^+e^- \to 2(\pi^+\pi^-)\pi^0$ cross section has been measured by BaBar
from threshold up to 4.5 GeV~\cite{Aubert:2007ef}. A large coupling of the $J/\psi$ and $\psi(2S)$ to this channel is observed. 
The systematic error of the measurement is about 7\% around the peak of the mass spectrum. 
In the $\pi^+\pi^-\pi^0$ mass distribution the $\omega$ and 
$\eta$ peaks are observed; the rest of the events have a $3\pi\rho$ structure.
\\
BaBar performed also the first measurement of the $e^+e^-\to 2(\pi^+\pi^-)\eta$ cross section. 
A peak value of about 1.2 nb at about 2.2 GeV is observed, followed by a monotonic decrease towards 
higher energies. Three intermediate states are seen: $\eta\rho(1450)$, $\eta^{\prime}\rho(770)$ and
$f_1(1285)\rho(770)$.
\begin{figure}[t]
\begin{center}
\epsfig{file=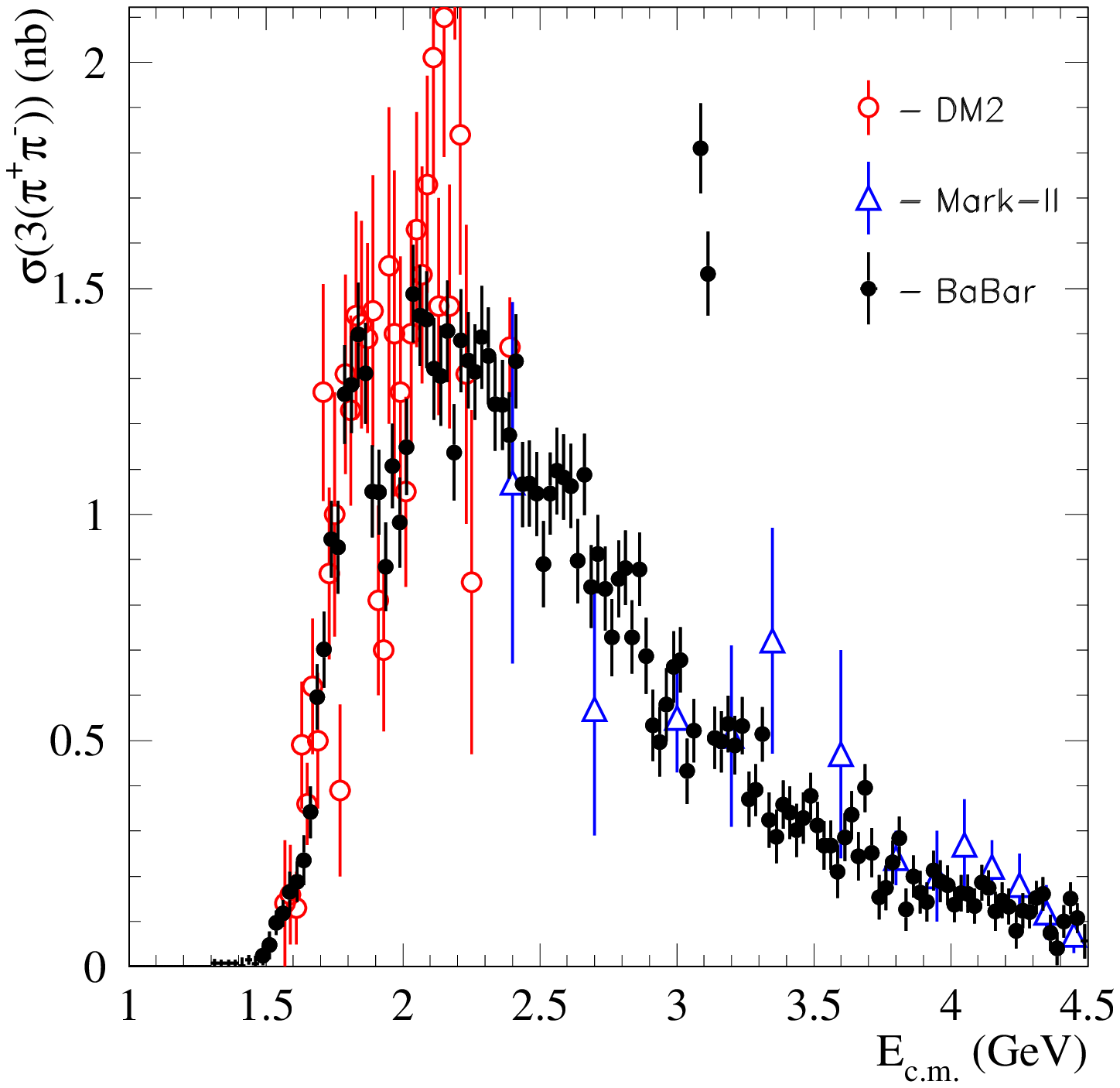, width=6.cm}
\epsfig{file=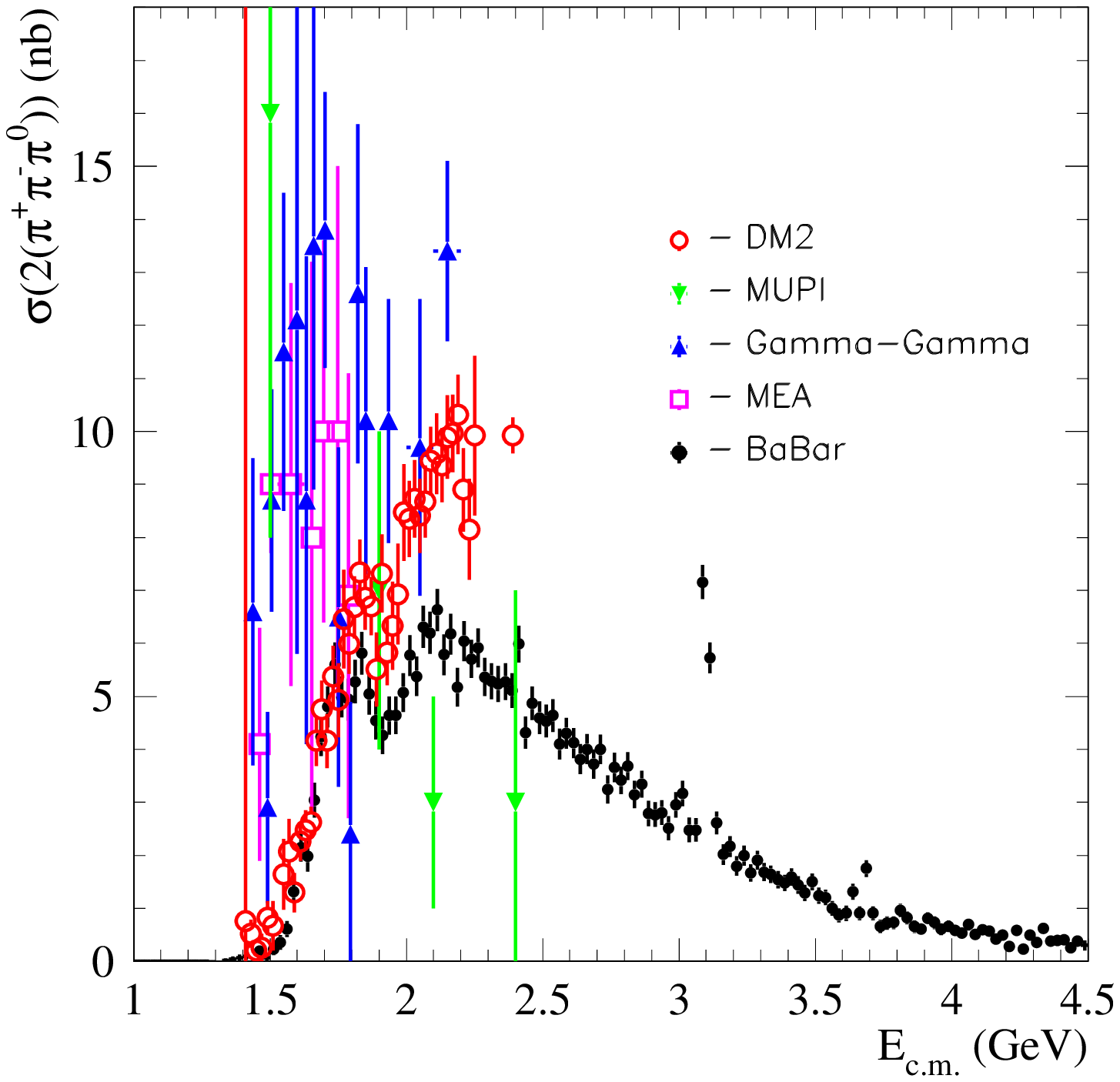, width=6.cm}
\caption{\small{The energy dependence of the cross sections for
$e^+e^-\to 3(\pi^+\pi^-)$ (upper plot) and $e^+e^-\to
2(\pi^+\pi^-)2\pi^0$ (lower plot), obtained by BaBar (filled circles)
by radiative return, in comparison with previous data.}}
\label{fig.6pi}
\end{center}
\end{figure}   
\\
\\
{\bf \it $e^+e^-\to 6$ hadrons} 
\\
The $6$ hadron final state has been measured in the exclusive channels 
$3(\pi^+\pi^-)$, $2(\pi^+\pi^-)2\pi^0$ and $K^+K^-2(\pi^+\pi^-)$ \cite{Aubert:2006jq}. 
The cross section in the last case has never been measured before; the precision
in the first two cases is $\sim 20\%$, which is a large improvement with respect to existing data.  
Again, the entire energy range from threshold up to $4.5$ GeV is measured
in a single experiment. 
The distributions for the final states $3(\pi^+\pi^-)$ and $2(\pi^+\pi^-)2\pi^0$
are shown in Fig.~\ref{fig.6pi}.
A clear dip is visible at about $1.9$ GeV in both pion
modes. A similar feature was already seen 
by FOCUS~\cite{Frabetti:2001ah}  in the diffractive photo-production of six charged pions.
The spectra are fitted by BaBar using the sum of a Breit-Wigner resonance function 
and a Jacob-Slansky continuum shape. For the
$3(\pi^+\pi^-)$ ($2(\pi^+\pi^-)2\pi^0$) mode, BaBar obtains values of $1880 \pm 30$ MeV
($1860 \pm 20$ MeV) for the resonance peak, $130 \pm 30$ MeV
($160 \pm 20$ MeV) for the resonance width  and $21^{\rm o} \pm 14^{\rm o}$
($-3^{\rm o} \pm 15^{\rm o}$) for the phase shift between the resonance and continuum.
\\
\\
{\it \bf $e^+e^-\to K^+K^-\pi^0, K^+K^-\eta$, $K_S K^{\pm}\pi^{\mp}$}
\\
A recent BaBar ISR-analysis is dedicated to three hadrons in the final state, including a pair of 
kaons ($K^+K^-\pi^0$, $K K_S \pi$); a peak near 1.7 GeV, which is mainly due to 
the $\phi^{\prime}$(1680) state, is observed. 
A Dalitz plot analysis shows that the $KK^{*}$(892) and $KK^{*}_2$(1430)
intermediate states are dominating the $K\bar{K}\pi$ channel. 
A fit to the $e^+e^-\to K\bar{K}\pi$ cross section assuming
the expected contributions from the 
$\phi,\phi^{\prime},\phi^{\prime\prime},\rho^{0},\rho^{\prime},\rho^{\prime\prime}$
states was performed. The parameters of the $\phi^{\prime}$ and other excited 
vector meson states are compatible with PDG values.
\\
\\
{\bf \it Time-like proton form factor $e^+e^-\to p \bar{p}$, 
hyperon form factors $e^+e^-\to\Lambda^0\bar{\Lambda^0},\Lambda^0\bar{\Sigma^0},\Sigma^0\bar{\Sigma^0}$} 
\\
\begin{figure}[t]
\begin{center}
\epsfig{file=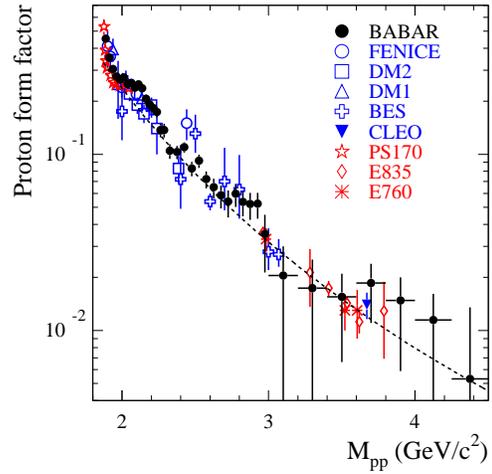, width=6.8cm}
\caption{\small{The $e^+e^-\to p\bar{p}$ cross section measured by BaBar (filled circles) 
in comparison with data from other $e^+e^-$ colliders (blue points) and from $\bar{p}p$
experiments (red points).}}
\label{fig.ppbar}
\end{center}
\end{figure} 
BaBar has also performed a measurement of the $e^+e^-\to p\bar{p}$ cross section~\cite{Aubert:2005cb}.
This time-like process is parametri\-sed by the electric 
and magnetic form factors, $G_E$ and $G_M$:
\begin{eqnarray*}
\sigma_{e^+e^- \to p \bar{p}} & = & \frac {4 \pi \alpha^2 \beta C}{3s} \\
                                 & \times & (|G_M|^2 + \frac{2m_p^2}{s} |G_E|^2),
\end{eqnarray*}
where $\beta=\sqrt{1-4m_p^2/s}$ and the 
factor $C=y/(1-e^{-y})$ (with $y=\pi \alpha m_p/(\beta \sqrt{s})$) accounts for the Coulomb interaction of the final state particles.
The proton helicity angle $\theta_p$ in the $p\bar{p}$ rest frame can be used to
separate the $|G_E|^2$ and $|G_M|^2$ terms. Their respective variations are approximately 
$\sim \sin^2\theta_p$ and $\sim (1+\cos^2\theta_p)$.
By fitting the $\cos\theta_p$ distribution to a sum of the two terms,
the ratio $|G_E|/|G_M|$ can be extracted. This is done separately in six bins 
of $M_{p\bar{p}}$. The results disagree significantly
with previous measurements from LEAR~\cite{Bardin:1994am} above threshold. BaBar observes a ratio
$|G_E|/|G_M|>1$ above threshold, while at larger values of
$M_{p\bar{p}}$ the BaBar measurement finds $|G_E|/|G_M| \approx 1$. LEAR data, on the
contrary, show a behaviour $|G_E|/|G_M|<1$ above threshold.
\\
In order to compare the cross section measurement with previous data ($e^+e^-$ and $\bar{p}p$ 
experiments), the {\it effective} form factor $G$ is introduced: 
$G = \sqrt{|G_E|^2 + 2m_p^2/s |G_M|^2}$. The BaBar measurement of 
$G$ is in good agreement with existing results, as can be seen in Fig.~\ref{fig.ppbar}. The structure of
the form factor is rather complicated; the following observations can be made:
(i) BaBar confirms an increase of $G$ towards threshold as seen before by other experiments; 
(ii) two sharp drops of the spectrum at $M_{p\bar{p}}=2.25$ and $3.0$ GeV are observed; 
(iii) data at large values
$M_{p\bar{p}}>3$ GeV are in good agreement with the prediction from perturbative QCD. 
\\
A continuation of the ISR program with baryon final states is the 
measurement of the $e^+e^-\to\Lambda\bar{\Lambda}$ cross section~\cite{Aubert:2007uf}. 
So far only one data point from DM2~\cite{Bisello:1990rf} was existing for this channel, which is in good
agreement with BaBar data. About 360 $\Lambda\bar{\Lambda}$ events could be selected 
using the $\Lambda\to p\pi$ decay.
In two invariant mass bins an attempt has been made to extract the
ratio of the electric to magnetic form factor $|G_E|/|G_M|$. 
In the mass range below $2.4$ GeV this ratio is above unity --
as in the proton case -- with
a significance of one standard deviation ($|G_E|/|G_M|=1.73^{+0.99}_{-0.57}$). 
Above $2.4$ GeV the ratio is consistent with unity ($|G_E|/|G_M|=0.71^{+0.66}_{-0.71}$).
Also the $\Lambda$ polarisation and the phase between $G_E$ and $G_M$ was studied 
using the slope of the angle between the polarisation axis
and the proton momentum in the $\Lambda$ rest frame.
The following limit on $\Lambda$ polarisation is obtained: $-0.22 < \zeta < 0.28$; the relative phase
between the two form factors is measured as $-0.76 < \sin(\phi) < 0.98$, which is
not yet significant due to limited statistics.
\\
Finally, the first measurements of the 
$e^+e^-\to\Sigma^0\bar{\Sigma^0}$ and $e^+e^-\to\Sigma^0\bar{\Lambda}(\Lambda\bar{\Sigma^0})$ cross 
sections were performed. For the detection of the $\Sigma^0$ baryon, the decay 
$\Sigma^0\to\Lambda\gamma\to p\pi\gamma$ was used.
About 40 candidate events were selected for the reaction
$\Sigma^0\bar{\Sigma^0}$ and about 
20 events for $\Lambda\bar{\Sigma^0}$. 
All baryon form factors measured by BaBar have a similar size and mass shape, namely
a rise towards threshold. The reason for this peculiar behaviour is not
understood.

\subsubsection{Radiative return at Belle}
\label{rr:belle}

{\noindent \it ISR studies at Belle\\}

Until now most of the Belle analyses using radiative return 
focused on studies of the charmonium and charmo\-ni\-um-like 
states. They can be subdivided into final states with open and
hidden charm.\\   

{\noindent \it Final states with open charm\\}

Belle performed a systematic study of various exclusive channels of
$e^+e^-$ annihilation into charmed mesons and baryons using ISR, 
often based on the so called partial reconstruction to increase 
the detection efficiency.

In Ref.~\cite{Abe:2006fj} they measured the cross sections of the processes
$e^+e^- \to D^{*\pm}D^{*\mp}$  and $e^+e^- \to D^+D^{*-} + c.c.\,$.
The shape of the former is complicated and has several local maxima
and minima. The first two maxima are close to the $\psi(4040)$ and
$\psi(4160)$ states. The latter shows significant excess of events 
near the $\psi(4040)$.

The cross sections of the processes  $e^+e^- \to D^+D^-$ and
 $e^+e^- \to D^0\bar{D}^0$ show a signal of the $\psi(3770)$, 
as well as hints of the $\psi(4040)$, $\psi(4160)$ and 
$\psi(4415)$~\cite{Pakhlova:2008zza}. There is
also an enhancement  near 3.9~GeV, which qualitatively agrees with 
the prediction of the coupled channel model \cite{Eichten:1979ms}. 

The cross section of the process $e^+e^- \to D^0D^-\pi^+$ has a
prominent peak at the energy corresponding to the
$\psi(4415)$~\cite{Pakhlova:2007fq}. From a study of the resonant 
substructure in
the decay $\psi(4415) \to D^0D^-\pi^+$ they conclude that it is
dominated by the intermediate $D\bar{D}^*_2(2460)$ mechanism.

In contrast to expectations of some hybrid models predicting
$Y(4260) \to D^{(*)}{\bar D^{(*)}}\pi$ decays, no clear structures were 
observed in the cross section of the process 
$e^+e^- \to D^0D^{*-}\pi^+$~\cite{Pakhlova:2009jv}. There is only 
some evidence ($\sim 3.1\,\sigma$) for the $\psi(4415)$. 

\begin{center}
\begin{figure}
\begin{center}
\includegraphics[height=0.43\textheight]{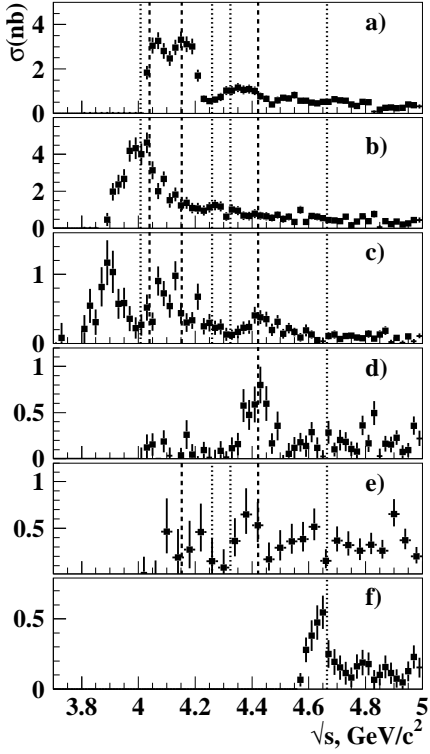}
\caption{Cross sections of various exclusive processes measured by
Belle: a)  $e^+e^- \to D^{*\pm}D^{*\mp}$, b) $e^+e^- \to D^+D^{*-} + c.c.$,
c) $e^+e^- \to D\bar{D}$,
d) $e^+e^- \to D^0D^-\pi^+ + c.c.$, e)  $e^+e^- \to D^0D^{*-}\pi^+ + c.c.$, and
f) $e^+e^- \to \Lambda^+_c\Lambda^-_c$. The dashed lines show the
position of the $\psi$ states, while the dotted lines correspond
to the $Y(4008),~Y(4260),~Y(4360)$, and $Y(4660)$ states.}
\label{fig:allcc}
\end{center}
\end{figure}
\end{center}

Finally, they measure the cross section of the reaction
$e^+e^- \to \Lambda^+_c\Lambda^-_c$ and observe a significant peak near 
threshold that they dub $X(4630)$~\cite{Pakhlova:2008vn}. 
Assuming that the peak 
is a resonance, they find that its mass and width are compatible
within errors with those of the $Y(4660)$ 
state found by Belle in the $\psi(2S)\pi^+\pi^-$ final state 
via ISR~\cite{:2007ea}. However, interpretations other than 
$X(4630) \equiv Y(4660)$ cannot be excluded. For example, peaks 
at the baryon-antibaryon threshold are observed in various 
processes.
According to other assumptions, 
the $X(4630)$ is
a $\psi(5S)$~\cite{Badalian:2008dv} or $\psi(6S)$~\cite{Li:2009zu} 
charmonium state, 
or, for example, a threshold effect which is due to the $\psi(3D)$, 
slightly below the $\Lambda^+_c\Lambda^-_c$ 
threshold~\cite{vanBeveren:2008rt}. 
Figure~\ref{fig:allcc} shows all cross sections mentioned above, with
the vertical lines showing positions of both well established states
like 
$\psi(4040)$, $\psi(4160)$
and $\psi(4415)$, and new charmonium-like 
states $Y(4008)$, $Y(4260)$, $Y(4360)$ and $Y(4660)$ discussed below.

Summing the measured cross sections and taking into account not yet
observed final states on base of isospin symmetry they find that 
the sum of exclusive cross sections almost saturates the 
total inclusive cross section measured by BES~\cite{Bai:2001ct}.\\

{\noindent \it Final states with hidden charm\\}

\begin{figure}
\begin{center}
\includegraphics[width=0.25\textwidth,angle=-90]{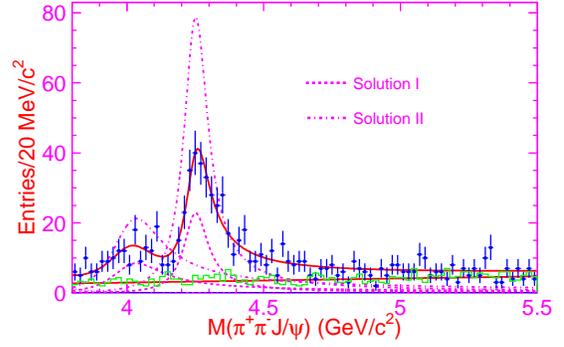}
\caption{The $J/\psi\pi^+\pi^-$ invariant mass distribution.}
\label{fig:bel42}
\end{center}
\end{figure}

Studying the $J/\psi\pi^+\pi^-$ final state, Belle confirmed the
$Y(4260)$ discovered by BaBar and in addition observed a new structure
dubbed $Y(4008)$~\cite{:2007sj}, see Fig.~\ref{fig:bel42}.
They also observe the reaction $e^+e^- \to J/\psi K^+K^-$
and find first evidence for the reaction  
$e^+e^- \to J/\psi K^0_SK^0_S$~\cite{:2007bt}.

\begin{figure}
\begin{center}
\includegraphics[width=0.25\textwidth,angle=-90]{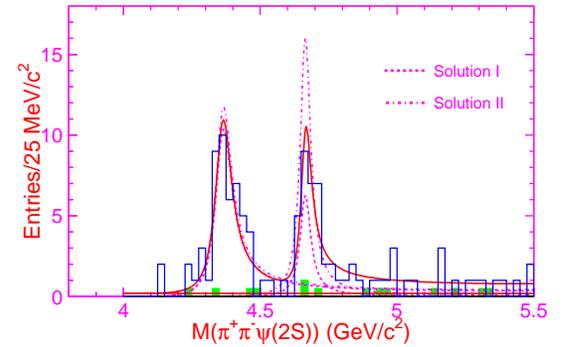}
\caption{The $\psi(2S)\pi^+\pi^-$ invariant mass distribution.} 
\label{fig:bel43}
\end{center}
\end{figure}

Studying the $\psi(2S)\pi^+\pi^-$ final state, Belle confirmed the
$Y(4360)$ discovered by BaBar and in addition observed a new structure
dubbed $Y(4660)$~\cite{:2007ea}, see Fig.~\ref{fig:bel43}.

It is worth noting that the resonance interpretation of various
enhancements discussed above is not unambiguous and can be strongly affected
by close thresholds of different final states and rescattering
effects.
 
Various ISR studies performed at the Belle detector in the charmonium
region are summarised in Table~\ref{tab:belisr}.\\

\begin{table}
\begin{center}
\caption{Summary of ISR studies in the $c\bar{c}$ region at Belle.}
\label{tab:belisr}
\begin{tabular}{lll}
\hline
Final state & $\int{L~{\rm d}t}$, fb$^{-1}$ & Ref. \\
\hline
$D^{*+}D^{*-}$ & 547.8 &  \cite{Abe:2006fj} \\
\hline
$D^{\pm}D^{*\mp}$ & 547.8 &  \cite{Abe:2006fj} \\
\hline
$D^0\bar{D}^0,~D^+D^-$ & 673 & \cite{Pakhlova:2008zza} \\
\hline
$D^0D^-\pi^+$ & 673 &  \cite{Pakhlova:2007fq} \\
\hline
$D^0D^{*-}\pi^+$ & 695 &  \cite{Pakhlova:2009jv} \\
\hline 
$\Lambda^+_c\Lambda^-_c$ & 695 &  \cite{Pakhlova:2008vn} \\
\hline
$J/\psi\pi^+\pi^-$ & 548 &  \cite{:2007sj} \\
\hline
$\psi(2S)\pi^+\pi^-$ & 673 &  \cite{:2007ea} \\
\hline
$J/\psi K^+K^-$ & 673 & \cite{:2007bt} \\
\hline
\end{tabular}
\end{center}
\end{table}

{\noindent \it ISR studies of  light quark states\\}

\begin{figure}
\begin{center}
\begin{tabular}{cc}
\begin{minipage}{4.2cm}
\includegraphics[width=1.\textwidth]{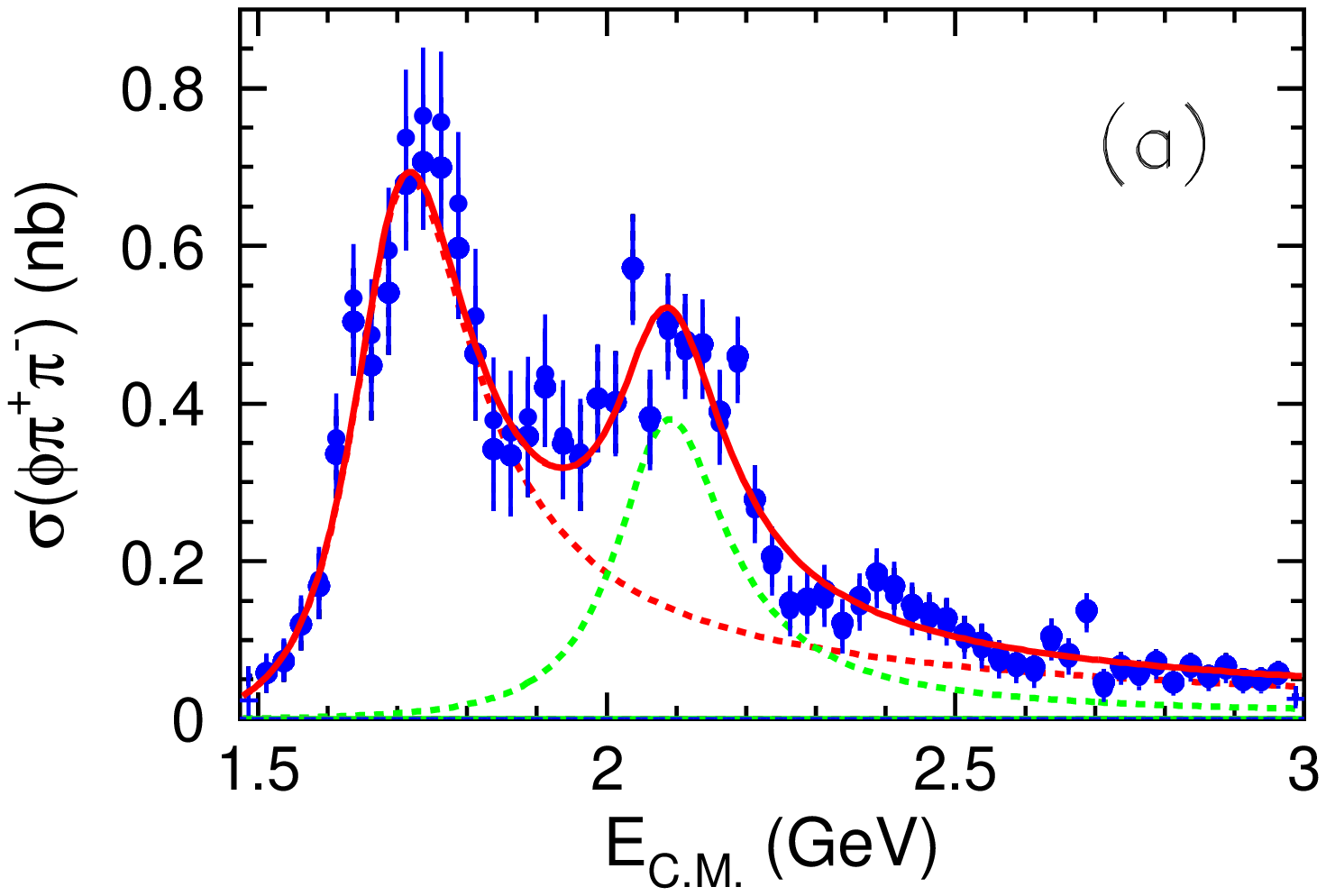}
\end{minipage} &
\begin{minipage}{4.2cm}
\includegraphics[width=1.\textwidth]{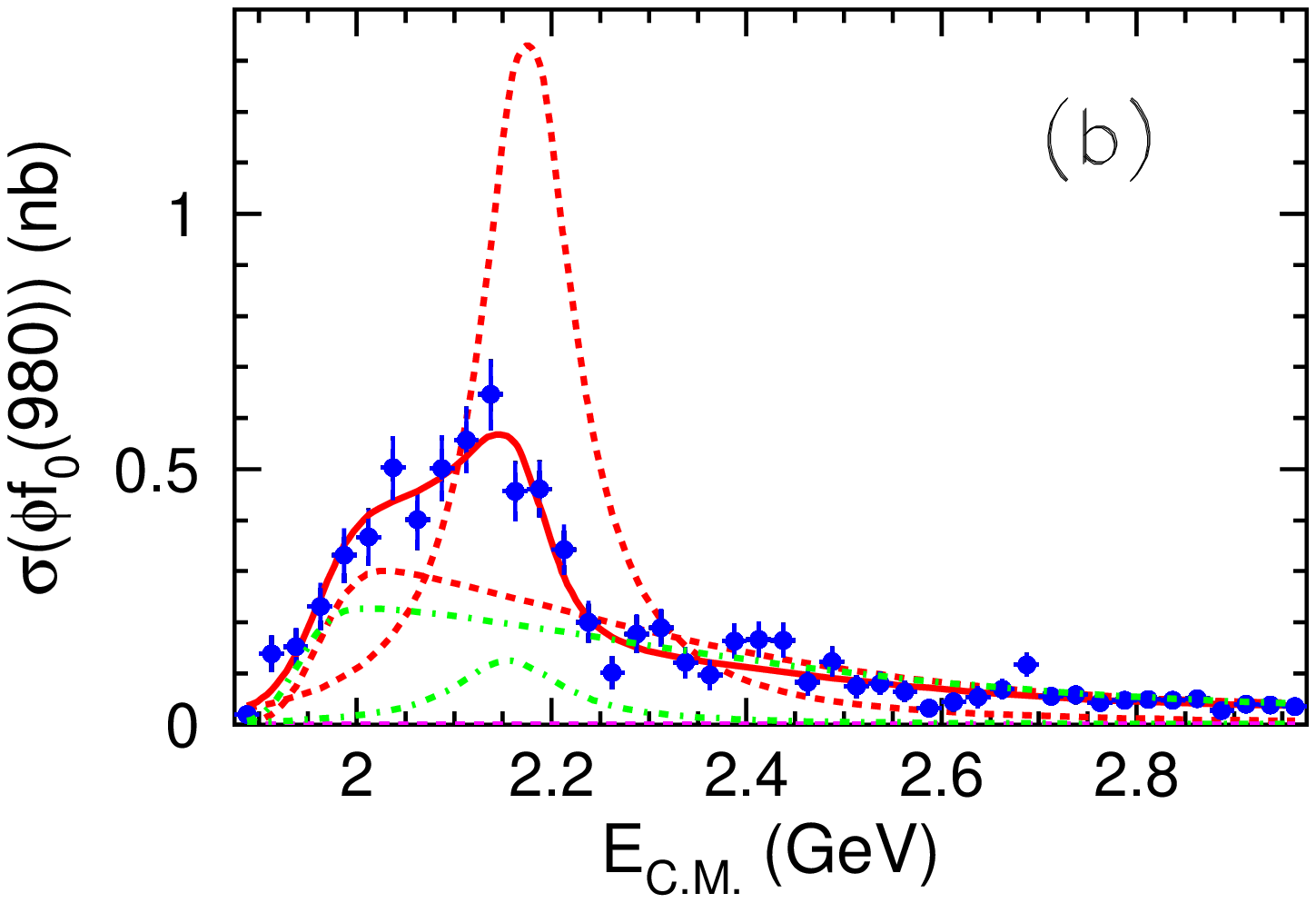}
\end{minipage}
\end{tabular}
\caption{Cross sections of the processes 
$e^+e^- \to  \phi\pi^+\pi^-$ (a) and $e^+e^- \to \phi f_0(980)$ (b).}
\label{fig:belphi}
\end{center}
\end{figure}

In one case the ISR method was used to study the light quark 
states~\cite{:2008ska}. In this analysis the cross sections of the
reactions $e^+e^- \to \phi\pi^+\pi^-$ and  $e^+e^- \to \phi f_0(980)$
are measured from threshold to 3~GeV, using a data sample of 673~fb$^{-1}$,
see Fig.~\ref{fig:belphi} (a, b).
In the $\phi\pi^+\pi^-$ mode the authors observe and measure for
the first time the parameters of the $\phi(1680)$; they also observe
and measure the parameters of the $\phi(2170)$. Also selected in this
analysis is the $\phi f_0(980)$ final state, which shows a clear
signal of the $\phi(2170)$. For Monte Carlo simulation they use a
version of PHOKHARA in which the produced resonance decays into  
$\phi\pi^+\pi^-$ or  $\phi f_0(980)$ with the subsequent decays
$\phi \to K^+K^-$ and $f_0(980) \to \pi^+\pi^-$. The $\pi^+\pi^-$ 
system is in the $S$-wave, the $\pi^+\pi^-$ system
and the $\phi$ are also in a relative 
$S$-wave.  The $\pi^+\pi^-$ mass distribution is generated according
to phase space. They assign 0.1\% as  the systematic uncertainty
of the ISR photon radiator.

In all the ISR studies the Monte Carlo simulation is performed as
follows. First, the kinematics of the initial state radiation is
generated using the PHOKHARA v5.0 package for simulation of the
process $e^+e^- \to V\gamma_{\rm ISR}(\gamma_{\rm ISR})$ \cite{Czyz:2005as}.
Then a $q\bar{q}$ generator is used to generate $V$ decays.

\subsubsection{Prospects for radiative return at VEPP-2000}
\label{rr:vepp2000}

As discussed above, the major hadronic leading-order contribution to 
$a_{\mu}^{\rm had}$ comes from the energy range below 1 GeV, where in turn
the $\pi^+\pi^-$ channel gives the dominant contribution. Direct scan at VEPP-2000 will
deliver huge statistics at the experiments CMD-3 and SND, but the accuracy of the cross
sections will be determined by systematic errors. Therefore, any other
possibility to measure the pion form factor, for example with ISR,
will be a valuable tool to provide a cross check for better understanding
the scale of systematic effects.

The design luminosity of $\sim 10^{32}$cm$^{-2}$c$^{-1}$ is expected at 
$\sqrt{s}$ = 2 GeV. 
The luminosity recalculated to the $\rho$-peak will be close to the one obtained 
with CMD-2.
Let us recollect that the ISR method provides a continuous ``low energy scan'',  while
taking data at fixed high energy. The threshold region, $2m_{\pi}$ -- 0.5 GeV,
gives about 13\% of the total contribution to the muon anomaly. As a rule, 
the collider luminosity dramatically decreases at low energies. To overcome the lack of
data in the threshold region, the ISR method can serve as a very efficient
and unique way to measure the pion form factor inside this energy region.

Today, the theoretical precision for the cross section of the process $e^+e^-
\to \pi^+\pi^-\gamma$ is dominated by the uncertainty of the radiator function
(0.5\%), and there is hope to reduce it to a few per mill in the future.
 In the case of the pion form factor extraction
from the $\pi^+\pi^-\gamma / \mu^+\mu^-\gamma$
ratio, the dependence on theory will be significantly reduced, since the
main uncertainty of the radiator function and vacuum polarisation effects
cancel out in the ratio. With the integrated luminosity of several inverse
femtobarn at 2 GeV, one can reach a fractional accuracy on the total error
smaller than 0.5\%.

In direct scan experiments the data are collected at fixed
energy points. Thus, some ``empty'' gaps without data naturally arise.
The experiments with ISR will cover the whole energy scale, filling any existing gaps.
Trigger and reconstruction efficiencies, detector imperfections and many other factors
will be identical for all data in the whole energy range. Therefore, some systematic 
errors will be cancelled out in part. Comparison of cross sections for the process 
$e^+e^- \to \mu^+\mu^-$, measured both with ISR and direct scan, can serve as a 
benchmark 
to study and control systematic effects. It should confirm the validity of this 
method and 
 help to determine the energy scale. A fit of the $\omega$ and $\phi$ resonances
will also provide a calibration of the energy scale -- an important feature to achieve a
systematic accuracy of a few per mill for the pion form factor.


\subsubsection{Prospects for radiative return at BES-III}
\label{rr:besiii}

The designed peak luminosity of BEPC-II is $1\times 10^{33}$
cm$^{-2}s^{-1}$ at $\sqrt{s}=3.77$~GeV, i.e. the $\psi(3770)$
peak. It has reached 30\% of the design luminosity now and is
starting to deliver luminosity to BES-III for physics. Although the
physics programs at BES-III are rather rich~\cite{Asner:2008nq}, most of
the time, the machine will run at $\sqrt{s}=3.77$~GeV and 4.17~GeV
for charm physics, since the cross sections of $J/\psi$ and
$\psi(2S)$ production are large and the required statistics can be
accumulated in short time, say, one year at each energy point. The 
estimated running time of BEPC-II at $\sqrt{s}=3.77$ and
4.17~GeV is around eight years, which corresponds to an integrated
luminosity of about 20~fb$^{-1}$ at each energy point.

Data samples at $\sqrt{s}=3.77$ and 4.17~GeV can be used for
radiative return studies, for the c.m. energies of the
hadron system between the $\pi^+\pi^-$ threshold to above 2.0~GeV.
This will allow for measurements of the pion, kaon and proton form 
factors, as well as of cross sections for some multi-hadron final states. 
The good coverage of
the muon detector at BES-III also allows the identification of
the $\mu^+\mu^-$ final state, thus supplying a normalisation
factor for the other two-body final states.

\begin{figure}
  \centerline{\psfig{file=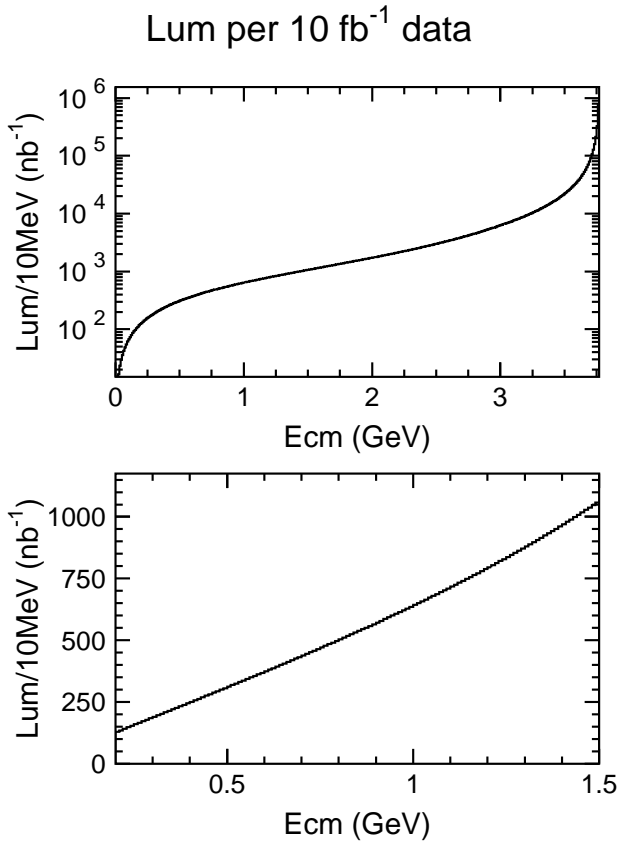,width=0.45\textwidth}}
\caption{Expected luminosity at low energies due to ISR for
10~fb$^{-1}$ data accumulated on the $\psi(3770)$
peak.}\label{lum3773}
\end{figure}

Figure~\ref{lum3773} shows the expected luminosity at low energies
in 10~MeV bins for 10~fb$^{-1}$ data accumulated on the
$\psi(3770)$ peak. In terms of luminosity at the $\rho^0$ peak,
one can see that 10~fb$^{-1}$ of data at $\sqrt{s}=3.77$~GeV is
equivalent to 70~fb$^{-1}$ at 10.58~GeV, i.e. at the $B$ factories.

\begin{figure}[htb]
  \centerline{\psfig{file=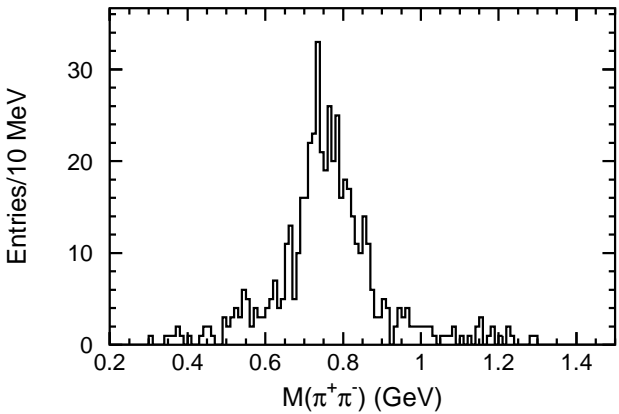,width=0.45\textwidth}}
\caption{Detected $\gamma_{\rm ISR}\pi^+\pi^-$ in 10000 produced
events at the $\psi(3770)$ peak. The sample is generated with
PHOKHARA.} \label{ff_bes3}
\end{figure}

With Monte Carlo generated $e^+e^-\to \gamma_{ISR}\pi^+\pi^-$ data
using PHOKHARA~\cite{Rodrigo:2001kf}, after a fast simulation and
reconstruction with the BES-III software, one found the efficiency
for events at the $\rho^0$ peak to be around 5\% if one requires the
detection of the ISR photon. This is higher than the efficiency at 
BaBar \cite{wangll}. Figure~\ref{ff_bes3} shows the
signal for 10,000 generated $\pi^+\pi^-$ events. One estimates the
number of events in each 10~MeV bin to be around 20,000 at the
$\rho^0$ peak, for 10~fb$^{-1}$ of data at $\sqrt{s}=3.77$~GeV. This
is comparable with the recent BaBar results based on 232~fb$^{-1}$ of 
data at the $\Upsilon(4S)$ peak~\cite{wangll}.

The most important work related to the pion form factor
measurement is the estimate of the systematic error. Since the
cross section of good events at the $\psi(3770)$ peak is not large
(around 30~nb for the total hadronic cross section, with about 400~nb
cross section for the QED processes) 
compared to the highest
trigger rates at $J/\psi$ and $\psi(2S)$ peak energies, a loose
trigger is mandatory to allow the ISR events to be recorded. 
In principle, the trigger rate for these events could reach 100\%, 
with an allowed trigger purity of less than 20\%.

With enough $D\bar D$ events accumulated at the same energy, the
tracking and particle ID efficiencies can be measured with 
high precision (as has been done at CLEO-c~\cite{cleoc:2007zt}). In
addition, a huge data sample at the $\psi(2S)$ and the well measured
large branching fraction of $\psi(2S)$ transition modes, such as
$\pi^+\pi^-J/\psi$, $J/\psi\to \mu^+\mu^-$, can be used to study
the tracking efficiency, $\mu$-ID efficiency and so on. All this 
will greatly help to understand the detector performance and
to pin down the systematic errors in the form factor
measurement.

The kaon and proton form factors can be measured as well since
they are even simpler than the measurement of the pion form factor.
This will allow us to better understand the structure close
to threshold and possible existing high-mass structures.

Except for the lowest lying vector states ($\rho$, $\omega$, $\phi$), 
the parameters of other vector states are poorly known, 
and further investigations are needed. BES-III ISR analyses may
reach energies slightly above 2~GeV, while beyond that BEPC-II can run
by adjusting the beam energy. This allows BES-III to study the full range of
vector mesons between the $\pi^+\pi^-$ threshold and 4.6~GeV, which is
the highest energy BEPC-II can reach, thus covering the $\rho$,
$\omega$ and $\phi$, as well as the $\psi$ sector. One will have the chance
to study the excited $\rho$, $\omega$ and $\phi$ states between 1
and about 2.5~GeV. The final states include $\pi^+\pi^-\pi^0$,
$K\bar K$, 4 pions, $\pi\pi K K$, etc. Final states with more than
four particles will be hard to study using the ISR method, since the
$D \bar D$ decay will contribute as background.

\section{Tau decays}
\label{sec:5}
\hyphenation{author another created financial paper re-commend-ed Post-Script}


\subsection{Introduction}

After discovery of the $\tau$ lepton, which is a fundamental lepton,
heavy enough to decay not only into leptons, but also into
dozens of various hadronic final states, it became clear 
that corresponding Monte Carlo (MC) event generators are needed for various 
purposes:

\begin{itemize}
\item
{To calculate detector acceptance, efficiencies 
 and various distributions for signal event selection and comparison
to data. In general the acceptance is small (a few percent) and depends on
the model; in principle, it is a complicated function of invariant
masses, angles, and resolutions. 
Analysis of publications shows that effects of MC signal modelling 
are almost always neglected.}
\item
{To estimate the number of background (BG) events
$N^{\rm BG}_{\rm ev}$ and their distributions;
in addition to background coming from $\tau^+\tau^-$ pairs (so called
cross-feed), there might be BG events from $q\bar{q}$ continuum,
$~\gamma\gamma$ collisions etc.}
\item
To unfold observed distributions 
to get rid of detector effects, 
important when extracting resonance parameters.
\end{itemize}
 
Various computer packages like, e.g.,  
KORALB~\cite{Jadach:1984iy}, KKMC~\cite{kkcpc:1999}, 
TAUOLA~\cite{Jadach:1990mz,Jezabek:1991qp,Jadach:1993hs} and 
PHOTOS~\cite{photos2:1994} 
were developed to generate events for $\tau$ lepton production
in $e^+e^-$ annihilation and their subsequent decay, taking into
account the possibility of photon emission.
These codes became very important tools 
for experiments at LEP, CLEO, Tevatron and HERA.

Simulation of hadronic decays requires the knowledge of hadronic form 
factors.  Various hadronic final states were
considered 
in the 90's, resulting in 
a large number of specific hadronic currents~\cite{Kuhn:1992nz}.


However, already experiments at LEP and CLEO show\-ed that with increase
of the collected data sets a more precise description is necessary.
Some attempts were made to improve the parametrisation of various
hadronic currents. One should note the serious efforts of the ALEPH
and CLEO Collaborations, which created their own parametrisations of TAUOLA
hadronic currents already in the late 90's, or 
a  parametrisation of the hadronic current in the $4\pi$ 
decays~\cite{Bondar:2002mw}, based on the experimental information on 
$e^+e^- \to 2\pi^+2\pi^-,~\pi^+\pi^-2\pi^0$ from Novosibirsk~\cite{Akhmetshin:1998df}, which is now implemented in the presently distributed TAUOLA code \cite{Golonka:2003xt}.

\subsection{Current status of data and MC generators}
In this section we will briefly discuss the most precise recent
experimental data on $\tau$ lepton decays, showing, wherever possible, 
their comparison with the existing MC generators and discussing the 
decay dynamics.
 \begin{center}
\begin{figure}
\includegraphics[width=0.37\textwidth]{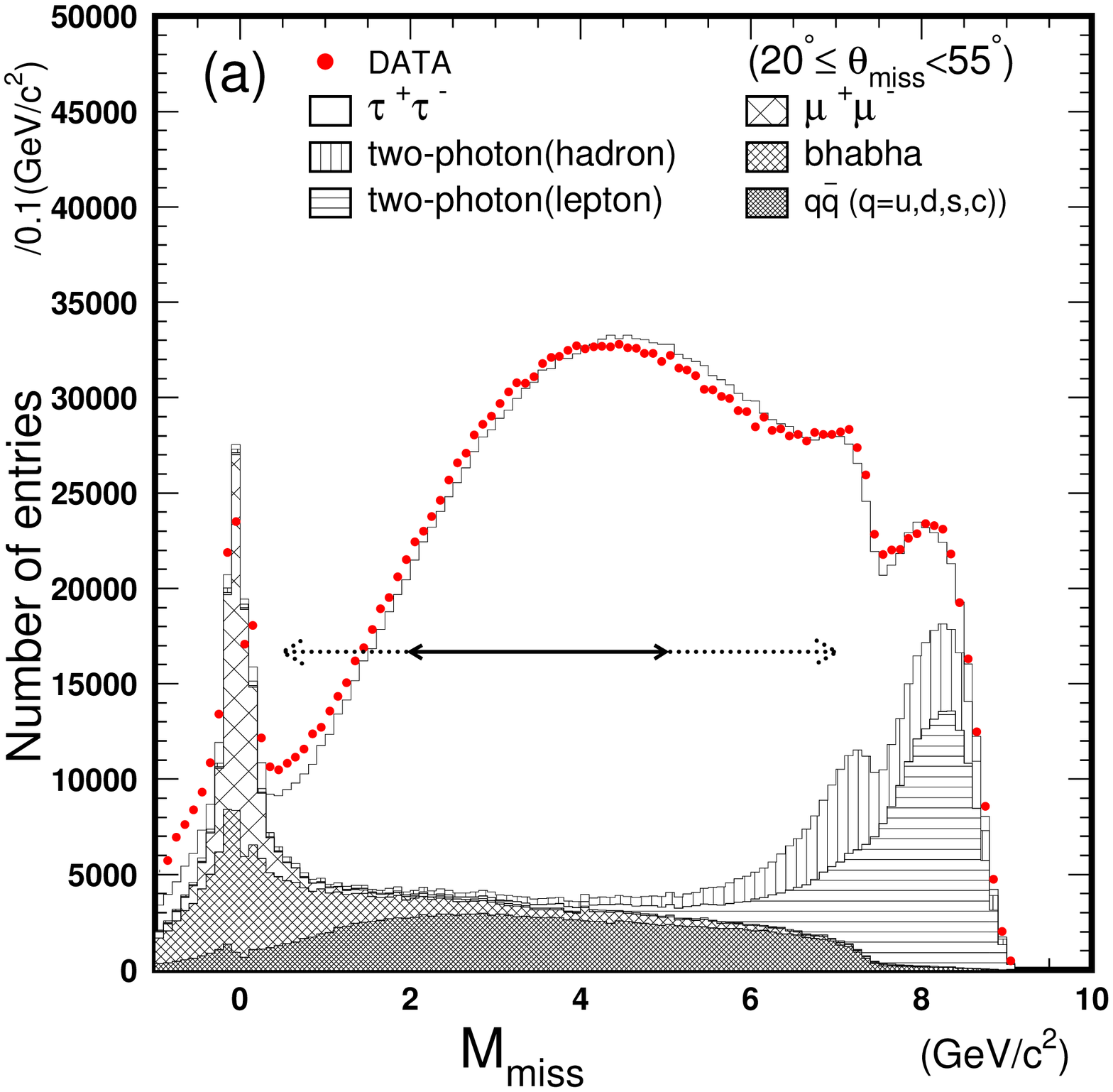}
\includegraphics[width=0.37\textwidth]{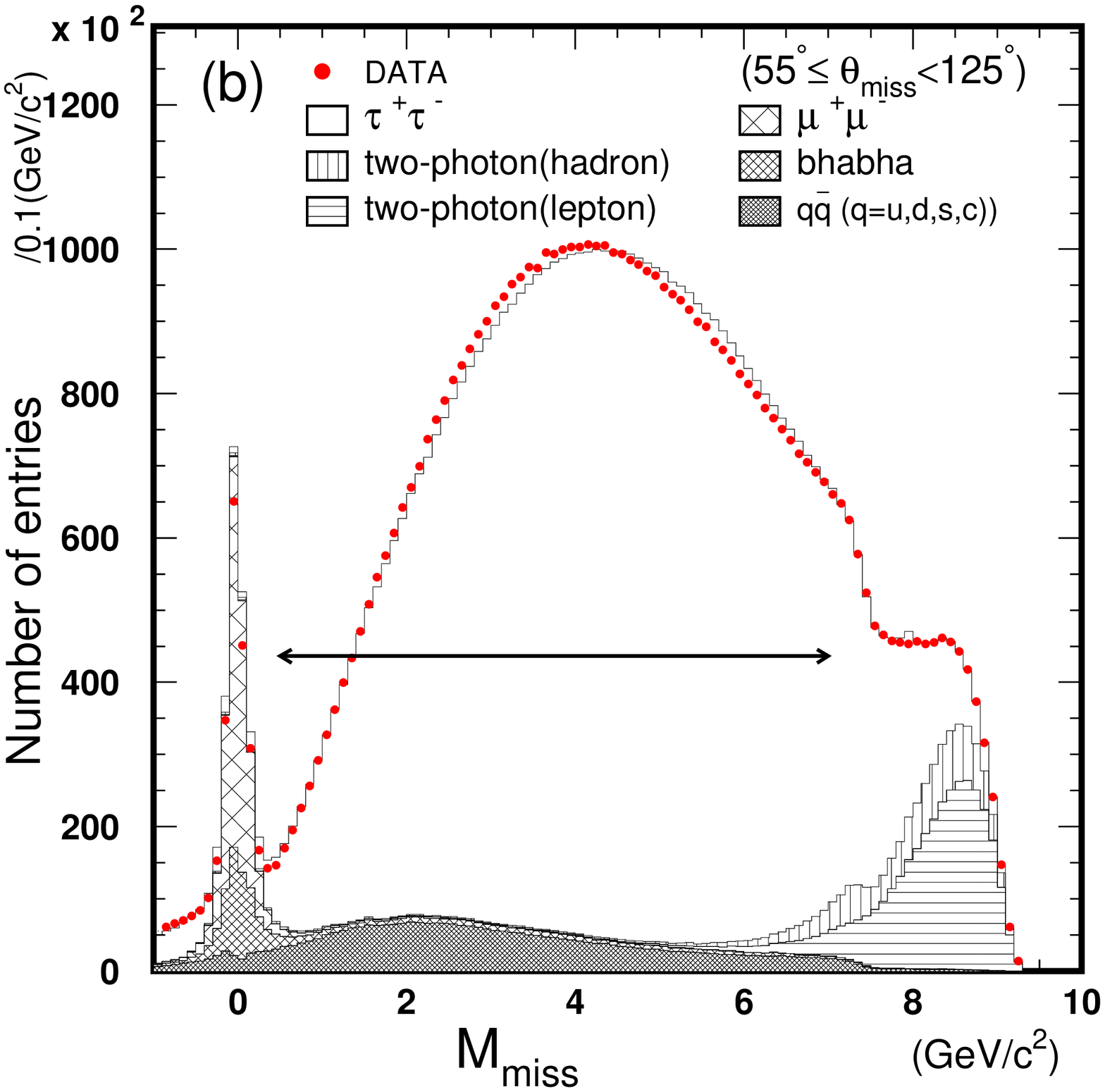}
\includegraphics[width=0.37\textwidth]{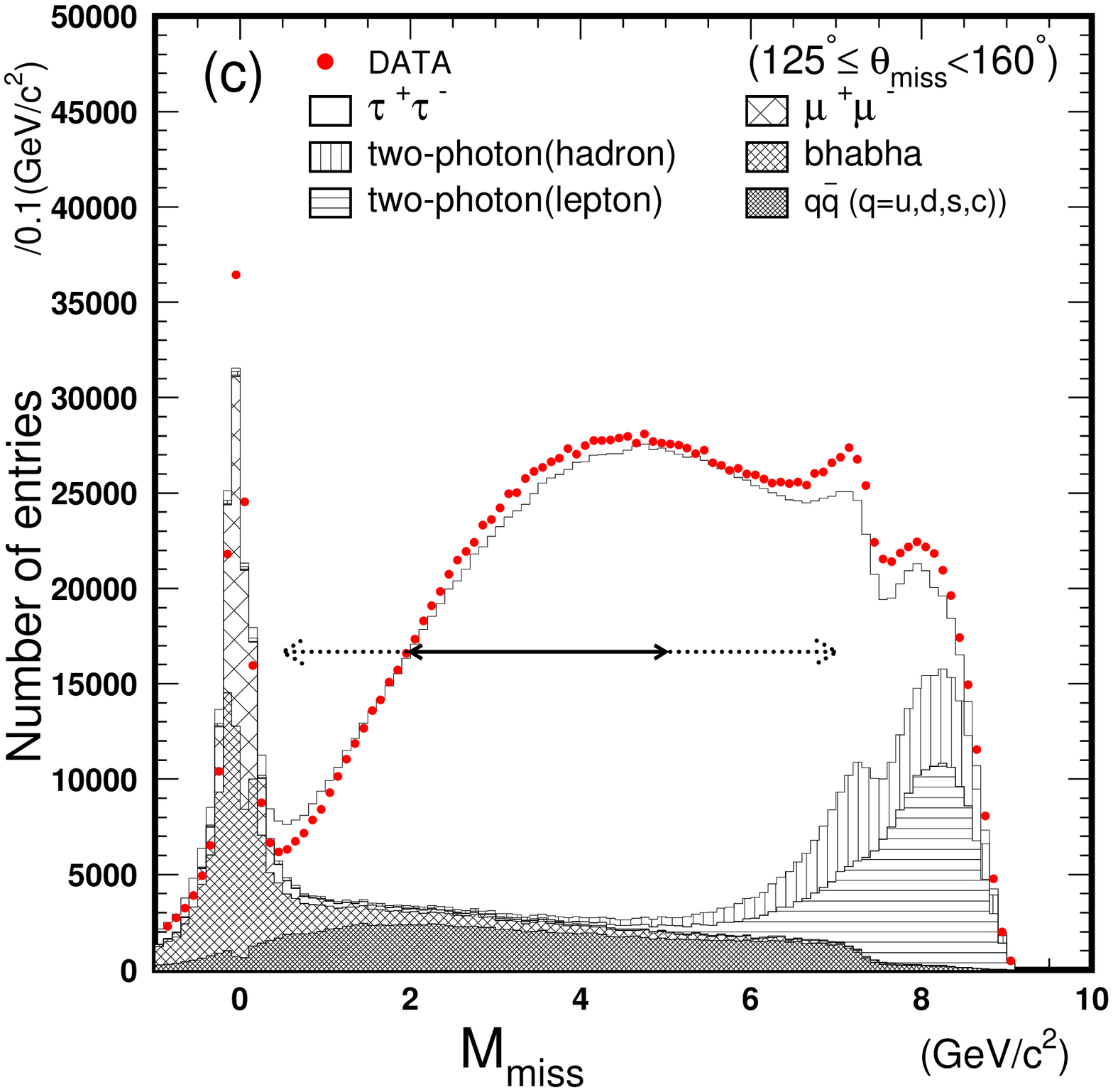}
\caption{Projections to the missing mass
and missing direction for 
$\tau^- \to  \pi^-\pi^0\nu_{\tau}$ decays at Belle:
(a)--(c) correspond to different ranges of the missing polar angles.
The solid circles represent the data and the histograms the MC
simulation (signal + background). The open histogram shows 
the contribution from $\tau^+\tau^-$ pairs, the vertical
(horizontal) striped area shows that from two-photon leptonic
(hadronic) processes; the wide (narrow) hatched area 
shows that from Bhabha ($\mu^+\mu^-$), and the shaded area that from the 
$q\bar{q}$ continuum.} 
\label{fig:tau21}
\end{figure}
\end{center}
 
\begin{center}
\begin{figure}
\includegraphics[width=0.40\textwidth]{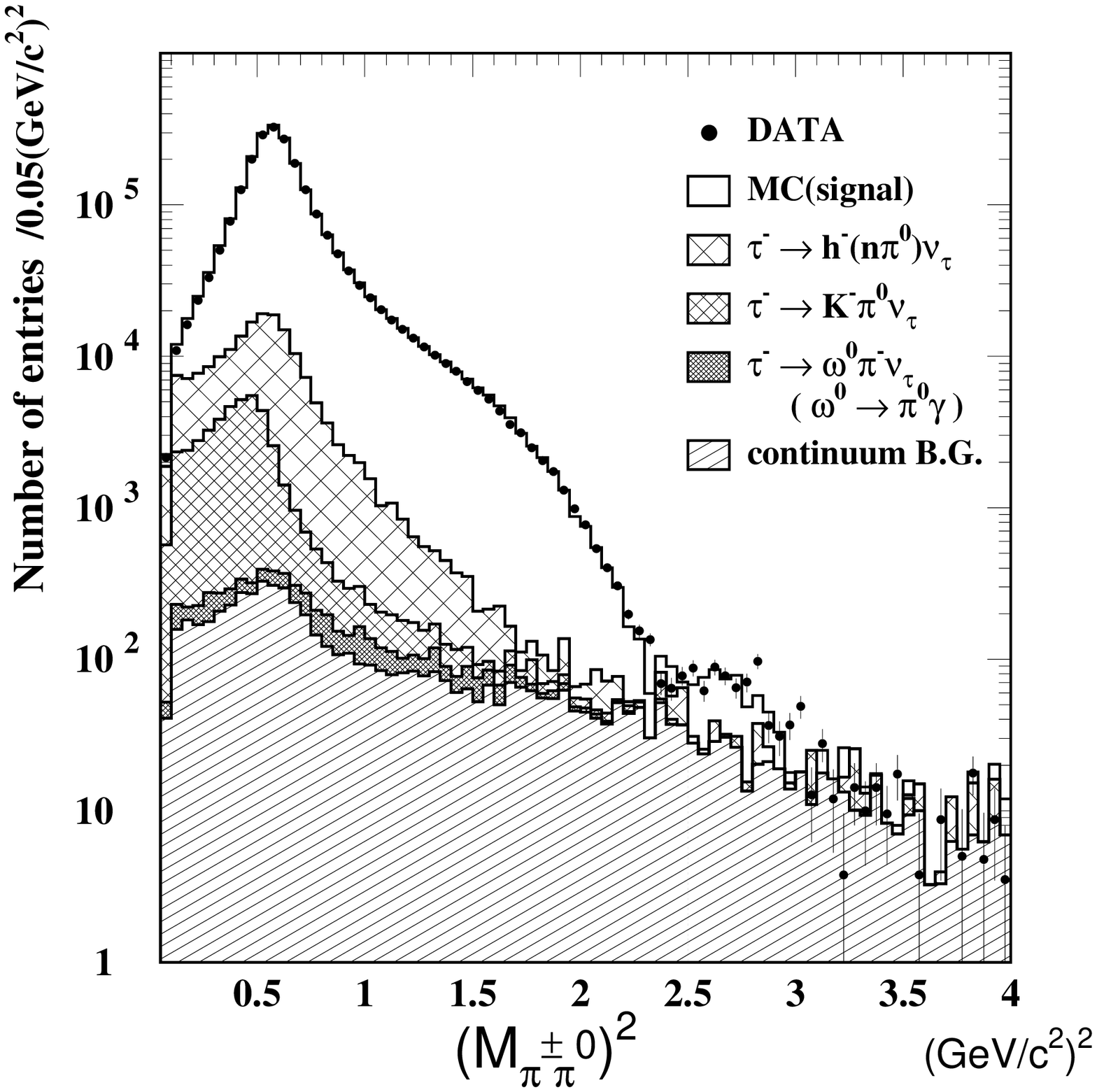}
\\
\includegraphics[width=0.36\textwidth]{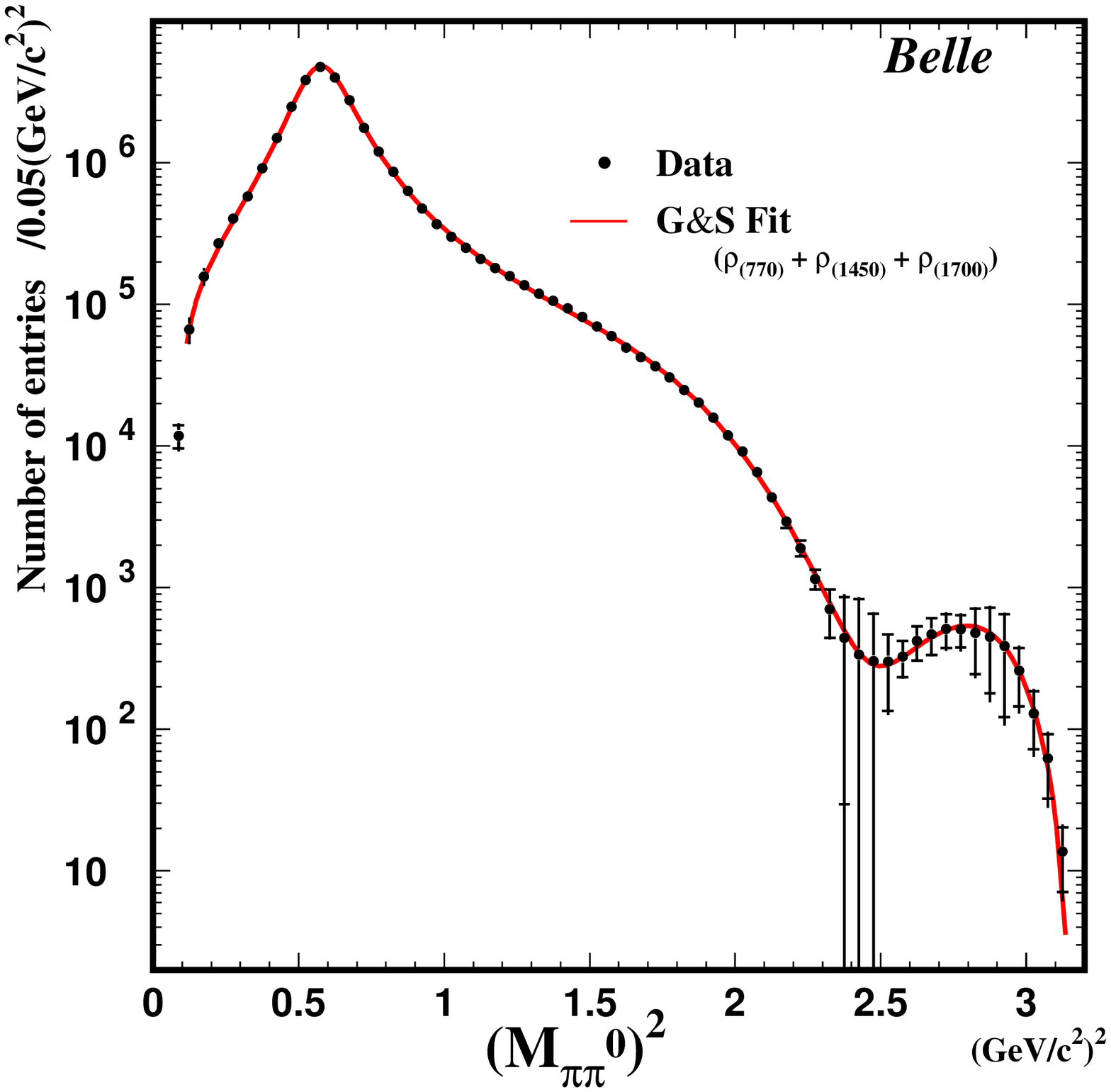}
\caption{Invariant-mass-squared distribution for
$\tau^- \to  \pi^-\pi^0\nu_{\tau}$ decay at Belle.
(a) Contributions of different background sources. The solid circles 
with error bars represent the data,
and the histogram represents the MC simulation 
(signal + background).
(b) Fully corrected distribution. The solid curve is the result
of a fit to the Gounaris-Sakurai model with the $\rho(770)$,
$\rho(1450)$ and $\rho(1700)$ resonances.}
\label{fig:tau22}
\end{figure}
\end{center}

\subsubsection{$\tau^- \to  \pi^-\pi^0\nu_{\tau}$ at Belle}
Recently results of a study of the
$\tau^- \to  \pi^-\pi^0\nu_{\tau}$ decay by the Belle Collaboration
were published~\cite{Fujikawa:2008ma}.
From less than 10\% of the dataset available the authors selected a
huge statistics of 5.4M events, about two orders of magnitude larger 
than in any previous experiment, determined the branching fraction and 
after the unfolding 
obtained the hadronic mass spectrum, in which for the first time
three $\rho$-like resonances were observed together: 
$\rho(770),~\rho(1450)$ and $\rho(1700)$. Their parameters were also
determined.


The comparison of the obtained missing mass distributions 
with simulations for different polar angle ranges
(Fig.~\ref{fig:tau21}) shows that there exist small discrepancies 
between MC and data.

Figure~\ref{fig:tau22} shows various background contributions to
the di-pion mass distribution (upper panel) and underlying dynamics
(lower panel), clearly demonstrating a pattern of the three interfering
resonances $\rho(770),~\rho(1450)$ and $\rho(1700)$.

\subsubsection{$\tau^- \to
  \bar{K}^0\pi^-\nu_{\tau},~K^-\pi^0\nu_{\tau}$}

\begin{center}
\begin{figure}
\includegraphics[width=0.35\textwidth]{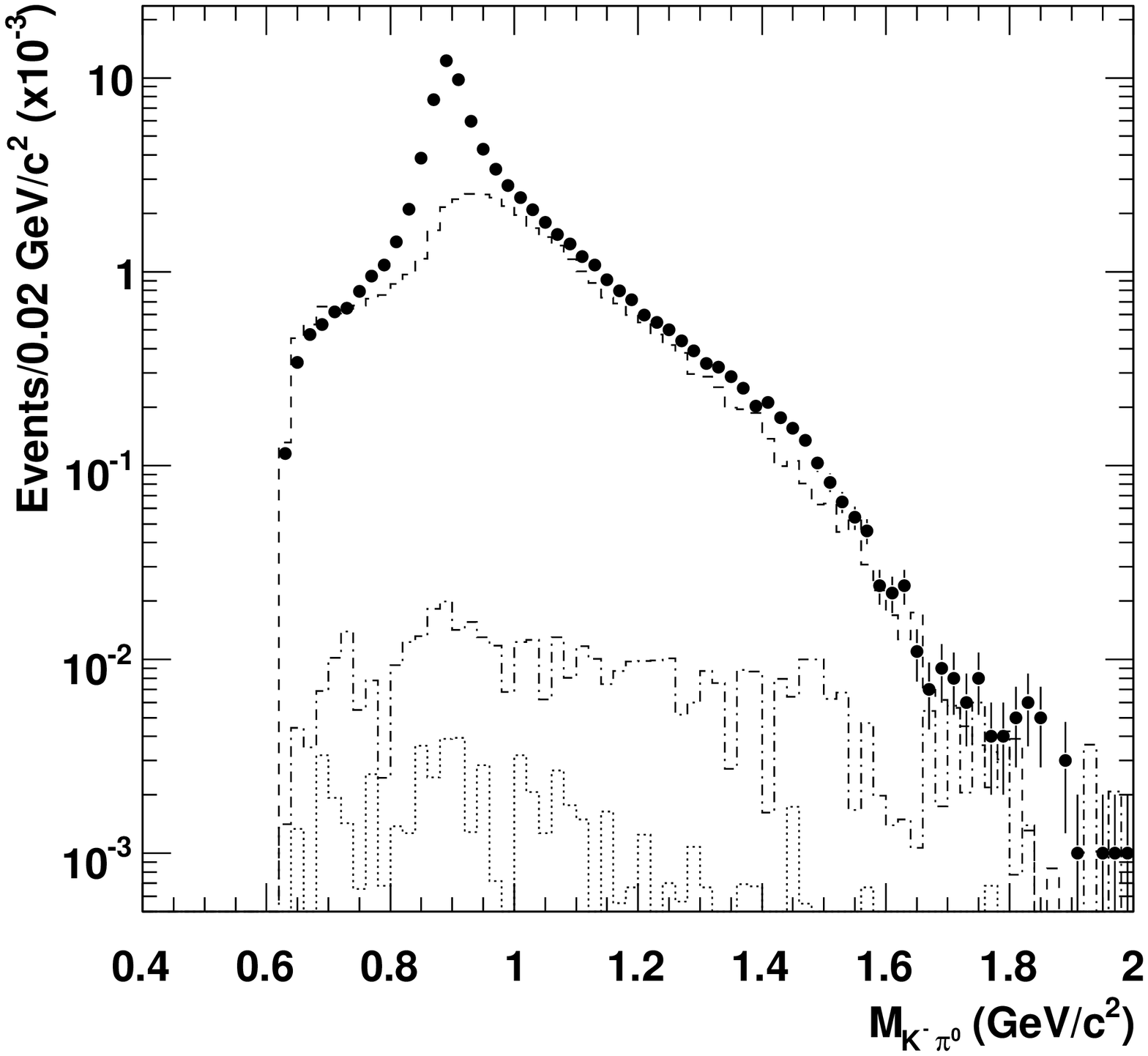}
\caption{The $K\pi$ invariant mass distribution for the decay 
$\tau^- \to K^-\pi^0\nu_{\tau}$ at BaBar. The dots are the data,
while the histograms are background MC events with selection and 
efficiency corrections: $\tau$ background (dashed line), $q\bar{q}$
(dash-dotted line), $\mu^+\mu^-$ (dotted line).}
\label{fig:tau23}
\end{figure}
\end{center}

Two high-precision studies of the $\tau$ decay into the $K\pi\nu_{\tau}$ 
final state were recently published.  The BaBar Collaboration
reported  a measurement of the branching fraction of the
$\tau^- \to K^-\pi^0\nu_{\tau}$ decay~\cite{Aubert:2007jh}. They do not study
in detail the $K\pi$ invariant mass distribution, noting only that the
$K^*(892)^-$ resonance is seen prominently above the simulated
background, see Fig.~\ref{fig:tau23}. Near 1.4 GeV$/c^2$ decays 
to higher $K^*$ mesons are
expected, such as the $K^*(1410)^-$ and  $K^*_0(1430)^-$, but their
branching fractions are not yet measured well. These decays are not
included in the BaBar simulation of $\tau$ decays, but seem to be
present in the data around  1.4 GeV$/c^2$. It is also worth noting
that this decay mode is heavily contaminated by cross-feed backgrounds
from other $\tau$ decays. For example, below 0.7 GeV$/c^2$ the
background is dominated by $K^-\pi^0\pi^0\nu_{\tau}$ and  
$K^-K^0\pi^0\nu_{\tau}$ events,
for which the branching fractions are only known with large 
relative uncertainties of $\approx 37\%$ and $\approx 13\%$,
respectively. Non-negligible background may also come from
the $\tau^- \to \pi^-\pi^0\nu_{\tau}$ decay, which has a large
branching fraction and thus should be simulated properly.
    
\begin{center}
\begin{figure}
\includegraphics[width=0.35\textwidth]{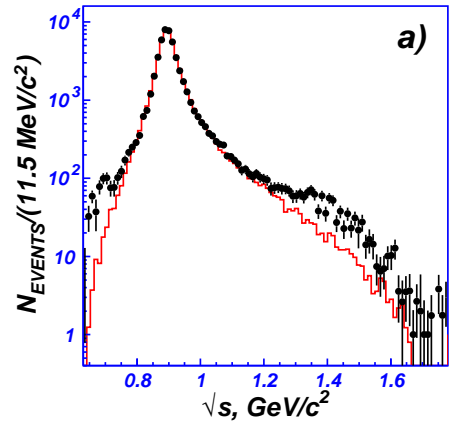} 
\\
\includegraphics[width=0.35\textwidth]{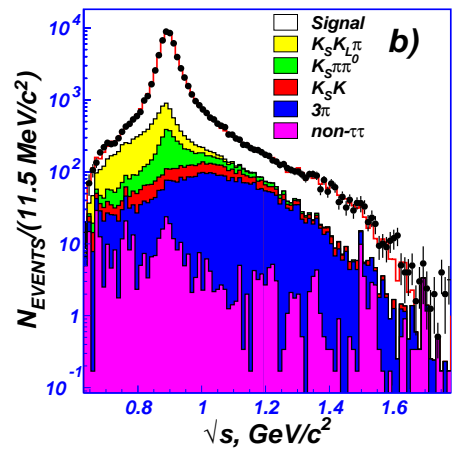}
\caption{The $K\pi$ invariant mass distribution for the decay 
$\tau^- \to K^-\pi^0\nu_{\tau}$ at Belle. Points are experimental 
data, histograms are spectra expected for different models. 
(a) shows the fitted result in the model with the $K^*(892)$ alone.
(b) shows the fitted result in the  $K^*(892)+ K^*_0(800) + K^*(1410)$
model. Also shown are different types of background.} 
\label{fig:tau24}
\end{figure}
\end{center}
{}
Another charge combination of the final state particles, i.e.,
$K^0_S\pi^-\nu_{\tau}$, was studied in the Belle
experiment~\cite{:2007rf}. In this case a detailed analysis of the
$K\pi$ invariant mass distribution has been performed. The authors   
also conclude that the decay dynamics differs from pure $K^*(892)$: 
the best fit includes $K^*_0(800)+K^*(892)+K^*(1410)/K^*_0(1430)$,
see Fig.~\ref{fig:tau24}.
\begin{center}
\begin{figure}
\includegraphics[width=0.5\textwidth]{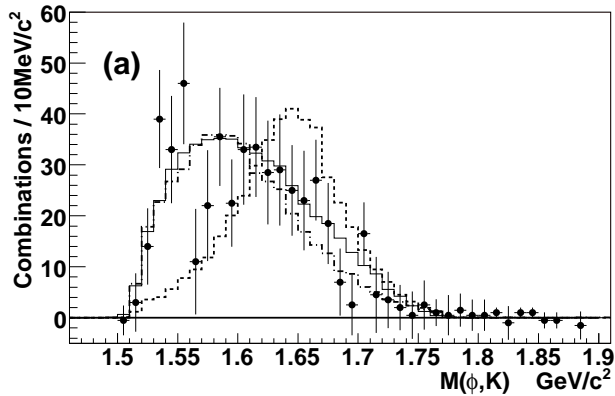}
\caption{The $\phi K$ invariant mass distribution for the decay 
$\tau^- \to \phi K^- \nu_{\tau}$ at Belle. Points with error
bars are the data. The open histogram is the phase-space distributed 
signal MC, and dotted and dot-dashed histograms indicate the signal MC
mediated by a resonance with mass and width of 1650 MeV and 100 MeV,
and 1570 MeV and 150 MeV, respectively.}
\label{fig:tauphik}
\end{figure}
\end{center}

\begin{figure*}
\begin{center}
\includegraphics[width=0.75\textwidth]{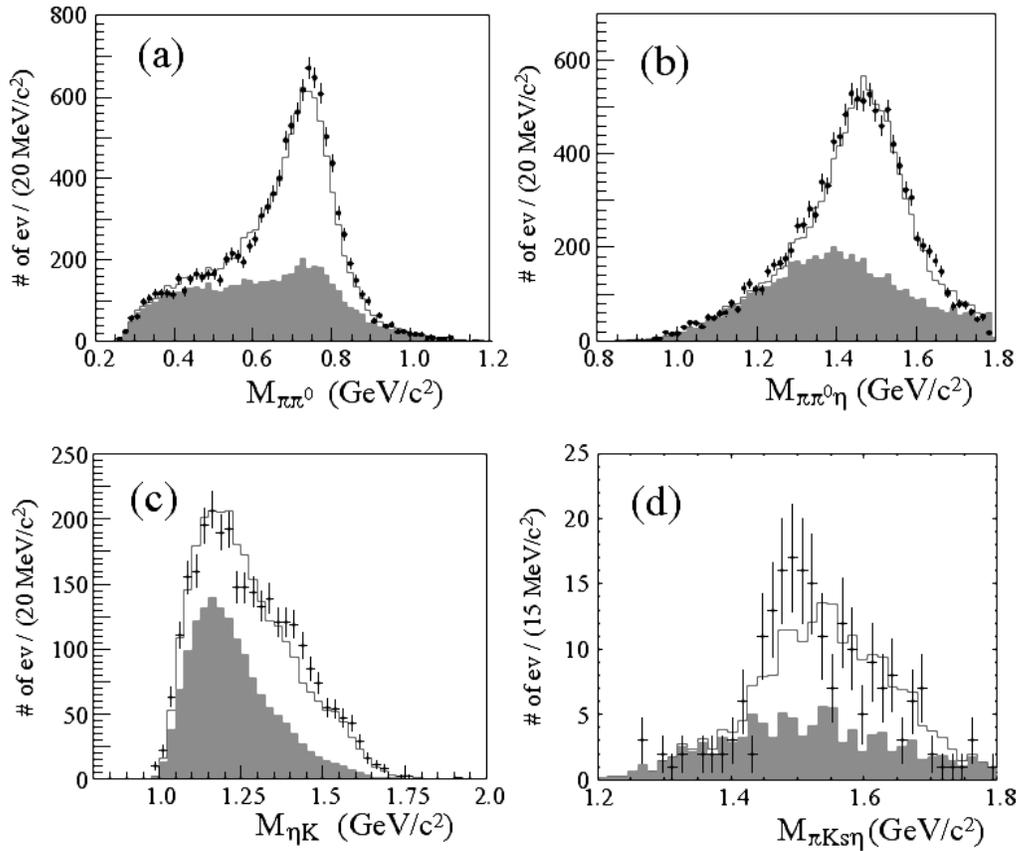}
\caption{Invariant mass distributions: 
(a) $\pi\pi^0$ and (b) $\pi\eta\pi^0$ for 
$\tau \to \pi\pi^0\eta\nu_{\tau}$; (c) $\eta K$ for 
$\tau \to K\eta\nu_{\tau}$ and (d) $\pi K^0_S\eta$ for 
$\tau \to \pi K^0_S\eta\nu_{\tau}$ at Belle. The points with error 
bars are the data. The normal and filled histograms indicate the signal and
$\tau^+\tau^-$ background MC distributions, respectively.}
\label{fig:beleta}
\end{center}
\end{figure*}


\subsubsection{$\tau$ decays into three pseudoscalars}

Recently a measurement of the branching fractions of
various particle combinations in the decay to three charged
hadrons (any combination of pions and kaons) was reported by the 
BaBar Collaboration~\cite{Aubert:2007mh}.
A similar study was also performed by the Belle group~\cite{:2008sg}.
However, both groups have not yet analysed the mass spectra in detail.
In the $K^-K^+K^-\nu_{\tau}$ final state BaBar~\cite{Aubert:2007mh} 
and Belle~\cite{Inami:2006vd} reported 
the observation of the decay mode $\phi K^-\nu_{\tau}$, while in the 
$K^-K^+\pi^-\nu_{\tau}$ final state BaBar observed the 
$\phi \pi^- \nu_{\tau}$ decay mode~\cite{Aubert:2007mh}. Belle analysed 
the spectrum of the $\phi K^-$ mass and concluded that it might 
have a complicated dynamics, see Fig.~\ref{fig:tauphik}.

The most detailed previous study of the mass spectra was done by the
CLEO group~\cite{Asner:1999kj}. With the statistics of about 8,000 events 
they conclude that the $3\pi$ mass spectrum is dominated by the
$a_1(1260)$ meson, and confirmed that the decay of the latter 
is not saturated by the $\rho\pi$ intermediate state, having in
addition a significant $f_0(600)\pi^-$ component observed earlier in
$e^+e^-$ annihilation into four charged pions~\cite{Akhmetshin:1998df}.

Recently the Belle Collaboration performed a detailed study of
various decays with the $\eta$ meson in the final
state~\cite{Inami:2008ar}. 
They measured
the branching fractions of the following decay modes:
$\tau^- \to K^-\eta\nu_{\tau}$, $\tau^- \to K^-\pi^0\eta\nu_{\tau}$, 
$\tau^- \to \pi^-\pi^0\eta\nu_{\tau}$, 
$\tau^- \to \pi^-K^0_S\eta\nu_{\tau}$, and 
$\tau^- \to K^{*-}\eta\nu_{\tau}$. They also set upper limits on the
  branching fractions of the  decays into $K^-K^0_S\eta\nu_{\tau}$,
$\pi^-K^0_S\pi^0\eta\nu_{\tau}$, $K^-\eta\eta\nu_{\tau}$,
 $\pi^-\eta\eta\nu_{\tau}$, and non-resonant  $K^-\pi^0\eta\nu_{\tau}$
final states.

Figure~\ref{fig:beleta} shows that there is reasonable agreement for
$\eta\pi^-\pi^0\nu_{\tau}$ (a, b) and a worse one for
$\eta K^- \nu_{\tau}$ (c) and $\eta K^{*-}\nu_{\tau}$ (d). 

\subsubsection{$\tau$ decays to four pions}

There are two possible isospin combinations of this hadronic final state,
$2\pi^-\pi^+\pi^0$ and $\pi^-3\pi^0$. Both have not yet been studied
at $B$ factories, so the best existing results are based on 
ALEPH~\cite{Buskulic:1996qs} and CLEO~\cite{Edwards:1999fj} results.
    
The theoretical description of such decays is based on the CVC relations 
and the available low energy $e^+e^-$ 
data~\cite{Decker:1994af,Czyz:2000wh,Bondar:2002mw,Czyz:2008kw}.

\subsubsection{$\tau^- \to  3h^-2h^+\nu_{\tau}$ at BaBar}

A new study of the $\tau^- \to  3h^-2h^+\nu_{\tau}$ decay ($h=\pi,~K$)
has been performed by the BaBar Collaboration~\cite{Aubert:2005wa}. A large
dataset of over 34,000 events (two orders of magnitude larger than in the
best previous measurement at CLEO~\cite{Gibaut:1994ik}) allows one 
a first search for resonant structures and decay dynamics. 

The invariant mass distribution of the five charged particles
in Fig.~\ref{fig:tau51} shows a clear discrepancy between the data 
and the MC simulation, which uses the phase space distribution
for $\tau^- \to  3\pi^-2\pi^+\nu_{\tau}$.

The mass of the $h^+h^-$ pair combinations in Fig.~\ref{fig:tau52}
(upper panel), 
with a prominent shoulder at 0.77 GeV$/c^2$, suggests a strong
contribution from the $\rho$ meson. Note that there are three allowed
isospin states for this decay, of which two may have a $\rho$ meson. 
The mass of the $2h^+2h^-$ combinations in Fig.~\ref{fig:tau52} (lower
panel) also shows a structure at 1.285 GeV$/c^2$ coming from the 
$\tau^- \to f_1(1285)\pi^-\nu_{\tau}$ decay.         

The first attempt to take into account the dynamics of this decay was
recently performed in Ref.~\cite{Kuhn:2006nw}.

\subsubsection{$\tau$ decays to six pions}

The six-pion final state was studied by the CLEO 
Collaboration~\cite{Anastassov:2000xu}. Two charge combinations,
 $3\pi^-2\pi^+\pi^0$ and $2\pi^-\pi^+3\pi^0$, 
were observed and it was found that the decays are saturated by
intermediate states with $\eta$ and $\omega$ mesons. Despite the rather
limited statistics (about 260 events altogether), it became clear that
the dynamics of these decays is rather rich.

\subsubsection{Lepton-Flavour Violating Decays}

More than 50 different Lepton-Flavour Violating  (LFV)
decays have been studied by the CLEO, BaBar and Belle
Collaborations.
Publications rarely describe how the simulation of such decays is performed.
Moreover, theoretical papers suggesting LFV in new models usually 
do not provide differential cross sections. In some experimental
papers the authors claim that the production of final state
hadrons with a phase space distribution is assumed.
However, the real meaning of this statement is not very clear 
since LFV assumes New Physics and, therefore, matrix elements are not 
necessarily separated into weak and hadronic parts. 

However, there exist a few theoretical papers considering 
differential cross sections. For example,   
angular correlations for
$\tau^- \to \mu^-\gamma,~\mu^-\mu^+\mu^-$ and $\mu^-e^+e^-$ decays
were studied in Ref.~\cite{Kitano:2000fg}. An attempt to classify different
types of operators entering New Physics Lagrangians for $\tau$ decays 
to three charged leptons was made in~\cite{Dassinger:2007ru}.

\subsection{Status of Monte Carlo event generators for 
$\tau$ production and decays}

High-statistics and high-precision experiments, as well as  
searches for rare processes, result in a new challenge: 
Monte Carlo generators based on
an adequate theoretical description of energy and angular distributions.
In the following we will 
describe the status of the Monte Carlo programs used by experiments. 
We will review the building blocks used in the simulation 
with the goal in mind to localise the
points requiring most urgent attention.

At present, for the production of $\tau$ pairs, the Monte Carlo programs
KORALB ~\cite{Jadach:1984iy} and
KKMC ~\cite{kkcpc:1999} are the standard
codes to be used. 
For the generation of brems\-strah\-lung in decays, 
the Monte Carlo PHOTOS~\cite{photos2:1994} is used. Finally, $\tau$ decays 
themselves are simulated 
with the program TAUOLA ~\cite{Jadach:1990mz,Jezabek:1991qp,Jadach:1993hs}. The 
EvtGen code was written and maintained for simulation of $B$ meson decays,
see \newline\texttt{www.slac.stanford.edu/\~{}lange/EvtGen/}\ .
It offers a \newline unique opportunity to specify, at run time, a list of the
final state particles\footnote{E.g. $\tau$ lepton decay products 
including neutrinos.},
without having to change and/or compile the underlying code.  In
a multi-particle final state dominated by
phase space considerations, this generator provides an adequate
description of the final state momenta,
for which the underlying form factor calculation is more involved and
not presently available in a closed form. That is why it is used by 
experiments measuring $\tau$ decays too.


So far,  our discussion has been based on the comparison of experimental data 
and theory embodied into Monte Carlo 
programs treated as a black box. 
One could see that a typical signature
of any given $\tau$ decay channel is matching rather poorly 
the publicly available Monte Carlo predictions. This 
should be of no surprise as efforts to compare data with
predictions were completed  for the last time 
in late 90's by the ALEPH and CLEO collaborations.
The resulting hadronic currents were afterwards implemented 
in~\cite{Golonka:2003xt}.
Since that time no efforts to prepare a complete parametrisation of $\tau$ 
decay simulation for the public use were undertaken seriously.

There is another important message which can be drawn from these comparisons.
Starting from a certain precision level, the study of a given decay mode can 
not be separated from the discussion of others. In the distributions aimed 
at representing the given decay mode, a contribution from the other 
$\tau$ decay modes can be large, up to even 30\%. 

It may be less clear that experiments differ significantly in the way how 
they measure individual decay modes. For instance, ALEPH 
produced $\tau$ samples free of the non-$\tau$ backgrounds,
but, on the other hand, strongly boosted, making the
reconstruction of some angles in the hadronic system more difficult. 
This is important and affects  properties of the decay models 
which will be used for a parametrisation. In particular, when the 
statistics is small, possible fluctuations may affect the picture 
and there are not enough data to complete an estimate of the systematic errors.
In this case, details of the description of the hadronic current,
as the inclusion of intermediate resonances, are not important.
Let us consider, as an example, 
$\tau^- \to K_S^0 \pi^- \pi^0 \nu_{\tau}$. 
The matrix element in the ALEPH parametrisation is saturated by 
 $\rho^-\to \pi^- \pi^0 $ and
$K^{*0} \to K_S^0 \pi^0$, and a 
similar parametrisation is used for $K^{*-}\to K_S^0 \pi^-$. 
In practice, the contribution of the $\rho$ is more significant in 
the ALEPH parametrisation in contrast to the CLEO one where 
the  $K^*$ dominates.
One has to admit that at the time when both collaborations were 
preparing their parametrisations  to be used in 
TAUOLA, the data samples of both experiments were rather
small and the differences were not of much significance. This can, however, 
affect possible estimates of backgrounds for searches of rare decays, 
e.g. of $B$ mesons at LHCb.\footnote{LHCb performed MC studies 
for $B_s^0 \to \mu^+ \mu^-$ and the 
radiative decays $B^0 \to K^* \gamma$ and $B_s^0 \to \phi \gamma$, 
but $\tau$ decays have not yet been taken into account. 
These results are not public and exist only as internal documents
LHCB-ROADMAP1-002 and LHCB-ROADMAP4-001.} 

Let us now go point by point and discuss examples of Monte Carlo programs 
and fitting strategies. We will focus on subjects requiring most 
attention and future work. We will  review the theoretical 
constraints which are useful  in the construction of the models 
used for the data description. 

\begin{center}
\begin{figure}
\includegraphics[width=0.48\textwidth]{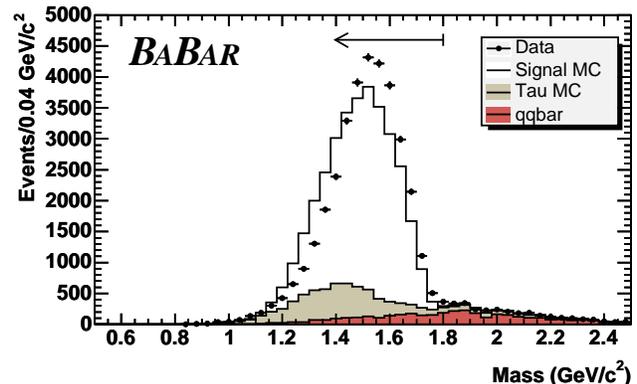}
\caption{Invariant mass of five charged particles for 
$\tau^- \to  3h^-2h^+\nu_{\tau}$ at BaBar.}
\label{fig:tau51}
\end{figure}

\begin{figure}
\includegraphics[width=0.48\textwidth]{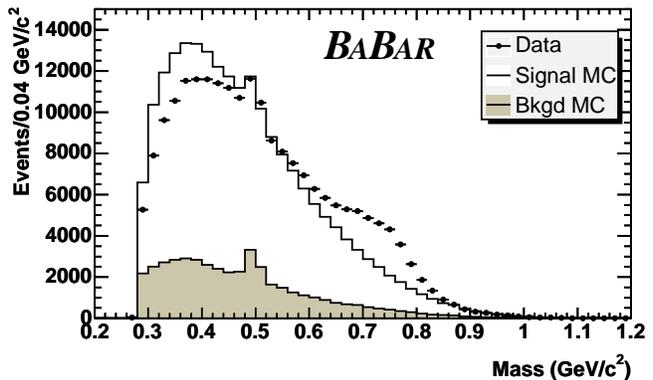} 
\\
\includegraphics[width=0.48\textwidth]{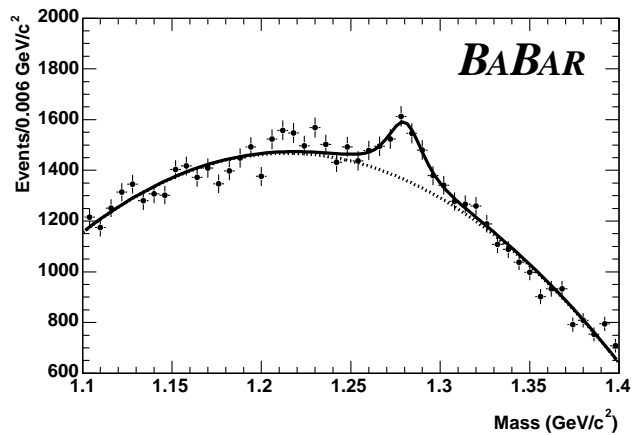}
\caption{Invariant mass distributions
 for $\tau^- \to  3h^-2h^+\nu_{\tau}$ at BaBar. Points with 
error bars are the data: Upper panel -- $h^+h^-$; the unshaded and shaded 
histograms are the signal and background predicted by MC.  
Lower panel -- $2\pi^+ 2\pi^-$; the solid line is a fit to the data
using a second-order polynomial (dashed line) for the background and 
a Breit-Wigner convoluted with a Gaussian for the peak region.} 
\label{fig:tau52}
\end{figure}
\end{center}

\subsection{Phase space}
Because of the relatively low multiplicity of final state particles, 
it is possible
to separate the description of $\tau$ production and decay into 
segments describing the matrix elements and the phase space.
In the phase space no approximations are used, contrary to the
 matrix elements where all 
approximations and assumptions reside. 
The description of the phase space 
used in TAUOLA is given in detail in~\cite{Jadach:1993hs}. The description 
of the phase space for $\tau$ production is given in~\cite{kkcpc:1999}. Thanks 
to conformal symmetry it is exact for an arbitrary number of 
photons. Using exponentiation, 
see, for example, Yennie-Frautchi-Suura~\cite{yfs:1961}, 
the phase space description 
can be exact and the matrix element can be refined order by order. 
For radiative corrections in the decay PHOTOS can be 
used. Its phase space is described, for example, 
in the journal version of~\cite{Nanava:2006vv} and is exact.
Approximations are made in the matrix element only.
Benchmark comparisons\footnote{The purpose of this type of tests may
vary. If two 
programs differ in their physics assumptions, it may help to control
the physics precision. If the physics assumptions are identical, but
the technical constructions differ, then the comparison checks the
correctness of the implementation of the algorithm. Finally, the
comparison of results from the same program, but installed on
different computers, may check the correctness of the code's
implementation in new software environments. Such comparisons, or just
the data necessary for comparisons, will be referred to as physical,
technical and installation benchmarks, respectively. They are
indispensable for the reliable use of Monte Carlo programs.} 
with other calculations, which are actually 
based on second-order matrix elements and exponentiation, found
excellent agreement \cite{Golonka:2005pn,Golonka:2006tw}. 

\subsection{Spin effects}
The lifetime of the $\tau$ lepton is 
orders of magnitude larger not only than its formation time in high
energy experiments, but also than the time scale of all phenomena
related to higher-order corrections such as bremsstrahlung. 

The separation of $\tau$ production and decay is excellent due to the small width 
of the  $\tau$ lepton. Its propagator can be well approximated by
a delta function for 
phase space and matrix elements.
%
The cross section for the process 
$ f \bar f \to \tau^+\tau^-  Y; 
\tau^+ \to X^+ \bar{\nu}_{\tau}; \tau^- \to l^- \nu_l \nu_{\tau}$ reads
 \[ 
{\rm d} \sigma = \sum_{spin }|{\cal M}|^2 
{\rm d}\Omega= \sum_{spin }|{\cal M}|^2 {\rm d}\Omega_{\rm prod} \; 
{\rm d}\Omega_{\tau^+} \; {\rm d}\Omega_{\tau^-}\,,\]
where $Y$ and  $X^{+}$ stand for particles produced together with
 the $\tau^+\tau^-$ and in the $\tau^+$ decay, respectively;
${\rm d}\Omega$,  ${\rm d}\Omega_{\rm prod}$, ${\rm
d}\Omega_{\tau^+}$, 
${\rm d}\Omega_{\tau^-}$ denote
the phase space in the original process, in production and
decay, respectively.

This formalism looks simple,
 but because of the over 20 $\tau$ decay channels 
there are more than 400 distinct processes.

Let us write the spin amplitude separated into the parts for 
$\tau$ pair production and decay:
\[ 
{\cal M}=\sum_{\lambda_1\lambda_2=1}^2{\cal
      M}_{\lambda_1\lambda_2}^{\rm prod} \; 
      {\cal M}_{\lambda_1}^{\tau^+}{\cal M}_{\lambda_2}^{\tau^-}.
\]
After integrating out the $\tau$ propagators, the formula for the 
cross section can be rewritten as
 \[ 
{\rm d} \sigma = \Bigl(\sum_{spin }|{\cal M}^{\rm prod}|^2 \Bigr)
 \Bigl(\sum_{spin }|{\cal M}^{\tau^+}|^2 \Bigr)
 \Bigl(\sum_{spin }|{\cal M}^{\tau^-}|^2 \Bigr) 
\]
\[
\hskip 7 mm \times\, wt \; {\rm d}\Omega_{\rm prod} \; {\rm
d}\Omega_{\tau^+} \; {\rm d}\Omega_{\tau^-},
\]
where
\[ 
wt=  \Bigl( \sum_{i,j=0,3} R_{ij} h_+^i h_-^j \Bigr),
\]
\[  
R_{00} =1,~~~ <wt>=1,~~~ 0 \le wt \le 4.
\]  

$R_{ij}$ can be calculated from  ${\cal M}_{\lambda_1\lambda_2}$, $h_+^i$ and $h_-^j$ from $ {\cal M}^{\tau^+}$ 
and $ {\cal M}^{\tau^-}$, respectively. 
Bell inequalities (related to the Einstein-Rosen-Podolsky 
paradox~\cite{Einstein:1935rr}) tell us that in general it is
impossible to rewrite $wt$ in 
 the following factorised form, $wt^{\rm factorized}$:
\[ 
 wt \ne wt^{\rm factorized}= \Bigl( \sum_{i,j=0,3} R^A_{i} h_+^i \Bigr)\Bigl( \sum_{i,j=0,3} 
R^B_{j}  h_-^j \Bigr),
\]
where $R^A_{i}$ and $R^B_{j}$ are four-component objects 
calculated from variables of the process of 
$\tau$ pair production.
In the Monte Carlo construction it is thus impossible to generate a
$\tau^+$ $\tau^-$ pair, where each of the two is in some quantum
state, and later to perform the decays of the $\tau^+$ and the
$\tau^-$ independently. 
This holds at all orders of the perturbative expansion. 
$\tau$ production and decay 
are correlated through spin effects, 
which can be represented by the well-behaved factor $wt$ introduced
previously. 
The above formulae do not lead to any loss of precision and hold in
presence of radiative corrections as well. Different options for the
formalism, based on these expressions, are used in Monte Carlo
programs and are basically well founded.
This should be confronted with processes where instead of $\tau$
leptons short-lived intermediate states are considered. 
Then, in general, ambiguities appear
and corrections proportional to the ratio of the resonance 
width to its mass (or other energy scales of the process resulting,
for example, from cut-offs) must be included. 
Interfering background diagrams may cause additional 
problems.  For details we refer 
to~\cite{koralb:1985,Jadach:1990mz,kkcpc:1999}.

\subsection{$\tau$ lepton production}
KORALB was 
published~\cite{Jadach:1984iy,Jadach:1985ac} more than 
{\it twenty years ago}. 
It included first-order QED corrections and complete  mass and spin 
effects. It turned out to be very useful, 
and still remains in 
broad use. On the other hand, some of its ingredients
are outdated
and do not match the present day requirements, even for technical tests.
For example the function PIRET(S), which describes the
real part of the photon hadronic vacuum polarisation as measured by the data
collected until the early 80's should be replaced by one of the 
new precise codes (see Section~\ref{sec:4} for details).

Unfortunately, this replacement
does not solve all
normalisation problems of KORALB. For example, it is well known 
that the one-loop corrections are not sufficient.
The two major improvements which were developed  
during the LEP era are the introduction 
of higher-order QED corrections into Monte Carlo simulation and a better way
to combine loop corrections with the rest of the field theory calculations.
For energies up to  10 GeV  (typical of the $B$ factories), 
the KKMC
Monte Carlo~\cite{kkcpc:1999} provides a realisation of the above 
improvements. This program includes higher-order QED
matrix elements with the help of exclusive exponentiation, and 
explicit matrix elements up to the second order. Also in this case 
the function calculating the 
vacuum polarisation must be replaced by a version appropriate for low energy
(see Section~\ref{sec:4}).
 
Once this is completed, and if the 
two-loop photon vacuum polarisation can be neglected, 
KORALB and KKMC can form a base for tests and studies 
of systematic errors for cross section normalisations 
at low energies. Using a strategy 
similar to the one for Bhabha scattering~\cite{Jadach:1991pj},
the results obtained in~\cite{Banerjee:2007is,Jadach:2000ir}
allow to expect a precision  of 0.35--0.45\% using KKMC
at Belle/BaBar energies.
 Certainly, a precision tag similar to that for 
linear colliders can also be achieved for lower energies. 
Work beyond ~\cite{Banerjee:2007is} and explained in that paper
would then be necessary.

\subsection{Separation into leptonic and hadronic current}

 The matrix element used in TAUOLA for semi-leptonic decays, 
 $\tau(P,s)\rightarrow\nu_{\tau}(N)X$,
\begin{equation}
{\cal M}=\frac{G}{\sqrt{2}}\bar{u}(N)\gamma^{\mu}(v+a\gamma_{5})u(P)J_{\mu}
\end{equation}
requires the knowledge of the hadronic current  $J_{\mu}$. The 
expression is easy to manipulate. One obtains:
\begin{eqnarray}
|{\cal M}|^{2}&=& G^{2}\frac{v^{2}+a^{2}}{2}
( \omega + H_{\mu}s^{\mu} ), \nonumber\\
\omega&=&P^{\mu}(\Pi_{\mu}-\gamma_{va}\Pi_{\mu}^{5}), \hskip 5 mm, \nonumber\\
H_{\mu}&=&\frac{1}{M}(M^{2}\delta^{\nu}_{\mu}-P_{\mu}P^{\nu})(\Pi_{\nu}^{5}-
\gamma_{va}\Pi_{\nu}), \nonumber\\
\Pi_{\mu}&=&2[(J^{*}\cdot N)J_{\mu}+(J\cdot N)J_{\mu}^{*}-(J^{*}\cdot J)
N_{\mu}], \nonumber\\
\Pi^{5\mu}&=&2~ {\rm Im} ~\epsilon^{\mu\nu\rho\sigma}
J^{*}_{\nu}J_{\rho}N_{\sigma}, \hskip 5 mm \nonumber\\
\gamma_{va}&=&-\frac{2va}{v^{2}+a^{2}}.
\end{eqnarray}
If the $\tau$ coupling is 
$v+a\gamma_{5}$ and $m_{\nu_{\tau}} \neq 0$ 
is allowed, one has
to add  to $\omega$ and $H_{\mu}$:
\begin{eqnarray}
\hat{\omega}&=&2\frac{v^{2}-a^{2}}{v^{2}+a^{2}}
m_{\nu}M(J^{*} \cdot J), \nonumber \\
\hat{H}^{\mu}&=&-2\frac{v^{2}-a^{2}}
{v^{2}+a^{2}}m_{\nu}~ {\rm Im}~\epsilon^{\mu\nu\rho\sigma}
J_{\nu}^{*}J_{\rho}P_{\sigma}.
\end{eqnarray}
The expressions are useful for Monte Carlo applications and 
are also calculable from first principles. The resulting
expression can be used to the precision level of the order of 0.2--0.3\%. 

In contrast to other parts, the hadronic current $J_{\mu}$ still
can not be calculated reliably from first principles. Some theoretical 
constraints need to be fulfilled, but in general it has to be obtained 
from experimental data. We will return to this point later 
(see Section \ref{hadroniccurrents}). 
 
\subsection{Bremsstrahlung in decays}
The PHOTOS Monte Carlo is widely used for generation
 of radiative corrections 
in cascade decays, starting from the early 
papers~\cite{Barberio:1990ms,Barberio:1994qi}. With time the precision 
of its predictions improved significantly, but the main principle 
remains the same.
Its algorithm is aimed to modify the content of the event record 
filled in with 
complete cascade decays at earlier steps of the generation.  {\tt PHOTOS} 
modifies the content of the event record; it adds additional photons 
to the decay vertices and at the same time modifies the kinematic 
configuration of other decay products.

One could naively expect that this strategy is bound 
to substantial approximations. 
 However, the algorithm is compatible with NLO calculations,
 leads to a complete coverage of the phase space for multi-photon 
final states and provides correct distributions in soft photon
limits. For more details of the program organisation and its phase 
space generation we address the reader to~\cite{Nanava:2006vv}.

The changes introduced over the last few years 
into the PHOTOS Monte Carlo program itself were rather small and
 the work concentrated on its theoretical foundations.
 This wide and complex subject goes beyond the scope of this Review
 and the interested reader can consult~\cite{Was:2008zz}, where  
 some of the topics are discussed.  
Previous tests  of two-body decays of the $Z$  
into a pair of charged leptons~\cite{Golonka:2006tw}
and a pseudoscalar $B$  into a pair of scalars~\cite{Nanava:2006vv} 
were recently supplemented~\cite{Photos_tests} with the study 
of $W^\pm \to l^\pm \nu \gamma$. 
The study of the process $\gamma^* \to \pi^+\pi^-$ is on-going~\cite{Xu}.
In all of these cases a universal kernel of PHOTOS was replaced 
with the one matching the exact first-order matrix element. 
In this way terms for the NLO/NLL 
level are implemented. The algorithm covers the full multi-photon 
phase space and it is exact in the infrared region of the phase space. 
 One should
not forget that PHOTOS generates weight-one events.

The results of all tests of PHOTOS with an NLO kernel are
at a sub-per mill level. No differences with 
benchmarks were found, even for samples of $10^9$ events. 
When simpler physics assumptions were used,
differences  between total rates at sub-per mill level 
 were observed  
 or they were matching 
a precision of the programs used for tests. 

This is very encouraging and points to the possible extension of the
approach beyond (scalar) QED, and in particular to QCD and/or models
with phenomenological Lagrangians for interactions of photons with
ha\-drons. For this work to be completed, spin amplitudes have 
to be further studied~\cite{vanHameren:2008dy}.
 
The  refinements discussed above
affect the practical side of simulations for $\tau$ physics only indirectly. 
Changes  in the kernels necessary for NLO  may remain as 
options  for tests only. They are available from the PHOTOS 
web page~\cite{Photos_tests}, but are not recommended for wider use. 
The corrections are small, and distributions visualising their size are available.
On the other hand, their use could be perilous, as it requires control of the
decaying particle spin state. 
It is known (see, e.g., ~\cite{Was:2004bk}) that this is not easy
because of technical reasons.

We will show later that radiative corrections do not provide a limitation 
in the quest for  improved precision of matching theoretical models 
to experimental data until issues discussed in subsection \ref{Thechallenges} 
are solved.

\subsection{Hadronic currents}
\label{hadroniccurrents}
So far all discussed contributions to the predictions were found 
to be controlled to the precision level of 0.5\% with respect to the decay 
rate under study.\footnote{This $0.5\%$ uncertainty is for QED 
radiative effects.  One should bear in mind other mechanisms involving 
the production of
photons, like, for example, the decay channel 
$\omega\to \pi \gamma$, which occurs with a probability of 
$(8.28 \pm 0.28)$\% and  does not belong to 
the category of radiative corrections. }

This is not the case for the hadronic current, which is the main source 
of our difficulties.
It can not be obtained from perturbative QCD as the energy scales involved are too
small. On the other hand, for the low energy limits the scale is too
large. 
Despite these difficulties one can obtain a theoretically clear object
if enough effort is devoted. This may lead to a better understanding of
the boundaries of the perturbative domain of QCD as well.

The unquestionable property which hadronic currents must fulfil 
is Lorentz invariance. For example, if the final state consists of
three scalars with momenta $p_1$, $p_2$, $p_3$, respectively, it must
take the form 
{\small{
\begin{eqnarray}
\label{fiveF}
&J^\mu & =N \bigl\{T^\mu_\nu \bigl[ c_1 (p_2-p_3)^\nu F_1  + c_2 (p_3-p_1)^\nu
 F_2  \nonumber \\
 &+& c_3  (p_1-p_2)^\nu F_3 \bigr] + c_4  q^\mu F_4  -{ i  c_5\epsilon^\mu_{.\ \nu\rho\sigma} \over 4 \pi^2 f_\pi^2}     
 p_1^\nu p_2^\rho p_3^\sigma F_5      \bigr\}\,,
\end{eqnarray}
}}
where  $T_{\mu\nu} = g_{\mu\nu} - q_\mu q_\nu/q^2$ is the transverse
projector and $q=p_1+p_2+p_3$.
The functions $F_i$ depend on three variables that can be chosen 
as $q^2=(p_1+p_2+p_3)^2$
and two of the following three, $s_1=(p_2+p_3)^2$, $s_2=(p_1+p_3)^2$, 
$s_3=(p_1+p_2)^2$.
 This form is obtained  from Lorentz invariance only. 

Among the first four  hadronic structure functions 
($F_1$, $F_2$, $F_3$, $F_4$),   only three  are independent.
We leave the structure function $F_4$ in the basis because, 
neglecting the pseudoscalar resonance production mechanism, 
the contribution due to $F_4$ is negligible ($\sim
m_{\pi}^2/q^2$)~\cite{GomezDumm:2003ku} and (depending on the decay
channel) one of $F_1$, $F_2$ and $F_3$ drops out, exactly as it is in
TAUOLA since long. 

In each case, 
the number of independent functions  is  four (rather than  five) 
and not larger 
than the dimension of our space-time.
That is why  
the projection operators can be defined, for two- and three-scalar 
final states.
Work in that direction has already been done in 
Ref.~\cite{Kuhn:1992nz} and then 
implemented in tests of TAUOLA too.
Thanks to such a method, hadronic currents can be obtained from data 
without any 
need of phenomenological assumptions. Since long such methods  were useful for 
data analysis, but only in part. Experimental samples were simply too small.

At present, for high statistics and precision the method may  be revisited. 
That is why it is of great interest to verify whether 
detector deficiencies will invalidate the method or if adjustments due
to incomplete phase space coverage are necessary. We will return to that 
question later. In the mean time let us return to other theoretical 
considerations which  constrain the form of hadronic currents, 
but not always  to the precision of today's data.

\subsection{The resonance chiral approximation and its result for the currents}
Once the allowed Lorentz structures are determined and a proper 
minimal set of them is chosen, one should impose the QCD symmetries 
valid at low energies. The chiral symmetry of massless QCD allows
 to develop an 
effective field theory description valid for momenta much smaller 
than the $\rho$ mass, $\chi PT$~\cite{Gasser:1983yg,Gasser:1984gg}.\\ 
Although $\chi PT$ cannot provide predictions valid over the full $\tau$ 
decay phase space, it constrains the form and the normalisation of the 
form factors in such limits.\\
The model, proposed in~\cite{Kuhn:1990ad} for $\tau$ decaying to pions
  and used also 
 for extensions to other 
decay channels, employs weighted products of Breit-Wigner functions to 
take into account resonance exchange. The form factors
used there have the right chiral limit at LO. However, as it 
was demonstrated in~\cite{GomezDumm:2003ku},
they  do not reproduce the  NLO chiral limit.
 \\
The step towards incorporating the right low-energy limit up to  NLO
 and the contributions from meson resonances which reflect the experimental
  data was
  done within Resonance Chiral Theory 
($R\chi T$~\cite{Ecker:1988te,Ecker:1989yg}). 
 The current state-of-the-art for the 
hadronic form factors ($F_i$) appearing in the $\tau$ decays is 
described in~\cite{Dumm:2009kj,Dumm:2009va}. Apart from the correct low energy
properties, 
it includes the right falloff~\cite{Brodsky:1973kr,Lepage:1980fj} 
at high energies.\\ 
The energy-dependent imaginary parts
in the propagators 
of the vector and the axial-vector mesons, 
$~1/(m^2-q^2- i m\Gamma(q^2))$, were
  calculated in ~\cite{GomezDumm:2000fz}  at one-loop,
exploiting the optical theorem that relates the appropriate hadronic 
matrix elements of $\tau$ decays and the cuts with on-shell mesons 
in the (axial-) vector-(axial-) vector correlators.\\
This formalism has been shown to successfully describe the invariant
mass spectra of
experimental data in $\tau$ decays for the following hadronic 
systems: $\pi\pi$ \cite{Guerrero:1997ku,Pich:2001pj,Pich:2002ne}, 
$\pi K$ \cite{Jamin:2006tk,Jamin:2008qg}, 
$3\pi$ \cite{GomezDumm:2003ku,Dumm:2009kj,Dumm:2009va,Roig:2008xt} and 
$KK\pi$\cite{Dumm:2009kj,Roig:2008xt}. 
Other channels will be worked out along the same lines.
\\
It has already been checked that the $R\chi T$ results 
provide also a good description of the three-meson processes 
$\Gamma(\tau \to 3\pi \nu_\tau)$~\cite{Barate:1998uf} and 
$\sigma(e^+e^-\to KK\pi)_{I=1}$~\cite{Aubert:2007ym}.\\
Both the spin-one resonance widths and the form factors of 
the decays $\tau^- \to (\pi\pi,\ \pi K,\ 3\pi,\ KK\pi)^- \nu_\tau$ 
computed within $R\chi T$ are being implemented in TAUOLA only now.

Starting from a certain precision level, the predictions, 
like  the ones presented above,
may turn out to be not sufficiently precise.
Nonetheless, even in such a case they can provide some essential 
constraints on the form of the functions $F_i$. 
Further refinements will  require  large and combined efforts of 
experimental and theoretical physicists. We will elaborate on possible 
technical solutions later in the review.
Such attempts turned out to be difficult in the past
and a long time was needed 
for parametrisations given 
in~\cite{Golonka:2003xt} to become
public. Even now they are semi-official and are not based on the
final ALEPH and/or CLEO data.

\subsection{Isospin symmetry of the hadronic currents}

If one neglects quark masses, QCD is invariant under a 
transformation replacing quark flavours.
As a consequence, hadronic currents describing vector $\tau$ decays
($2\pi,\ 4\pi,\ \eta\pi\pi,\ \ldots$) and low energy
$e^+e^-$ annihilation into corresponding iso\-vector final states are
related and can be obtained from one another~\cite{Tsai:1971vv,Thacker:1971hy}.
This property, often referred to as conservation of the vector current
(CVC) in $\tau$ decays, results in the possibility to predict invariant
mass distributions of the hadronic system, as well as the corresponding
branching fractions in $\tau$ decays using $e^+e^-$ data.
 A systematic check of these predictions showed
that at the (5--10)\% level they work rather well~\cite{Eidelman:1990pb}.

In principle, the corrections due to mass  and
charge differences between $u$ and $d$ quarks are not expected 
to provide significant and impossible to control 
effects~\cite{Cirigliano:2001er,Cirigliano:2002pv}. However,
the high-precision data of the CLEO \cite{Anderson:1999ui},
 ALEPH \cite{Schael:2005am}, 
 OPAL \cite{Ackerstaff:1998yj},  Belle \cite{Fujikawa:2008ma},
 CMD-2 \cite{Akhmetshin:2003zn,Aulchenko:2006na,Akhmetshin:2006wh,Akhmetshin:2006bx}, SND \cite{Achasov:2006vp} and KLOE \cite{:2008en}
 collaborations in the $2\pi$ channel challenged this statement, and
 as it was shown 
 in~\cite{Davier:2002dy,Jegerlehner:2003qp,Jegerlehner:2003rx,Ghozzi:2003yn,Davier:2003pw,Jegerlehner:2008zza,Jegerlehner:2009ry} that 
 the
spectral functions for $\tau^- \to \pi^-\pi^0\nu_{\tau}$ significantly differ
 from those obtained using $e^+e^- \to \pi^+\pi^-$ data.
 Some evidence for a similar
discrepancy is also observed in the  $\tau^- \to 2\pi^-\pi^+\pi^0\nu_{\tau}$
decay  \cite{Davier:2005xq,Druzhinin:2007cs,Czyz:2008kw}.
 This effect remains unexplained. The magnitude of the isospin-breaking
corrections has been updated recently, making the discrepancy in the
$2\pi$ channel smaller~\cite{Davier:2009ag}.

These CVC based relations were originally used in the TAUOLA
form factors parametrisation,
but they were often modified to improve fits to the data.
Let us point here to an example where experimental $e^+e^-$ data were
used for the model of the $\tau \to 4\pi \nu_{\tau}$ decay 
channels~\cite{Bondar:2002mw}.
In this case, only a measurement of the distribution
  in the total invariant mass
of the hadronic system was available. 
This is not enough to fix the distribution over the multidimensional phase space.
For other dimensions one had to rely on
phenomenological models or other experiments.
In the future, this may not be necessary, but will always 
remain as a method of benchmarks construction.

\subsection{The challenges}
\label{Thechallenges}

As we have argued before, refined techniques for fits,
 involving simultaneous fits to many $\tau$ decay channels,  are necessary
 to improve the phenomenological description 
of $\tau$ decays. Complex backgrounds (where each channel contributes 
to signatures of other decay modes as well),
different sensitivities of experiments for measurements of some 
angular distributions within the same hadronic system, and sometimes even 
an incomplete reconstruction of final states, 
are the main cause of this necessity.
Moreover, theoretical models based on the Lagrangian approach
simultaneously describe more than one  $\tau$ decay channel with the same
set of parameters, and only simultaneous fits allow to establish
 their experimental constraints in a consistent way.
Significant efforts are thus necessary and
  close collaboration between phenomenologists and 
experimental physicists is indispensable.
 As a result,
techniques of automated calculations of hadronic currents may become 
 necessary~\cite{Korchin}. 

\subsection{Technical solutions for fits}

For the final states of up to three scalars, 
the use of projection operators~\cite{Kuhn:1992nz} is popular since 
long~\cite{Davier:2005xq}. 
It enables, at least in principle, to obtain form factors used in
hadronic currents directly from the data, for one scalar function
defined in Eq.~(\ref{fiveF}) at a time. Only recently experimental samples became sufficiently large.
However, to exploit this method one may have to 
improve it first by systematically including the effects of a limited detector 
acceptance.  
Implementation of the projection operators into packages like
MC-TESTER~\cite{Davidson:2008ma} may be useful. Efforts
in that direction are being pursued now\footnote{ This may help to
embed the method in the modern software for fits, see, e.g.,
\cite{Brun:2008zza}.}~\cite{OlgaArtur}.

On the theoretical side one may need to choose  
predictions from many  models, before a sufficiently good
agreement with data will be achieved. 
Some automated methods of calculations may then become 
useful~\cite{Automated}. This is especially important 
for hadronic multiplicities larger than three,  when 
projector operators have never been defined.

Certain automation of the methods is
thus advisable. To discriminate from the broad spectrum of choices, 
new methods of data analysis may become useful~\cite{Hocker:2007ht}. 
Such methods may require simulating samples
of events where several options for the matrix element calculation 
are used simultaneously.\footnote{Attempts to code such methods into TAUOLA, 
combined with programs for $\tau$ pair production and experimental 
detector environment, were recently
performed~\cite{VladimirTomasz}, but they were applied so far 
as prototypes only, see Fig. 1  of Ref.~\cite{Was:2009iy}.}

The neutrino coming from $\tau$ decays escapes detection 
and as a result the $\tau$ rest frame 
can not be reconstructed. Nevertheless, as was 
shown in Ref.~\cite{Kuhn:1992nz}, angular distributions can be used 
for the construction of projection operators, which allow the extraction
 of the hadronic structure functions from the data.
 This is possible as they
depend on $s_1$, $s_2$ and $q^2$ only.

A dedicated module for the MC-TESTER~\cite{Davidson:2008ma}, 
implementing the moments of different angular functions defined in 
Eqs.~(39)--(47) of Ref.~\cite{Kuhn:1992nz}, is under development.
The moments are proportional to combinations of the type 
$\alpha |\mathrm{F_{i}}|^{2} + \beta |\mathrm{F_{j}}|^{2} + 
\gamma \mathrm{Re(F_{i} F_{j}^{*}})$,
where the coefficients $\alpha$, $\beta$ and $\gamma$ are functions 
of hadron four-momentum components in the hadronic rest frame.
Preliminary results obtained with large statistics of five million 
$\tau \rightarrow a_{1} \nu_{\tau} \rightarrow 3\pi \nu_{\tau}$ decays, 
and assuming vanishing $\mathrm F_{3}$ and $\mathrm F_{5}$ form factors,
show that it is possible to extract 
$\mathrm |F_{1}|^{2}$, $\mathrm |F_{2}|^{2}$ and 
$\mathrm |F_{1}\cdot F_{2}^{*}|^{2}$ as functions of 
$s_{1}$, $s_{2}$ and $Q^{2}$. 
This extraction requires solving a set of equations. Since the
solution is sensitive to the precision of the estimation of 
the moments entering the equation, large data samples of the order of 
$O (10^{6}-10^{7})$ are necessary. The calculation of the moments 
also requires the knowledge of the initial $\sqrt{s}$ of the 
$\tau$ pair, which makes the analysis sensitive to 
initial state radiation (ISR) effects. The same studies show 
that the analysis is easier if one, instead of extracting
 the form factors $\mathrm |F_{i}|^{2}$, compares the 
moments obtained from the experimental data with theoretical
predictions. Such a comparison does not require repetition of the Monte Carlo
simulation of $\tau$ decays with different form factors, and only 
the calculation of combinations
of  $|\mathrm{F_{i}}|^{2}$ and $\mathrm{Re(F_{i} F_{j}^{*}})$ is necessary. 
This is much simpler than 
comparing the kinematic distributions obtained from data
with distributions coming from Monte Carlo simulations
  with various theoretical models. 
Further complications, for example, due to the presence 
of an initial state bremsstrahlung or an incomplete acceptance of decay 
phase space, were not yet taken into account. 

\subsection{Prospects}

Definitely the improvements of $\tau$ decay simulation packages and 
fit strategies
are of interest for phenomenology of low energy.
As a consequence, their input for such domains 
like phenomenology of the muon $g-2$ or $\alpha_{\rm QED}$,
$\alpha_{\rm QCD}$ 
and their use in constraints of new physics would improve. 

In this section, let us argue if  possible benefits for LHC 
phenomenology may arise from a better 
understanding of $\tau$ decay channels in measurements as well. In 
the papers~\cite{Bower:2002zx,Desch:2003rw} it was shown that spin effects 
can indeed be useful to measure 
properties of the Higgs boson such as parity. Moreover, 
such methods were verified to work well when detector effects 
as proposed for a future linear collider 
were taken into account.
Good control of the decay properties is helpful. For example,
in Ref.~\cite{Privitera:1993pr}  it was shown that for
the $\tau \to \ a_1 \nu_{\tau} \to 3\pi\nu_{\tau}$ decay the
sensitivity to the $\tau$ polarisation increases about four times when
all angular variables are used compared with the usual ${\rm
d}\Gamma/{\rm d}q^2$, see also~\cite{Kuhn:1995nn}. 

Even though $\tau$ decays provide some of the  most prominent signatures
for the LHC physics program, see, e.g., Ref.~\cite{RichterWas:2009wx},
for some time it was  expected that methods exploiting detailed
properties of $\tau$ cascade decays are not practical
for LHC studies. Thanks to
efforts on reconstruction of $\pi^0$ and  $\rho$ 
invariant mass peaks, this opinion evolves. 
Such work was done for studies of the CMS ECAL
detector inter-calibration~\cite{Bayatian:2006zz}, and 
in a relatively narrow $p_T$ range (5--10 GeV) some potentially 
encouraging results were obtained.
Some work in context of searches for new particles 
started recently~\cite{Nattermann:2009gh}. There, improved knowledge of
distinct $\tau$ decay modes may become important at a certain point.

One can conclude that the situation is  similar to that at the start
 of LEP, and
 some control of all $\tau$ decay channels is  important.
Nonetheless,  only if detector  studies of $\pi^0$ and  $\rho$ reconstruction
will provide positive results, the gate to improve the 
sensitivity of $\tau$ spin measurements with most of its decay modes, 
as at LEP~\cite{:2008zzm,Aad:2009wy,Heister:2001uh}, will be open. 
At this moment, however, it is difficult to judge about the importance
of such improvements in the description of $\tau$ decays for LHC perspectives.
The experience of the first years of LHC must be consolidated first.
In any case such an activity is important for the physics of future 
Linear Colliders.

\subsection{Summary}

We have shown  that the most urgent challenge in the quest
for a better understanding 
of $\tau$ decays is the development of efficient techniques for fitting 
multidimensional distributions, which take into account
 realistic detector conditions.
  This includes cross 
contamination of different $\tau$ decay modes, their respective signatures 
 and detector acceptance effects, which have to be simultaneously 
taken into account
 when fitting experimental data. 
 Moreover, at the current experimental precision, theoretical concepts have 
to be reexamined. In contrast to the
past, the precision of predictions based on chiral
Lagrangians and/or isospin symmetry 
can not be expected to always match the precision of the data. 
The use of model-independent data analyses
should be encouraged whenever possible in 
realistic conditions.

Good understanding of $\tau$ decays is crucial for understanding 
the low energy regime of strong interactions and
the matching between the non-perturbative and the perturbative domains. 
Further work on better simulations of $\tau$ decays at the LHC is needed 
 to improve its potential
to study processes of new physics, especially in the Higgs sector.
In addition, an accurate simulation of $\tau$ decays is important   
for the control 
of backgrounds for very rare decays.
 For the project to be successful, this should lead  
 to the encapsulation of our  knowledge on $\tau$ decays
  in form of 
a Monte Carlo library to be used by  low-energy as well as  
high-energy applications.

\section{Vacuum polarisation}
\label{sec:4}

\def\lapproxeq{\lower .7ex\hbox{$\;\stackrel{\textstyle <}{\sim}\;$}}
\def\Re{\mathop{\rm Re}\nolimits} \def\Im{\mathop{\rm Im}\nolimits}

%

\subsection{Introduction}
The vacuum polarisation (VP) of the photon is a quantum effect which
leads, through renormalisation, to the scale dependence (`running')
of the electromagnetic coupling, $\alpha(q^2)$. It therefore plays an
important role in many physical processes and its knowledge is crucial
for many precision analyses. A prominent example is the precision
fits of the Standard Model as performed by the electroweak working
group, where the QED coupling $\alpha(q^2 = M_Z^2)$ is the least well
known of the set of fundamental parameters at the $Z$ scale, $\{G_{\mu}, M_Z,
\alpha(M_Z^2)\}$. Here we are more concerned about the VP at lower
scales as it enters all photon-mediated hadronic cross sections. 
These are used, e.g., in the determination of the strong coupling
$\alpha_s$, the charm and bottom quark masses from $R_{\rm had}$ as
well as in the evaluation of the hadronic contributions to the muon
$g-2$ and $\alpha(q^2)$ itself.
It also appears in Bhabha scattering in higher orders of perturbation
theory needed for a precise determination of the luminosity. 
It is hence clear that VP also has to be included in the corresponding
Monte Carlo programs.

In the following we shall first define the relevant notations, then
briefly discuss the calculation of the leptonic and hadronic VP
contributions, before comparing available VP parametrisations. 

\begin{figure}[h]
\begin{center}
\includegraphics[width=5cm]{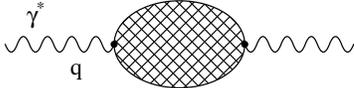}
\end{center}
\vspace{-0.2cm}
\caption{\label{fig:VP} Photon vacuum polarisation $\Pi(q^2)$.} 
\end{figure}
Conventionally the vacuum polarisation function is denoted by $\Pi(q^2)$
where $q$ is a space- or time-like momentum. The shaded blob in
Fig.~\ref{fig:VP} stands for all possible one-particle irreducible leptonic or hadronic
contributions. The full photon propagator is then the sum of the bare
photon propagator and arbitrarily many iterations of VP insertions,
\begin{eqnarray}
\mbox{full photon propagator} \ \sim\ \frac{-i}{q^2}\,\cdot
\qquad\qquad\qquad\qquad & &\nonumber\\
\quad\left( 1\,+\,\Pi\,+\,\Pi\cdot\Pi\,+\,\Pi\cdot\Pi\cdot\Pi\,+\,\ldots\right)\,.&&
\label{eq:photonprop}
\end{eqnarray}
The Dyson summation of the real part of the one-particle irreducible
blobs then defines the effective QED coupling
\begin{equation}
\qquad\alpha(q^2)\ =\ \frac{\alpha}{1-\Delta\alpha(q^2)}\ =
\ \frac{\alpha}{1-{\rm Re} \Pi(q^2)}\,,
\label{eq:defalphaqed}
\end{equation}
where $\alpha \equiv \alpha(0)$ is the usual fine structure constant,
$\alpha \sim 1/137$.
It is determined most precisely through the anomalous magnetic moment
of the electron, $a_e$, as measured by the Harvard group to an amazing $0.24$
ppb~\cite{Hanneke:2008tm}, in agreement with less precise
determinations from caesium and rubidium atom experiments. The most
precise value for $\alpha$, which includes the updated calculations
of $O (\alpha^4)$ contributions to $a_e$~\cite{Aoyama:2007mn},
is given by $1/\alpha = 137.035\, 999\, 084\, (51)$. 

By using Eq.~(\ref{eq:defalphaqed}) we have defined $\Pi$ to include
the electric charge squared, $e^2$ for leptons, but note that
different conventions are used in the literature, and sometimes $\Pi$
is also defined with a different overall sign. 

Equation~(\ref{eq:defalphaqed}) is the usual definition of the running
effective QED coupling and has the advantage that one obtains a real
coupling. However, the imaginary part of the VP function $\Pi$ is
completely neglected, which is normally a good approximation as the
contributions from the imaginary part are formally suppressed. This
can be seen, e.g., in the case of the `undressing' of the experimentally
measured hadronic cross section $\sigma_{\rm had}(s)$. 
The measured cross section $e^+e^- \to \gamma^* \to
hadrons$ contains $|\mbox{full photon propagator}|^2$, i.e. the
modulus squared of the infinite sum (\ref{eq:photonprop}). Writing
$\Pi = e^2 (P+iA)$ one easily sees that 
\begin{eqnarray}
& & |1 + e^2 (P+iA) + e^4 (P+iA)^2 + \ldots|^2 \ =\ \nonumber\\
& & \ 1\, +\, e^2\, 2 P\, +\, e^4\, ( 3 P^2 - A^2)\, +\, e^6\, 4P(P^2
- A^2)\, +\, \ldots \nonumber
\end{eqnarray}
and that the imaginary part $A$ enters only at order $O (e^4)$
compared to $O (e^2)$ for the leading contribution from the real
part $P$. To account for the imaginary part of $\Pi$ one may therefore
apply the summed form of the `(un)dressing' factor with the relation
\begin{equation}
\sigma_{\rm had}(s) = \frac{\sigma^0_{\rm had}(s)}{|1-\Pi|^2}
\label{eq:defsigma0}
\end{equation}
instead of the traditionally used relation with the real effective coupling,
\begin{equation}
\sigma_{\rm had}(s) = \sigma^0_{\rm had}(s) \left(\frac{\alpha(s)}{\alpha}\right)^2\,.
\label{eq:defsigma0re}
\end{equation}
We shall return to a comparison of the different approaches below for
the case of the hadronic VP.

It should be noted that the summation breaks down and hence can not be
used if $|\Pi(s)| \sim 1$. This is the case if $\sqrt{s}$ is very
close to or even at narrow resonance energies. In this case one can
not include the narrow resonance in the definition of the effective
coupling but has to rely on another formulation, e.g. through a
Breit-Wigner propagator (or a narrow width approximation with a
delta-function). For a discussion of this issue see~\cite{hmnt}.
Also note that the VP summation covers only the class of one-particle
irreducible diagrams of factorisable bubbles depicted in
Fig.~\ref{fig:VP}. This includes photon radiation within and between
single bubbles, but clearly does not take into account
higher-order corrections from initial state radiation or
initial-final state interference effects in $e^+e^-\to hadrons$.

As will be discussed in the following, leptonic and hadronic
contributions to $\Delta\alpha$ are normally calculated separately and
then added, $\Delta\alpha(q^2) = \Delta\alpha_{\rm lep}(q^2) +
\Delta\alpha_{\rm had}(q^2)$. While the leptonic contributions can be
predicted within perturbation theory, the precise determination of the
ha\-dronic contributions relies on a dispersion relation using
experimental data as input.

\subsection{Leptonic contributions}
The leptonic contributions $\Delta\alpha_{\rm lep}$ have been
calculated to sufficiently high precision. The leading order (LO) and
next-to-leading order (NLO) contributions are known as analytic
expressions including the full mass dependence~\cite{Kallen:1955fb},
where LO and NLO refer to the expansion in terms of $\alpha$. The
next-to-next-to-leading order (NNLO) contribution is available as
an expansion in terms of $m_{\ell}^2/q^2$~\cite{Steinhauser:1998rq},
where $m_{\ell}$ is the lepton mass. To evaluate $\Delta \alpha_{\rm
  lep}(q^2)$ for $|q^2| \lapproxeq m_\tau^2$, this expansion is not
appropriate, but this is exactly the region where the hadronic
uncertainties are dominant. Also from the smallness of the NNLO
contribution, we conclude that we do not need to further improve the
leptonic contributions beyond this approximation. 

The evaluation of the LO contribution is rather simple, and we briefly
summarise the results below. Hereafter, it is understood that we
impose the renormalisation condition $\Pi(0)=0$ on $\Pi(q^2)$. For
$q^2 < 0$, the VP function reads 
\begin{eqnarray}
 \Pi(q^2)
& = &
 -\frac{e^2}{36\pi^2} 
 \Big( 5-12\eta\\
& & + 3 ( -1 + 2 \eta ) \sqrt{1+4 \eta} 
      \,\ln \frac{\sqrt{1+4 \eta}+1}{\sqrt{1+4 \eta}-1}
 \Big)\,,\nonumber
\label{eq:LOleptonic} 
\end{eqnarray} 
where $\eta \equiv m_{\ell}^2/(-q^2)$. For $0\le q^2 \le 4 m_{\ell}^2$
one obtains
\begin{eqnarray}
 \Pi(q^2)
& = &
 -\frac{e^2}{36\pi^2} 
 \Big( 
     5-12\eta\\
& & + 3 ( -1 + 2 \eta ) \sqrt{- 1 - 4 \eta}
      \,\arctan \frac{\sqrt{-1-4 \eta}}{-1-2 \eta}
 \Big)\,,\nonumber
\end{eqnarray}
and for $q^2 \ge 4 m_{\ell}^2$
\begin{eqnarray}
 \Pi(q^2)
& = &
 -\frac{e^2}{36\pi^2} 
 \Big( 
     5-12\eta + 3 ( -1 + 2 \eta ) \sqrt{1+4 \eta}\\
 & & 
      \cdot\ln \frac{1+\sqrt{1+4 \eta}}{1-\sqrt{1+4 \eta}}
 \Big) 
- \frac{i\,e^2}{12\pi} ( 1 - 2 \eta ) \sqrt{1+4 \eta}\,.\nonumber
\end{eqnarray}
An easily accessible reference which gives the NLO contributions is,
for instance, Ref.~\cite{Djouadi:1993ss,PhysRevD.53.4111}. As
mentioned above, the NNLO contribution is given in
Ref.~\cite{Steinhauser:1998rq}. For all foreseeable applications the
available formulae can be easily implemented and provide a sufficient
accuracy. While the uncertainty from $\alpha$ is of course completely
negligible, the uncertainty stemming from the lepton masses is only
tiny. Therefore the leptonic VP poses no problem.

\subsection{Hadronic contributions}
In contrast to the leptonic case, the hadronic VP $\Pi_{\rm had}(q^2)$
can not be reliably calculated using perturbation theory. This is
clear for time-like momentum transfer $q^2 > 0$, where, via the
optical theorem $\Im\Pi_{\rm had}(q^2) \sim \sigma(e^+e^- \to
hadrons)$ goes through all the resonances in the low energy
region. However, it is possible to use a dispersion relation to
obtain the real part of $\Pi$ from the imaginary part. The dispersion
integral is given by
\begin{equation}
\Delta\alpha_{\rm had}^{(5)}(q^2)\ =\ 
  -\frac{q^2}{4\pi^2\alpha}\, {\rm P}\int_{m_{\pi}^2}^\infty
\frac{\sigma_{\rm had}^0(s)\,{\rm d}s}{s-q^2}\,,
\label{eq:defdelalhad5}
\end{equation}
where $\sigma_{\rm had}^0(s)$ is the (undressed) hadronic cross
section which is determined from experimental data. Only away from
hadronic resonances and (heavy) quark thresholds one can apply
perturbative QCD to calculate $\sigma_{\rm had}^0(s)$. In this region
the parametric uncertainties due to the values of the quark masses and
$\alpha_s$, and due to the choice of the renormalisation scale, are
small. Therefore the uncertainty of the hadronic VP is dominated by
the statistical and systematic uncertainties of the experimental data
for $\sigma_{\rm had}^0(s)$ used as input in (\ref{eq:defdelalhad5}). 

Note that the dispersion integral (\ref{eq:defdelalhad5}) leads to a
smooth function for space-like momenta $q^2 < 0$, whereas in the
time-like region it has to be evaluated using the principal value description
and shows strong variations at resonance energies, as demonstrated
e.g. in Fig.~\ref{fig:dalphaFJ03}.
In Eq.~(\ref{eq:defdelalhad5}) $\Delta\alpha_{\rm had}^{(5)}$ denotes the
five-flavour hadronic contribution. At energies we are interested in,
i.e. far below the $t\bar t$ threshold, the contribution from the top
quark is small and usually added separately. The analytic expressions
for $\Delta\alpha^{\rm top}(q^2)$ obtained in perturbative QCD are
the same as for the leptonic contributions given above, up to
multiplicative factors taking into account the top quark charge and the
corresponding SU(3) colour factors, which read $Q_t^2 N_c$ at LO and
$Q_t^2 \frac{N_c^2-1}{2N_c}$ at NLO. 

Contributions from narrow resonances can easily be treated using the
narrow width approximation or a Breit-Wigner form. For the latter one
obtains
\begin{equation}
\Delta\alpha^{\rm Breit-Wigner}(s) =  \frac{3\Gamma_{ee}}{\alpha
  M}\,\frac{s(s-M^2-\Gamma^2)}{(s-M^2)^2+M^2\Gamma^2}\,, 
\label{eq:BW}
\end{equation}
with $M$, $\Gamma$ and $\Gamma_{ee}$ the mass, total and electronic
width of the resonance. For a discussion of the undressing of
$\Gamma_{ee}$ see~\cite{hmnt}. 

Although the determination of $\Delta\alpha_{\rm had}^{(5)}(q^2)$ via the
dispersion integral~(\ref{eq:defdelalhad5}) may appear straightforward,
in practice the data combination for $\sigma_{\rm had}^0(s)$ is far from
trivial. In the low energy region up to about $1.4 - 2$ GeV many data sets
from the different hadronic exclusive final states (channels) from
various experiments have to be combined, before the different channels
which contribute incoherently to $\sigma_{\rm had}^0(s)$ can be
summed. For higher energies the data for the fast growing number of possible
multi-hadronic final states are far from complete, and instead
inclusive (hadronic) measurements are used. For the details of the
data input, the treatment of the data w.r.t. radiative corrections,
the estimate of missing threshold contributions and unknown
subleading channels (often via isospin correlations) and the
combination procedures we refer to the publications of the different
groups cited below.

In the following we shall briefly describe and then compare the
evaluations of the (hadronic) VP available as para\-me\-tri\-sations or
tabulations from different groups.

\subsection{Currently available VP parametrisations}
For many years Helmut Burkhardt and Bolek Pietrzyk have been providing
the Fortran function named REPI for the leptonic and hadronic
VP \cite{Burkhardt:1981jk,Burkhardt:1982kr,Burkhardt:1989ky,Burkhardt:1995tt,Burkhardt:2001xp}. While the leptonic VP is coded in analytical form
with one-loop accuracy, the hadronic VP is given as a very compact
parametrisation in the space-like region, but does not cover the
time-like region. For their latest update
see~\cite{Burkhardt:2005se}. The code can be obtained from Burkhardt's
web-pages which contain also a short introduction and a list of older
references, see \newline{\tt http://hbu.web.cern.ch/hbu/aqed/aqed.html}.

\begin{figure} \begin{center}
\includegraphics[bb=0 0 382 245,width=8.9cm]{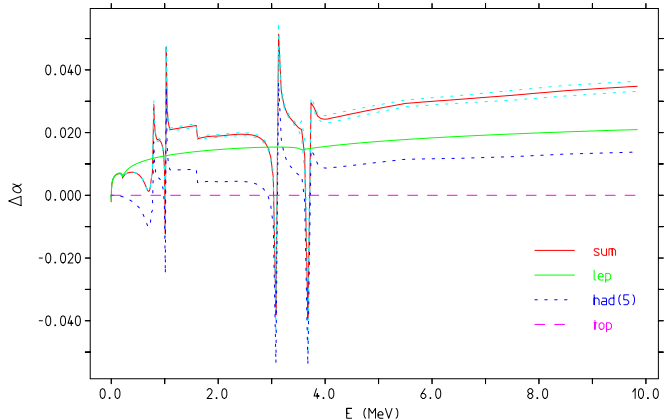}
\end{center}
\vspace{-0.1cm}
\caption{Different contributions to $\Delta\alpha(s)$ in the time-like
  region as given by the routine from Fred Jegerlehner. Figure
  provided with the package {\tt alphaQED.uu} from his homepage.} 
\label{fig:dalphaFJ03}
\end{figure}

Similarly, Fred Jegerlehner has been providing a package of
Fortran routines for the running of the effective QED
coupling \cite{Jegerlehner:1985gq,Eidelman:1995ny,Jegerlehner:2003ip,Jegerlehner:2003qp,Jegerlehner:2006ju,Jegerlehner:2008rs}.
It provides leptonic and hadronic
VP both in the space- and time-like region. For the leptonic VP the
complete one- and two-loop results and the known high energy
approximation for the three-loop corrections are included. The
hadronic contributions are given in tabulated form in the subroutine
HADR5N. The full set of routines can be downloaded from
Jegerlehner's web-page \newline{\tt
  http://www-com.physik.hu-berlin.de/$\sim$fjeger/}. The version available
from there is the one we use in the comparisons below and was last
modified in November 2003. It will be referred to as J03 in the
following. An update is in progress and other versions
may be available from the author upon request. Note that for quite
some time his routine has been the only available code for the time-like
hadronic VP. Fig.~\ref{fig:dalphaFJ03} shows the leptonic and hadronic
contributions together with their sum as given by Jegerleh\-ner's routine.

\begin{figure} \begin{center}
\includegraphics[bb=0 0 530 380,width=8.8cm]{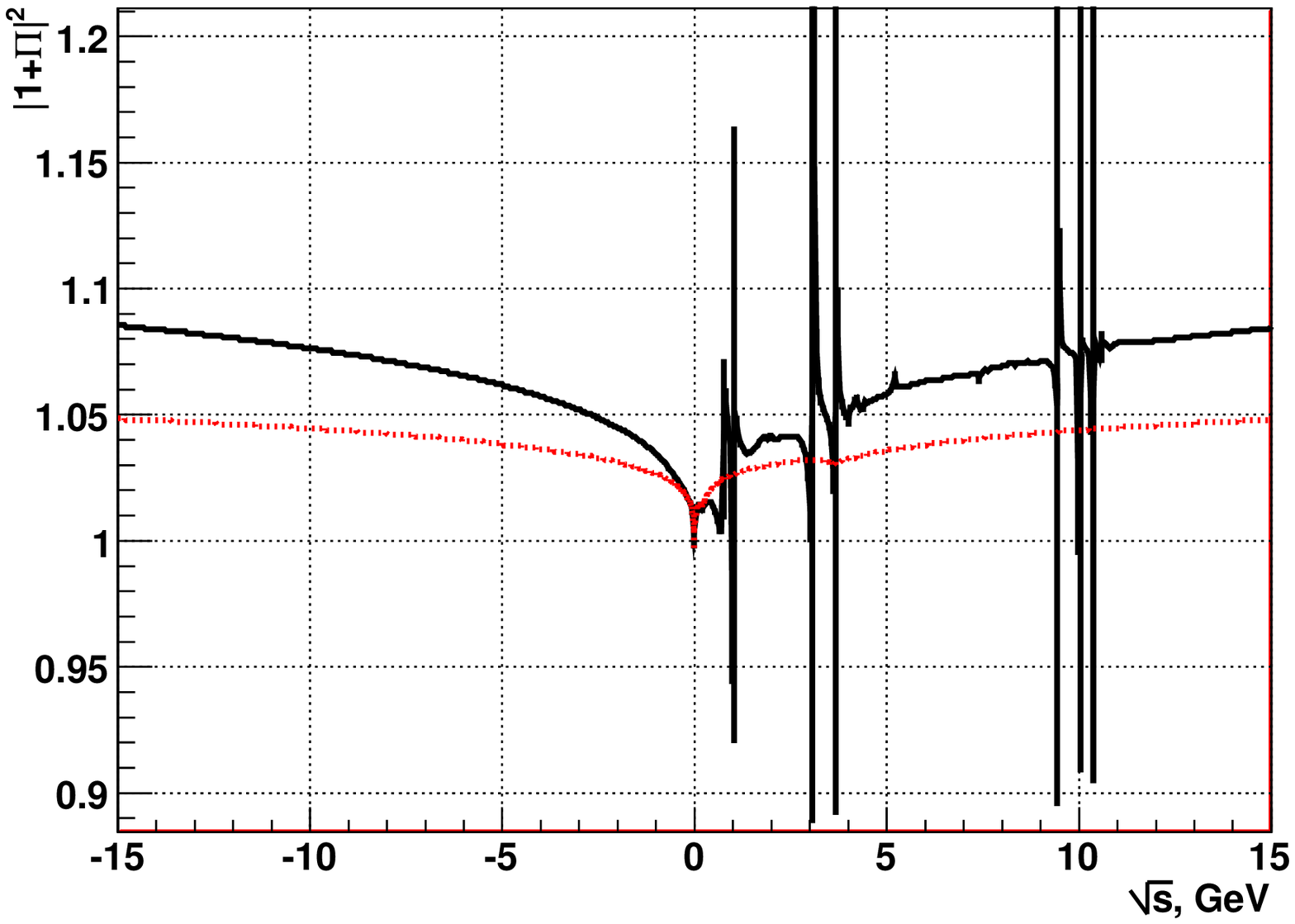}\\
\includegraphics[bb=0 0 530 380,width=8.8cm]{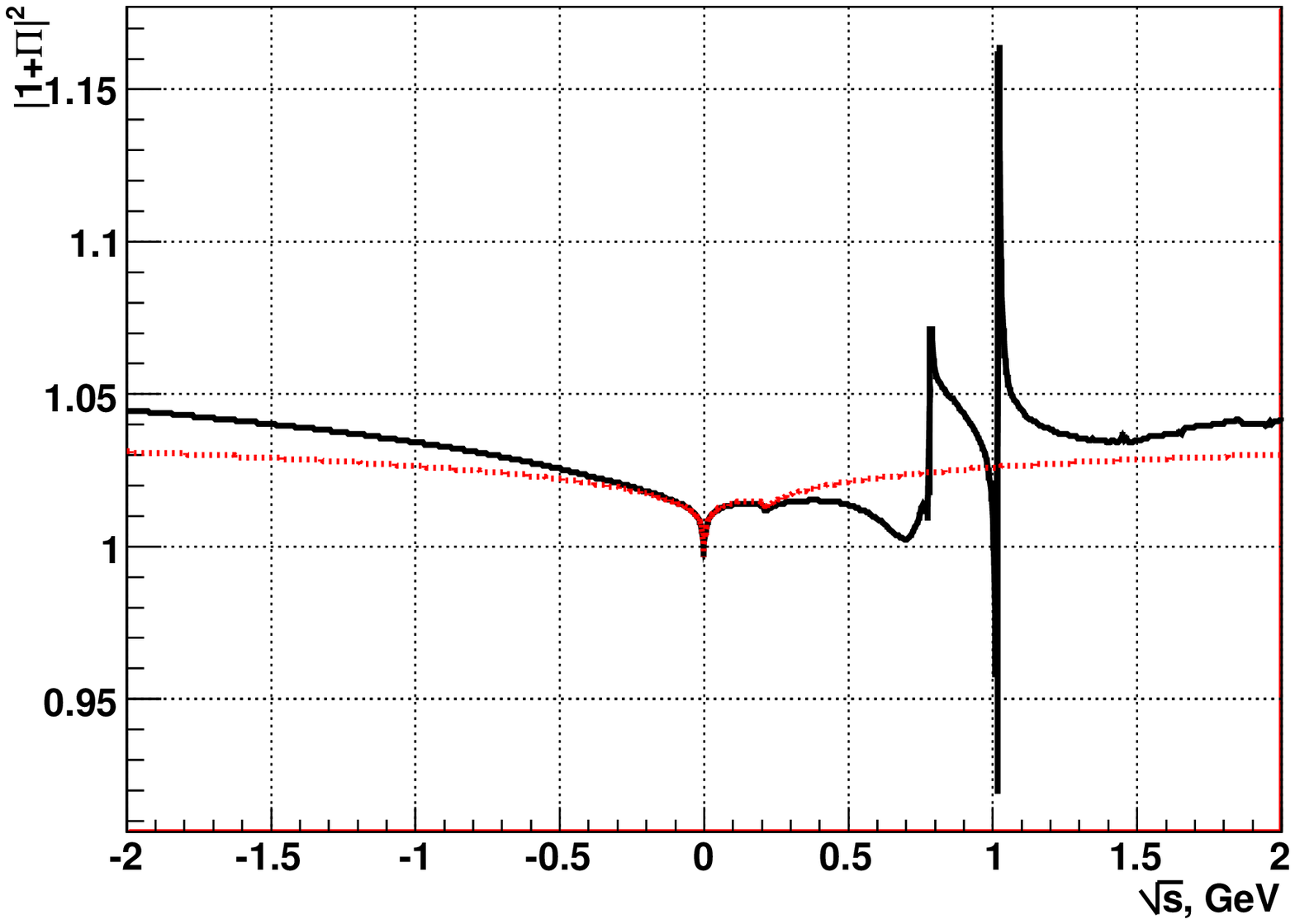}
\end{center}
\vspace{-0.1cm}
\caption{$|1+\Pi|^2$ from CMD-2's compilation for space- and
  time-like momenta (labelled $\sqrt{s}$); solid (black) lines: leptonic plus hadronic
  contributions, dotted (red) lines: only leptonic
  contributions. Upper panel: $-(15\ {\rm GeV})^2 < q^2 < (15\ {\rm GeV})^2$. Lower panel:
  $-(2\ {\rm GeV})^2 < q^2 < (2\ {\rm GeV})^2$. Figures provided by Fedor Ignatov.} 
\label{fig:cmd2vp}
\end{figure}

The experiments CMD-2 and SND at Novosibirsk are using their own VP
compilation to undress hadronic cross sections, and the values used
are given in tables in some of their
publications. Recently CMD-2 has made their compilation publicly
available, see Fedor Ignatov's web-page {\tt
  http://cmd.inp.nsk.su/$\sim$ignatov/vpl/}.\ \ There links are given to a
corresponding talk at the `4th meeting of the Working Group on
Radiative Corrections and Monte Carlo Generators for Low Energies'
(Beijing 2008), to the thesis of Ignatov (in Russian) and to a file
containing the tabulation, which can be used together with a
downloadable package. The tabulation is given for the real and
imaginary parts of the sum of leptonic and hadronic VP, for both
space- and time-like momenta, and for the corresponding
errors. Fig.~\ref{fig:cmd2vp}, also displayed on their web-page,
shows the results from CMD-2 for $|1+\Pi|^2$ both for the space- and
time-like momenta in the range $-(15\ {\rm GeV})^2 < q^2 < (15\ {\rm
  GeV})^2$ (upper panel) and for the important low energy region $-(2\
{\rm GeV})^2 < q^2 < (2\ {\rm GeV})^2$. The solid (black) lines are
the sum of leptonic and hadronic contributions, while the dotted (red)
lines are for the leptonic contributions only. 

Another independent compilation of the hadronic VP is available from
the group of Hagiwara et al.~\cite{hmnt} (HMNT), at present upon
request from the authors. They provide tabulations (with a simple
interpolation routine in Fortran) of $\Delta\alpha^{(5)}_{\rm had}(q^2)$ both in
the space- and time-like region, and also a compilation of
$R_{\rm had}(s)$. Currently available routines are based on the
analysis~\cite{Hagiwara:2003da,Hagiwara:2006jt}. Two
different versions are provided, one including the narrow resonances
$J/\psi, \psi^{\prime}$ and the Upsilon family,
$\Upsilon(1S)-\Upsilon(3S)$, in Breit-Wigner form, one excluding them.
However, for applications of $\Delta\alpha$ it should be remembered
that close to narrow resonances the resummation of such large
contributions in the effective coupling breaks down. 
In this context, note that the compilation from Novosibirsk contains
these narrow resonances, whereas the routine from Jegerlehner does contain
$J/\psi$ and $\psi^{\prime}$, but seems to exclude (or smear over) the
Upsilon resonances. When called in the charm or bottom resonance
region Jegerlehner's routine gives a warning that the ``results may
not be reliable close to J/Psi and Upsilon resonances''. 

In the following we shall compare the parametrisations from the
different groups.

\subsection{Comparison of the results from different groups}
\begin{figure}
\begin{center}
\includegraphics[bb=80 280 480 540,width=8.8cm]{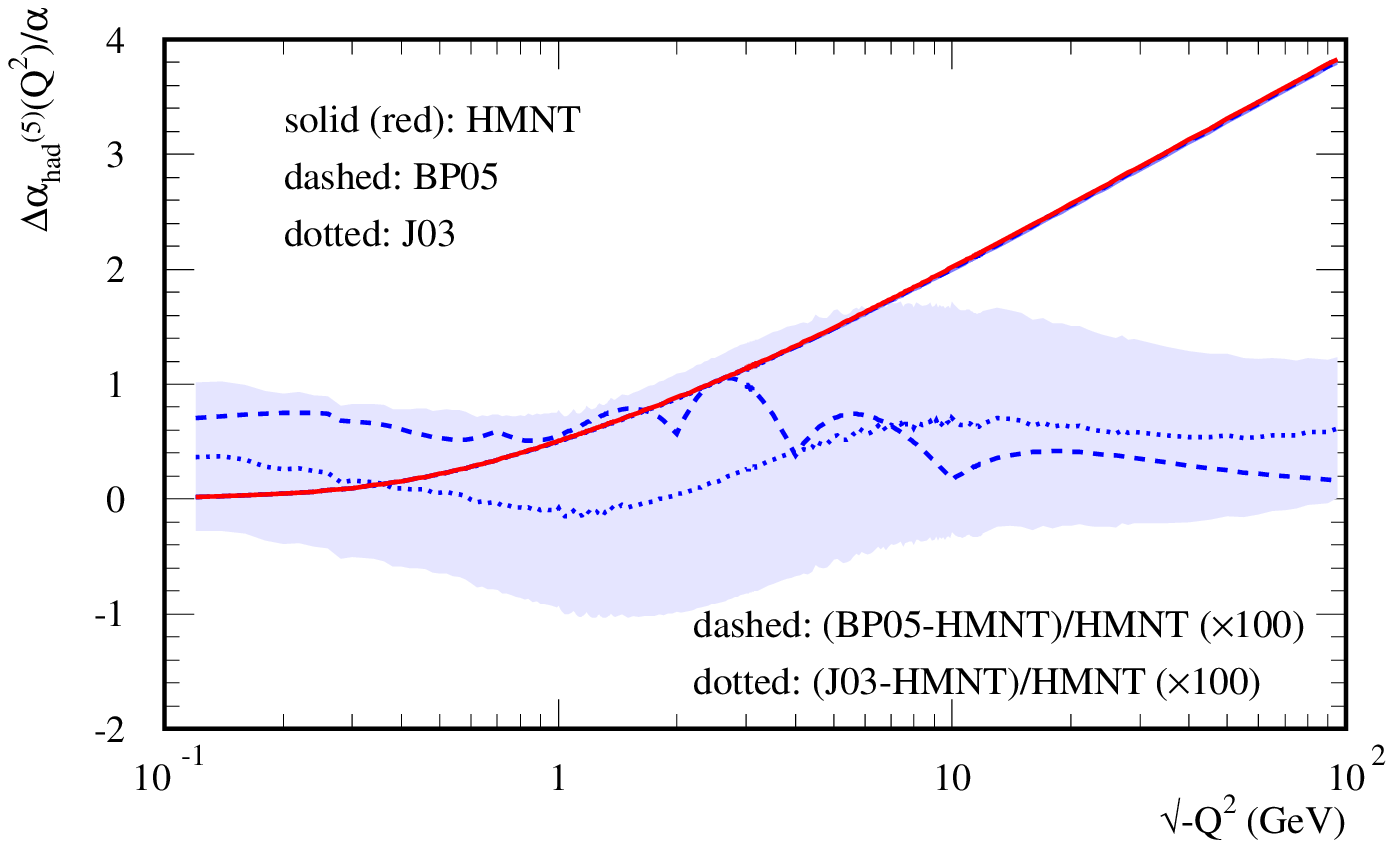}\\
\includegraphics[bb=80 280 480 540,width=8.8cm]{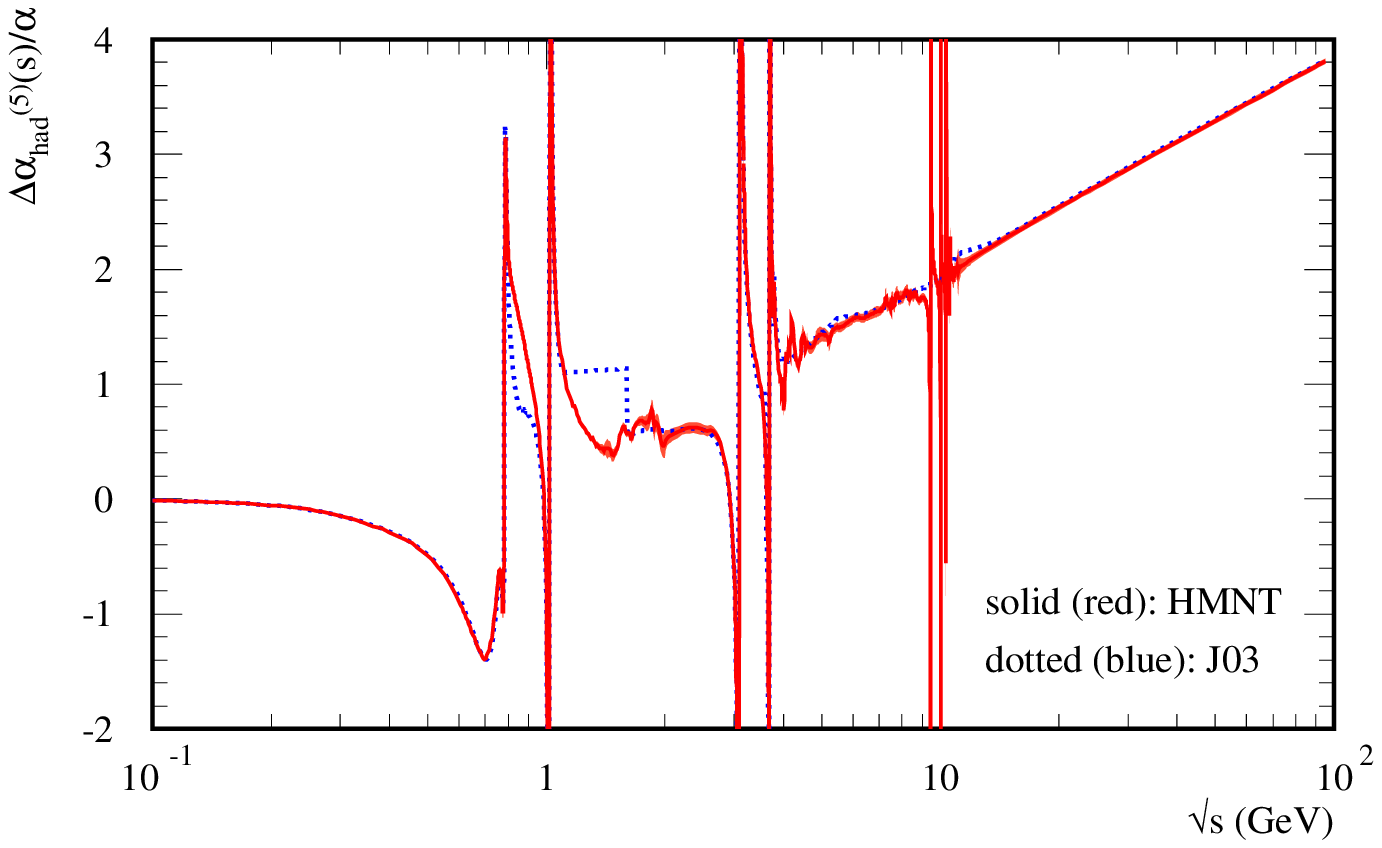}
\end{center}
\vspace{-0.3cm}
\caption{Comparison of the results from Hagiwara et
  al. (HMNT~\cite{hmnt}) for $\Delta\alpha_{\rm had}^{(5)}(q^2)$ in
  units of $\alpha$ with parametrisations from Burkhardt and Pietrzyk
  (BP05~\cite{Burkhardt:2005se}) and Jegerlehner (J03). Upper panel:
  $\Delta\alpha_{\rm had}^{(5)}(Q^2)/\alpha$ for space-like momentum transfer
  ($Q^2<0$), where the three parametrisations are indistinguishable.
  The differences (normalised and multiplied by 100) are highlighted
  by the dashed and dotted curves; the wide light (blue) band is
  obtained by using the error band of HMNT in the normalised
  difference to J03, labelled `(J03-HMNT)/HMNT ($\times 100$)'. Lower
  panel: $\Delta\alpha_{\rm had}^{(5)}(s)/\alpha$ from J03 and HMNT (as
  labelled) for time-like momenta ($q^2=s$). For readability, only the
  error band of HMNT is displayed.} 
\label{fig:delalfcomp}
\end{figure}

\begin{figure} \begin{center}
\includegraphics[width=8.9cm]{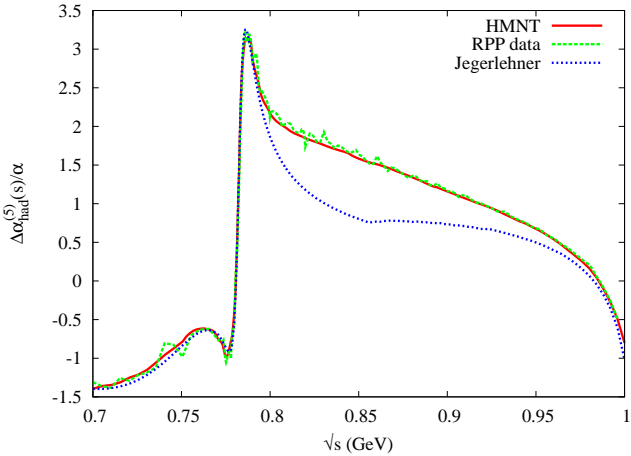}
\end{center}
\vspace{-0.1cm}
\caption{\label{fig:compzoom}
  Comparison of the results from Hagiwara et al. (HMNT, solid (red)
  line) for $\Delta\alpha_{\rm had}^{(5)}(s)/\alpha$ with the
  parametrisation from Jegerlehner (J03, dotted
  (blue) line) in the time-like region in the range $\sqrt{s} = 0.7 -
  1$ GeV. The dashed (green) line shows the result if the data
  compilation from the PDG~\cite{Amsler:2008zzb} is used.} 
\end{figure}

In Fig.~\ref{fig:delalfcomp}, we compare the parametrisations from
Burk\-hardt and Pietrzyk (BP05), Jegerlehner
(J03) and Hagiwara et al. (HMNT) in
the space-like (upper) and time-like region (lower panel). For the
space-like region the differences among the three parametrisations are
roughly within one standard deviation in the whole energy range
shown. However, for the time-like region, there is disagreement
between HMNT and J03 at several energy regions, most notably at $1\
{\rm GeV} \lapproxeq \sqrt{s} \lapproxeq 1.6$ GeV, and at $0.8$ GeV
$\lapproxeq \sqrt{s}$ $\lapproxeq 0.95$ GeV. As for the
discrepancy at $1$ GeV $\lapproxeq \sqrt{s} \lapproxeq 1.6$ GeV,
checking the routine from Jeger\-leh\-ner, one finds that a too sparsely
spaced energy grid in this region seems to be the reason. The
discrepancy at $0.8$ GeV $\lapproxeq \sqrt{s} \lapproxeq$ $0.95$
GeV is further scrutinised in Fig.~\ref{fig:compzoom}, where in
addition to the two parametrisations HMNT (solid (red) line) and J03
(dotted (blue) line), the result for $\Delta\alpha_{\rm
  had}^{(5)}(s)/\alpha$ obtained by integrating over the $R$-data as
compiled by the PDG~\cite{Amsler:2008zzb}\footnote{The actual
  compilation of the data is available in electronic form from {\tt
    http://pdg.lbl.gov/2008/hadronic-xsections} {\tt
    /hadronicrpp\_page1001.dat}\,.} is shown as the dashed (green)
line. While the results from HMNT and the one based on the PDG $R$-data
agree rather well, their disagreement with the J03 compilation in the
region $0.8$ GeV $\lapproxeq \sqrt{s} \lapproxeq 0.95$ GeV is
uncomfortably large compared to the error but may be due to a
different data input of the J03 parametrisation.

In the following we shall compare the parametrisation from HMNT with
the one from the CMD-2 collaboration which has become available very
recently. Note that for undressing their experimentally measured hadronic
cross sections, CMD-2 includes the imaginary part of the VP function
$\Pi(q^2)$ in addition to the real part. Before coming to the
comparison with CMD-2, let us discuss some generalities
about $\Im\Pi(q^2)$. If we are to include the imaginary part, then the
VP correction factor $\alpha(q^2)^2$ should be replaced as
\begin{eqnarray}
& &\left(\frac{\alpha}{1 - \Delta \alpha(q^2)}\right)^2
= \left(\frac{\alpha}{1 - \Re \Pi(q^2)}\right)^2
\to\\ 
& &\left| \frac{\alpha}{1 - \Pi(q^2)} \right|^2
= \frac{\alpha^2}{(1 - \Re\Pi(q^2))^2 + (\Im \Pi(q^2))^2}\,.\nonumber
\end{eqnarray}
Note that, as mentioned already
in the introduction, the contribution from the real part appears at
$O (e^2)$ in the denominator, while that from the imaginary part
starts only at $O (e^4)$. Because of this suppression we expect
the effects from the imaginary part to be small. Nevertheless we would
like to stress two points. First, field-theoretically, it is more
accurate to include the imaginary part which exists above
threshold. Including only $\Re\Pi(q^2)$ in the VP correction is an
approximation which may be sufficient in most cases. Second, it is
expected that the contribution from the imaginary part is of the order
of a few per mill of the total VP corrections. While this seems
small, it can be non-negligible at the $\rho$ meson region where the
accuracy of the cross section measurements reaches the order of (or even
less than) 1\%. Similarly, in the region of the narrow $\phi$
resonance, the contributions from the imaginary part become
non-negligible and should be taken into account.

\begin{figure}[htb]
\begin{center}
\includegraphics[bb=60 55 405 300, width=8.9cm]{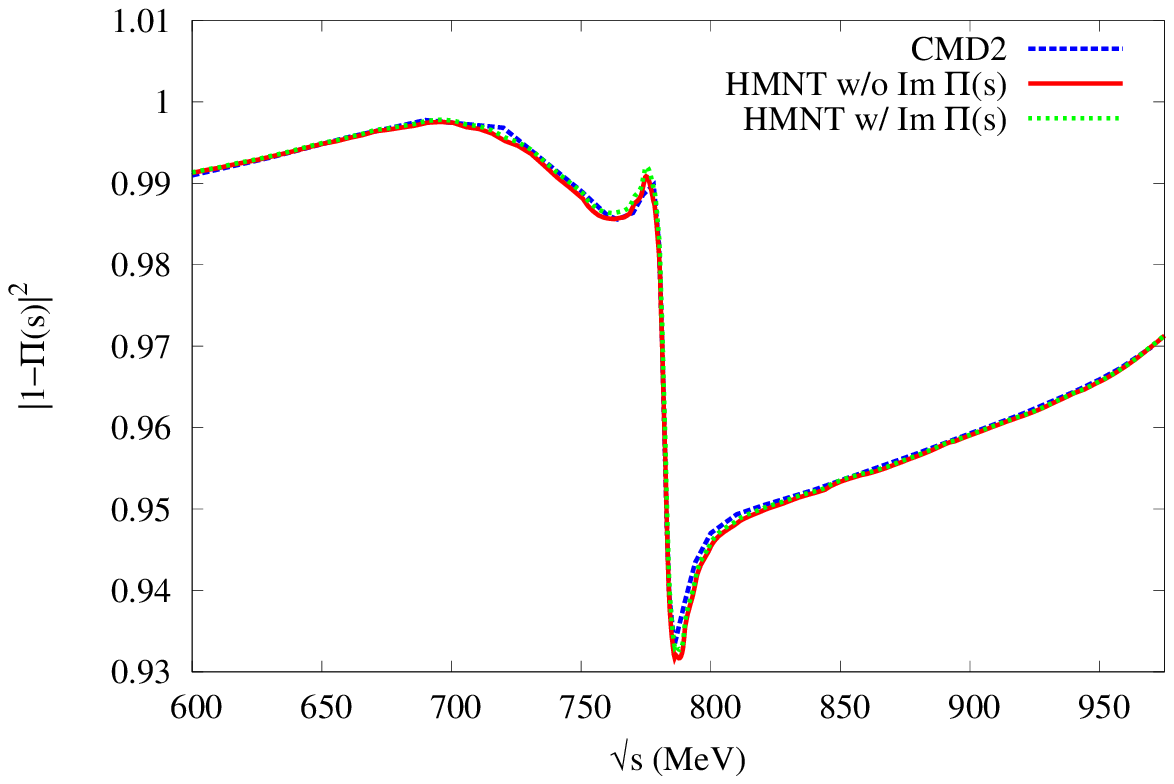}\\
\includegraphics[bb=60 55 405 300, width=8.9cm]{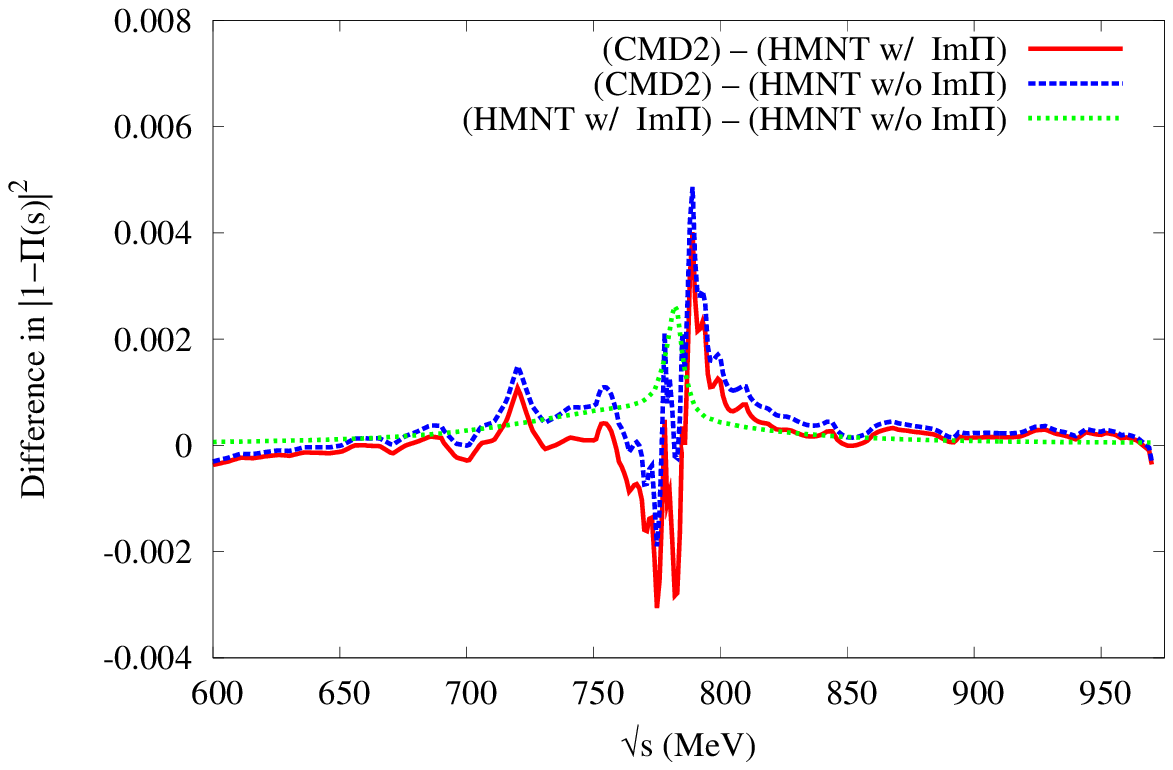}
\end{center}
\vspace{-0.2cm}
\caption{Upper panel: Correction factor $\left | 1 - \Pi(s) \right
  |^2$ as used for `undressing' by the CMD-2 collaboration in~\cite{Akhmetshin:2006bx}
  (dashed line) compared to the same quantity using the HMNT
  compilation for the $e^+e^- \to hadrons$ data (solid line). Also
  shown is the correction factor $(1 - \Re \Pi)^2 =
  (\alpha/\alpha(s))^2$, based on $\alpha(s)$ in the time-like region
  from HMNT (dotted line). Lower panel: Differences of the quantities
  as indicated on the plot.} 
\label{fig:cmd2comp}
\end{figure}
In Fig.~\ref{fig:cmd2comp} the VP correction factor, based on the
compilation from HMNT, with and without $\Im\Pi(q^2)$ is compared to
$|1-\Pi(s)|^2$ as used by the CMD-2 collaboration in their recent
analysis of the hadronic cross section in the $2\pi$ channel in the
$\rho$ central region~\cite{Akhmetshin:2006bx}.\footnote{We thank
  Gennadiy Fedotovich for providing us with a table including the VP
  correction factors not included in~\cite{Akhmetshin:2006bx}.} In the
\begin{figure}[htb]
\begin{center}
\includegraphics[bb=75 280 480 535, width=8.9cm]{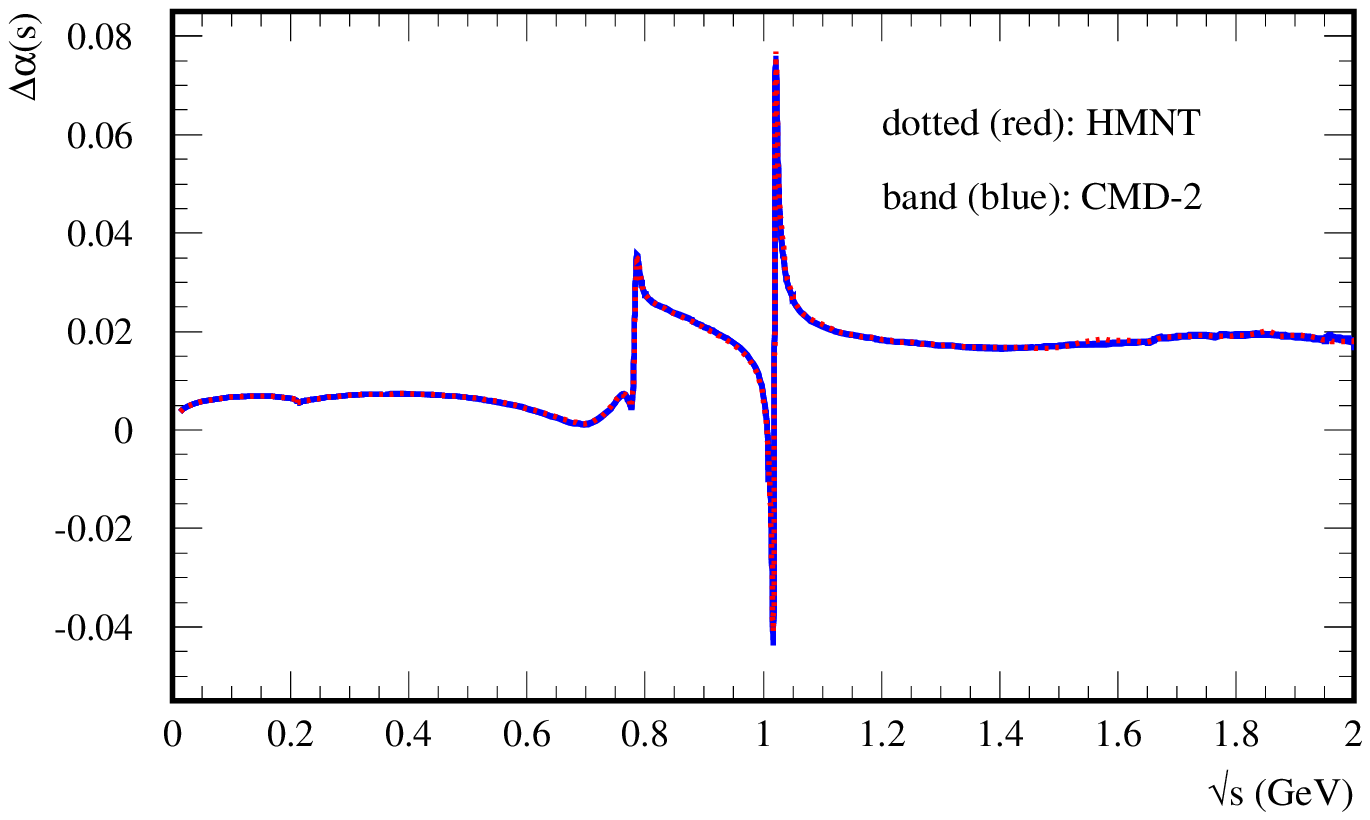}\\
\includegraphics[bb=75 280 480 535, width=8.9cm]{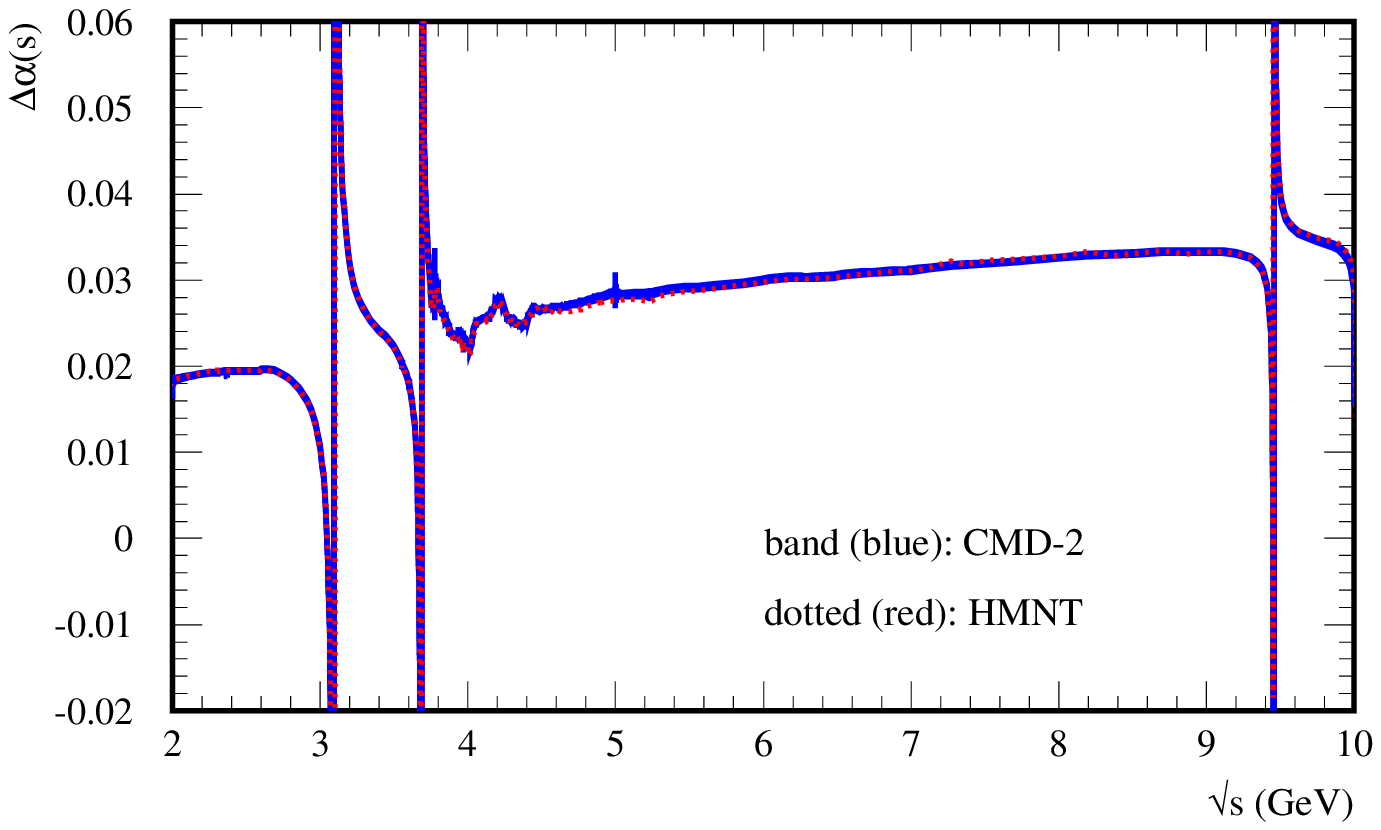}
\end{center}
\vspace{-0.3cm}
\caption{$\Delta\alpha(s)$ in the time-like region as given by the
  parametrisation from CMD-2 (solid (blue) band) compared to the same quantity
  from HMNT (dotted (red) line). Upper panel: $0 < \sqrt{s} < 2$ GeV, lower
  panel: $2$ GeV $< \sqrt{s} < 10$ GeV.}
\label{fig:cmd2mnhtda}
\end{figure}
upper panel the VP correction factors are given, whereas in the lower
panel the differences are shown. As expected, the differences between
the three are visible, and are about a few per mill at most. The
difference between the CMD-2 results and the one from HMNT including
$\Im\Pi(q^2)$ (solid (red) curve in the lower panel of
Fig.~\ref{fig:cmd2comp} shows a marked dip followed by a peak in the
$\rho - \omega$ interference region where the $\pi^+\pi^-$ cross
section falls sharply. This is most probably a direct consequence of
the different data input used. However, in most applications such a
difference will be partially cancelled when integrated over an energy
region including the $\rho$ peak. 

\begin{figure}[htb]
\begin{center}
\includegraphics[bb=75 280 480 535, width=8.9cm]{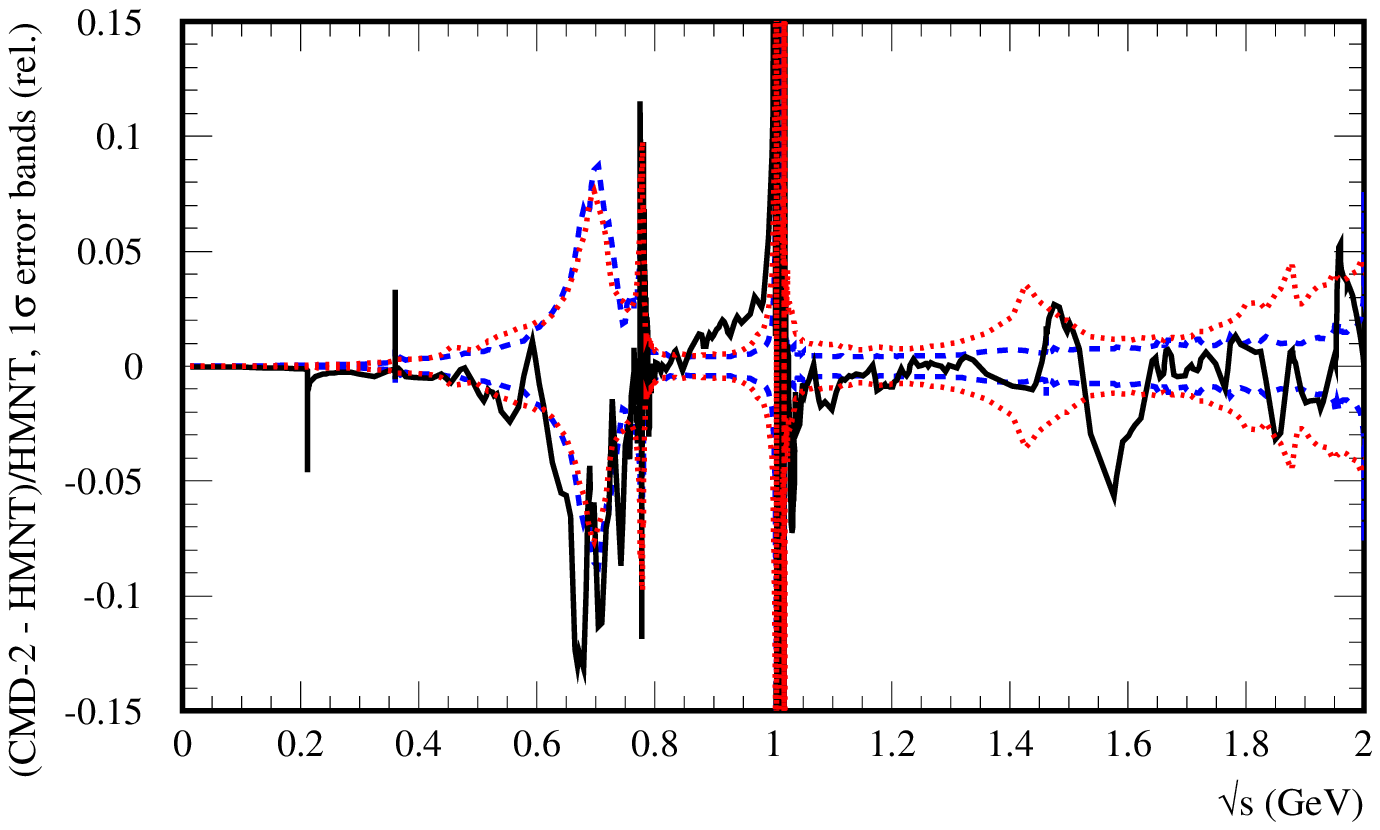}\\
\includegraphics[bb=75 280 480 535, width=8.9cm]{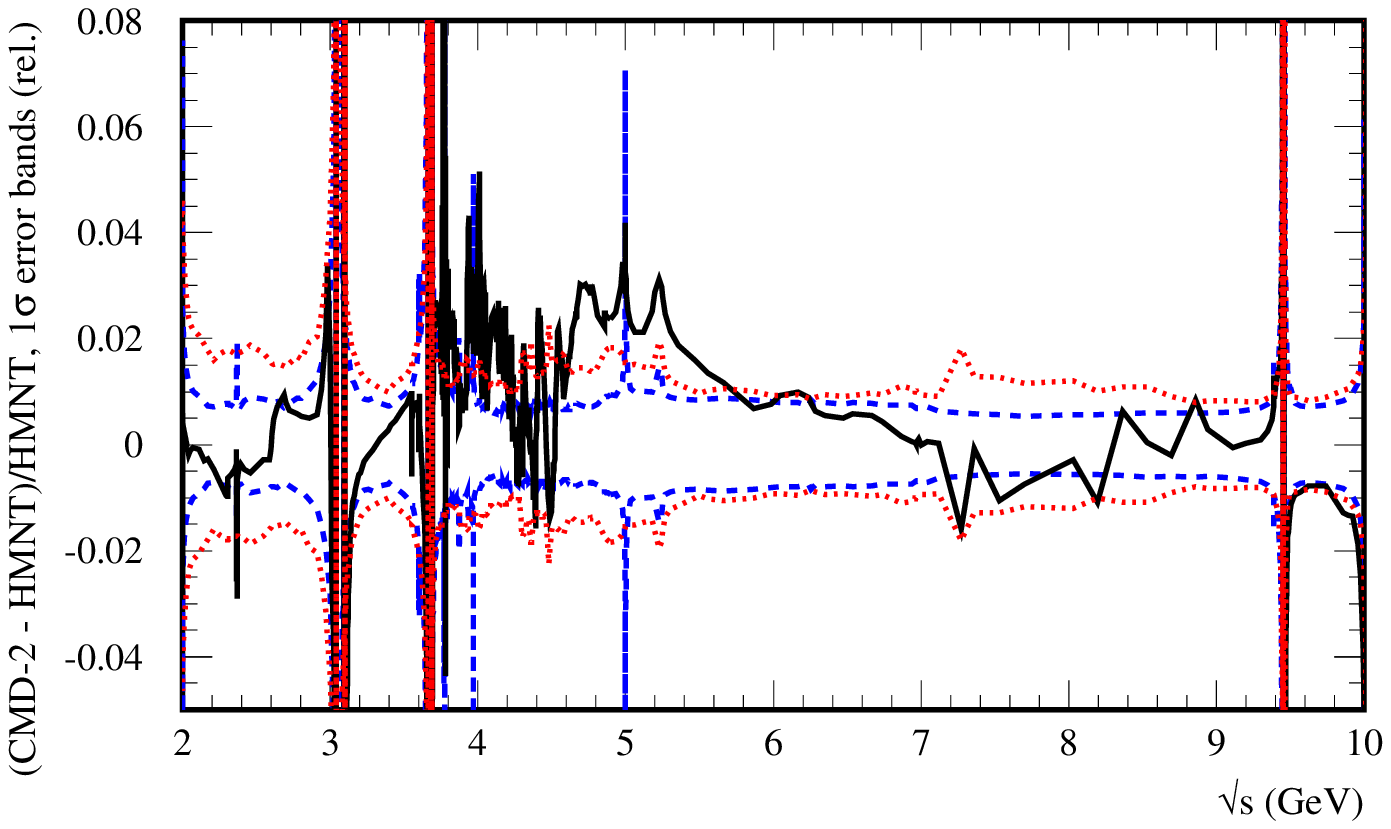}
\end{center}
\vspace{-0.3cm}
\caption{Solid (black) lines: Normalised difference $(\Delta\alpha^{\rm
    CMD-2}(s)-\Delta\alpha^{\rm HMNT}(s))/\Delta\alpha^{\rm HMNT}(s)$ in the
  time-like region. The dashed (blue) and dotted (red) lines indicate
  the relative error for the CMD-2 and HMNT parametrisations. Upper
  panel: $0 < \sqrt{s} < 2$ GeV, lower panel: $2$ GeV $< \sqrt{s} < 10$ GeV.}
\label{fig:cmd2mnhtdae}
\end{figure}
In Figs.~\ref{fig:cmd2mnhtda} and~\ref{fig:cmd2mnhtdae} we compare
$\Delta\alpha(s)$ in the time-like region as given by the
parametrisation from CMD-2 with the one from HMNT, where for HMNT we
have calculated the leptonic contributions (up to including the NNLO
corrections) as described above. The two panels in
Fig.~\ref{fig:cmd2mnhtda} (upper panel: $0 < \sqrt{s} < 2$ GeV, lower
panel: $2$ GeV $< \sqrt{s} < 10$ GeV) show $\Delta\alpha(s)$ with the 
$1\sigma$ error band from CMD-2 as a solid (blue) band, whereas for
HMNT the mean value for $\Delta\alpha(s)$ is given by the dotted (red)
line, which can hardly be distinguished. To highlight the differences
between the two parametrisations, Fig.~\ref{fig:cmd2mnhtdae} displays
the normalised difference $(\Delta\alpha^{\rm CMD-2}(s) -
\Delta\alpha^{\rm HMNT}(s))/\Delta\alpha^{\rm HMNT}(s)$ as a so\-lid
(black) line, and also shows the relative errors of CMD-2 and HMNT as
dashed (blue) and red (dotted) lines, respectively. As visible in
Fig.~\ref{fig:cmd2mnhtdae}, the error as given by the CMD-2
parametrisation is somewhat smaller than the one from HMNT. Both
parametrisations agree fairly well, and for most energies the
differences between the parametrisations are about as large or smaller
than the error bands. Close to narrow resonances the estimated
uncertainties are large, but as discussed above, there the
approximation of the effective coupling $\alpha(s)$ breaks down and
resonance contributions should be treated differently.

\subsection{Summary}
Vacuum polarisation of the photon plays an important role in many
physical processes. It has to be taken into account, e.g., in Monte Carlo
generators for hadronic cross sections or Bhabha scattering. When
low energy data are used in dispersion integrals to predict the
hadronic contributions to muon $g-2$ or $\Delta\alpha(q^2)$, undressed data
have to be used, so VP has to be subtracted from measured cross
sections. The different VP contributions have been discussed,
and available VP compilations have been briefly described and
compared. Until recently only one parametrisation has been available
in the time-like region, now three routines in the space- and time-like
regions exist,
from Jegerlehner, CMD-2 and HMNT, and a fourth from Burkhardt and
Pietrzyk in the space-like region. While the accuracy of the hadronic cross
section data themselves is the limiting factor in the precise
determination of $g-2$ and $\Delta\alpha(M_Z^2)$, the error of the VP
(or $\Delta\alpha(q^2)$) is not the limiting factor in its current
applications. With the ongoing efforts to measure $\sigma_{\rm
  had}(s)$ with even better accuracy in the whole low energy region,
further improvements of the various VP parametrisations are foreseen.

\section{Summary}
\label{sec:6}
 In this Report we have summarised the achievements of the last years of the
experimental and theoretical groups working on hadronic cross section
 measurements and tau physics. In addition we  have sketched 
the prospects in this  field for the years to come.
  We have emphasised the importance of continuous
 and close collaboration between the experimental and theoretical groups
 which is crucial in the quest for precision in hadronic physics. The platform
 set to simplify this collaboration is a  
{\it Working Group on Radiative Corrections  and  Monte Carlo Generators
for Low Energies (Radio MontecarLow)}, for the better understanding of
 the needs and limitations of both experimental and theoretical communities
 and to facilitate the information flow between them. This Review is a result
 of the Working Group.

 The Report was divided
 into five  Sections covering the luminosity
 measurements at low energies (up to the energy of $B$ factories)
 (Section \ref{sec:1}),
 $R$ measurement by energy scan (Section \ref{sec:2}), $R$ measurement 
 using radiative return (Section \ref{sec:3}), tau physics (Section \ref{sec:5}), and the calculation
 of the vacuum polarisation with emphasis on the hadronic contributions
 (Section \ref{sec:4}).
  In all the Sections, with the exception of Section \ref{sec:4},
  we gave an overview of the experimental results
 and the status of the Monte Carlo event generators used in the experimental
 analyses with emphasis on their accuracy and tests.

Concerning the work done on the topic of precision luminosity measurement
(Section \ref{sec:1}),
 a particular effort was paid to arrive at an up-to-date estimate
 of the accuracy of the most precise MC tools used by the experimentalists. 
Several tuned comparisons between the predictions of independent
 generators were presented, considering the large-angle Bhabha 
process with realistic event selection criteria and at different
 c.m. energies. It turned out that the three most precise luminosity tools,
 i.e. the programs BabaYaga@NLO, BHWIDE and MCGPJ, agree within 0.1\% for the
 integrated cross sections and within less than 1\% for the differential 
distributions. 
Therefore the main conclusion of the work on tuned comparisons is
 that the technical precision of MC programs is well under control,
 the (minor) discrepancies still observed being due to slightly
 different 
details in the treatment of radiative corrections 
and their implementation. 
 The theoretical accuracy of the generators with regard to  radiative corrections
not fully taken into account was assessed 
by performing detailed comparisons between the results of the generators
 and those of exact perturbative calculations. In particular, explicit
 cross-checks with the predictions of available NNLO QED calculations
 and with new exact results for lepton and hadron pair corrections led
 to the conclusion that the total theoretical uncertainty is at the one
 per mill level for the large-angle Bhabha process at different c.m.
 energies. Albeit this error estimate could be put on firmer grounds
 thanks to further work in progress, it appears to be already quite
 robust and sufficient for a precise determination of the luminosity.

In Section  \ref{sec:2} 
we presented the current status of the studies of $e^+e^-$ annihilation 
into hadrons and muons at the energies up to a  few GeV.  Accurate 
measurements of the ratio $R$, {\it i.e.} the ratio of the cross sections
of hadron and muon channels, are crucial 
for the evaluation of the hadronic contribution to vacuum polarisation and
subsequently for various precision tests of the Standard Model.
Results of several experimental collaborations have been reviewed for the most
important processes with the final states $\mu^+\mu^-$, $\pi^+\pi^-$, 
$\pi^+\pi^-\pi^0$, $\pi^+\pi^-2\pi^0$, $\pi^+2\pi^-$, two kaons and 
heavier mesons.
In particular, $R$ scans at the experiments CMD-2, SND, CLEO and BES experiments have been
 discussed.
Analytic expressions for the Born level cross sections of the main processes 
have been presented.
 First-order QED radiative corrections have been given explicitly 
for the case of muon, pion and kaon pair production.
The two latter cases are computed using scalar QED to describe interactions 
of pseudoscalar mesons with photons in the final state.
Matching with higher-order QED corrections evaluated in the leading logarithmic 
approximation have been discussed. Good agreement between 
different Monte Carlo codes for the muon channel has been shown.
The theoretical uncertainty in the description of these processes has been
 evaluated.
For the two main channels, $e^+e^-\to\mu^+\mu^-$ and $e^+e^-\to\pi^+\pi^-$, 
this uncertainty has been estimated to be of the order of $0.2\%$.

In Section  \ref{sec:3} 
we have given an overview of experimental measurements via radiative
return and described the Monte Carlo generators used in the
analyses. Special emphasis has been put on the modelling of the meson-photon 
interaction, crucial for reaching an accuracy below  1\%.
Radiative return has been applied successfully at the 
experiments KLOE in Frascati, BaBar in Stanford and 
Belle in Tsu\-ku\-ba, obtaining important results for the 
measurement 
of precise hadronic cross sections as well as in the field of hadron 
spectroscopy. In all three experiments, the ISR physics 
programme is still going on. New experiments like the BES-III detector at 
BEPC-II in Beijing and the experiments at the VEPP-2000 machine in Novosibirsk 
will use radiative return 
to complement their standard physics programme of energy scanning
in the regions of  2 -- 4.6 GeV (BEPC-II) and 1 -- 2 GeV (VEPP-2000). 
The success of this programme was possible only through close collaboration
between experimental and theoretical groups. Dedicated Monte Carlo
generators (PHOKHARA, EKHARA, FEVA, FASTERD) were developed to
make
the experimental analyses possible. 
The physics programme allowed for better modelling
of the photon-meson interaction which is crucial for a precise determination of the pion form factor. The measurements
 of the hadronic cross sections by means of radiative return 
 allowed to reduce the error of the  hadronic
 contribution to the anomalous magnetic moment of the muon and to the
 running of the fine structure constant.
 Ongoing and forthcoming
 measurements will aim at an even better modelling of the hadron-photon
 interaction and the inclusion of those QED radiative corrections not yet
 accounted for in the Monte Carlo generators. 
 This ongoing physics programme will lead
 to further improvements in the precision of the calculation of the hadronic
 contribution to the anomalous magnetic moment of the muon and to the
 running of the fine structure constant, which in turn
 is crucial for tests of the Standard Model and searches for New
 Physics.

In Section  \ref{sec:5} we described the 
present status of the simulation programs for the
production and decay of $\tau$ leptons. The available programs
have been discussed in the context of the required accuracy to match 
current high-statistics experimental data. After a review of the
existing programs used in the data analysis we have emphasised the
topics which will require particular attention in the future.
We have elaborated on the efforts which are going on at present
 and focused on the necessary
improvements. The techniques for fitting $\tau$ decay
currents require particular attention.
The observed spectra and angular distributions are a convolution of
theoretical predictions with experimental effects which should
be taken into account in the fitting procedures.
Background contributions also play an important role 
if high precision is requested. 
We have also commented on the impact of these efforts for forthcoming high
energy experiments (like at LHC), where $\tau$ decays are used to
constrain hard processes rather than  to measure properties of $\tau$
decays. 

 In Section \ref{sec:4}
the different vacuum polarisation (VP) contributions have been discussed, and available
parametrisations have been compared.
VP forms a universal part of radiative
corrections and as such is an important ingredient in Monte Carlo
programs. In addition, to evaluate the hadronic contributions to the muon $g-2$
and $\Delta\alpha(q^2)$ via dispersion relations, one has to use the 
`undressed' hadronic cross section, {\it i.e.} data with the VP effects
removed. 
Therefore the precise knowledge of VP is required. While in the space-like region
the VP is a smooth function and the parametrisations are in excellent
agreement, in the time-like region the VP is a fast varying function
and differences exist between different parametrisations, especially
around resonances. However, the accuracy which is typically of the
order of or below a few per mill and the agreement of the more
recent compilations indicate that the current precision of VP is
sufficient for the envisaged applications. In the future better
hadronic cross section data will lead to further improved accuracy.


\begin{acknowledgement}
 \noindent
{\bf \large  Acknowledgements}

 \noindent
  This work was supported in part by:
\begin{itemize}

\item
   European Union  Marie-Curie Research Training Networks
   MRTN-CT-2006-035482 ``FLAVIAnet'' and MRTN-CT-2006-035505
   ``HEPTOOLS'';

\item
European Union Research Programmes at LNF, FP7,
Trans\-na\-tio\-nal Access to Research Infrastructure (TARI), Hadron
Physics2-Integrating
Activity, Contract No. 227431;

\item
  Generalitat
Valenciana \\ under Grant No. PRO\-ME\-TEO/2008/069;

\item
 German Federal Ministry of Education and Research (BMBF)  grants 05HT4VKA/3,
06-KA-202 and 06-MZ-9171I;

\item
German Research Foundation (DFG): 'Emmy Noether Programme',
contracts DE839/1-4, 'Heisenberg Programme'
  and  Sonderforschungsbereich/Transregio SFB/TRR 9;

\item
Initiative and Networking Fund of the Helmholtz Association, contract HA-101 ("Physics at the Terascale"); 

\item
   INTAS project Nr 05-1000008-8328
  ``Higher-order effects in $e^+e^-$ annihilation and muon anomalous
  magnetic moment'';

\item
Ministerio de Ciencia e
Innovaci\'on under Grant No. FPA2007-60323, and CPAN
(Grant No. CSD2007-00042);

\item
National Natural
Science Foundation of China under Contracts
Nos. 10775142, 10825524 and 10935008;

\item
Polish Government grant N202 06434 (2008-2010);

\item
PST.CLG.980342

\item
Research Fellowship of the Japan Society
    for the Promotion of Science for Young Scientists;

\item
 RFBR grants 
 03-02-16477, 04-02-16217, 04-02-1623, 04-02-16443, 04-02-16181-a,
 04-02-16184-a, 05-02-16250-a, 06-02-16192-a, 07-02-00816-a,
 08-02-13516, 08-02-91969 and 09-02-01143;

\item
Theory-LHC-France initiative of CNRS/IN2P3;

\item
  US DOE contract DE-FG02-09ER41600.

\end{itemize}

 \noindent
We thank J.~Libby for useful correspondence about the luminosity measurement at CLEO-c, and A.~Pich, J.~Portol\'es, D.~G\'omez-Dumm,  
M.~Jamin and Z.H.~Guo for fruitful collaborations and useful  
suggestions related to the Tau Physics section.
S.~Eidelman and V.~Cherepanov are grateful to the Cracow Institute of Nuclear Physics where part of this work has been performed.
M.~Gunia acknowledges a scholarship from the UPGOW project co-financed
by the European Social Fund. F.~Jegerlehner acknowledges support by the Foundation for Polish Science.

\end{acknowledgement}

%
\bibliographystyle{epj}
\bibliography{pub_intro,pub_lumi,pub_scan,pub_radret,pub_vacpol,pub_tau}
%
%
%

\end{document}